\documentclass[sn-basic]{sn-jnl}
\setcitestyle{aysep={}} 

\usepackage[utf8]{inputenc}
\usepackage[acronym,nogroupskip,nonumberlist,nopostdot,toc]{glossaries}
\usepackage{graphicx}%
\usepackage{multirow}%
\usepackage{amsmath,amssymb,amsfonts}%
\usepackage{amsthm}%
\usepackage{mathrsfs}%
\usepackage[title]{appendix}%
\usepackage{xcolor}%
\usepackage{textcomp}%
\usepackage{manyfoot}%
\usepackage{booktabs}%
\usepackage{algorithm}%
\usepackage{algorithmicx}%
\usepackage{algpseudocode}%
\usepackage{listings}%
\usepackage{xspace}%
\usepackage{subfig}%
\usepackage{lineno}
\usepackage[defaultcolor=red]{changes}

\definecolor{ForestGreen}{RGB}{34, 139, 34}

\newcommand{\newglossaryacronym}[3]{
\newglossaryentry{#1}
{
    name={#1},
    long={#2},
    description={#2: #3},
    first={#2 (#1)},
    firstplural={#2\glspluralsuffix\ (#1\glspluralsuffix)}
}
}

\newglossaryacronym{ANN}{Artificial Neural Network}{A machine learning algorithm that is based on the structure and function of biological neural networks in the human brain.}
\newcommand{\ann}{\gls{ANN}\xspace}
\newcommand{\anns}{\glspl{ANN}\xspace}

\newglossaryacronym{BBH}{Binary Black Hole}{A binary composed of two black holes, typically of interest to us as a source of a compact binary coalescence signal. Purely described by general relativity without matter effects. The prototypical example was GW1501914 \citep{Abbott:2016blz}.}
\newcommand{\bbh}{\gls{BBH}\xspace}
\newcommand{\bbhs}{\glspl{BBH}\xspace}

\newglossaryacronym{BNS}{Binary Neutron Star}{A binary composed of two neutron stars, typically of interest to us as a source of a compact binary coalescence signal, of longer duration inside the detector band than a binary black hole and with more complicated waveforms due to matter effects. First multi-messenger detection with GW170817 \citep{LIGOScientific:2017vwq}.}
\newcommand{\bns}{\gls{BNS}\xspace}
\newcommand{\bnss}{\glspl{BNS}\xspace}

\newglossaryacronym{CBC}{Compact Binary Coalescence}{A binary of two compact objects radiating away orbital energy in the form of gravitational waves, getting increasingly closer, producing chirp-like gravitational-wave signals with the three characteristic phases of inspiral, merger and ringdown.}
\newcommand{\cbc}{\gls{CBC}\xspace}
\newcommand{\cbcs}{\glspl{CBC}\xspace}

\newglossaryacronym{CCSN}{Core-Collapse SuperNova}{The explosion at the end of a massive star's life, which due to asymmetries can emit gravitational-wave bursts.}
\newcommand{\ccsn}{\gls{CCSN}\xspace}
\newcommand{\ccsne}{\glspl{CCSN}\xspace}

\newglossaryacronym{CGAN}{Conditional Generative Adversarial Networks}{A variant of generative adversarial networks (see below) where the networks are provided with additional information, e.g. labels of the training data. See \citet{2014arXiv1411.1784M}.}
\newcommand{\cgan}{\gls{CGAN}\xspace}

\newglossaryacronym{CNN}{Convolutional Neural Network}{Neural networks that convolve the input data with kernels representing local perceptive fields. A popular method for image recognition tasks and hence also often used on intermediate gravitational-wave data products such as spectrograms. See e.g. \citet{venkatesan2017convolutional} for a general introduction.}
\newcommand{\cnn}{\gls{CNN}\xspace}
\newcommand{\cnns}{\glspl{CNN}\xspace}

\newglossaryacronym{CSC}{Constrained Subspace Classifiers}{A class of algorithms for classification.}

\newcommand{\CSCs}{\glspl{CSC}\xspace}

\newglossaryacronym{CVAE}{Conditional Variational Auto-Encoder}{A version of auto-encoders (see above) where ``variational'' refers to considering distributions as the central objects of interest, and the network is ``conditional'' on specific observations. see \citet{2019arXiv190406264T,2018arXiv181204405P}.}
\newcommand{\cvae}{\gls{CVAE}\xspace}

\newglossaryacronym{CW}{Continuous Gravitational Wave}{Persistent or at least very long-duration gravitational-wave signals from individual spinning deformed neutron stars or other, more exotic sources. see \citet{Riles:2022wwz} for a comprehensive review.}
\newcommand{\cw}{\gls{CW}\xspace}
\newcommand{\cws}{\glspl{CW}\xspace}

\newglossaryacronym{DL}{Dictionary Learning}{A machine learning method that sparsely represents data in a higher-dimensional space by using an optimal set of basis functions.}
\newcommand{\DL}{\gls{DL}\xspace}

\newglossaryacronym{EM}{electromagnetic}{Physical interactions mediated by photons. In our context also used as a shorthand for the methods and observatories of classic astronomy ranging from radio to $\gamma$-ray wavelengths.}
\newcommand{\elmag}{\gls{EM}\xspace}

\newglossaryacronym{EMRI}{Extreme Mass Ratio Inspiral}{Binary mergers with very high mass ratios, e.g. higher than $10^4$ following \citep{LISAConsortiumWaveformWorkingGroup:2023arg}. They are mainly accessible to space-based gravitational-wave detectors.}
\newcommand{\emri}{\gls{EMRI}\xspace}
\newcommand{\emris}{\glspl{EMRI}\xspace}

\newglossaryacronym{ET}{Einstein Telescope}{A planned third-generation gravitational-wave detector in Europe, see \citet{0264-9381-27-19-194002}.}

\newglossaryacronym{CE}{Cosmic Explorer}{A planned third-generation gravitational-wave detector in the United States, see \citet{2017CQGra..34d4001A}.}

\newglossaryacronym{GAN}{Generative Adversarial Network}{An unsupervised machine learning method where two networks are designed to train through competing with each other, a generator trying out new signals and a discriminator trying to classify them. see \citet{goodfellow2020generative}.}
\newcommand{\gan}{\gls{GAN}\xspace}

\newglossaryacronym{GMM}{Gaussian Mixture Model}{The approach of approximating arbitrary data sets or probability distributions by a sum of a variable number of Gaussian functions. See section 16 of \citet{press2007numerical}. In the limit of a large and adaptively chosen number of Gaussians, this becomes comparable to Gaussian Kernel Density Estimators (KDEs).}
\newcommand{\gmm}{\gls{GMM}\xspace}

\newglossaryentry{curric}
{
name={Curriculum Learning},
text={curriculum learning},
description={A procedure to improve the training of, e.g., a neural network, by presenting it with two or more separate training sets of increasing difficulty. see \citet{Bengio2009}.}
}

\newglossaryentry{gaussianproc}
{
name={Gaussian Process},
plural={Gaussian Processes},
description={Nonparametric supervised learning method. For a review see \citet{Rasmussen:2006gpr}.}
}
\newcommand{\gp}{\gls{gaussianproc}\xspace}
\newcommand{\gps}{\glspl{gaussianproc}\xspace}

\newglossaryacronym{GPU}{Graphical Processing Unit}{Computer hardware originally designed for 3D graphics, with massively parallel processing capabilities, which also makes them ideal for machine learning applications.}
\newcommand{\gpu}{\gls{GPU}\xspace}
\newcommand{\gpus}{\glspl{GPU}\xspace}

\newglossaryacronym{GRU}{Gated Recurrent Unit}{A type of recurrent neural network architecture with a simpler architecture and fewer parameters than long short-term memory networks.}
\newcommand{\gru}{\gls{GRU}\xspace}

\newglossaryacronym{GW}{Gravitational Wave}{Ripples in space-time, i.e. propagating solutions of Einstein's equations. The types of astrophysical signals we are considering.}
\newcommand{\gw}{\gls{GW}\xspace}
\newcommand{\gws}{\glspl{GW}\xspace}
\newcommand{\GW}{\gls{GW}\xspace}

\newglossaryacronym{GWTC}{Gravitational-Wave Transient Catalog}{The catalogs of high-probability gravitational-wave candidates published by LIGO, Virgo and KAGRA, with all entries in the four releases from observing runs O1--O3 \citep{PhysRevX.9.031040,PhysRevX.11.021053,LIGOScientific:2021usb,KAGRA:2021vkt} so far being compact binary coalescences. These catalogs are considered cumulative, e.g. GWTC-3 contains events from O1--O3, even though the individual publications typically focus on the most recent subset of events.}

\newglossaryacronym{LIGO}{Laser Interferometer Gravitational-wave Observatory}{Two 4\,km armlength gravitational-wave detectors in the US. Upgraded to Advanced LIGO by 2015, as described in \citet{LIGOScientific:2014pky}.}
\newcommand{\ligo}{\gls{LIGO}}

\newglossaryacronym{LISA}{Laser Interferometer Space Antenna}{A space-based gravitational-wave detector scheduled for launch in the 2030s by ESA and NASA, targeting lower frequencies than current terrestrial detectors.}
\newcommand{\lisa}{\gls{LISA}}

\newglossaryacronym{LSTM}{Long Short Term Memory}{Recurrent neural network family.}
\newcommand{\lstm}{\gls{LSTM}\xspace}
\newcommand{\lstms}{\glspl{LSTM}\xspace}

\newglossaryacronym{LVK}{LIGO--Virgo--KAGRA}{The current global network of interferometric gravitational-wave detectors. Also used as a joint acronym for the three collaborations who build and operate them and analyze their data.}
\newcommand{\lvk}{\gls{LVK}}

\newglossaryacronym{KNN}{k-nearest-neighbor}{A non-parametric supervised classification/regression method, see \citet{shakhnarovich2005nearest}.}

\newglossaryacronym{MCMC}{Markov Chain Monte Carlo}{A stochastic sampling method, often used for parameter estimation of gravitational-wave signals, using one or more chains exploring the parameter space with a jump probability depending only on the current position, not the previous history (Markovian). See e.g. Chapter 12 of \citet{gregory2005bayesian}.}
\newcommand{\mcmc}{\gls{MCMC}\xspace}

\newglossaryentry{MF}
{
name={Matched Filter},
text={matched filter},
description={The optimal linear solution to finding weak signals in a noise background of known properties, by ``filtering'' the full (noise+signal) input data with a model for the expected signal. See \citet{wienerbook,Wainstein:1962vrq} for the general concept, \citet{Allen:2005cz} for the compact binary coalescence case and \citet{Jaranowski_Krolak_2009} for continuous waves.}
}
\newcommand{\mf}{\gls{MF}\xspace}

\newglossaryacronym{ML}{Machine Learning}{The set of data-driven analysis methods we review here.}
\newcommand{\ml}{\gls{ML}\xspace}
\newcommand{\ML}{\gls{ML}\xspace}

\newglossaryacronym{MLP}{Multi-Layer Perceptron}{Another term for neural networks made out of multiple layers, each including multiple neurons. Traditionally, a perceptron is a neuron with various binary inputs and one binary output.}
\newcommand{\mlp}{\gls{MLP}\xspace}

\newglossaryentry{NNOISE}{
name={Newtonian Noise},
text={Newtonian noise},
description={Environmental noise that is produced by tiny fluctuations of the Earth's gravitational field, such as those generated by the motion of air and soil. Newtonian noise couples to ground-based interferometric gravitational-wave detectors limiting their sensitivity at low frequencies. It is expected to become the dominant source of noise in the low-frequency band in third-generation detectors.}
}

\newglossaryentry{NF}
{
name={Normalizing Flow},
text={normalizing flow},
description={A technique to build up representations of complex probability distributions by learning the necessary transformations from a simpler base distribution (e.g. a Gaussian). see \citet{Papamakarios:2019fms}.}
}
\newcommand{\nf}{\gls{NF}\xspace}
\newcommand{\nfs}{\glspl{NF}\xspace}

\newglossaryacronym{NR}{numerical relativity}{Methods for solving the Einstein equations with computers, most importantly for cases such as compact binary coalescences where no full analytical solutions exist. NR simulation results are one of the main ingredients for inspiral-merger-ringdown waveform models. see \citet{Palenzuela:2020tga} for an introduction.}
\newcommand{\nr}{\gls{NR}\xspace}

\newglossaryacronym{NS}{Neutron Star}{The extremely dense remnant of a massive star, but not massive enough to collapse to a black hole. Promising gravitational-wave emitters either in a compact binary {binary neutron stars or neutron star -- black hole binaries} or individually for continuous-wave signals.}
\newcommand{\ns}{\gls{NS}\xspace}
\newcommand{\nss}{\glspl{NS}\xspace}

\newglossaryacronym{NSBH}{Neutron Star-Black Hole}{A mixed compact binary consisting of one black hole and one neutron star. A possible multi-messenger source. First detected during the O3 observing run \citep{LIGOScientific:2021qlt}.}
\newcommand{\nsbh}{\gls{NSBH}\xspace}
\newcommand{\nsbhs}{\glspl{NSBH}\xspace}

\newglossaryacronym{PBH}{Primordial Black Holes}{Black Holes that have not formed through stellar evolution, but are remnants from early universe physics.}
\newcommand{\pbh}{\gls{PBH}\xspace}
\newcommand{\pbhs}{\glspl{PBH}\xspace}

\newglossaryacronym{PCA}{Principal Component Analysis}{A popular dimensionality reduction algorithm, see e.g. \citet{jolliffe2013principal}.}
\newcommand{\pca}{\gls{PCA}\xspace}

\newglossaryacronym{PE}{Parameter Estimation}{Determining the physical properties of the sources. Usually Bayesian.}
\newcommand{\pe}{\gls{PE}\xspace}

\newglossaryacronym{PINN}{Physics-Informed Neural Network}{Neural networks used to solve differential equations dwhich describe physical laws. see \citet{2017arXiv171110561R,2021arXiv210309655M}.}

\newglossaryacronym{PSD}{Power Spectral Density}{A frequency-domain description of the fluctuation power in a time series, e.g. for detector noise or signal strength, with units 1/Hz. This is the square of the also commonly used Amplitude Spectral Density (ASD). For gravitational-wave applications, see \citet{LIGOScientific:2019hgc}.}
\newcommand{\psd}{\gls{PSD}\xspace} 

\newglossaryacronym{RELU}{Rectified Linear Unit}{A type of activation function often used for neural networks, defined as $\max(0,x)$.}
\newcommand{\relu}{\gls{RELU}\xspace}

\newglossaryacronym{ResNet}{Residual neural network}{A deep learning architecture in which the neural network layers learn residual functions based on their inputs. This technique enables the model to bypass one or more layers, allowing the network to be trained on a large number of layers without compromising performance.}
\newcommand{\resnet}{\gls{ResNet}\xspace}

\newglossaryentry{RF}
{
name={Random Forest},
text={Random Forest},
description={An ensemble learning method for classification and regression based on a set of decision trees. see \citet{Breiman2001}.}
}
\newcommand{\RF}{\gls{RF}\xspace}
\newcommand{\RFs}{\glspl{RF}\xspace}

\newglossaryentry{RL}
{
name={Reinforcement Learning},
text={Reinforcement Learning},
description={Deep learning approach based on trial-and-error and objective-based reward/penalty feedback. see \citet{8103164}.}
}
\newcommand{\rl}{\gls{RL}\xspace}

\newglossaryacronym{RNN}{Recurrent Neural Network}{Deep learning structure with information exchange between layers.
It is frequently used to process and model speech, text, and time series in general.
}

\newcommand{\rnns}{\glspl{RNN}\xspace}

\newglossaryacronym{ROQ}{Reduced Order Quadrature}{A technique to speed up likelihood-based inference in gravitational-wave parameter estimation by precomputing ingredients to the signal likelihood for a given waveform model.}
\newcommand{\roq}{\gls{ROQ}\xspace}

\newglossaryacronym{SBI}{Simulation Based Inference}{A version of likelihood-free inference that is useful when explicit parametrized models are computationally prohibitive or difficult to write down, while forward-simulating noise+signal realizations is easy and hence neural network methods, such as normalizing flows, can be trained to perform the analysis. see \citet{Cranmer:2019eaq}.}
\newcommand{\sbi}{\gls{SBI}\xspace}

\newglossaryacronym{SGWB}{Stochastic Gravitational-Wave Background}{A non-deterministic gravitational-wave signal corresponding to a universal background made of many overlapping signals, either from astrophysical processes or early-universe physics. see \citet{Renzini:2022alw,vanRemortel:2022fkb} for recent reviews.}

\newglossaryacronym{SNR}{Signal-to-Noise Ratio}{A common measure for signal detectability. Note that this can be defined quite differently for different applications, including different conventions on expectation values in pure noise and scaling with amplitude or with power/energy (amplitude squared). For example, see \citet{Allen:2005cz} for the standard definition in the context of compact binary coalescences and \citet{Riles:2022wwz} for the continuous waves case.}
\newcommand{\snr}{\gls{SNR}\xspace}
\newcommand{\snrs}{\glspl{SNR}\xspace}

\newglossaryacronym{SOAP}{Officially not an acronym but can be assumed to stand for Snakes On A Plane}{An algorithm for detecting continuous-wave signals as time-frequency tracks, introduced by \citet{Bayley:2019bcb}. It involves a discrete number of time steps, where the states at each step are computed with the highest probability given the data. The algorithm got its name as a joke on the observation that the signal tracks, with their Doppler modulation, have snake-like shapes in the time-frequency plane, and the 2006 American action thriller film ``Snakes on a Plane'' directed by David R.~Ellis and starring Samuel L.~Jackson \citep{soap}.}
\newcommand{\soap}{\gls{SOAP}\xspace}

\newglossaryacronym{SVD}{Singular Value Decomposition}{A standard matrix decomposition of linear algebra that can also be used for dimensionality reduction, by chosing a reduced subset of the obtained basis set.}
\newcommand{\svd}{\gls{SVD}\xspace}

\newglossaryacronym{SVM}{Support Vector Machine}{A class of algorithms for classification, see e.g. chapter 5 of \citet{vapnik2013nature}.}
\newcommand{\svm}{\gls{SVM}\xspace}

\newcommand{\svms}{\glspl{SVM}\xspace}

\newglossaryentry{autoencoder}
{
name={Auto-encoder},
text={auto-encoder},
description={A type of machine learning algorithm that combines two networks, one for encoding and one for decoding, to transform between inputs/outputs and a lower-dimensional latent representation space. See e.g. \citet{2022arXiv220103898M} for an introduction.}
}

\newglossaryentry{bayesian}
{
name={Bayesian statistics},
description={One approach to statistics, where probability is interpreted as degrees of belief. In the gravitational-wave context, most parameter estimation techniques are Bayesian. See \citet{Jaynes_2003,gregory2005bayesian}.}
}

\newglossaryentry{burst}
{
name={Burst},
text={burst},
description={A short transient gravitational-wave signal, typically those for which no explicit waveform model is available.}
}
\newcommand{\burst}{\gls{burst}\xspace}
\newcommand{\bursts}{\glspl{burst}\xspace}

\newglossaryentry{deeplearning}
{
name={Deep learning},
text={deep learning},
description={Machine-learning methods based on deep, i.e. many-layered, neural networks. see \citet{Goodfellow-et-al-2016} for an introduction.}
}

\newglossaryentry{frequentist}
{
name={Frequentist statistics},
description={one approach to statistics, where probability is interpreted as the limiting fraction in infinitely repeated experiments. Often referred to as the opposite of Bayesian.}
}

\newglossaryentry{genetica}
{
name={Genetic Algorithm},
description={An algorithm that uses rules inspired by reproductive processes and natural selection to improve a population of candidate solutions.}
}
\newcommand{\genetica}{\gls{genetica}\xspace}

\newglossaryentry{geneticp}
{
name={Genetic Programming},
description={An approach to problem solving based on evolving an initial population of computer programs.}
}
\newcommand{\geneticp}{\gls{geneticp}\xspace}

\newglossaryentry{KAGRA}
{
name={KAGRA},
description={A 3\,km armlength underground gravitational-wave detector in Japan, described in \citet{10.1093/ptep/ptaa125}.}
}
\newcommand{\kagra}{\gls{KAGRA}}

\newglossaryentry{k-means-clustering}
{
name={k-means clustering},
description={A popular clustering algorithm that iteratively refines an initial set of clusters, see \citet{Jin2010}.}
}

\newglossaryentry{multimessenger}
{
name={Multi-messenger},
text={multi-messenger},
description={Astronomical observations that use signals from at least two of the following categories: electromagnetic, gravitational waves, and/or astroparticles (neutrinos, cosmic rays, \dots). see \citet{Corsi:2024vvr} for a recent review.}
}

\newglossaryentry{nestedsampling}
{
name={Nested sampling},
text={nested sampling},
description={A stochastic sampling technique often used as an alternative to {Markov Chain Monte Carlo algorithms} in gravitational-wave parameter estimation, introduced by \citet{Skilling:2006gxv}. It is particularly well suited for evidence calculation, as needed in model selection. See also \citet{Ashton:2022grj} for a pedagogical review.}
}

\newglossaryentry{transformer}
{
name={Transformer},
text={transformer},
description={A deep learning architecture based on the multi-head attention mechanism, proposed by \citet{NIPS2017_3f5ee243}.}
}

\newglossaryentry{Virgo}
{
name={Virgo},
description={A 3\,km armlength gravitational-wave detector in Italy, later upgraded to Advanced Virgo as described in \citet{Acernese_2015}.}
}
\newcommand{\virgo}{\gls{Virgo}}

\newglossaryentry{Viterbi}
{
name={Viterbi},
description={A dynamic programming algorithm that can efficiently find the highest-probability signal path through an observed data set, e.g.\ a gravitational-wave spectrogram. Originally introduced for communications decoding by \citet{Viterbi1967}.}
}

\newglossaryentry{waveform}
{
name={Waveform},
text={waveform},
description={A signal model for the time-domain strain time series or frequency-domain amplitude and phase evolution of gravitational waves from a specific source. For the case of compact binary coalescences, these are either taken directly from numerical relativity simulations or combine these with analytical relativity information. see \citet{Schmidt:2020ekt} for a brief review and \citet{LISAConsortiumWaveformWorkingGroup:2023arg} for a more in-depth overview, though focused on the case of the space-based LISA observatory.}
}

\makeglossaries 

\begin{document}

\title{Applications of machine learning in gravitational wave research with current interferometric detectors}

\author*[1,2]{\fnm{Elena} \sur{Cuoco}}\email{elena.cuoco@unibo.it}
\author[3]{\fnm{Marco} \sur{Cavagli\`a}}\email{cavagliam@mst.edu}
\equalcont{These authors contributed equally to this work.}
\author[4]{\fnm{Ik Siong} \sur{Heng}}\email{ik.heng@glasgow.ac.uk}
\equalcont{These authors contributed equally to this work.}
\author[5]{\fnm{David} \sur{Keitel}}\email{david.keitel@ligo.org}
\equalcont{These authors contributed equally to this work.}
\author[4]{\fnm{Christopher} \sur{Messenger}}\email{christopher.messenger@glasgow.ac.uk}
\equalcont{These authors contributed equally to this work.}

\affil*[1]{\orgdiv{Physics and Astronomy Department (DIFA)}, \orgname{Alma Mater Studiorum- Universit\`a di Bologna}, \orgaddress{\street{Via Zamboni, 33}, \city{Bologna}, \postcode{40126}, \state{Bologna}, \country{Italy}}}
\affil*[2]{\orgdiv{INFN Sezione di Bologna}, \orgaddress{\street{Viale C. Berti Pichat, 6/2}, \city{Bologna}, \postcode{40126}, \state{Bologna}, \country{Italy}}}

\affil[3]{\orgdiv{Institute of Multi-messenger Astrophysics and Cosmology}, \orgname{Missouri University of Science and Technology}, \orgaddress{\street{1315 N.~Pine St.}, \city{Rolla}, \state{MO}, \postcode{65409}, \country{USA}}}

\affil[4]{SUPA, School of Physics and Astronomy, University of Glasgow, Glasgow, United Kingdom}

\affil[5]{\orgdiv{Departament de F{\'i}sica}, \orgname{Universitat de les Illes Balears, IAC3--IEEC}, \orgaddress{\street{Carretera de Valldemossa, km 7.5}, \postcode{E-07122} \city{Palma}, \country{Spain}}}


\abstract{This article provides an overview of the current state of machine learning in gravitational-wave research with interferometric detectors. 
Such applications are often still in their early days, but have reached sufficient popularity to warrant an assessment of their impact across various domains, including detector studies, noise and signal simulations, and the detection and interpretation of astrophysical signals.

In detector studies, machine learning could be useful to optimize instruments like LIGO, Virgo, KAGRA, and future detectors. Algorithms could predict and help in mitigating environmental disturbances in real time, ensuring detectors operate at peak performance. Furthermore, machine-learning tools for characterizing and cleaning data after it is taken have already become crucial tools for achieving the best sensitivity of the LIGO--Virgo--KAGRA network.

In data analysis, machine learning has already been applied as an alternative to traditional methods for signal detection, source localization, noise reduction, and parameter estimation.
For some signal types, it can already yield improved efficiency and robustness, though in many other areas traditional methods remain dominant.

As the field evolves, the role of machine learning in advancing gravitational-wave research is expected to become increasingly prominent. This report highlights recent advancements, challenges, and perspectives for the current detector generation, with a brief outlook to the next generation of gravitational-wave detectors.}

\clearpage




\maketitle

\tableofcontents

\clearpage
\printglossary[type=\acronymtype]
\printglossaries

\section{Introduction}

The first direct detection of \gws, the \bbh coalescence event GW1501914 \citep{Abbott:2016blz},
was made on September 14, 2015, by the twin detectors of the \ligo, located in Livingston, Louisiana, and Hanford, Washington, USA.
In 2017, the \virgo\ detector in Italy joined the global network of \gw detectors,
and August 2017 brought the first \gls{multimessenger} event, the \bns inspiral GW170817 \citep{LIGOScientific:2017vwq} with rich \elmag counterparts \citep{LIGOScientific:2017ync}.
Since then, the \lvk\ Collaboration has published over 90 \gw signals caused by \cbcs, with most of them due to \bbh mergers \citep{PhysRevX.9.031040,PhysRevX.11.021053,LIGOScientific:2021usb,KAGRA:2021vkt}, but also including several \bns \citep{LIGOScientific:2017vwq,LIGOScientific:2020aai} and \nsbh \citep{LIGOScientific:2021qlt} events.
As of the submission of this review, \ligo~\citep{LIGOScientific:2014pky}, \virgo~\citep{Acernese_2015}, and \kagra~\citep{10.1093/ptep/ptaa125} are performing their fourth observing run O4\footnote{\url{https://observing.docs.ligo.org/plan/}} after undergoing further upgrades to improve their sensitivity.
We anticipate that many more events will be detected, including some multi-messenger events like GW170817 \citep{KAGRA:2013rdx}.

Besides the construction, commissioning and characterization of the detectors, \gw astronomy relies on a variety of signal modeling and data analysis techniques \citep{LIGOScientific:2019hgc,LVK:OBSWP}. Most of these have been developed in the frameworks of \gls{frequentist} or \gls{bayesian} statistics.
For some signal types, such as \cbcs and \cws from spinning neutron stars,
predictive \gls{waveform} models of varying accuracy are available.
In these cases, \mf techniques \citep{wienerbook} are typically used,
though model mismatch can be a concern and continued development of these models is of crucial importance, see e.g. \citet{Dhani:2024jja}.
On the other hand, a variety of cross-correlation and pattern-recognition techniques
are used for less well-understood sources, such as transient \gw \glspl{burst} from a variety of sources, as well as \glspl{SGWB} \citep{doi:https://doi.org/10.1002/9783527636037.ch7}.

In recent years, several novel \ml approaches have been explored as alternatives for noise mitigation, source modeling, signal detection and characterization in \gw astronomy. A first survey of these developments was presented a few years ago \citep{Cuoco:2020ogp}.
Now with this review article, we aim to provide an updated comprehensive summary of where the field stands. We focus on applications for the current \lvk\ network and the near future, though we also briefly comment on future ground- and space-based \gw detectors towards the end of the article (Sect.~\ref{sec:nextgeneration}).

\ml for \gw astronomy fits within a wider context of such methods gaining ground in physics and astronomy.
Various papers and other resources have been dedicated to making concepts, methods and software from the \ml world more accessible to astronomers,
including both general reviews and introductory articles \citep{2019arXiv190407248B,2022ExA....53....1S,2023RSOS...1021454S} as well as those about individual methods, e.g. \glspl{transformer} \citep{2023arXiv231012069T} -- the reader is also kindly referred to our glossary for short definitions and additional basic references on methods and concepts.
A useful set of recommendations for reliable and impactful
\ml applications in astronomy are made by \citet{2023arXiv231012528H}.
Criticisms to the indiscriminate application of \ml in the physical sciences, and possible answers, are discussed in \citet{2024arXiv240518095H}.

In this review, we mainly focus on concrete applications of \ml in the \gw context. 
However, we also provide a few introductory sections that briefly summarize the status of both fields.
These do not aim to be comprehensive didactic treatments, but just to offer sufficient context for the later sections and pointers for the reader to study the rest of the literature.
In this understanding, in Sect.~\ref{section:GW-data} we provide a brief general introduction to \gw detector data and the various standard methods both for preprocessing it and for detecting astrophysical signals in it.
In Sect.~\ref{sec:MLintro}, we provide a concise overview of the principal \ml methods employed within the context of our review. This includes, but is not limited to, supervised learning, unsupervised learning, and reinforcement learning techniques. 
The goal of Sect.~\ref{sec:MLworkflow} is to present a conceptional and visual summary of the domains where \ml is presently utilized in \gw astronomy, or could potentially find application in the near future.

These different domains are then discussed in detail in the remaining sections of this review. 
In Sect.~\ref{sec:noisemitigation} we will demonstrate the considerable focus of numerous studies on utilizing \ml to clean data and mitigate noise contributions. 
Section \ref{sec:datagen} discusses the application of \ml techniques for efficiently simulating astrophysical \gw sources and the emitted \glspl{waveform}, as well as simulating realistic detector noise. In Sect.~\ref{sec:searches} we explore \ml applications for \gw signal searches and in Sect.~\ref{sec:interpretation} we address aspects related to source interpretation, including the \pe of individual \gw signals as well as wider inference tasks such as multi-messenger, population and cosmology studies, which are crucial areas where \ml can significantly impact outcomes. Faster PE algorithms can open the way for discoveries in new physics, facilitate real-time studies, and enhance alert systems. In particular, in Sect.~\ref{sec:multimessenger}, we outline the preliminary framework for a multimodal machine learning application in multimessenger astrophysics, where faster PE will have a fundamental role.
Throughout these sections, we mainly aim to give a general overview of the historical development and current state of the art of \ml applications in the \gw context, rather than detailed comparisons against other established techniques, though performing such comparisons case-by-case is a crucial step for the adoption of \ml methods as tools for routine practical use.

In Sect.~\ref{sec:citizenAI}, we explore the links of the realm of citizen science with \ml strategies, motivated by the growing need for labeled datasets. This demand is being addressed through the engagement of volunteers in citizen science initiatives. Furthermore, these activities could potentially connect the expertise of data scientists outside the \gw community.
A concise outlook towards applications of \ml to next-generation \gw detectors is provided in Sect.~\ref{sec:nextgeneration}. Like much of this review, this section in particular will continue to evolve significantly in the coming years, as we face new challenges arising from the upgraded sensitivity of \gw detectors.

We provide our overall perspective on the role and significance of \ml for \gw astronomy in the summary and outlook of Sect.~\ref{sec:summary}.

\section{A brief primer on GW data and searches}
\label{section:GW-data}

Advanced \ligo\ \citep{LIGOScientific:2014pky}, Advanced \virgo\ \citep{Acernese_2015} and \kagra\ \citep{10.1093/ptep/ptaa125} are second-generation laser-interferometric detectors specifically designed for the detection of \gw signals. They utilize the Michelson interferometer technique, which allows them to be highly sensitive to the strain in space-time caused by passing \gws \citep{Saulson_book_2017}.
This strain causes fluctuations in the relative lengths of the interferometer arms, which in turn lead to corresponding power variations in the interferometer's output. To precisely capture these fluctuations, photo-diodes are used to measure changes in the intensity of the laser light \citep{LIGOScientific:2019hgc}. The signal obtained from these photo-diodes, after calibration \citep{Viets:2017yvy,Virgo:2018gxa}, serves as the readout for detecting \gws.

However, alongside astrophysical \gw signals, these detectors are also susceptible to a variety of environmental noises: terrestrial forces can also directly cause time-varying variations in the lengths of the interferometer arms. These variations pose a challenge in accurately identifying astrophysical signals amidst the noise \citep{Abbott_2016,LIGOScientific:2019hgc,LIGO:2021ppb,Acernese_2023,Covas:2018oik}.
Hence, the detector output also functions as an error signal for controlling the relative lengths of the arms. This continuous monitoring and adjustment of the arm lengths based on the control signal allow the detectors to optimize their sensitivity to \gws while minimizing the impact of unwanted noise sources. In addition to the control signal, a vast array of auxiliary monitoring channels is employed to track and monitor noise sources, aiming to reduce their influence on the \gw strain.

\subsection{Preprocessing techniques}

The photo-diode output in the interferometer yields time-series data, recorded at a designated sampling rate, covering a wide spectrum of emissions originating from diverse astrophysical sources. \ligo, \virgo, and \kagra\ commonly sample their data at a rate of 16384\,Hz. For most astrophysical analyses, the frequency range of interest typically spans only from 20\,Hz to 2000\,Hz, where sensitivity is best. (Calibration is typically valid from 10\,Hz on, but due to the steep sensitivity losses at lower frequencies, few analyses go that low in practice \citep{LIGOScientific:2019hgc}.) Thus, in practice, when it comes to analyzing this wealth of data, most \gw analysis pipelines employ a down-sampling process, reducing the data to a lower sampling rate, e.g. $4096$\,Hz. This down-sampling approach effectively condenses the information in the time series while retaining essential details, enabling researchers to focus on \gw signals within a frequency range that extends up to half the chosen sampling rate, also known as the Nyquist frequency \citep{Shannon1949}.

This down-sampling strategy serves several crucial purposes in the field of \gw research. Firstly, it helps manage the computational demands of data analysis, as the original high-frequency data can be extremely data-intensive and challenging to process. 
If the signal has no contributions above the Nyquist frequency, by reducing the sampling rate, analysis pipelines become more computationally manageable.

Secondly, this down-sampling aligns with the fact that many astrophysical sources of \gws fall within the frequency range preserved by the lower sampling rate. The down-sampling strategy can be applied with different sampling rates, depending on the type of astrophysical signal search being conducted or if the data analysis involves detector characterization procedures.

This strategy of down-sampling achieves a balance between data efficiency and the capability to detect and investigate \gw signals emitted by diverse astrophysical phenomena.

Another step that is often useful is to reduce the bandwidth of interest for analysis. This is done by using digital filters with different purposes, such as high-pass, low-pass or band-pass filters, which allow greater efficiency on the area of interest for searching gravitational signals and can reduce the computational cost of subsequent operations.

Another important aspect to improve the sensitivity and robustness of \gw analyses is to mitigate the impact of certain spectral lines \citep{Covas:2018oik} due to the presence of noise of a persistent nature at certain frequencies, such as those created by the power grid frequency, mirror suspension resonances, and other sources. This can be done in a variety of ways best suited to each analysis pipeline, e.g. by simple notch filters on the frequencies to be removed \citep{10.5555/516039},
veto techniques at analysis time (see e.g. \citealt{Leaci:2015iuc} for an overview for the case of \cw searches),
or the \ml noise subtraction methods discussed in Sect.~\ref{denoising}.

The extraction of \gw signals from the background noise requires sophisticated signal processing techniques due to the relatively low amplitudes of the gravitational signals of interest, since the signals usually have a signal-to-noise ratio close to one. These techniques involve advanced algorithms and analyses to separate and identify the desired signals from the noise, ensuring accurate detection and characterization of \gws, and are linked to the kind of signals we are looking for \citep{LIGOScientific:2019hgc}.

One important preprocessing procedure that is applied in some \gw searches for transient signals is the whitening algorithm. 
The purpose of the whitening procedure is essentially to remove the contribution of stationary noise associated with the second-order statistics of the data, information encapsulated in the \psd of the data. Thus, whitening transforms colored noise, which includes stationary and Gaussian noise contributions, into white noise, meaning it becomes delta-correlated.

This is done by applying a filter that compensates for the frequency-dependent response of the detector, and can be accomplished through techniques applied either in the frequency domain \citep{LIGOScientific:2019hgc} or in the time domain \citep{Cuoco_2001}.
Whitening is an intrinsic part of modeled \mf techniques
while for most unmodeled detection algorithms an explicit whitening preprocessing step significantly improves pipeline efficiency. 

The data in the time domain is frequently subjected to various transformations to gain deeper insights and extract valuable information useful for the signal detection or characterization algorithms.
Two common domains in which these transformations occur are the frequency domain and the time-frequency domain. The frequency domain is typically achieved through the Fourier transform, a mathematical technique that decomposes a signal into its constituent frequencies, allowing for the analysis of its frequency components.
The time-frequency domain provides a representation of the signal that captures both time and frequency information simultaneously. This is particularly useful for analyzing non-stationary signals. Techniques used in this domain include the Short-Time Fourier Transform (STFT), which divides the signal into short segments and applies the Fourier transform to each segment; the Q-transform, which provides a time-frequency representation with variable resolution; and wavelet transforms, which use wavelet functions to analyze signals at different scales and positions, offering a more flexible approach to time-frequency analysis. In \citet{PhysRevD.102.124038}, the various methodologies for time-frequency representation in the field of \gw\ are described.
The choice between these time-frequency analysis techniques depends on the specific characteristics of the data and the research objectives. They all offer unique advantages in terms of revealing temporal and spectral information that may be hidden in the original time series.
Transforming time-series data into different domains, such as frequency or time-frequency representation, can help uncover hidden patterns, track changes over time, and gain deeper insights into the underlying dynamics of the data. Very often the time-frequency transformation is used before \ml applications based on image classification. 
 
\begin{figure}[ht]
    \centering
    \subfloat[Time domain]{\label{a}\includegraphics[width=.33\linewidth ]{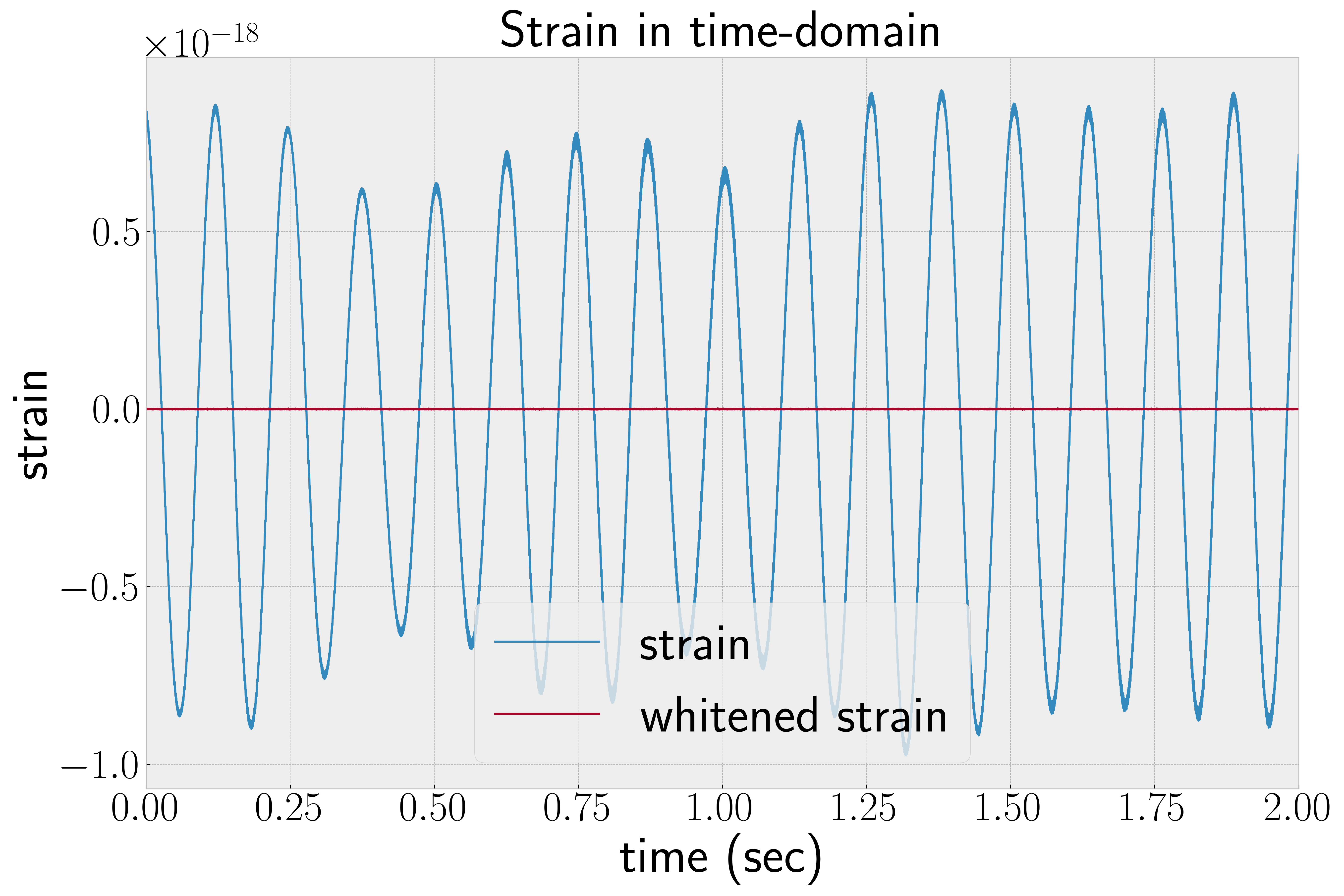}}
    \subfloat[Frequency domain]{\label{b}\includegraphics[width=.33\linewidth]{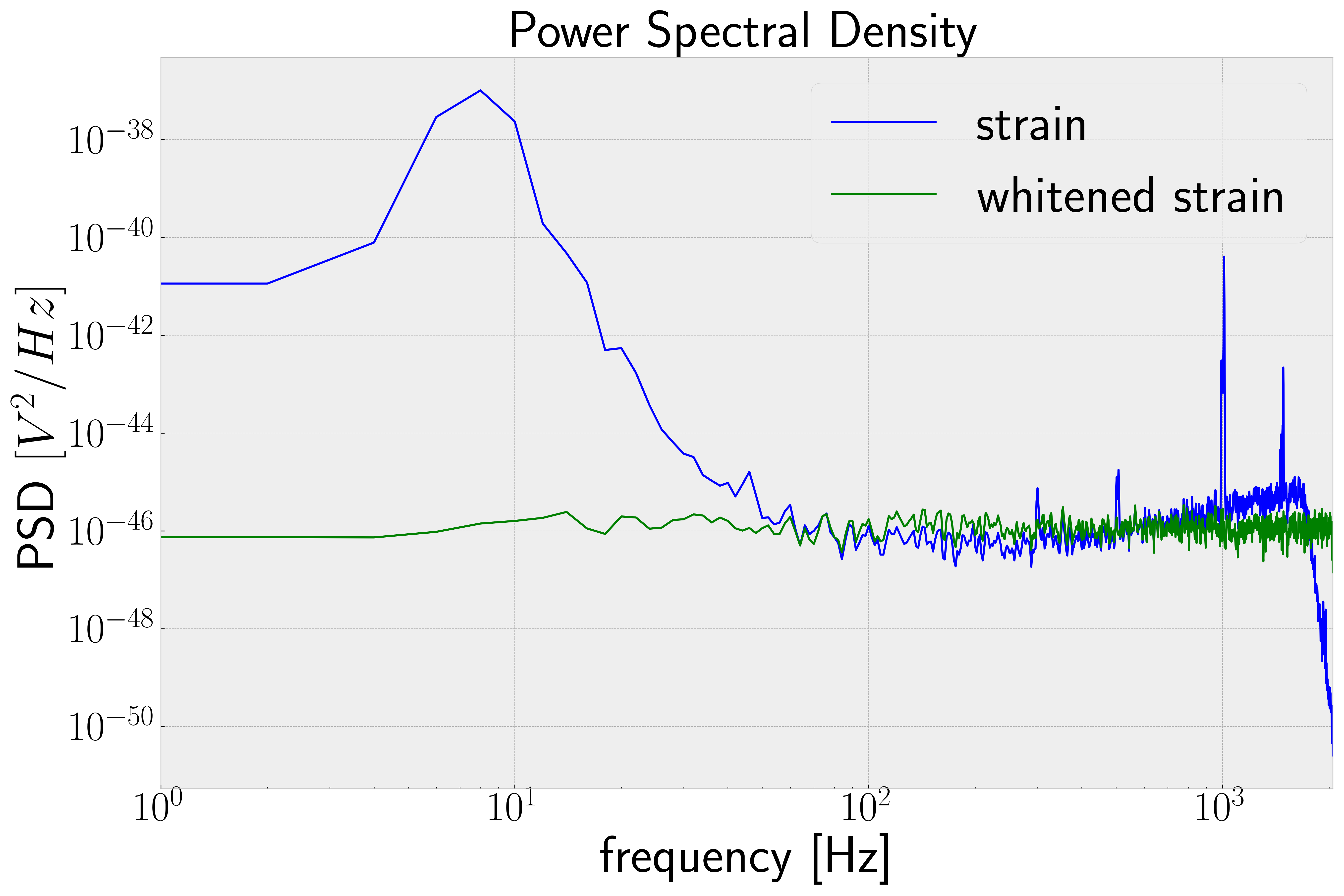}}
    \subfloat[Time-Frequency domain]{\label{c}\includegraphics[width=.33\linewidth]{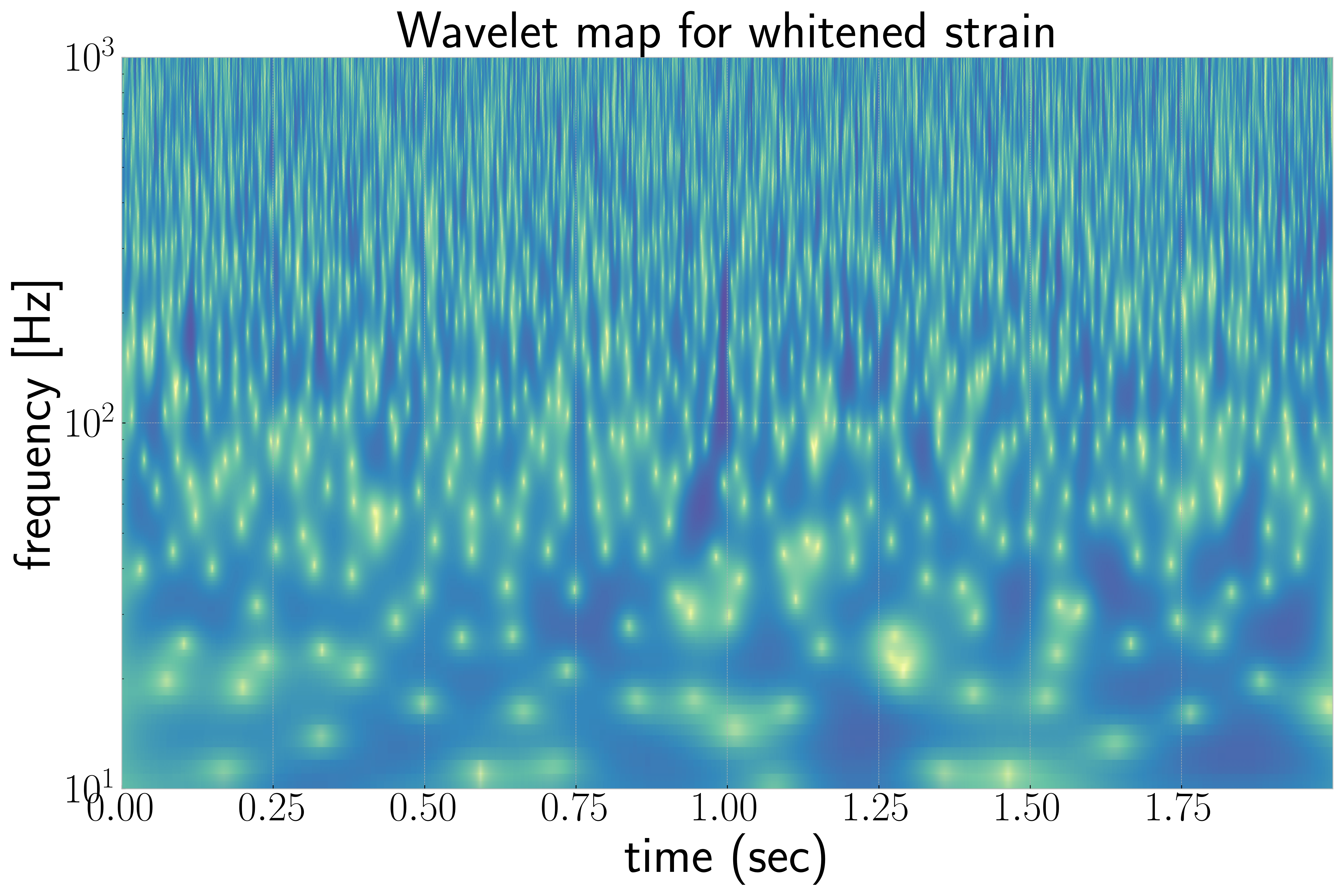}} 
    \caption{The image shows the identical not whitened and whitened strain data represented in the time domain (a) and frequency domain (b). In (c) we illustrated, as time-frequency representation, the mapping to a Morlet continuous wavelet basis of the whitened strain. }
    \label{fig:data-mapping}
\end{figure}

In Fig.~\ref{fig:data-mapping} we show different representations of a \gw signal in the time domain, the frequency domain and in time-frequency domain, as a wavelet map. The wavelet transform was utilized after signal whitening to improve the visibility of transient signals.

\subsection{GW searches} 
The exploration of detector data in search of astrophysical gravitational signals involves distinct methodologies, depending on the specific nature of the signals under investigation. Broadly speaking, \gw signals are categorized into three main types: continuous signals, transient signals, and stochastic signals \citep{doi:https://doi.org/10.1002/9783527636037.ch5}. A more nuanced classification arises from our understanding of the signal's source and our ability to model its \gls{waveform} accurately. Consequently, research efforts are further stratified into two overarching domains: the examination of signals characterized by known waveforms and those characterized by unknown waveforms. 

The \mf technique (see \citealt{wienerbook,Wainstein:1962vrq} for general mathematical background and \citealt{LIGOScientific:2019hgc} for a modern, \gw-focused review) is designed for scenarios where a known signal waveform and known noise background can be assumed, with the noise typically characterized as a stationary and Gaussian (but possibly colored) distribution It represents the optimal choice under these constraints in the Neyman--Pearson sense \citep{neyman1933} of maximizing detection probability at fixed false-alarm probability. In the \lvk\ different pipelines implement the \mf for \cbc signals in real time (also called ``online'' mode) \citep{Nitz:2018rgo, CANNON2021100680,Aubin:2020goo,Chu:2020pjv}. Many more implementations exist for ``offline'' analyses, where the data is reanalyzed at higher latency (also called ``archival'' searches), for both \cbcs and \cws. In this review however, we will mostly focus on \ml alternatives to this kind of approach.

When dealing with signals characterized by unknown \glspl{waveform}, the requirements shift towards more generic algorithms capable of identifying signal excess over a background of noise, with coherence across multiple detectors valuable to take into account. One example of this kind of pipeline is coherent WaveBurst \citep{Klimenko:2008fu,Drago:2020kic}, based on wavelet transform. Another important class of algorithms is based on cross-correlation techniques \citep{Dhurandhar:2007vb}, used mainly for unmodeled transients (\glspl{burst}) and stochastic signals, but also for complicated cases of \cws such as those from \nss in binary orbits.

We will not go further into the details of these existing pipelines in the rest of this review. Instead, our focus is to outline various solutions based on \ml approaches that have been researched and implemented in the community. These solutions aim to enhance existing search pipelines, either by complementing them or by serving as potential new pipelines integrated into the standard framework. 

Observations of transient signals from the first three observing runs have been reported in, so far, four releases of the \gls{GWTC} \citep{PhysRevX.9.031040,PhysRevX.11.021053,LIGOScientific:2021usb,KAGRA:2021vkt},
all corresponding to \cbcs.
The full cumulative catalog is also available online at \url{https://gwosc.org/eventapi/html/GWTC/}.
An overview of all \lvk\ observational results, including more detailed papers on individual events as well as upper limits on other \gw signal types, can be found at \url{https://pnp.ligo.org/ppcomm/Papers.html} and a live catalog of detection candidates from O4 and future runs is provided at \url{https://gracedb.ligo.org/}.
Data releases of archival \lvk\ data \citep{LIGOScientific:2019lzm,KAGRA:2023pio} can be found in the Gravitational Wave
Open Science Center, \url{https://gwosc.org/},
along with pointers to the software used in obtaining \lvk\ results.

\section{A brief overview of machine learning techniques}
\label{sec:MLintro}
\ml offers a diverse range of approaches for solving complex problems across numerous domains, each tailored to specific types of problems and data characteristics.
For a general introduction to machine learning, the reader may consider reading \citet{theobald2017machine} or,
with more of a focus on astronomy applications, \citet{2019arXiv190407248B,2022ExA....53....1S,2023RSOS...1021454S}.
\Gls{deeplearning} \citep{Goodfellow-et-al-2016} is a subfield of machine learning that focuses on training \anns with multiple layers to learn complex patterns and representations from data. Deep learning algorithms automatically learn hierarchical representations of the data directly from raw inputs. This ability to automatically discover intricate, nonlinear patterns makes deep learning particularly well-suited for tasks involving large amounts of data, such as image and speech recognition, natural language processing, and reinforcement learning \citep{nielsen2015neural}.

The rise of \ml and, in particular, \gls{deeplearning} methods has been closely linked with the use of highly-parallelized high-performance \gpu hardware for general computation and data processing applications.
Their ability to perform numerous calculations simultaneously allows for faster training times and the execution of sophisticated models that were previously impractical with traditional CPU-based systems.
On the software side, ML applications in \gw astronomy are often built on standard libraries from other fields, which we however do not review in detail here.
The further embedding of \gw applications into existing online computing infrastructures optimized for \ml is also a promising approach,
as discussed e.g. by \citet{Gunny:2021gne}.

Below, we present a concise overview of the primary approaches utilized in \gw machine learning applications.
The glossary of this article also collects the main definitions and references.

\subsection{Supervised methods}
Supervised learning involves training a model on a labeled dataset, where the input data is paired with corresponding output labels.
The goal of supervised learning is to learn a mapping from input data to output labels, such that the algorithm can make accurate predictions on new, unseen data. Examples of supervised learning algorithms include linear regression, Decision Trees \citep{Fürnkranz2010}, Random Forests \citep{Breiman2001}, and Support Vector Machines \svm \citep{Cristianini2008}.
A comparison among the different techniques for supervised methods is reported in \citet{osisanwo2017supervised}.
Deep learning neural network models like \cnns and \rnns are also often used in supervised modes, though they can also form building blocks for unsupervised or reinforcement learning approaches.
Many \ml applications in \gw science are based on supervised methods, which will be discussed in the following sections.

Neural networks and other supervised methods with explicit training stages can depend a lot, both in the ease and speed of training and in their eventual performance, on the design of the training data set. Besides general rules like ensuring that the training set is large enough and representative of (but statistically independent from) the eventual evaluation and application cases, specific tricks have been developed in the \ml community to enhance training. For example, ``\gls{curric}'' \citep{Bengio2009} first presents the algorithm or model to be trained with an ``easy'' training set (e.g., in the \gw contest, simulated data with high signal-to-noise ratios or with simple Gaussian noise only), and then moves onto another training stage with a second, more difficult set. More than two stages are also possible. This is, in principle, also applicable to the unsupervised methods discussed next. For another popular enhancement, data augmentation, see Sect.~\ref{sec:augmentation} below.

\subsection{Unsupervised methods}
Unsupervised learning involves training a model on an unlabeled dataset, where the model must learn the underlying structure or patterns in the data without explicit guidance. Unlike supervised learning, there are no explicit output labels provided during training. Instead, the model identifies relationships or clusters among the input data points. Common tasks in unsupervised learning include clustering, where the algorithm groups similar data points together, and dimensionality reduction, where the algorithm reduces the number of features while preserving the essential information. Examples of unsupervised learning algorithms include \gls{k-means-clustering} \citep{Jin2010}, hierarchical clustering, \pca \citep{MACKIEWICZ1993303}, and \glspl{autoencoder} \citep{LI2023110176}.

\subsection{Semi-supervised methods}
Semi-supervised learning techniques \citep{vanEngelen2020} work with both labeled and unlabeled data for training. Its goal is to recover the information present in both labeled and unlabeled data to improve the performance of the learning algorithm. The availability of unlabeled data can be advantageous in scenarios where data labeling is expensive or time-consuming, as it allows the model to learn from a larger pool of examples. Semi-supervised learning algorithms aim to exploit the underlying structure or relationships within the data to make predictions on both labeled and unlabeled instances. One prevalent strategy in semi-supervised learning involves utilizing the unlabeled data to generate additional training examples or to regularize the learning process. Semi-supervised learning has applications in various domains, including natural language processing, computer vision, and speech recognition. 

\subsection{Reinforcement learning}
Reinforcement Learning (RL) \citep{8103164} involves training an agent to interact with an environment to achieve a specific goal. The agent receives feedback in the form of rewards or penalties based on its actions, and the goal is to learn a policy that maximizes cumulative reward over time. Unlike supervised and unsupervised learning, reinforcement learning is focused on learning through trial and error rather than explicit input-output mappings. \rl has applications in areas such as robotics, game playing, autonomous driving, and resource management.
A reinforcement learning control system strategy for non-linear systems is discussed in \citet{1200173}, which could have applications for \gw detector control (see Sect.~\ref{sec:control}).

\subsection{Data augmentation}
\label{sec:augmentation}
Data augmentation is not a specific kind of \ml algorithm, but a technique used in various approaches to increase the diversity and size of a training dataset by applying various transformations to the existing data samples. The objective of data augmentation is to improve the robustness and generalization capability of \ml models by exposing them to a wider range of variations in the input data. Data augmentation is particularly useful in scenarios where the training data set is limited or when the data distribution is unbalanced. By introducing variations to the training data, data augmentation helps prevent over-fitting \citep{Ying_2019} and improves the model's ability to generalize to unseen data.
In \citet{MUMUNI2022100258} the authors report a comprehensive review of data augmentation methods specifically tailored to computer vision domains, with a focus on recent and advanced techniques, including a comparative analysis of several state-of-the-art augmentation methods.

\subsection{Simulation-based inference and neural posterior estimation}
\label{sec:sbi-npe}
Estimating the parameters of \gw detections, and other more general inference tasks such as those at a population level, are traditionally done by Bayes' theorem and stochastic sampling techniques \citep{Jaynes_2003,gregory2005bayesian}, requiring an explicit likelihood model for the data containing noise and signal contributions. Since realistic detector noise can be difficult to model (see Sect.~\ref{sec:noisegen}) and signal models can be expensive to evaluate, or even too complicated to fully express, alternative likelihood-free inference schemes are attractive.
One class of such alternatives is referred to as \sbi, which inverts the usual approach of Bayesian inference:
instead of evaluating a large number of models on each observation,
large numbers of fake data realizations are generated in advance and then each observation is compared to this ensemble.
One popular implementation is neural posterior estimation,
where neural networks (often \nfs, and variational \glspl{autoencoder}, which are designed to deal with continuous distributions such as needed for Bayesian posteriors) are trained to learn the mapping from data to a posterior distribution in parameter space and can then typically be evaluated extremely quickly for each new observation, as long as it falls within the training space.
A brief overview of the \sbi approach is given by \citet{Cranmer:2019eaq}.
For an early introduction of the ideas of neural posterior estimation,
see e.g. \citet{2016arXiv160506376P},
though we discuss \gw applications more concretely in Sect.~\ref{sec:interpretation-transientPE}.

\subsection{Explainable/interpretable machine learning}
\label{sec:explainableML}
\ml methods are often perceived as ``black boxes'', where the researcher applying a trained model has little insight into how it comes to a certain result given certain inputs, and sometimes the same may even be true for the developer who trained a neural network. Particularly in fundamental physics with its traditional emphasis on deep understanding of the underlying processes and the ability to forward-model many of the systems being studied, a real or perceived lack of interpretable algorithm behavior can be a hurdle to the adoption of \ml solutions. A broad trend in the wider \ml landscape are so-called explainable or interpretable algorithms and models. These can range from additional tools that allow to study the responses of networks trained in standard ways to novel architectures designed from the ground up to ease the understanding of their inner workings. See e.g. \citet{2019PNAS..11622071M,2020Entrp..23...18L} for a review and some clarifications on the related concepts and nomenclature.

\section{Machine learning applications in the GW data analysis workflow}
\label{sec:MLworkflow}

\begin{figure}[ht]
    \centering
    \includegraphics[width=\linewidth]{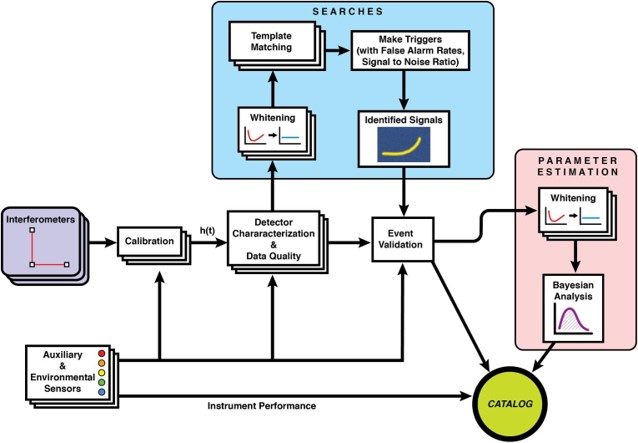}
    \caption{Typical analysis workflow for the data from \gw detectors, taken from \citet{LIGOScientific:2019hgc}.
    \label{fig:DA-workflow}
    }
\end{figure}

Figure \ref{fig:DA-workflow} illustrates the standard \GW data analysis workflow as outlined in \citet{LIGOScientific:2019hgc}, delineating the principal areas involved. Our objective in this paper is to showcase the latest advancements in \ml applications across the diverse domains depicted within the same figure. We aim to outline the ever-growing significance of \ml methodologies in the \gw data analysis framework.

As discussed in Sect.~\ref{section:GW-data}, the output from the detection photodiodes in \gw detectors comprises a time series that may contain \gw signals, but also many sources of instrumental and environmental noise. Alongside the primary channel (strain) which contains the \gw signals, numerous other channels are acquired with varying sampling rates to facilitate detector operation. Some channels serve as control signals for machine operation, while others monitor environmental conditions. All of these channels contain valuable information that can be utilized for data quality checks and cleaning procedures.

Following an initial stage involving calibration, cleaning, and vetting through data quality checks, the data is prepared for detection algorithms. The approaches employed vary, as elaborated in Sect.~\ref{section:GW-data}. For the specific case of applying a \mf, it is also necessary to input a template bank of signals generated in a parameter space suitable to cover the possible \gls{waveform} parameters.
After detecting the signal, a comprehensive parameter estimation procedure is initiated, including sky localization. 

\begin{figure}[ht]
    \centering
    \includegraphics[width=1.0\linewidth]{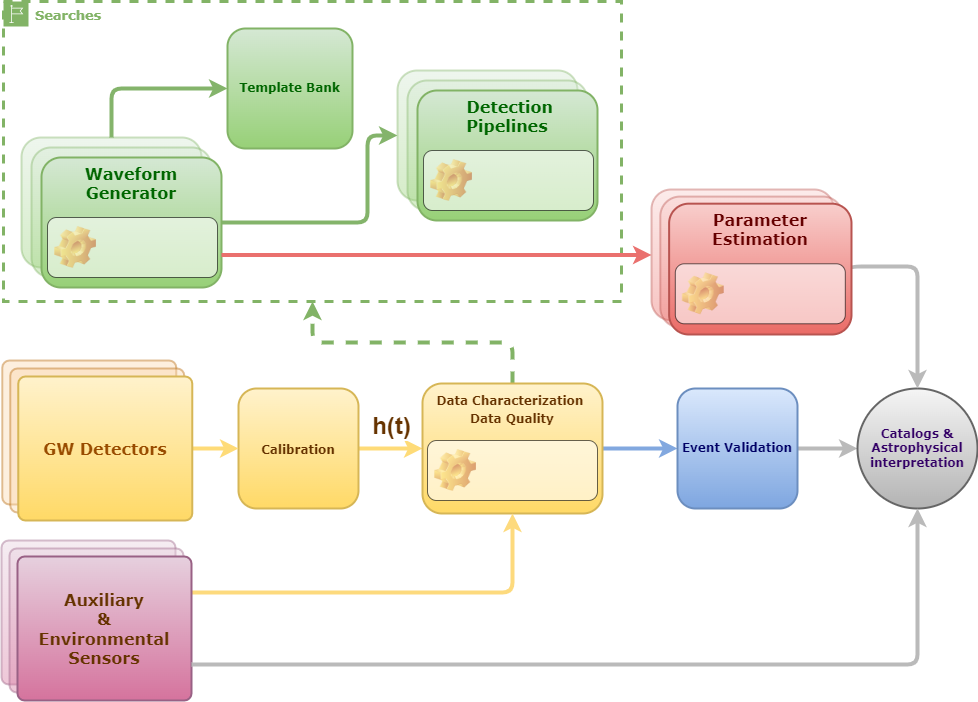} 
    \caption{Example of where machine learning fits in the workflow for \gw detectors and data analysis. The gear in the picture indicates the parts of the data analysis chain where \ml could be used. It is not the full picture covering all the \ml-based solutions. Most of the kinds of \ml applications studied will be described in the following sections of the paper.
    \label{fig:ML-workflow}
    }
\end{figure}

At each stage outlined above, there exists the opportunity to either adopt an equivalent \ml approach or to utilize \ml techniques to enhance the speed and efficiency of existing processes. This concept is illustrated in Fig.~\ref{fig:ML-workflow}, where we have incorporated typical \ml solutions that have already been explored in \gw science. These \ml solutions will be elaborated upon in greater detail in the subsequent sections of this paper. By integrating \ml methods into our workflow, we aim to optimize various aspects of \gw data analysis and enhance our ability to extract meaningful insights from the acquired data.

In this review article, our aim is not to show a detailed quantitative comparison with the standard techniques widely used in \gw data analysis. We mainly aim to report a general overview of the state of the art of \ML algorithms that have been studied and tested as alternatives or complementary to the techniques already in use.

\section{Strategies for noise mitigation}
\label{sec:noisemitigation}

The background noise in the instrument band of interest of a \gw detector determines its sensitivity. Ideally, this background noise is Gaussian, stationary, and determined by the detector's design \citep{LIGOScientific:2019hgc}. For instance, the quantum noise of the laser light, the thermal noise of the mirror coatings and optic suspensions, the electronic and feedback control system designs, and the inevitable seismic noise all limit the background floor of current ground-based \gw interferometric detectors \citep{LIGOScientific:2014pky,Acernese_2015,KAGRA:2022qtq}.
However, the noise floor of \gw detectors is not actually Gaussian nor stationary in practice. Most noise sources that couple to the detector, whether they be environmental or instrumental, vary over various time scales \citep{AdvLIGO:2021oxw}. The detector may also experience non-linear coupling from noise sources, which can for example cause sidebands, i.e. frequencies that appear on both sides of a carrier frequency during modulation, carrying the actual information of the signal, at frequencies close to other known disturbances \citep{LIGO:2021ppb}. These extra noise artifacts result in a decrease in a \gw detector's {duty cycle and in its design} sensitivity at specific frequencies or times \citep{Acernese_2023,LIGO:2021ppb}. Even if such artifacts are correctly identified, the simplest approach of dealing with them by excising certain time or frequency ranges from analysis will lead to a reduced duty cycle (fraction of usable data over the total run duration) or frequency coverage.
Hence, more advanced mitigation strategies are highly valuable.

Among transient noise phenomena, shifts in the noise floor are usually referred to as ``non-stationarities'', while the term ``glitches'' refers to various types of excess noise that are more-or-less well-localized in the time domain. These have an impact particularly on searches for burst signals and \cbcs because their presence increases the false alarm rate in both modeled and unmodeled searches \citet{LIGOScientific:2016gtq,2018CQGra..35f5010A,Nitz:2017lco,Davis:2020nyf,Mozzon:2020gwa,Mozzon:2021wam,Kumar:2022tto}. A glitch overlapping with a signal, or in close vicinity to it, can also significantly impact the accuracy of low-latency alerts to astronomer partners for EM follow-up \citep{Macas:2022afm} and the fidelity of parameter estimation, both in terms of sky localization and estimates of intrinsic parameters. The GW170817 \bns detection \citep{LIGOScientific:2017vwq} is a well-known example, where an improved parameter estimation was obtained after subtracting the time-frequency wavelet reconstruction of an instrumental glitch from the \ligo-Livingston data. Such instances become more common as the detectors improve, reducing their noise background and increasing the rate of detectable astrophysical signals, but also uncovering new types of transient noise artifacts.

On the other hand, excess localized noise in the frequency domain, e.g., narrow spectral lines or other narrow-band features, are the main contaminants in searches for continuous \gws, long-lived \gw transients, and the \gls{SGWB}, where they can also massively increase the false-alarm rate if not adequately accounted for through advance mitigation or post-processing strategies.

\begin{figure}
 \centering
 \includegraphics[width=0.6\linewidth]{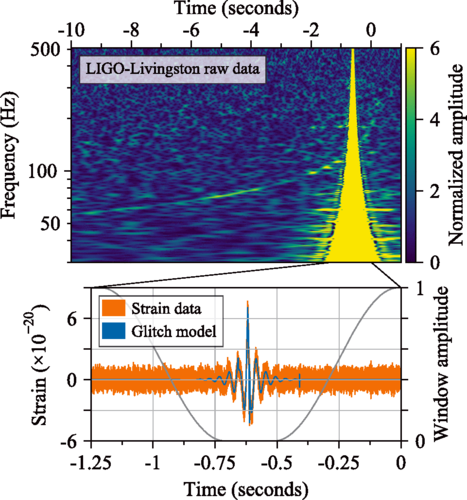}
 \caption{\label{fig:glitch_gw170817} Top panel: Time-frequency representation \citep{2004CQGra..21S1809C} of the \ligo-Livingston data at the time of the GW170817 binary neutron star merger. The time-frequency track of GW170817 shows the typical chirp-like shape. A loud glitch occurs 1.1 seconds before the coalescence time. Bottom panel: Strain data in the time domain (orange curve) band passed between 30 Hz and 2 kHz. To mitigate the glitch, the raw detector data were initially multiplied by an inverse Tukey window represented by the gray curve. A model of the glitch based on a wavelet reconstruction (blue curve) was later subtracted from the data to mitigate the glitch in the measurement of the source's properties. Image reproduced with permission from \citet{LIGOScientific:2017vwq}; copyright by the author(s).
}
\end{figure}

For all these reasons, a critical step in any real-time and follow-up \gw detection workflow is the understanding and mitigation of detector non-astrophysical excess noise in the instrument band of sensitivity \citep{Abbott_2016}. 
Hence, characterizing non-astrophysical noise, enhancing the quality of \gw search data, and commissioning detectors are major areas of focus for the \gw collaborations \citep{LIGO:2021ppb,Acernese_2023,Klimenko:2008fu}. These tasks are all completed by the detector characterization team, which is part of the joint collaboration operation division. Here, spectral features and unwanted glitches that taint \gw searches are recognized, categorized, and reduced by a collaborative effort between instrumentalists, commissioners, and data analysts.

The data from the \gw detector output and numerous instrumental and environmental auxiliary data streams are typically mined to complete these tasks. To accomplish these tasks, \gw researchers have created and applied a variety of signal processing techniques and algorithms over time \citep{Slutsky:2010ff,LIGO:2021ppb,F_Acernese_2023}. A few of the data quality tools are available online and are capable of rapidly assessing the interferometers' state for low-latency searches \citep{Chaudhary:2023vec}. For follow-up with \gw candidates and deeper searches, an alternative set of tools is utilized offline. The findings of these studies are then shared with data analysts operating \gw searches, detector commissioners, and operators to help mitigate undesired non-astrophysical disturbances.

However, current data quality methods based on conventional signal processing techniques will probably prove inadequate for the tasks ahead due to the increase in detections of \cbc and the discovery of \gws from other types of astrophysical sources. Investigating the use of sophisticated techniques based on computational learning theory as a complement to conventional signal processing methods is one way to enhance the way \gw collaborations perform in the area of data quality.

Machine learning algorithms are effective instruments for identifying patterns and analyzing large volumes of data. They are made to carry out specified tasks and, through the use of adaptive techniques and iterative procedures, automatically enhance their performance. Furthermore, because models created to address one problem can be readily modified to address another, they are adaptable, robust, and portable to a variety of scenarios. Because of these features, machine learning techniques are ideal for developing novel approaches to non-astrophysical noise reduction \citep{Cuoco:2020ogp}.

Under this framework, a crucial step towards implementing new techniques and improving \gw detector operations is the creation of self-contained algorithms that can be incorporated into already-existing searches or noise investigation processes. 

\subsection{Noise characterization with detector strain data (transient and continuous)}
\label{sec:noisecharact}
Thus far, the majority of \ml algorithm applications to the characterization of \gw interferometric detector noise have concentrated on the identification and classification of glitches in the time domain. \ml applications to spectral line characterization have mainly focused on noise line subtraction. These studies will be described in \ref{denoising}. 

Developing and implementing algorithms for the identification and predictive modeling of detector noise is the ultimate goal of detector characterization. If the source of the excess noise is found, it might be possible to design software or hardware upgrades that remove it. Various methods can be used to accomplish this aim.

Most \ml algorithms that are publicly accessible have been explored throughout the years, covering the primary categories outlined in section~\ref{sec:MLintro}. 
As a result, a substantial body of literature has been written on this topic (see \citealt{Cuoco:2020ogp,Benedetto:2023jwn} and references therein). The classification of glitches in the detector output, as well as the separation of astrophysical signals from detector glitches, were the primary goals of early \ml applications to detector characterization \citep{Powell:2015ona,Powell:2016rkl,Cuoco:2017evl,PhysRevD.88.062003,2017PhRvD..96j4015K}. Several of these notable methods that have advanced to the point of being implemented in \gw searches and/or helped characterize real interferometric data in various observing runs include Gravity Spy \citep{Zevin_2017}, GWSkyNet \citep{Cabero:2020eik,Abbott:2021cuf} and deep learning algorithms.

Gravity Spy \citep{Zevin_2017,Zevin:2023rmt,Wu:2024tpr,Bahaadini:2017dqg} is a project that combines \gw science with machine learning and citizen science. Utilizing citizen scientists' voluntary efforts, Gravity Spy aims to categorize \ligo's glitches and produce an ever-expanding labeled data set that can be used as input for machine learning algorithms \citep{Glanzer:2022avx}. A similar effort, GWitchHunters, has been developed in the Virgo community, with citizens assisting in labeling glitch data sets \citep{RAZZANO2023167959}.

Time-frequency images, specifically Omega Scans \citep{Chatterji:2004qg}, are used in the Gravity Spy framework to represent glitches. Volunteers manually label these images using a predetermined list of glitch classes \citep{JACKSON2020106198} and perform their own investigations. One of the useful tools for this is the Similarity Search, which finds similar images using distance in the feature space \citep{Coughlin:2019ref}. 

Gravity Spy employs a deep learning model with \cnn layers, a \relu activation
function, and a fully connected final layer to train the algorithm over the images. For glitches that do not fall into any of the established categories, a ``None of the Above'' class is supplied. This makes it possible for citizen scientists to identify new glitch classes \citep{Soni_2021}. The machine learning algorithms are retrained to include these newly discovered glitch categories in the data set. As a result, the Gravity Spy product becomes a dynamic set that incorporates variations in the detectors' noise characteristics over time.

\begin{figure}[ht]
 \centering
 \includegraphics[width=\linewidth]{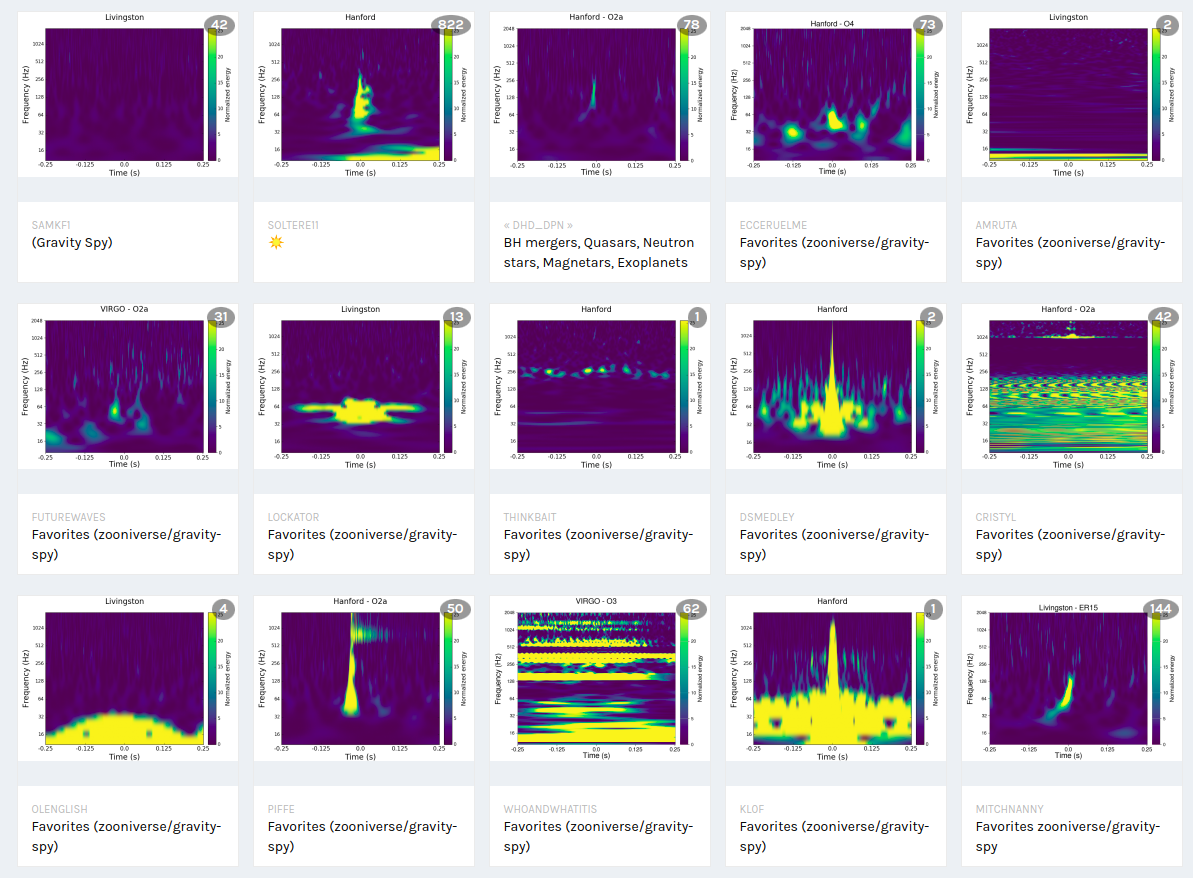}
 \caption{\label{fig:gravity_spy_snapshot} A snapshot from the Gravity Spy website with a few examples of citizen science classifications of glitches. Retrieved February 20, 2024 from \url{https://www.zooniverse.org/projects/zooniverse/gravity-spy/favorites}.
}
\end{figure}

Neural network algorithms are designed to extract features from two-dimensional matrices and use their features for classification purposes. Thus, like GravitySpy, the majority of neural network-based techniques rely on time-frequency transform images for the classification of glitches. 

One of the first attempts at glitch classification using deep learning is described in \citet{Razzano_2018}, where the authors present a classification pipeline utilizing \cnns to categorize glitches based on their time-frequency evolution, represented as images. This earlier application demonstrated that deep learning can accurately identify glitches, paving the way for real-time detector characterization and advanced algorithm implementation.

A recent application of deep learning to glitch classification was reported in \citet{Fernandes_2023}. Here, the \resnet architecture is used to classify glitches in both supervised and unsupervised modes. The supervised algorithm is directly trained on Gravity Spy public images with the Fastai library \citep{fastai}. In unsupervised mode, the algorithm is pre-trained with automatically generated labels before being fine-tuned with Gravity Spy labels.

In recent years, there has been a great deal of interest in developing complementary strategies to detection pipelines for discriminating \gw signals from noise artifacts \citep{Schafer:2022dxv}.
This issue has been addressed using a variety of \ML architectures, from \gp regression \citep{Lopez:2021ikt} and \RF \citep{Shah:2023twc} to \geneticp \citep{Cavaglia:2020qzp} and deep learning \citep{Andres-Carcasona:2022prl,Chatterjee:2022ggk,Jadhav:2023mqx,Trovato:2023bby,Baltus:2021nme}, and targeting both modeled and unmodeled \gw sources as well as signals of varying durations \citep{Boudart:2022apz,Boudart:2022xib,Skliris:2020qax}.
See Sect.~\ref{sec:searches} for more discussion of search pipelines incorporating such techniques to increase robustness against noise artifacts.

Two promising recent examples of deep learning applications are SiGMa-Net \citep{PhysRevD.107.024030} and GSpyNetTree \citep{alvarezlopez2023gspynettree,alvarezthesis}. Both these algorithms use \cnn architectures and input from Gravity Spy to distinguish \gw signals from noise artifacts.

SiGMa-Net uses sine-Gaussian projections as the deep learning neural network's input to distinguish binary black hole signals from short-lived ``blip'' glitches for potential applications in low latency.

GSpyNetTree uses spectrograms with varying durations; the same input feature set as Gravity Spy. Based on a \gw candidate's estimated merger mass, these spectrograms are then sent to one of a set of classifiers with multi-label architecture, tuned to identify \gws and glitches of different characteristic durations \citep{Jarov:2023qpt}. The classifier returns a series of scores indicating the likelihood that the data contains a \gw and/or noise artifacts. GSpyNetTree is currently used in the validation of \ligo-\virgo\ event candidates.

Beyond glitches and other time-domain non-stationarities, spectral lines are of concern to all \gw searches, but particularly to those for persistent stochastic backgrounds or \cw.
Many line investigation tools have been developed \citep{Covas:2018oik,LIGO:2021ppb,Accadia:2012zz} that can then feed either into instrumental mitigation interventions or the creation of lists of affected frequencies that need to be vetoed in astrophysical searches.
Computing cross-correlations or coherence measures with additional instrumental and environmental monitoring channels is often crucial for safe line identification, as the methods should not falsely dismiss true astrophysical narrow-band signals like \cws.
The NoEMI (Noise Frequency Event Miner) framework \citep{Accadia:2012zz} was developed for Virgo detector characterization.
It automatically monitors Fourier-transformed data for spectral features and creates a database of identified features,
but does not include advanced \ml techniques.
Most tools developed on the LIGO side \citep{Covas:2018oik,LIGO:2021ppb} rely heavily on human interaction and visual inspection.
One exception is the \cnn implemented by \citet{Bayley:2020zfa} for three-way candidate classification as background noise, spectral lines or astrophysical \cws in the \soap pipeline \citep{Bayley:2019bcb}, discussed further in sections \ref{sec:searches-cw} and \ref{sec:interpretation-CWPE}.

\subsection{Noise characterization with auxiliary channels}

All the above implementations discussed in the previous section have as a common denominator the classification of short-lived glitches (either time series or time-frequency transforms) from the detector's main output that may impact burst or \cbc searches.

\begin{figure}
 \centering
 \includegraphics[width=1\linewidth]{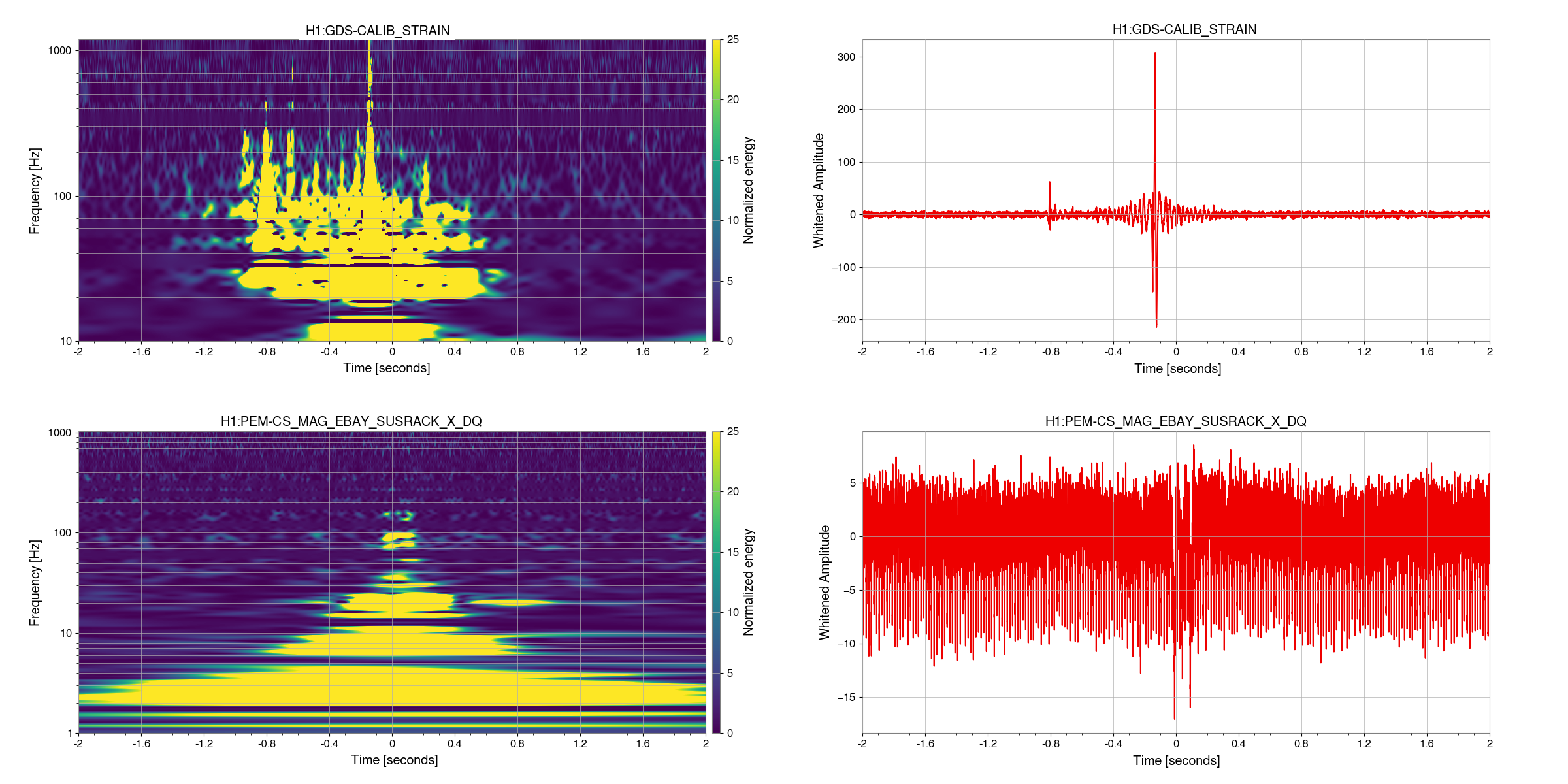}
 \caption{\label{fig:aux_channel_glitch}
  Example of a glitch appearing in the main output of the \ligo-Hanford detector ({\tt H1:GDS\_CALIB\_STRAIN}) as well as in one of the environmental magnetometer auxiliary channel ({\tt PEM-CS\_MAG\_EBAY\_SUSRACK\_X\_DQ}). The left panels show Omega Scan \citep{Chatterji:2004qg} representations of the glitch. The right panels are the corresponding whitened time series. The plots were obtained with the \textit{GWDetChar} Python package \citep{GWDetChar}. \textit{GWDetChar} is distributed under the GNU General Public License, Copyright 2023, Alex Urban, Duncan Macleod. 
 }
\end{figure}

Although the upstream classification of glitches in the detector output can provide some insight into their origins and relationships to the environment and the instrument, this process rarely makes it possible to identify how these glitches are coupled to the different detector components and ultimately come up with solutions to remove them. Utilizing the data supplied by auxiliary channels monitoring the different detector subsystems is the obvious next step.

Different \ml algorithms have been investigated to uncover correlations between the upstream interferometer output and auxiliary channels. Again, investigations have used a variety of algorithms, from tree-based classifiers to neural network architectures, for real-time or offline identification of excess noise. 

iDQ \citep{Essick_2021} is a supervised learning algorithm that can autonomously identify noise artifacts in \gw detectors in real time by leveraging data from thousands of detector auxiliary data streams. During the first four advanced-detector observing runs, iDQ has been running in real-time and has produced probabilistic statements on the presence of excess noise in \ligo\ data. iDQ uses wavelet-based extractors to generate feature tables from auxiliary channels. Different supervised classifier models are then applied to these high-dimensional representations of the detector's state to evaluate the probability of observing a specific instance of excess noise. The classifiers are regularly retrained to detect excess noise without human involvement. Supporting multiple modes of operation and accommodating any supervised learning algorithm that works with tabular data are two of iDQ's key advantages.

A significant amount of work has gone into creating offline, on-demand techniques for the identification and characterization of glitches in real time, in addition to iDQ and other online algorithms. For the sake of brevity, we will just describe a few examples of methods developed over the past decade. 

The first use of an \ml algorithm for offline noise excess identification with auxiliary detector channels dates back to the initial \gw detector era \citep{PhysRevD.88.062003}. \ann, \svm, and \RF classifiers were used to identify and remove noise artifacts during four weeks of \ligo's fourth science run and one week of \ligo's sixth science run, demonstrating the feasibility of using \ml for the detection of instrumental and environmental excess glitches with auxiliary channels. 

Subsequent studies examined ready-to-use algorithms for the identification and classification of excess noise. In this case, having straightforward, adaptable algorithms at hand is essential so that new noise artifacts in the interferometer data can be quickly and easily addressed. One of the desired features of this approach is the ability to train the algorithm with small datasets.

Examples of \ml algorithms to identify the origin of the noise artifacts and infer the relevant mechanical couplings in the detector were presented in \citet{CiCP-25-963} and \citet{PhysRevD.101.102003}. In these methods, event trigger generators operating on auxiliary channels provide input to \ml binary classification algorithms like \RFs, \geneticp, and logistic regression to either rank the channels according to their correlation to the \gw channel \citep{CiCP-25-963} or produce a probability estimate to classify data periods as glitchy or clean \citep{PhysRevD.101.102003}. In the former approach, several instances of classifiers are run with a fixed number of estimators, and the channels with importance below a pre-defined threshold are removed iteratively from the results. In the latter approach, a predictive model is trained by iteratively minimizing the residual error between predicted class probability and ground truth via gradient descent.

Unsupervised \ml methods have recently been used to learn the underlying distribution of glitches. Unsupervised algorithms have the advantage of not making prior assumptions about data distribution. Catalogs of glitches can be constructed using tensor and matrix factorization techniques \citep{Gurav2022IdentifyingWT}. Deep-learning methods, such as \glspl{autoencoder}, can be used to find anomalous time periods that deviate significantly from the general trend as they can discover structures and patterns in unlabeled datasets. Their performance is typically evaluated by comparing their output to that of supervised classifiers, which serve as the benchmark. For example, the unsupervised algorithm in \citet{Laguarta:2023evo} is benchmarked against Gravity Spy high-confidence classifications. In that case, the information from safe auxiliary channel time series is first encoded in their fractal dimension \citep{Cavaglia:2022vlu} to reduce the dimensionality of the data set. Then, similar to neural network algorithms acting on time-frequency transforms, a convolutional \gls{autoencoder} algorithm is applied to a ``time-fractalgram'' to identify the anomalous periods.

As detectors evolve and become more sensitive, the ability to characterize their noise in real time will become increasingly important. In this context, it is expected that machine learning methods, particularly unsupervised methods, will continue to gain traction as tools for the noise characterization of detectors. These methods are highly general and are likely to be used in the characterization of the upcoming generation of \gw detectors.

\subsection{Interferometer control}
\label{sec:control}

One area of experimental \gw physics where machine learning might be useful is detector control. Although the detector is intrinsically nonlinear, most detector control techniques rely on linear controllers. Machine learning algorithms may be able to approximate instrumental nonlinear behavior around the planned mode of operation.

Deep learning approaches have recently been applied to the problem of \ligo\ lock acquisition \citep{ma2023deep} and lock loss prediction \citep{Coughlin_2017,Biswas_2020}. Interferometer lock is usually obtained by controlling the detector's longitudinal translational degrees of freedom \citep{Staley_2014}. This requires knowing the state of the mirrors and then making use of that knowledge to drive their motion to the operational point using an appropriate model. This method has traditionally been handled with case-by-case procedures that are difficult to scale to more complicated systems. Lock acquisition can take a few minutes to tens of minutes due to the variability of initial boundary conditions.

 A \gru was used in \citet{ma2023deep} to provide an accurate non-linear state assessment of \ligo's mirror locations. The approach is precise enough, according to simulations, to allow quick lock acquisition of the interferometer's power-recycled cavity. This approach has two advantages: it can obtain a detector lock without requiring an expert's understanding of the instrument, and it may be scalable to other detector designs. Additional hardware testing is planned to validate the simulated results.

Noise reduction by machine learning non-linear control is another use of machine learning techniques that has been investigated. The complexity of the mathematical equations required in modeling the system typically hampers detector non-linear control. As a result, only a few non-linear systems can be analytically characterized. This could be avoided with neural networks. Noise reduction via detector control is still in its early stages. Proof-of-concept studies are currently being conducted at the \ligo\ laboratories on simple setups such as seismometers used to measure seismic motion.

\subsection{Methods for denoising}
\label{denoising}
The removal of excess noise is the next desirable step in the \gw detection workflow after noise identification and characterization. Machine learning techniques can help with this task by creating algorithms for modeling and removing non-astrophysical noise.

Machine learning algorithms may be especially effective at removing instrumental and technical noise that couples to the detector in nonlinear or nonstationary ways without previous knowledge of the physical mechanisms of the noise. Because the majority of excess noise in the interferometer's output does not always result from linear couplings, standard signal processing techniques such as Wiener filtering may be incapable of removing this type of noise. Because of machine learning's ability to detect nonlinear patterns in data, it may be possible to develop machine learning-based ``transfer functions'' for the nonlinear components of excess noise. The algorithm can then be used to subtract those nonlinear couplings from the output data, thereby reducing the noise floor of the detector.

Machine learning-based denoising methods fall into two categories: methods for removing persistent noise, such as spectral lines like the 60 Hz line, and methods for removing glitches, either to reduce search backgrounds or to improve \gw signal parameter estimation. 

\subsubsection{Algorithms for denoising of persistent noise}

DeepClean \citep{PhysRevResearch.2.033066} and NonSENS are the two primary algorithms that have been developed for line noise subtraction of \gw interferometers. 

DeepClean \citep{PhysRevResearch.2.033066} employs a one-dimensional \cnn algorithm that takes the detector auxiliary channels as input and produces an estimate of the excess noise in the \gw strain data, which can then be filtered out. To deal with non stationary noise couplings, the algorithm is designed to be easily retrained on time scales smaller than the duration of the analyzed data. On a standard GPU, training on 300--1024 seconds of training data takes roughly 2-6 minutes (including data preprocessing). After training the data, the inference process takes a few seconds. This makes the algorithm suitable for use in both offline and real-time subtraction. DeepClean was tested on data collected by the \ligo\ detectors during their second and third observation runs, and it was shown to be capable of removing broadband beam jitter noise as well as the 60 Hz linear coupling and its sidebands \citep{Saleem:2023hcm}. A similar, \cnn-based approach has been used in \citet{Yu:2021swq} to mitigate noise in the aLIGO angular control system, one of the major noise sources limiting the sensitivity of the detector in the sub-30 Hz band.

NonSENS (NON-Stationary Estimation of Noise Subtraction) \citep{PhysRevD.101.042003}, like DeepClean, is a Deep Neural Networks-based algorithm designed to characterize non-stationary noise couplings in auxiliary witness data streams and perform time-domain noise subtraction in the target detector's strain. The algorithm can model noise coupling modulations sensed by slowly varying witness sensors and applies to both linear and stationary couplings. The algorithm determines the best Infinite Impulse
Response (IIR) filters to use for subtraction in the time domain. So far, it has been used in the third observing run to subtract the 60 Hz power line and sidebands, \ligo's Alignment Sensing and Control (ASC) dither lines, and ASC and Length Sensing and Control (LSC) control noise. After training a noise subtraction model, current NonSENS implementations can either generate subtracted frame file data with a script that runs on low-latency data or use \ligo's frontend model to subtract noise in real time. 
During the first six months of Advanced LIGO's O3 run, nonstationary subtraction of the 60 Hz line and its sidebands effectively enhanced astrophysical sensitivity, extending the detector's range for high-mass binary black hole systems by 25 Mpc and increasing the observable volume by 11\%.

\begin{figure}
 \centering
 \includegraphics[width=0.9\linewidth]{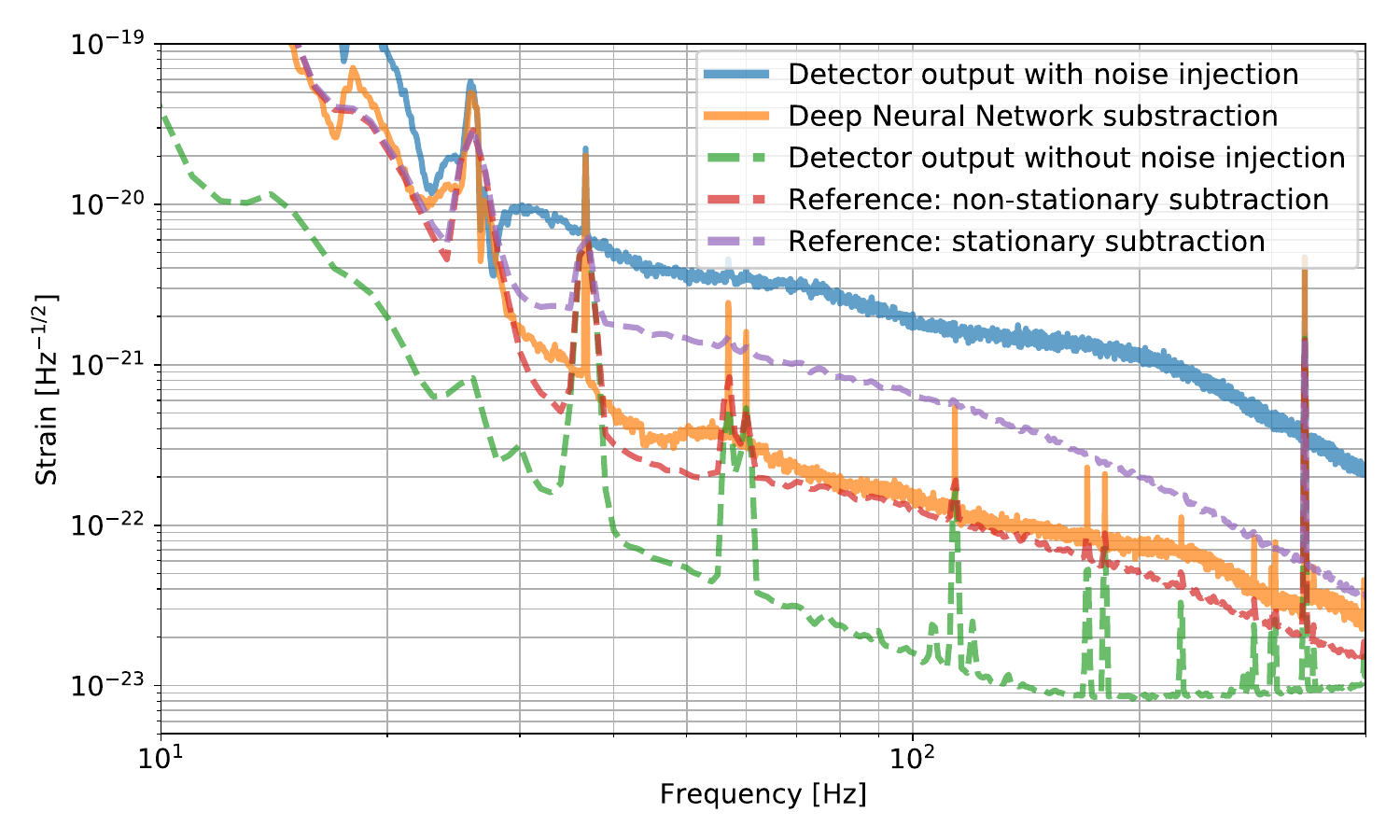}
 \caption{\label{fig:nonsens} Noise subtraction of persistent noise with the NonSENS algorithm \citep{PhysRevD.101.042003}. The curves show a comparison between the Deep Neural Network-based subtraction and the stationary noise subtraction. Image reproduced with permission from \citet{PhysRevD.101.042003}, copyright by APS 
 }
\end{figure}

\subsubsection{Algorithms for signal denoising }

The idea behind machine learning methods for glitch subtraction is to train the algorithm on a set of glitches in the interferometer's main output or auxiliary channels to reconstruct the excess noise in strain data.

 \DL \citep{PhysRevD.102.023011} and \anns \citep{Mogushi_2021} were two of the earliest machine learning techniques that were suggested for removing glitches from \gw signals. A similar approach based on a recurrent neural net denoising auto-encoder was reported in \citet{WEI2020135081}

To represent the input data, the \DL approach \citep{DL_book} uses a linear combination of basic elements known as atoms. Data are mathematically represented as a linear combination of a small set of basis functions (atoms) in a higher-dimensional space (the dictionary). The training process is optimized to identify the most effective dictionary that reduces reconstruction error while preserving sparsity. \DL has proven effective in a range of applications including image and signal processing, as well as data reduction.

In \citet{PhysRevD.102.023011}, \DL is used to identify and subtract ``Blip'' glitches, which are one of the most prevalent short-lived glitches detected in \ligo\ detector data and can interfere with transient \gw searches. In most of the detector frequency bands, the approach can remove the noise contribution of blip glitches.

\begin{figure}
 \centering
 \includegraphics[width=0.5\linewidth]{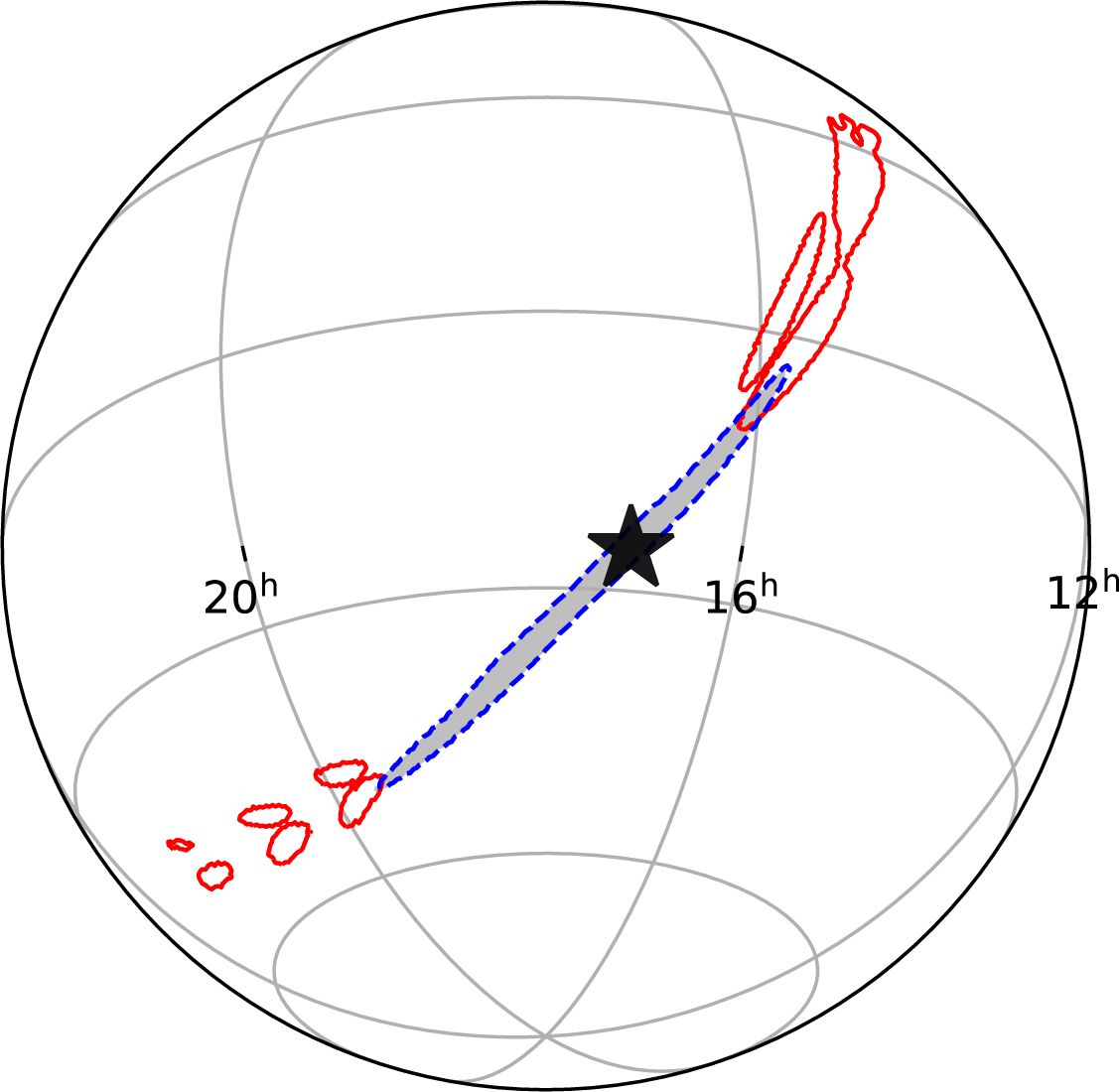}
 \caption{\label{fig:nnetfix} The 90\% sky localization error region of a simulated \bbh signal in a two-detector Advanced \ligo\ network (gray region). The black star indicates the true location of the injected signal. The red empty contours show the localization area when a 130 ms gate is applied 30 ms before the geocentric merger time in one of the two detectors. The dashed-blue contour indicates the signal localization area after the missing portion of the signal has been reconstructed with NNETFIX. Image reproduced with permission from \citet{Mogushi_2021}, copyright by the author(s).}
\end{figure}

NNETFIX \citep{Mogushi_2021} is an \ann technique used to reconstruct data with a binary black hole merging signal overlapping with a glitch as if the latter were not present. The algorithm is trained to predict the fraction of the signal that must be gated owing to the existence of excess noise. The neural network's output is a full-time series of the signal, which may subsequently be utilized as input to other algorithms to generate sky localization maps or parameter estimates. Figure \ref{fig:nnetfix} illustrates a comparison of a sky localization error region obtained using NNETFIX reconstructed data to one obtained with (incomplete) gated data.

\citet{Bacon_2023} recently investigated the use of a convolutional neural network in an encoder-decoder architecture to denoise merging binary black hole signals. The method is trained and evaluated on a population of several thousand synthetic astrophysical signals injected into interferometric noise, as well as real events from the first two \ligo-\virgo\ observation runs. The denoised output, like NNETFIX, might be used as an input for parameter estimate pipelines.

\section{Modeling and data generation}
\label{sec:datagen}

Many \gw data analysis methods, especially for \cbcs and \cws, rely on matched filtering of the detector data against predicted \gls{waveform} templates.
Modeling of astrophysical sources is an important contribution to \gw astrophysics, both to provide these waveform models and more generally to understand the physical behavior of these extreme physical systems, which can be useful even in guiding the design of template-free searches and in the astrophysical interpretation of any analysis results.

Due to the often complex physics of \gw sources, conventional modeling approaches tend to combine a large array of analytical approximations and numerical methods.
\ml methods are making inroads in these research areas, but with a few notable exceptions, they still have many challenges to overcome to become competitive.
We will split the discussion of these efforts into first source modeling on its own,
and then specific \gw signal models.

Another nontrivial problem is the realistic simulation of detector noise, which to first approximation is colored Gaussian noise, but in reality has many non-stationary aspects and contains various types of artifacts. See, again, \citet{LIGOScientific:2019hgc} for a general summary of \gw detector noise characteristics. 
Here, traditional approaches have been very limited, and \ml solutions are already considered the state of the art.
Generative noise modeling is also closely related to the noise mitigation and data characterization tasks discussed in Sect.~\ref{sec:noisemitigation},
as the ability to simulate large sets of realistic noise samples
is often a requirement for training those tools.

\subsection{Modeling of astrophysical sources}
\label{sec:modeling-astro}

\gw sources span a broad range of astrophysical systems, even if just focusing on the sensitive band of current detectors: \cbcs, individual \nss, \ccsne, exotic sources such as cosmic strings or dark matter condensates, and early-universe physics.

We will first focus on works regarding the modeling of the only type of \gw signals detected so far by the \lvk\ network, \cbcs.\footnote{
Evidence has also been found for a stochastic \gw background -- see \citet{InternationalPulsarTimingArray:2023mzf} and references therein -- but through the very different approach of pulsar timing arrays, which we do not cover in detail in this review.}
This requires solving the orbital dynamics of close binaries in the strong-field dynamic regime of general relativity, including gravitational back-reaction. The nominally ``simplest'', yet already enormously challenging case are \bbhs, which are pure gravity systems. On the other hand, \bnss and \nsbhs additionally require the modeling of matter and \elmag effects. Analytic approximations can be obtained for the inspiral phase from post-Newtonian theory, as well as the self-force and effective one-body approaches. see \citet{Blanchet:2013haa,Barack:2018yvs,Damour:2016bks} for reviews of these three frameworks, respectively. On the other hand,
the more complicated merger and postmerger phases require numerical simulations: pure \nr for \bbhs, and general relativistic magnetohydrodynamics (GRMHD) when \nss are involved. See e.g. \citet{Shibata:2016nr} for a textbook treatment and \citet{Duez:2018jaf} for a review of this field. 
Adequately covering the full parameter space of generic \bbhs (especially when including spin precession and orbital eccentricity) with high-fidelity \nr simulations remains a formidable challenge, and even more so for \bnss and \nsbhs.

Due to the physical and numerical complexity of these systems,
on the one hand \ml approaches appear very promising
as they could allow us to work around limitations of the classical approaches,
or at least to reduce the amount of algorithmic and initial data fine-tuning needed.
On the other hand,
decades of progress on the established methods
mean that \ml has a high standard to reach to be a valid alternative,
and the more abstract and intuitive aspects of a trained physicist's domain knowledge
can prove quite difficult to encode in a way
that benefits an algorithmic learning scheme.

\begin{figure}[ht]
 \centering
 \includegraphics[width=\linewidth]{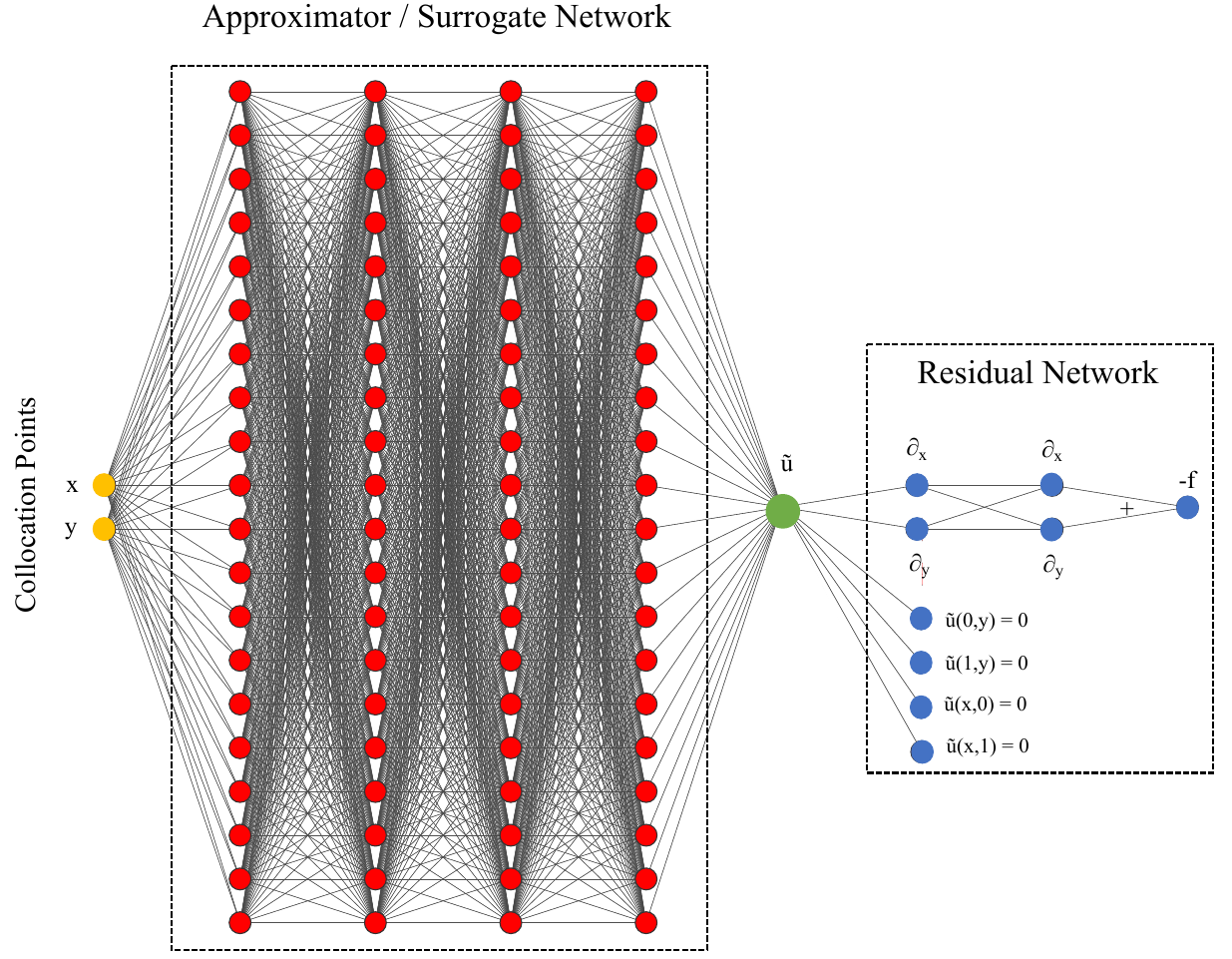}
 \caption{
  Example of the common two-block layout for a PINN.
  The approximator/surrogate network produces approximate solutions for the differential equation under study,
  with the cost function necessary for its training evaluated by the residual network. Image reproduced with permission from \citet{2021arXiv210309655M}, copyright by the author(s).}
  \label{fig:pinn-markidis}
\end{figure}

One class of approaches that have received attention across the broader physics community in recent years
are \glspl{PINN} \citep{2017arXiv171110561R,2021arXiv210309655M}
and the closely related physics-informed neural operators (PINOs) \citep{Rosofsky:2022lgb}.
In the most general sense,
this covers neural networks used to solve any type of differential equations
that describe a physical law.
The major advantage compared to traditional differential equation solvers
lies in using the power of automatic differentiation,
which is available in highly optimized form in all standard \ml packages,
as an alternative to other numerical differentiation schemes.
Training may work in supervised or unsupervised modes.
A common setup is to construct the PINN out of two building blocks,
a ''surrogate network`` that provides solutions to the equation
and a ''residual network`` that evaluates the cost function
associated with deviations from these solutions,
see Fig.~\ref{fig:pinn-markidis}.
For more detailed overviews of the concepts and sub-classes,
see \citet{2017arXiv171110561R,2021arXiv210309655M,Rosofsky:2022lgb}.

The first applications of PINNs with more or less direct relevance to \gw astrophysics include
\citet{Cornell:2022enn,Patel:2024wzo} for quasi-normal modes of non-rotating black holes,
\citet{Luna:2022rql} for the Teukolsky equation
(linear perturbations of the spinning Kerr metric),
\citet{Dieselhorst:2021zet} for relativistic hydrodynamics (without gravity or \elmag fields),
\citet{Rosofsky:2023dtc} for magnetohydrodynamics (without gravity),
\citet{2024MLS&T...5b5014A} for Newtonian self-gravitating hydrodynamic systems,
\citet{Urban:2023cfk,Stefanou:2023jxk} for pulsar magnetospheres
and \citet{Li:2023eys} for obtaining the Schwarzschild metric from the Einstein equations.
The promise of PINNs is that they could eventually solve broader
families of such differential equations for a variety of physical systems
more efficiently, more robustly, or in cases where no traditional solution is known.
But in practice, these examples are mostly still proofs of concept applied to cases for which long-established solution methods already exist.
Applications to more complicated \gw-emitting systems
without known solutions are still to be demonstrated.

Besides PINNs, other \ml methods can be used for emulating the solutions to differential equations. For example, \citet{Reed:2024urq} studied neural networks, \gls{gaussianproc} and a reduced-basis method for the Tolman--Oppenheimer--Volkoff equation describing neutron stars.
Yet another different approach was taken in another proof-of-concept study \citep{Keith:2021arn}
demonstrating the concept of ``inverse modeling'',
where real data is used as the main input to find an appropriate physical model.
The generic dynamics of \bbh systems were phrased
as a hyper-model of possible differential equations,
parameterized via feed-forward neural networks,
which are then trained on observed \gw data.

ML can also be used as a tool to improve \nr codes and catalogs.
Anomaly detection in \nr catalogs was considered in \citet{Pereira:2022kqn},
using a U-Net network \citep{2015arXiv150504597R}
to identify types of data quality issues in \nr waveforms.
in \citet{Ferguson:2022qkz},
a deep neural network (15 layers)
was used to optimize the placement of new simulations
based on mismatches between the entries of an existing catalog.
The same goal was pursued in \citet{Doctor:2017csx}
based on where a new simulation would most reduce
the uncertainties of a \gp surrogate waveform model
(see next subsection).
Another fruitful approach could be to use \ml tools
to improve internal aspects of \nr simulation codes,
such as initial data generation or grid setups.
But to the knowledge of the authors,
no such works had been published as of the writing of this article.

One of the key aspects of \ns physics,
the equation of state of dense nuclear matter \citep{Lattimer:2021emm}
(of relevance both to \cbc and \cw studies),
can -- besides many more classical interpolation approaches --
also be modeled
or inferred from observational data with \ml methods.
Many such works have been published following an initial study by \citet{Fujimoto:2017cdo} which used a 5-layer deep feed-forward network trained to infer the functional dependency from mock sets of mass--radius measurement.
These works have used a wide variety of algorithms,
with varying emphasis on the modeling or inference aspects.
For studies focusing on combining \gw and \elmag observations in a multi-messenger approach,
see also Sect.~\ref{sec:multimessenger}.
To highlight here just two more examples of how the equation of state can be modelled in a flexible way by \ml approaches,
the influential work by \citet{Essick:2019ldf} used a non-parametric Gaussian process model,
while \citet{Han:2022sxt} trained a predictive variational \gls{autoencoder} as a general representation of the equation of state.
This is a very active and rapidly evolving field,
and one where \ml-based or at least \ml-adjacent methods
(not all practitioners in the field agree whether Gaussian processes should be considered \ml) are already playing a highly significant role.
The flexibility of \ml approaches in this context also helps to extend the formalism to neutron stars in theories beyond general relativity,
as for example done with neural network surrogates in \citet{Liodis:2023adg,Biswas:2023ceq,Stergioulas:2024jgk}
which can help to efficiently analyze the higher-dimensional parameter space of neutron star properties and alternative gravity theories.

For \gw sources beyond \cbcs,
the physics can be even more complicated and poorly understood,
e.g. for \ccsne.
Various \ml approaches have been explored for supernova modeling in general,
and especially for the turbulent aspects of the involved stellar hydrodynamics
\citep{2021JOSS....6.3199K,Karpov:2022tro},
as well as for the prediction of \elmag lightcurves \citep{2023A&A...677A..16D}.

Furthermore,
ML methods can be helpful for simulating
complex astrophysical source populations, and there is fruitful interaction between pure simulation methods and \sbi approaches.
Not many dedicated \ml-based works on population simulations alone exist yet in \gw astronomy,
but similar approaches feed directly into the inference of source populations from \gw observations,
for which see Sect.~\ref{sec:interpretation-pop-cosmo}.

\subsection{Modeling of GW signals}

The \gw signal models needed for \mf methods, and many other data analysis tasks, have to be based on an understanding of astrophysical sources.
They can either be based purely on numerical simulations,
purely on analytical work,
or combine both kinds of inputs.
Simple expressions are typically used for \cws -- Taylor expansions, see the general \cw review of \citet{Riles:2022wwz} -- and \cw-like long-duration transients -- e.g power laws \citep{T1700408} or piece-wise expressions \citep{Grace:2023kqq} for long-lived \bns remnants.
For \cbcs the signals are more complicated. A basic introduction and review of waveform modeling for \bbhs is given in \citet{Schmidt:2020ekt} and a broader
 review of its history, current state and open challenges is given in the \lisa\ collaboration's white paper \citep{LISAConsortiumWaveformWorkingGroup:2023arg}.
While the latter focuses on the signals detectable at the mHz frequencies where LISA will be sensitive, much lower than for ground-based interferometers, it also covers many aspects relevant to the \lvk\ science case.
However, some aspects of waveform modeling are unique to the \lvk\ band, for example the matter effects in the late inspiral, merger and post-merger phases of binaries involving neutron stars. For the \bns and \nsbh cases, respectively, such effects and associated modeling approaches are reviewed in \citet{Dietrich:2020eud} and \citet{Kyutoku:2021icp}.

For \cbcs specifically,
the traditional modeling approaches for full inspiral-merger-ringdown waveforms
have long combined analytical approximations
with \nr simulation results.
A key tool in constructing these models are fitting (regression) methods.
such as the one developed in \citet{Jimenez-Forteza:2016oae,Keitel:2016krm} for the fourth generation of the ``IMRPhenom'' model family
\citep[][and other works building on it]{Pratten:2020fqn},
or as described for the latest ``SEOBNR'' model \citep{Ramos-Buades:2023ehm}
in \citet{Pompili:2023tna}
and for the ``TEOBResumS'' model in \citet{Nagar:2018zoe}.
Regression methods in this context have also been compared quantitatively in \citet{Setyawati:2019xzw},
including both traditional interpolation and fitting schemes
(e.g. least squares)
as well as two \ml frameworks: \gps and neural networks.
Especially for EOBNR models,
since these are typically computationally more expensive than IMRPhenom models,
reduced-order-modeling based on \svd
has often been used as an intermediate step
between the initial calibrated model
and practical applications in \gw searches and \pe,
starting from \citet{Purrer:2014fza,Purrer:2015tud}.

\begin{figure}[ht]
    \centering
    \includegraphics[width=\linewidth]{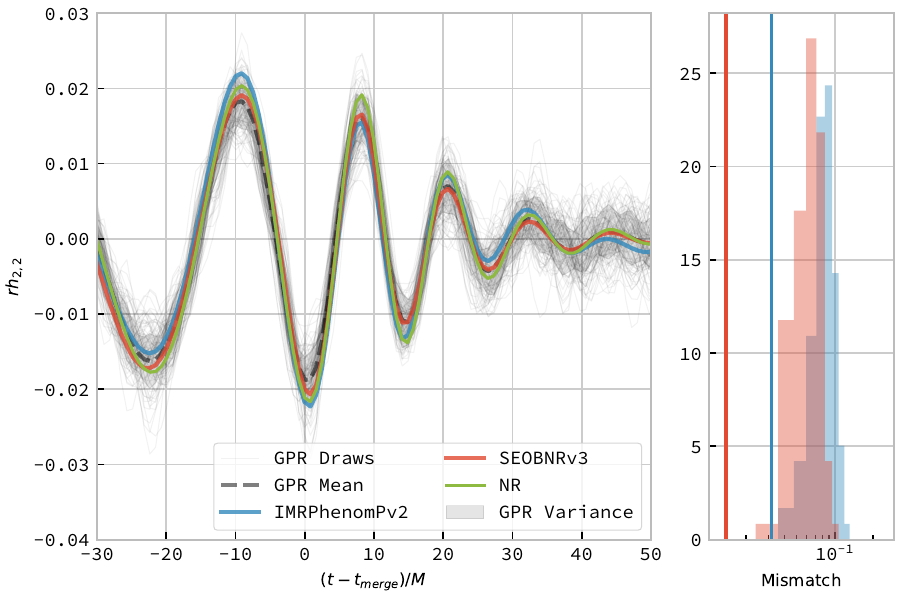}
    \caption{
     Example of a \gp waveform model, for a non-spinning binary,
     illustrating the variance over 100 draws from the GP
     and a comparison with full \nr and two other waveform models.
     Reproduced with permission from \citet{Williams:2019vub}, copyright by APS.}
     \label{fig:nrsur-gp-williams}
\end{figure}

These approaches have been joined by a more strongly data-driven approach to waveform modeling,
the NRSurrogate model family
(see, e.g., \citealt{Field:2013cfa} for the initial work and \citealt{Varma:2019csw} for the ``NRSur7dq4'' model most often used in data analysis up to now).
These use a variety of techniques
to construct a reduced basis that is sufficient to
interpolate an input set of waveforms from other methods
at discrete parameter-space points
for evaluation at any other point within the fitted domain.
Initially, mainly traditional parametric fits and spline interpolation \citep[e.g. in][]{Blackman:2017dfb,Blackman:2017pcm} were used.
However, \citet{Varma:2018aht} introduced a surrogate for the \bbh merger remnant properties based on \gp regression \citep{Rasmussen:2006gpr},
which is considered by many (though not universally) as a \ml technique.
This remnant surrogate then served as an additional ingredient to the full precessing waveform surrogate of \citet{Varma:2019csw}.
Other works that
have used a pure \gp approach for fitting and interpolation
include
\citet{Doctor:2017csx} who used spin-aligned IMRPhenomD waveforms \citep{Husa:2015iqa,Khan:2015jqa} as the input,
and \citet{Williams:2019vub,Khan:2024whs} who both applied a similar approach to constructing fully precessing \nr surrogates.
An example from \citet{Williams:2019vub} is shown in figure~\ref{fig:nrsur-gp-williams},
illustrating the probabilistic nature of a Gaussian process model
but also the level of match achieved with \nr and other models.
As an added bonus,
Gaussian process regression also provides uncertainty estimates on the waveforms,
which can then be marginalized over in \pe
\citep{Moore:2014pda,Moore:2015sza,Liu:2023oxw}
for more robust results.

Initial studies have also considered neural networks to further accelerate surrogate models.
\citet{Fragkouli:2022lpt} have used a 4-layer neural network for finding the coefficients in the parameterized interpolation scheme from \citet{Field:2013cfa}.
On the other hand, \citet{Nousi:2021arn} have again followed the basic surrogate approach from \citet{Field:2013cfa} but used \glspl{autoencoder}. They noticed a ``spiral'' pattern in the dependence of fitting coefficients on the mass ratio, which was exploited via the autoencoder's differentiable transformation from input to output parameters for a faster and more precise regression of the fitting coefficients.
See also \citet{Stergioulas:2024jgk} for a summary of these developments.
However, neither of these neural-accelerated surrogates has found practical adoption yet, and it remains to be seen whether these or similar techniques will become an important ingredient of future waveform model generations.

Besides full waveform models, surrogates are also available
to predict the remnant properties (final mass and spin, kick velocity)
of \bbh mergers
\citep{Varma:2018aht,Islam:2023mob}
and their peak luminosity \citep{Taylor:2020bmj,Islam:2023mob}.
These four works all use \gp regression.
An independent approach to \ml modeling of \bbh remnant properties
was introduced by \citet{Haegel:2019uop},
using a four-layer neural network,
with the eventual aim of improving the parameter-space fitting in ``IMRPhenom'' waveform models..

ML waveform models have also been trained on top of other inspiral-merger-ringdown models,
with the main goal of providing for accelerated evaluation.
The GP approach mentioned above for \nr simulation placement \citep{Doctor:2017csx}
used IMRPhenomD waveforms \citep{Husa:2015iqa,Khan:2015jqa}
as a proof of concept.
We provide here a representative, but not necessarily exhaustive, list of other combinations of architectures and input waveforms that have been the subject of at least proof-of-concept studies:
in \citet{Khan:2020fso}, a fully-connected neural network was trained on SEOBNRv4 \citep{Bohe:2016gbl} waveforms, and
\citet{Lee:2021isa} adapted a deep learning model previously developed for natural language processing to emulate the same waveform model.
SEOBNRv4 and TEOBResumS \citep{Nagar:2018zoe} waveforms were used in \citet{Schmidt:2020yuu}
with a ``mixture of experts'' \citep{Jacobs:1991moe} approach,
where the training is on the coefficients of a weighted combination of a number of linear regression functions.
in \citet{Thomas:2022rmc} SEOBNR4PHM waveforms \citep{Ossokine:2020kjp} were used to train fully-connected neural networks.
A \gls{transformer} network \citep{Vaswani:2017lxt} was built in \citet{Khan:2021czv} for the NRHybSur3dq8 surrogate model \citep{Varma:2018mmi}.
In the future, \ml-accelerated models like these could be used for \ml-accelerated Bayesian parameter estimation as discussed in Sect.~\ref{sec:interpretation-transientPE}.
However, for now, most applications rely instead on direct GPU acceleration of the original waveform models, see e.g. \citet{Edwards:2023sak}.

At the interface of waveform modeling and data analysis,
in \citet{Wong:2020wvd} explicit reproduction of the full waveform was skipped,
and instead a fully-connected network was trained to reproduce
optimal \snrs
as given by the IMRPhenomD \citep{Husa:2015iqa,Khan:2015jqa} and SEOBNRv4 models.
This differs from signal detection methods,
as discussed below in Sect.~\ref{sec:searches-cbc},
in that optimal SNRs do not take into account a particular data realization,
but solely depend on the source parameters
and a reference \psd for a detector.

Other areas in which \gls{waveform} modeling faces challenges
in physical and numerical complexity
are for example \gw signals from \bns remnants
\citep{Easter2019,Sarin:2020gxb}
and from ultralight dark matter particles forming clouds
around spinning black holes via the superradiance process \citep{Siemonsen:2022yyf}.
ML methods could conceivably contribute to studying these as well.
For example,
\citet{Whittaker:2022pkd} trained a \cvae on \nr simulations of hyper-massive \ns remnants.

In summary,
it is not always clear where to draw the dividing line between ``traditional'' and \ml methods in waveform modelling.
At least if considering surrogates as an \ml approach, then
waveform modeling has proven to be one of the areas where \ml methods are already very productively contributing to practical progress in \gw astronomy.
A healthy competition between different modeling approaches is driving the field forward towards meeting the challenges imposed by higher-sensitivity and lower-frequency future detectors \citep{LISAConsortiumWaveformWorkingGroup:2023arg} -- see also Sect.~\ref{sec:nextgeneration}.
Both pure \ml approaches and judicious combinations of such novel techniques and the strengths of traditional modelling expertise are likely to play an increasing role in these regimes.

\subsection{Tools for noise generation}
\label{sec:noisegen}

One key importance of having simulated datasets is the ability to understand and mitigate systematic errors or biases in data analysis pipelines. By generating synthetic data that mimics the characteristics of real observations, we can stress-test our detection or parameter estimation pipelines. In particular, transient noise perturbations can make it more difficult to detect signals or obtain good parameter estimation. Therefore, different groups work on creating tools to generate realistic data, which contains also non-stationary behavior and, in particular, loud transient noise signals (glitches).

\begin{figure}[ht]
    \centering
    \includegraphics[width=\linewidth]{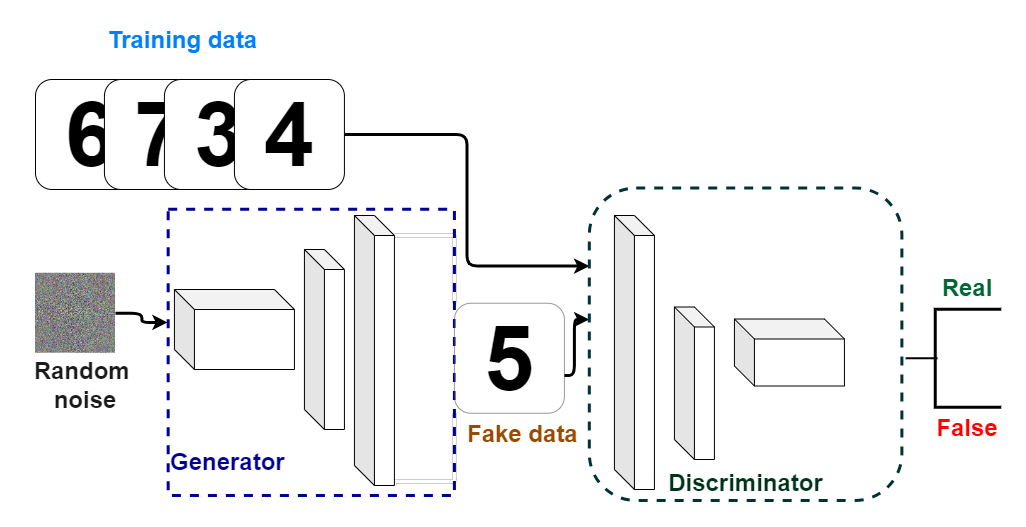}
    \caption{Example of basic GAN architecture.}
    \label{fig:GAN-arch}
\end{figure}

For example, \citet{PhysRevD.106.023027} addresses a significant challenge in the field of \GW searches: the so-called ``blip glitches'' which occur at a rate of approximately 1 per minute and hinder the detection of astrophysical signals. Due to the huge impact of these glitches on data quality, there is a pressing need for improved modeling and incorporation of glitches into large-scale studies. The authors employ a \gan to generate blip glitches in the time domain. They make the trained network available through an accessible open-source software package called ``gengli''\footnote{\url{https://git.ligo.org/melissa.lopez/gengli}} and provide practical examples of its application. 
In Fig.~\ref{fig:GAN-arch}, a typical GAN workflow is illustrated. 

Another work that addresses the issue of glitches in \GW data is \citet{Powell_2023} The authors propose a solution using GANs to simulate glitches. They create synthetic images representing the 22 most common glitch types observed in \ligo, \virgo, and \kagra\ detectors. These images are then converted into time series data, facilitating their integration into simulations and mock data challenges. Through neural network classification, the authors demonstrate that their artificial glitches closely resemble real glitches, achieving an average classification accuracy of 99.0\% across all 22 types. This work suggests that incorporating GAN-generated glitches could enhance the reliability of \gw searches and parameter estimation algorithms. 

In \citet{McGinn_2021}, the authors introduced the use of \cgan for the generation of generalized \gw bursts in the time domain. The CGAN in this work was conditioned on five classes of time-series signals commonly used in \GW burst searches: sine-Gaussian, ringdown, white noise burst, Gaussian pulse, and binary black hole merger. An example application was presented where a \cnn classifier was trained on burst signals generated by the CGAN. The results showed that a \cnn classifier trained solely on the standard five signal classes exhibits lower detection efficiency compared to one trained on a population of generalized burst signals drawn from the combined signal class space.

\section{Strategies for signal detection}
\label{sec:searches}

As discussed in Sect.~\ref{section:GW-data}, searches for \gw signals with the current ground-based \lvk\ detector network
have traditionally been separated into the four categories of
modeled transients (\cbcs),
unmodeled transients (\bursts),
\cws
and stochastic long-duration signals.
The observational science white paper of the \lvk\ collaboration
\citep{LVK:OBSWP}
can be used for an overview of efforts in all four areas.

This section covers the main \ml alternatives developed in the community for searches in these four areas.
However, there are also fairly generic \ml approaches that do not necessarily fall within any one search type.
For example, \citet{Morawski:2021kxv}
use convolutional \glspl{autoencoder} for generic detection of transient anomalies which could be of either noise or astrophysical origin.

\subsection{Methods for modeled transient searches (CBC, real-time and offline)}
\label{sec:searches-cbc}

The application of \ml to the problem of \cbc detection has arguably been the most popular area of \ml research in the \gw community since the recent acceleration of \ml technology through the use of \gpus. In most cases, the problem of signal detection is approached in terms of a classification problem in \ml. In the early examples and through to the most recent state-of-the-art applications, neural networks are trained via supervised learning to identify between the 2 main classes of data -- a piece of data containing detector noise and a piece of data containing detector noise plus a \cbc event. We will now discuss the main developments in the field and highlight the different approaches taken.

The first work on this topic \citep{2018PhLB..778...64G} laid the foundations for the analysis pipelines under development today. It consisted of a relatively basic, but highly appropriate, deep convolutional neural network architecture made of 2 convolutional layers and 2 fully connected output layers. These were interspersed with a series of max pooling layers used to reduce the feature space as the data passed through the network. The output was the result of a softmax layer applied to the pair of final layer neurons and was interpreted as a probability for each of the 2 classes ($0=$noise or $1=$noise plus signal). As is standard in binary classification problems, the loss function to be minimized during training was the binary cross entropy defined as
\begin{align}
    L_{\mathrm{BCE}} = \frac{1}{N}\sum_{i=1}^{N}\left[y_i\log\hat{y}_i + \left(1-y_i\right)\log\hat{y}_i\right]
\end{align}
where $y$ and $\hat{y}$ represent the true (0 or 1) and predicted labels respectively of each of $N$ training data items. The input data came in the form of 1-second long segments of whitened \gw strain time-series sampled at 8192~Hz and each sample labeled as 1 contained a signal whose mass and time of arrival parameters were sampled from a specified prior distribution. The whitening process used the average \psd of real noise measured from \ligo\ data during the O1 observing run. The trained network performance indicated that this emerging technology could already rival the sensitivity of the traditional matched filtering approach and set the bar for competing \ml approaches.

Soon after the initial work on this, \citet{PhysRevLett.120.141103} were able to independently reproduce the results and converged on a similar structure and complexity of network for the classification task. In this work, the emphasis was on providing a comparison with matched filtering that probed lower false alarm rates to indicate the performance of \cnn approaches in more realistic detection scenarios. One limitation of both studies \citep{2018PhLB..778...64G,PhysRevLett.120.141103} was the choice to train separate networks for different signal \snrs. The results presented indicated that \ml could achieve the same sensitivity as matched filtering under a range of conditions. However, as presented in both papers, in practical terms one would need to run each network separately on data since the \snr of an as-yet undetected signal was unknown. The authors did not provide a scheme for combining the outputs of each network into a single detection statistic.

In the years following the first work on this topic, there was much work placed on reproducing and enhancing the power of \ml for \cbc detection. One notable study \citep{Gebhard:2019ldz} made efforts to place limitations on the potential impact of \ml and \cnns specifically to the field of \cbc detection. They conclude that such approaches should not be used to quantify the statistical significance of \gw detections, due to the high false positive rate. However, they do note that networks such as the \cnn that they present, due to their computational efficiency, provide a useful and promising tool to produce real-time triggers for detailed analysis and follow-up searches. 

In \citet{PhysRevD.104.064051} the authors demonstrate possible improvements in search sensitivity by integrating deep learning with a conventional template based search pipeline. They construct a new coincident ranking statistic that incorporates information from an \ml model trained to identify multiple known transient noise features. The effect is to achieve a considerable reduction in background leading to elevated significance of events. Apart from recovering the GWTC-1 events \citep{PhysRevX.9.031040}, the search was also able to confidently detect an additional event, viz., GW151216, identified in the GWTC-1 catalogue as ``low significance'' and previously observed by \citet{PhysRevD.101.083030} and the OGC analyses \citep{Nitz_2019, Nitz_2020}. In \citet{Jadhav:2023mqx}, the problem of model reliability was addressed through stage-wise training of the network for \cbc detection. The authors also tested the fragility of the network and proposed a novel \gan based setup for improved robustness.

In 2021 the first mock data challenge for the detection of \cbc signals using \ml was launched through Kaggle and is discussed in Sect.~\ref{sec:kaggle-bbh}. This challenge was limited to semi-realistic cases and the aim was to cultivate interest and expertise in \gw detection from the \ml community. This was followed by a realistic data challenge \citep{Schafer:2022dxv} targeted at the \gw community itself to bring together the existing efforts within the field and calibrate them against one of the standard \cbc detection tools \citep{2016CQGra..33u5004U,pycbc-github, Aubin_2021}. The challenge consisted of 4 datasets, each representing different levels of difficulty. The first contained non-spinning \bbh signals with mass ranging from $10$--$50M_{\odot}$ in Gaussian noise with known detector \psd and the fourth extended the mass range down to $7\,M_{\odot}$, including spinning systems, and used real O3 \ligo\ noise. The challenge attracted 6 teams using independent analysis tools, 2 of which were existing non-\ml algorithms (PyCBC, \citealt{pycbc-github}, and coherent WaveBurst, \citealt{Klimenko:2008fu}) with the remainder being newly developed \ml applications. The primary metric used for comparison was the sensitive distance -- a chirp-mass weighted approximation to the range in the universe out to which each algorithm can detect signals.
This quantity is itself represented as a function of the false alarm rate and showed that when applied to the most challenging dataset (dataset 4) one of the existing non-\ml tools, PyCBC, achieved the highest sensitive distance at all false alarm rates (limited by the background to be $\ge 1$/month and exceeding the equivalent sensitive distance from the best \ml approach by a factor of $\approx 1.4$). However, for challenges 1--3 which did not include real detector noise, two \ml analyses \citep{2022PhRvD.105d3002S,2023PhRvD.108b4022N} using deep residual networks achieved almost equivalent results in terms of sensitive distance, e.g., obtaining sensitive distances of up to 95\% of those achieved by PyCBC at false alarm rates of 1 per month. In both cases, however, the duration of signals considered as input data for training was 1~sec, far shorter than the maximum length of signal contained within the data challenge ($\sim 12$~sec). We note that following the publication of the initial challenge results, two of the analyses used in the challenge have since been updated \citep{2024PhRvD.110b4024Z,2023PhRvD.108b4022N}. For the Virgo-AUTh approach the sensitive distances at false alarm rates $\ge 1\,\mathrm{month}^{-1}$ all exceed those of the traditional PyCBC implementation from the challenge. More recently studies into neural network approaches to the analysis of periods when only single detectors are operational \citep{Trovato:2023bby} have assessed the applicability of different network architectures. Additionally, two deep neural network analysis pipelines were developed \citep{2024arXiv240318661M,2024arXiv240707820K} that successfully detected gravitational wave signals in real, publicly available data, demonstrating their potential as pipeline detection tools.

The most compelling motivation for the use of \ml in \gw searches for \cbc events is the possibility to front-load the computational expense in training such that at run-time the search can be performed in real-time with relatively low computational cost. As an example we can take the mock data challenge of \citet{Schafer:2022dxv} where pre-trained \ml tools can run at $\mathcal{O}(1000)$ times faster than real time assuming 16 CPU cores. This can be compared to the run times of PyCBC within that challenge that were $\mathcal{O}(10)$ times faster than real time. The potential to detect signals with minimal computational cost and potentially before the binary stars merge (i.e., using only the information from the inspiral part of the signal) would be a powerful tool for multi-messenger astronomy. This need is most relevant for \cbc systems that contain one or more neutron stars since they, unlike \bbh systems, are expected to produce an electromagnetic counterpart signal that could be observed by ground or space-based telescopes \citep{2019NatRP...1..585M}. The challenge in such cases is the length of the expected signal in the \gw detector band, especially for \bns systems. The first \bns detection \citep{LIGOScientific:2017vwq} spent $\mathcal{O}(100)$~sec \citep{2019PhRvX...9a1001A} in band and as the low frequency sensitivity of ground based detectors improves, the duration will grow, e.g, for a detector with sensitivity at 2~Hz (e.g., the Einstein Telescope \citep{ET_design_study_2019}) a \bns signal could last for $\sim 1$ day. Recent efforts have attempted to address this issue and have made improvements in the signal duration considered within \ml detection algorithms. Work has extended the initial consideration of 1~sec long signal durations to 2~sec \citep{Fan:2018vgw}, 4~sec \citep{Lin:2019egc}, and 10~sec \citep{Krastev:2019koe,Krastev:2020skk}. It is also worth mentioning work specifically on early warning detection of \cbc events, where detection can be claimed before the merger \citep{Sachdev_2020, Magee_2021, Nitz_2020EW, Kovalam_2022}, with a neural network approach \citep{Yu:2021vvm}. In \citet{2021arXiv210513664B,2024arXiv240204589A} it is shown that both \cnns and \lstms can be very effective in the analysis of long time-series and could return confident detection statements after only analyzing a fraction of the \gls{waveform}. 

Often, \ml algorithms and neural networks for \cbc detection are treated as black boxes that are purely judged on how they compete with each other, and with existing non-\ml methods, in terms of high-level summary statistics such as detection performance at fixed false-alarm rate.
But understanding their detailed responses, e.g. in terms of the activation patterns of subnetworks or even individual neurons, can also be very enlightening to design better solutions -- following the general approach of explainable or interpretable \ml, for which see \citet{2019PNAS..11622071M,2020Entrp..23...18L} and also Sect.~\ref{sec:explainableML}.
For example, \citet{Gebhard:2019ldz} used activation maximization and feature visualization techniques \citep{zeiler2014visualizing,olah2017feature} to study the actual response patterns of the network they were using.
They also tested its behaviour to so-called ``adversarial attacks'' \citep{2013arXiv1312.6199S}, where the network is exposed to test cases deliberately designed to cause otherwise rare failure modes and elucidate the problematic associations it has learned.
The work of \citet{Safarzadeh:2022pij} focused on similar approaches to study the detailed responses of the two-branched architecture introduced in \citet{Huerta:2020xyq} for \bbh signals, by visualizing the layer-by-layer activation response of the networks and detailed neuron-level sensitivity maps.

A special case of \cbc searches is those for gravitationally lensed signals \citep{2023Univ....9..200G} where, as for \elmag radiation, \gws can also be lensed by massive celestial objects. Strong lensing magnifies signals, making the sources of binary merger signals appear closer and more massive, and can create multiple instances of the same event, separated by minutes to years, but appearing to come from similar sky locations. Microlensing introduces small time delays, causing overlapping waveforms and produces ``beating'' patterns in the signals. Besides many matched filtering methods developed so far and applied e.g. in two LVC/\lvk\ searches \citep{LIGOScientific:2021izm,LIGOScientific:2023bwz}, there are also some based on \ml methods.
Two works have proposed neural network spectrogram classifiers to identify microlensing beating patterns in individual \gw events \citep{Singh:2018csp,Kim:2020xkm}. For strongly lensed image pairs, another classifier \citep{Goyal:2021hxv} combines information from spectrograms and sky maps and was already used in \citet{LIGOScientific:2023bwz} as a complementary first-stage classifier for potentially lensed event pairs, together with the traditional KDE-based Bayesian posterior overlap method \citep{Haris:2018vmn}. Here the superimposed spectrograms of the two events are first passed through a separate \cnn for each detector (starting from pretrained DenseNet201 \citep{Huang:2016den} networks), and the three outputs then passed to the XGBoost algorithm \citep{Chen:2016btl}; while for the skymaps three feature statistics are defined which summarize the similarity and differences between the two maps and then XGBoost is applied to these. The final ranking statistic is the product of the two XGBoost outputs.
An alternative implementation \citep{Magare:2024wje} also combines two classifiers, one working on Q-transforms (time-frequency representations of detector data) and sine-Gaussian projections (which transform to a space characterized by the central frequency and quality factor of sine-Gaussian model functions), whose output in terms of probability for lensing are then simply multiplied with each other for the final ranking.
In all these cases, an input catalog of \gw events is used; direct searches for additional candidate lensed events from the full strain data sets have so far not been proposed with \ml methods.

XGBoost is also used in conjunction with a version of cWB that focuses on the detection of \bbh events \citep{PhysRevD.104.023014, PhysRevD.105.083018}. The results, reported in \citet{PhysRevD.105.083018}. show an improved sensitivity including the recovery of \bbh events previously missed by the standard cWB search. 

\subsection{Methods for GW searches associated with core-collapse supernovae}

\ccsne are extremely complex phenomena \citep{2021Natur.589...29B,2024Univ...10..148B} and modelling them is challenging, though great advances in numerical relativity models have managed to incorporate a wide range of the physics required for predicting the result \gw signal morphology. 
A recent review article by \citet{2024arXiv240111635M} provides a good overview of simulating and detecting \gw signals associated with core-collapse supernovae. 
\gw emissions from \ccsne are usually considered under the category of Burst \gw signals since \gw signal waveforms from \ccsne simulations are not yet suitable for generating template banks used by the matched filtering for \cbc sources. Furthermore, the stochastic nature of the processes involved in a \ccsn requires search pipelines that make minimal assumptions on the \gw signal waveform.

\citet{10.1088/2632-2153/ab7d31} explore the use of 1D and 2D \cnns and of \lstm networks for multilabel classification of \ccsne signals obtained from 3D simulations. The classification procedure is carried out after preprocessing and trigger generation by a wavelet based algorithm, the Wavelet Detection Filter \citep{Cuoco2018}, using real data from the second observing run. 

Similarly, \citet{2023PhRvD.108j3036D} simulated core-collapse \gw signals were processed by the coherent WaveBurst algorithm to produce a list of events with excess coherent energy in data from multiple detectors before the attributes of these events are then analyzed by a fully connected neural network. In this work, the training data used for the training network was augmented by a novel method of generation of expected long SASI \gw signal so as to allow the neural network to achieve competitive sensitivities.
\citet{2021PhRvD.103f3011L} performed a variation of this work which used a mini-inception \resnet on the time-frequency images of the data.

\citet{2022PhRvD.105h4054A} used linear discriminant analysis and support vector machines and characterized the effectiveness of these approaches as a supervised follow-up approach for coherent WaveBurst events originating from \ccsn signals. 
\citet{2020PhRvD.102d3022C} used a convolution neural network on time series data and demonstrated the ability of convolutional neural network to distinguish between \gw signals corresponding to different explosion mechanisms (rapidly-rotating vs neutrino-driven) and also characterized the sensitivity of their approach.
\citet{Cavaglia:2020qzp} proposed a single-detector \ccsn \gw signal search based on a combination of coherent WaveBurst and \geneticp.
\citet{2017PhRvD..96j4033M} introduced the harmonic regeneration noise reduction approach for reconstructing (denoising) supernova waveform signals and later proposed a convolutional neural network to improve the noise rejection ability of their proposed analysis \citep{2021PhRvD.103j3008M}. 

There have also been some efforts comparing and interpreting machine learning approaches for supernova. \citet{2023arXiv231118221P} compared different methods of classifying the \ccsn explosion mechanisms. Among the classification methods compared was a method based on dictionary learning and another method involving a convolutional neural network. All methodologies were able to correctly classify the corresponding explosion mechanism for the majority of the simulated \gw signals. The convolutional neural network approach was better at correctly classifying \gw signals from rapidly rotating stars while dictionary learning was better at classifying gravitational wave signals from non-exploding \ccsn simulations. Dictionary learning for \gws, in particular \ccsne, was first proposed by \citet{2016PhRvD..94l4040T} and later used to classify \ccsn signals \citep{10.1093/mnras/stac698}. Additionally, \citet{Sasaoka:2023kte} used the Class Activation Mapping approach to investigate and interpret how a \cnn classifies \gw signals from \ccsn.

\subsection{Methods for unmodeled transient searches (burst, real-time and offline)}
\label{sec:searches-burst}

Burst signals, by definition, do not have a well-modeled \gls{waveform}. Traditional burst search techniques rely on the principle that \gw signals are correlated between multiple, widely-spaced detectors while noise originating from the detector or the local environment is uncorrelated. The burst parameter space is defined by a range of proxy waveforms (eg. sine gaussians, ringdowns, white-noise bursts) which try to capture the main features of the wide parameter space. An overview of the burst search techniques and these proxy waveforms can be found in \citet{2012PhRvD..85l2007A} 

The lack of a well-defined signal waveform poses a challenge for many machine learning approaches which tend to be trained to detect specific signal morphologies that are simulated in the training data set. Machine learning use in Burst searches fall into two broad categories. The first category involves the direct application of machine learning techniques to the calibrated strain data, typically time series. These approaches are trained on one or more sets of simulated data with proxy waveforms injected (e.g. sine-gaussian signals). This class of machine learning approaches take additional steps to ensure generalisation of the approach and prevent overfitting on the training data set. 

The second category of burst machine learning techniques involve existing burst detection methodologies (e.g. coherent WaveBurst) being enhanced by machine learning techniques to improve the search sensitivity. These approaches ingest various event attributes with the goal of finding a mapping of these attribute values that best differentiates background noise events and simulated signal events. 

\subsubsection{Direct application of machine learning}

\citet{Li:2017chi} used a wavelet basis to decompose simulated \gw strain data as input to a classification CNN to detect simulated transient signals. The network was trained and characterized using only a decaying sinusoidal signal. Nonetheless, this is one of the first examples of CNN applications for Burst \gw searches.

MLy, developed by \citet{Skliris:2020qax}, is an \ml pipeline to search for Burst \gw signals. MLy consists of two CNNs, each trained to identify critical features of Burst \gw signals. One CNN is trained to detect the presence of transient signals in data from multiple \gw detectors. A second CNN uses both the whitened detector time series data and the corresponding Pearson correlation between data from pairs of detectors to distinguish between correlated \gw signals and uncorrelated spurious noise. A novel aspect of this approach is that the hyperparameters for this search pipeline were optimized using a \genetica. This approach has been shown to be competitive for online searches with false alarm rates of 1 per year.

The Gravitational Wave Anomalous Knowledge (GWAK) is a semi-supervised anomaly detection approach which uses deep recurrent \glspl{autoencoder} to encode the different signals and glitches into a latent space which captures the physical signatures of the different signal classes \citep{2023arXiv230911537R}. The encoding is informed by priors based on \gw signal features to allow for robust signal recovery of unmodeled transients. GWAK has been shown to have comparable burst signal sensitivity to MLy and can identify \cbcs.

Additionally, ALBUS \citep{Boudart:2022xib,Boudart:2022apz} is an approach which uses fourier time-frequency maps (spectrograms) as inputs to a \cnn to detect long-transient \gw signals which can last for many minutes.

\citet{10.1093/mnras/staa3550} have presented a search for unmodeled \gw signals using semi-supervised machine learning, processing first a set of labeled spectrograms and then searching for anomalies in the remaining dataset.

\subsubsection{Machine learning enhanced burst searches}

The Coherent WaveBurst (cWB) algorithm \citep{Klimenko:2008fu,Drago:2020kic} looks for signals using excess coherent energy between detectors in the time-frequency wavelet domain.
This algorithm outputs a set of events which correspond to time-frequency locations of excess coherent energy.
These triggers are characterized by a list of event attributes such as the event time, central frequency and strength of coherent energy.
Until the 3rd observing run (O3), the standard approach for optimising cWB's sensitivity for transient burst searches was to manually tune threshold for a set of trigger attributes
This approach relied on experience and intuition built up over many years of performing burst searches on data from \gw observatories \citep[for latest results see][]{KAGRA:2021vkt, 2021PhRvD.104l2004A}.
However, with the advent of machine learning and multi-variate approaches, new data-driven methodologies were developed to optimise search sensitivities. 

Gaussian Mixture Modeling (GMM) have been used as a supervised machine learning postprocessing analysis for the cWB pipeline \citep{2020PhRvD.102j4023G}.
GMMs are probabilistic models which use a sum of uni-modal Gaussian distributions to statistically model the multi-dimensional attribute space for background and signal data from cWB. These models are applied to data to calculate log-likelihood statistics, and a single detection statistic that distinguishes likely \gw triggers from noisy background glitches. The addition of GMM improves on the overall search sensitivity of standard signals in the all-sky short search, and removes the need for manual selection of triggers through binning and cuts which occurred in standard cWB post-production previously. 
The benefits of GMM to the cWB analysis are further exemplified in \citet{2022PhRvD.105f3024L}, where a comparison between standard cWB and cWB + GMM has been presented for the O3a all-sky short search. In this work it is shown that the addition of GMM enhances the detection efficiency for all standard injections, with improvements in detection efficiencies of between 5$\%$ and 10$\%$ for sine gaussian waveforms and about $100\%$ for gaussian pulses at a false alarm rate of 1 event per 100 years. The detection efficiencies for supernova waveforms were also improved by a few percent at false alarm rates of 1 event per 100 years.

Recently, the cWB pipeline was upgraded with XGBoost \citep{Chen:2016btl}, an ensemble based boosted decision-tree algorithm, to automate the signal-noise classification of cWB events \citep{PhysRevD.104.023014, PhysRevD.105.083018, 2023PhRvD.107f2002S}. Two types of input data are used: signal events from simulations and noise events from background estimations. For each event, a selected subset of cWB summary statistics/attributes is used by XGBoost as input features to train a signal-noise classification model. The output of XGBoost is incorporated into an overall detection statistic by multiplying the standard cWB ranking statistic with the XGBoost weighting factor. Therefore, this approach uses the XGBoost output as a penalty factor which applies a weight between 0 (noise) and 1 (signal) to the cWB ranking statistic. 
This enhanced cWB pipeline was tuned to search for generic \gw bursts in O3 data, where it demonstrated robustness as a model-agnostic search, and improved the all-sky search sensitivity across the broad spectrum of simulated signals, ranging from a few percent improvement for sine gaussian waveforms up to factors of about 3 for gaussian pulses \citep{2023PhRvD.107f2002S}]. The authors also report the most stringent constraints on isotropic emission of \gw energy from short-duration burst sources with the enhanced cWB pipeline, improving on previous constraints by about 5\% to 10\% depending on the frequency of the \gw signal. Moreover, Bini et al. trained a version of the cWB+XGBoost pipeline on \gw signals from hyperbolic encounters between compact objects \citep{PhysRevD.109.042009}. These scattering events are expected to occur in dense stellar environments releasing a \gw burst signal. The authors used O3b data to obtain the first observational upper limit on the rate density of hyperbolic encounters in the local universe.

In addition to using XGBoost, \citet{2023CQGra..40m5008B} combined cWB with an autoencoder which was trained on specific transient noise morphologies (blip glitches). The autoencoder allowed for better discrimination between known glitch classes and noise. The authors show that the sensitivity volume can be improved by up to 30$\%$ for signal morphologies similar to blip glitches at a false alarm rate of 1 event per 50 years.
In \citet{2018PhRvD..98l2002A}, time-frequency maps from cWB were studied with a \cnn classifier.

\subsection{Methods for Continuous-Wave searches}
\label{sec:searches-cw}

\gw signals are typically considered as \cws when they have longer duration than typical \cbc or burst transients,
with slow amplitude and frequency evolution.
For a broad review of \cw research see \citet{Riles:2022wwz},
and for other recent reviews with different focus areas
see \citet{Tenorio:2021wmz,Piccinni:2022vsd,Haskell:2023yrv,Wette:2023dom}.

A specific minimum duration for calling a signal a \cw can however not easily be given.
The classical case are
true, fully persistent \cws:
as long as, or longer than, a typical observing run of our \gw detectors.
A crucial aspect of searching for these are the time-varying Doppler shifts and antenna patterns from the Earth's diurnal rotation an orbital motion.
However, over the past decade data analysis methods from the \cw regime have increasingly also been applied to transient signals of varying duration.
and \cw-like transients
together as a single category
is that they are \emph{quasi-monochromatic}:
at any given time, the signal is limited to a narrow frequency band,
even if that frequency evolves slowly over time.
Such \cw-like transients can be days to months long, maintaining the importance of correcting for Earth's motion, but they can also be as short as seconds and hence the regime of interest overlaps with \cbc and burst analysis methods.

The prototypical sources for both fully-persistent and \cw-like transiemt signals, with ground-based detectors, are spinning \nss
with non-axisymmetric deformations or oscillation modes.
Other possible sources include
glitching pulsars or newborn strongly deformed NSs
(both yielding short- or long-duration \cw-like transients),
exotic compact objects,
boson clouds around spinning black holes,
and the early slowly varying inspiral phase of low-mass binaries,
such as primordial black holes.
Dark matter directly interacting with the interferometric detectors
could also be observed through \cw-like detection methods.
The more exotic cases will be discussed in section~\ref{sec:new-physics}.

\ml applications in the \cw context fall into two main categories:
(i) improvements to individual aspects of existing workflows,
typically in post-processing steps after an initial \mf search;
(ii) attempts at stand-alone \ml searches from input data that has been preprocessed to varying degrees.
For both categories, it is important to realize
that most current \cw searches
(except for targeted searches for known pulsars with well-constrained frequency evolution from \elmag observations)
are severely computationally limited:
The weak expected signals from typical astrophysical sources
make it necessary to precisely track the signal over long integration times
in order to achieve significant \snr.
For templated methods like \mf and cross-correlation,
this requires a very dense covering of the search space,
reaching over $10^{18}$ templates for the deepest all-sky blind searches
\citep{Steltner:2023cfk}.
Hence, all methods applied to wide-parameter space \cw searches
are by necessity statistically sub-optimal,
making trade-offs between sensitivity and computational efficiency.
While for \cbc searches, ``matching matched filtering'' \citep{PhysRevLett.120.141103}
can be considered the main benchmark,
for \cws there is a large gap between current practical algorithms
and the theoretical optimum of a fully-coherent \mf,
where \ml methods could potentially find better sensitivity-efficiency tradeoffs.
Thus, they could actually be crucial for opening up a new detection space with current \gw detectors.

Regarding improvements to existing workflows,
one promising area is candidate clustering.
Various supervised or unsupervised clustering algorithms
have been developed
\citep{Singh:2017kss,Beheshtipour:2020zhb,Beheshtipour:2020nko,Tenorio:2020cqm,Steltner:2022aze}
to reduce the number of candidates
that need to be followed up further \citep{Papa:2016cwb,Walsh:2019nmr,Tenorio:2021njf}.
These make use of the fact that both true \cw signals
and detector noise artifacts typically excite several nearby templates.
Some are based on nearest-neighbor techniques or graph theory \citep{Tenorio:2020cqm}.
On the other hand, the method from \citet{Beheshtipour:2020zhb,Beheshtipour:2020nko} used a neural network architecture called ``Mask R-CNN'' \citep{He2017:rcnn}, which is designed specifically to find and bound regions of interest. This method is so far limited to directed searches
(for sources with known sky location),
applied e.g. in \citet{Ming:2021xtz} to the supernova remnant G347.3,
while for the more challenging all-sky case
a more conventional binned approach \citep{Steltner:2022aze}
is still preferred by the same analysis group
\citep{Steltner:2023cfk}.

An intermediate approach combining traditional and \ml methods
is taken by \citet{Morawski:2019awi},
who trained \cnns
on the per-frequency-band outputs
of the \mf time-domain $\mathcal{F}$-statistic \citep{Jaranowski:1998qm} search
and then each band is classified as containing either
pure Gaussian noise,
line-like noise artifacts \citep{Covas:2018oik},
or a \cw signal,
with the latter two on top of Gaussian noise.

Taking one step further towards pure \ml-based searches,
\citet{Modafferi:2023nzt} used the so-called $\mathcal{F}$-statistic atoms \citep{Prix:2011qv},
which are intermediate short-duration \mf data products,
as inputs to a \cnn producing an emulated \snr-like
detection statistic for long-duration \cw-like transients
from glitching \nss.
This is a computationally limited search type,
with the CNN evaluating much faster even than a straight \gpu port \citep{Keitel:2018pxz}
of the pure MF method from \citet{Prix:2011qv}.
A \gls{curric} approach using simulated Gaussian data first and then adding real data
allowed for preserving performance in a search of real O2 data,
but the study was limited to a narrow frequency band.

From such hybrid solutions, it is a gradual but quite challenging step to fully stand-alone \ml searches.
One possible input are time-frequency maps with differing degrees of preprocessing.
An early study \citep{Mytidis:2015kea} used such maps as produced
by the STAMP pipeline \citep{Thrane:2010ri}
for transient \cw-like signals from neutron star r-mode oscillations,
and compared three \ml methods:
a shallow neural network with one hidden layer,
\svms and \CSCs.
Similarly, \citet{Miller:2019jtp} used a \cnn on time-frequency maps
to search for long-duration transients from \bns merger remnants,
obtaining results on real data competitive with other algorithms for the same target signal that had been developed until that point.
\citet{Attadio:2024otf} studied similar signals from newborn neutron stars,
again using time-frequency maps,
and found that combining a classifier \cnn with a previous denoising stage (see Sect.~\ref{denoising}) is beneficial to increase sensitivity towards weaker signals.

Working directly on the full detector strain data was pioneered by \citet{Dreissigacker:2019edy,Dreissigacker:2020xfr}
who trained deep neural networks
(modified versions of \resnet \citep{2016cvpr.confE...1H} and Inception-ResNet~\citep{2016arXiv160207261S})
for all-sky and directed wide-parameter space searches.
To overcome limitations encountered in these two works,
which we will also discuss below,
\citet{Joshi:2023hpx} took a step back and considered the
simpler case of targeted known-pulsar searches,
constructing a simpler neural network.
It uses spectrograms with well-chosen bandwidth as inputs
and various improvements such as
on-the-fly regeneration of Gaussian training data at each training epoch.
This allowed for nearly ``matching matched filtering'' sensitivity
at different points in parameter space,
up to signal durations of 10 days,
but still in stationary Gaussian noise
and not for full observing-run length data sets.

In follow-up work \citep{Joshi:2024bkj}, the same authors kept for the moment the limitation to 10 days of observation time but extended the method to be able to cover wide parameter spaces,
thus competing directly with traditional fully-coherent matched filters
for which a similar maximum duration applies in practice due to computational constraints (see, e.g., \citealt{Owen:2022mvu}) and which are more usually replaced with semi-coherent methods to cover a full observing run \citep{Tenorio:2021wmz,Riles:2022wwz,Wette:2023dom}.
in \citet{Joshi:2024bkj}, both directed (known sky position) and all-sky searches are considered,
and tested for 20--1000\,Hz.
The network architecture is made up of three blocks following the design principles from \citet{Joshi:2023hpx}.
Resulting sensitivity depths at 1\% false-alarm probability per 50\,mHz band and 10\% false-dismissal probability reach around \mbox{$\sqrt{S_\mathrm{n}}/h_0\approx30$\,Hz$^{-1/2}$} at 20\,Hz, while still falling off to $\approx20$\,Hz$^{-1/2}$ towards higher frequencies where the signal tracks are more challenging. For comparison, a fully-coherent matched filter reaches $\approx40$\,Hz$^{-1/2}$ across the frequency band.
This work demonstrates that a certain gap in sensitivity remains to be caught up, but that the proposed architecture can generalize already quite well in signal strength, frequency and sky position.

In an independent approach,
\citet{Yamamoto:2020pus}
combined excess-power detection on short-time Fourier transform data
with a \cnn for sky localization.
A similar network was used directly on Fourier-transformed data in \citet{Yamamoto:2022adl},
serving as a four-class classifier for
pure Gaussian noise, \cw signals, line artifacts, or \cws plus lines.
They found comparable sensitivity to established semi-coherent all-sky search methods
in quiet data,
but with significant degradation in the presence of (simulated) noise lines.

Occupying a space between traditional \cw search methods and pure \ml,
the class of \gls{Viterbi} searches
\citep{Viterbi1967,Suvorova:2016rdc,Suvorova:2017dpm,Sun:2017zge,Sun:2018owi,Bayley:2019bcb,Sun:2019bew,Melatos:2021mmz}
considers detector data as the output of a hidden Markov model,
with the true frequency evolution of the source as the hidden state.
The main attraction is robustness to non-deterministic source behavior,
such as \ns spin wandering.
These are essentially semi-coherent searches
that can be run either in a fully unmodeled mode
with short-time Fourier transforms as the input,
or on the outputs of coherent \mf analyses over limited coherence-time segments
(the $\mathcal{F}$-statistic and variants of it).
The Viterbi-based \soap pipeline \citep{Bayley:2019bcb} for all-sky searches
is in its latest iteration also being combined with two deep learning stages \citep{Bayley:2020zfa,Bayley:2022hkz}, one
for spectral line suppression and 
another for performing \pe on the most significant remaining candidates at the end of the search pipeline (after line suppression),
for which see Sect.~\ref{sec:interpretation-CWPE}.

\begin{figure}[ht]
    \centering
    \includegraphics[width=0.49\linewidth]{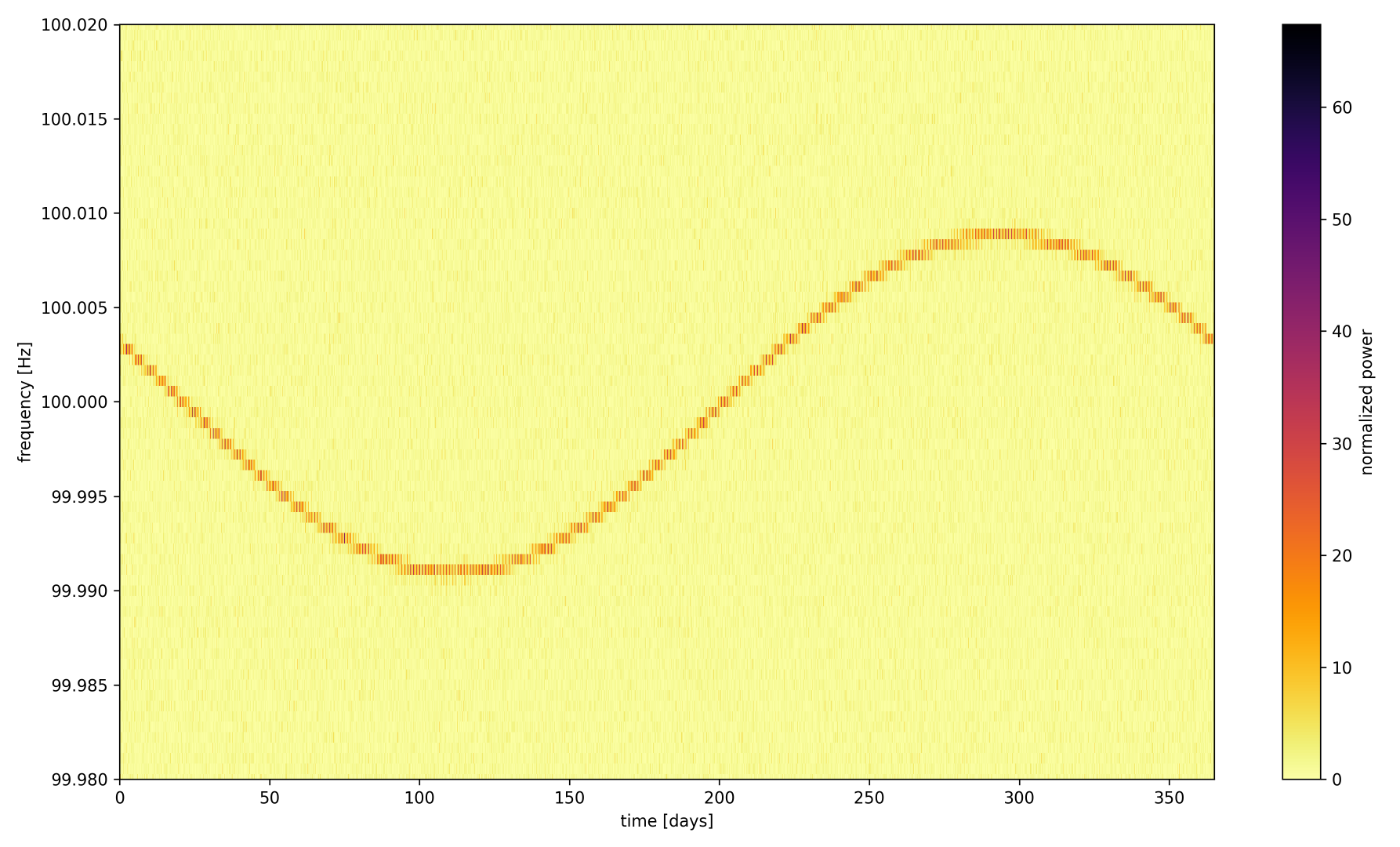}
    \includegraphics[width=0.49\linewidth]{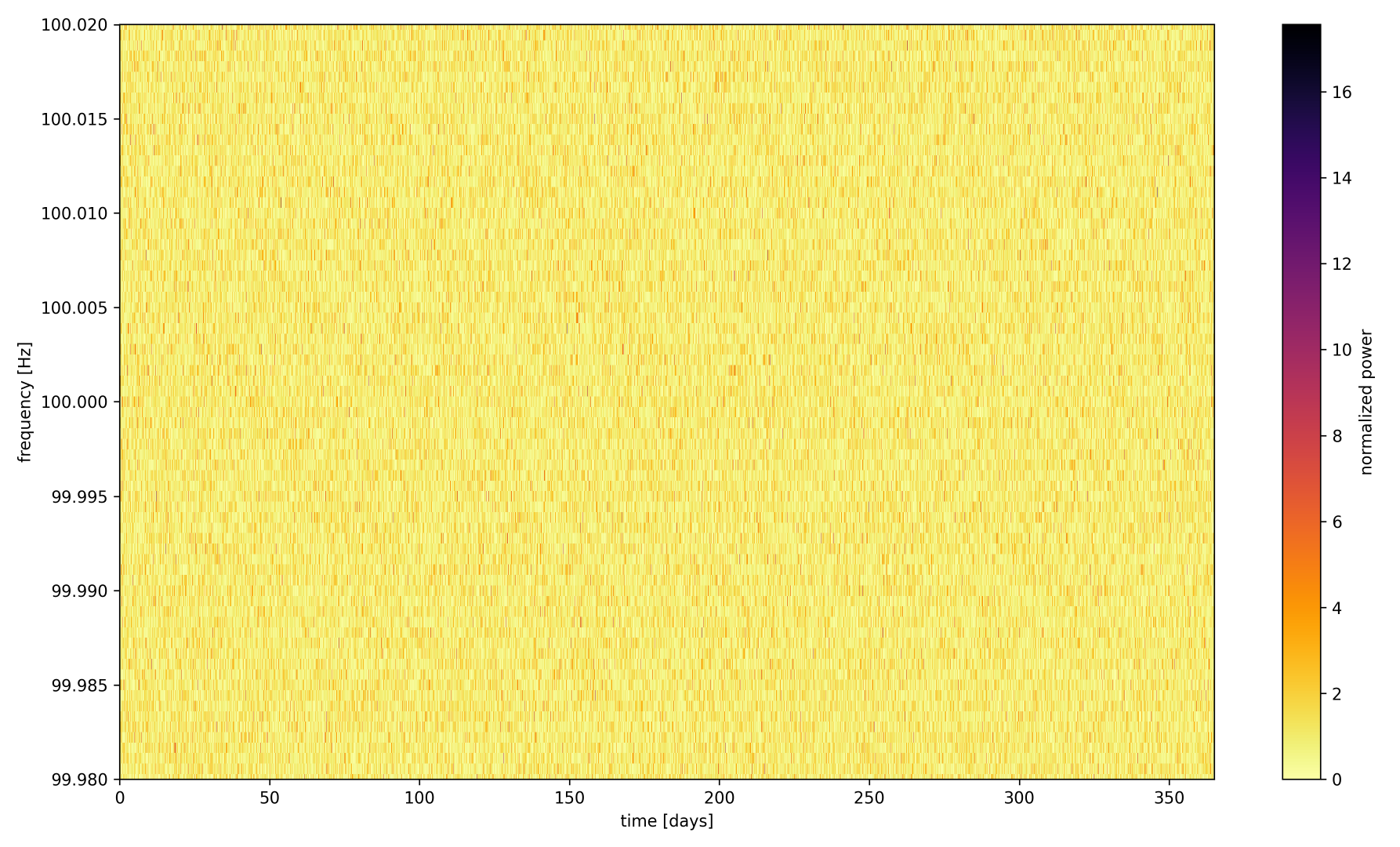}
    \caption{
    Example spectrograms of \cw signals, illustrating the difficulty of detecting these with pattern recognition algorithms focusing on local structure.
    The left panel shows a signal that is extremely strong for \cw standards
    (depth $\sqrt{S_\mathrm{n}}/h_0=5\,\mathrm{Hz}^{-1/2}$, see \citealt{Dreissigacker:2018afk} for a discussion of this quantity),
    already illustrating the extended and narrow structure in time-frequency space.
    The right panel with a depth of 50\,Hz$^{-1/2}$ (a realistic value for current semi-coherent all-sky \cw searches),
    the signal track is in addition not visible by eye above the noise, even though it is purely Gaussian.
    Graphs generated with PyFstat \citep{Keitel:2021xeq}.
    \label{fig:cw-spectrogram}
    }
\end{figure}

In summary, \ml has started to establish itself as a useful tool
for optimizing and generalizing various steps of typical \cw search workflows.
End-to-end \ml searches that could fully replace traditional algorithms
are however still in their infancy, with several crucial challenges identified so far \citep{Joshi:2023hpx,Yamamoto:2022adl}:
The defining characteristics of \cw signals--
being weak, long-duration and narrow-band--
makes them fundamentally more challenging for some popular \ml architectures,
especially \cnns and any other networks that learn the local structure of data sets.
This is illustrated in Fig.~\ref{fig:cw-spectrogram}.
Furthermore, the complicated Doppler modulations produced by the Earth's motion
over long observing times lead to very diverse signal patterns
across parameter space,
so that fully-coherent analyses of year-long data sets
by stand-alone neural network classifiers have remained out of reach so far.
On translating methods from simulated Gaussian noise to real data,
sensitivity also often takes a significant hit,
as nonstationarities in the detector noise floor
as well as stationary line detector artifacts \citep{Covas:2018oik}
make for very different characteristics.
The last point, though, is not a unique challenge of \ml methods,
as the outputs of traditional \cw searches are usually dominated
by candidates produced by noise lines too,
and much development has been put over the years
into making them more robust in this regard \citep{Keitel:2013wga,Leaci:2015iuc,Zhu:2017ujz,Intini:2020blc,Jones:2022fgs}.

See also Sect.~\ref{sec:interpretation-CWPE} for \pe methods
on \cw signal candidates.
Furthermore,
CW searches were also covered in a Kaggle challenge as described in Sect.~\ref{sec:kaggle-cw}.
The winning submissions turned out to be more closely related to established \cw analysis methods than to typical \ml approaches,
also indicating that finding competitive pure \ml solutions is still an open problem in this field.

\subsection{Methods for stochastic searches}
\label{sec:searches-stochastic}

Like \cws, searches for stochastic \gws are generally considered
to cover long-term persistent signals,
but in contrast lack deterministic signal models.
These cover both the \gls{SGWB}
(stochastic backgrounds --
either of cosmological nature, i.e. from early-universe physics,
or the stochastic superposition of unresolved signals
from individual astrophysical sources such as \cbcs)
and individual, but unmodeled signals.
The latter category includes searches for persistent point sources
without the strict model assumptions of \cw searches
and various types of long-duration transients.
Recent reviews of stochastic search methods
and the current observational status
can be found in \citet{Renzini:2022alw,vanRemortel:2022fkb}.

There are no mainstream analysis pipelines for stochastic backgrounds
based on \ml techniques in use by the \lvk\ yet.
A first exploration of different deep learning frameworks \citep{Utina:2021ipo}
has compared three architectures:
1D and 2D \cnns and a \lstm network.
Meanwhile, a good example for cases that might be considered as instances of \ml by some, but not by others, has appeared in the context of a proposed optimal Bayesian search \citep{Smith:2017vfk}
for astrophysical backgrounds
(superposed \cbc signals)The method itself is firmly based on traditional Bayesian \pe methods.
Additionally, a \gmm approach
has been used to predict detection prospects
and to simulate data realizations
for testing the analysis.
For the regime of intermittent, non-Gaussian backgrounds, \citet{Yamamoto:2022kuh} have compared three neural network architectures with different structures: two \cnns of varying depth and a residual network, which was found to perform best for detection purposes.
Another neural network was tested for signal classification and \pe (in the sense of estimating the duty cycle and \snr of the background).

Since stochastic background searches typically
rely on observing cross-correlations \citep{Allen:1997ad}
between the data of two or more \gw observatories,
it is crucial to control
possible correlated noise sources across sites
(e.g. through atmospheric magnetic channels),
as is the reduction of low-frequency noise components
such as \gls{NNOISE}.
ML can play an important role in this kind noise mitigation,
as explored e.g. in \citet{Badaracco:2020qmm}
with surrogate Wiener filtering.

Work is also ongoing on using \ml for stochastic background searches
in other \gw frequency bands,
e.g. with \lisa~\citep{Alvey:2023npw}
and Pulsar Timing Arrays~\citep{Chen:2020ehw,Shih:2023jme}.

Long-duration transient searches with stochastic-style methods,
for targets such as magnetar bursts \citep{Quitzow-James:2017zeq,LIGOScientific:2022sts}
or \bns post-merger remnants \citep{Abbott:2018hgk,Banagiri:2019obu}
typically involve pattern recognition tasks in \gw spectrograms,
similar to shorter-duration burst searches (Sect.~\ref{sec:searches-burst}).
As such, the STAMP pipeline,
first introduced in \citet{Thrane:2010ri}
and recently reimplemented in python \citep{Macquet:2021ttq},
uses either seeded or seedless clustering techniques \citep{Khan:2009tx,Thrane:2013bea}
for detecting \gw transients as time-frequency tracks.
At the shorter duration end ($\sim \mathcal{O}$(minutes)),
this overlaps with the regime
where neural networks are being applied
for burst-type searches \citep{Boudart:2022xib,Boudart:2022apz}.

\section{Strategies for source interpretation}
\label{sec:interpretation}

Once a \gw signal candidate has been detected with sufficient confidence,
additional methods are needed to characterize the parameters of its physical source in detail.
Here, we discuss how \ml methods are starting to make an impact on
the classification of candidates into different possible source types
and the detailed estimation of parameters,
for both the transient and continuous signal case.
We also discuss in this section applications for \gls{multimessenger} physics, population inference, cosmology and new physics beyond the standard model.

\subsection{Source classification (online/offline)}
\label{sec:interpretation-classification}

In the last two observing runs, the \lvk\ has been issuing prompt public alerts for \gw transient candidates crossing a pre-determined false alarm rate (FAR) significance \citep{Chaudhary:2023vec}. Preliminary alerts are issued seconds after the merger, with annotations released as additional and more accurate information becomes available. Since the fourth observing run, additional specialized searches \citep{Magee_2021,Sachdev_2020,Nitz_2020EW,Kovalam_2022} have been set up to issue ``early warning'' alerts when they detect sufficiently strong signal candidates before they reach the merger stage. In cases of positive confirmation of the candidate following human vetting, update notices are generally issued on time scales of a few hours with improved estimates of sky localization, detection significance, or source classification.

The classification of \gw candidate events and source property inference in real time present various issues, as the necessity for precision clashes with the need to release information as soon as feasible. These are situations where machine learning-based classifiers may be effective in obtaining accurate results that would otherwise be unachievable.

The \lvk\ low-latency pipeline \citep{Chaudhary:2023vec} for compact binary coalescence events \citep{Chatterjee:2019avs} currently uses two supervised learning algorithms: a \gls{KNN}~\citep{shakhnarovich2005nearest} algorithm is used to infer the presence of a neutron star as one of the binary components and of possible post-remnant matter, and a \RF algorithm~\citep{Breiman2001} to infer the presence of an object with mass in the ``gray'' region between the lowest-mass black holes and the heaviest neutron stars \citep{Chaudhary:2023vec}. Binary classification scores for these metrics are generated from the output of detection pipelines and equation of state models. The algorithms are typically trained using hundreds of thousands of simulated \cbc signals that are coherently inserted into real detector noise. The validity of an event's classification outcome is then evaluated by means of features of the algorithm's confusion matrix and the receiver operating characteristic curve. The benefit of this approach is its capacity to accommodate statistical and systematic errors in the search pipeline parameters. It also provides for a significant speed gain over the semi-analytic effective Fisher formalism technique used in the first two \lvk\ observing runs \citep{Chatterjee_2020}. Recent advancements have concentrated on improving this method and obtaining actual conditional Bayesian probabilities for the source property measures that are more simply interpretable than scores \citep{Berbel:2023vug}.

\subsection{Source parameter estimation (transient signals)}
\label{sec:interpretation-transientPE}

Our understanding of the compact binary systems that produce the \gws detected by the \lvk\ Collaboration hinges on our ability to perform Bayesian inference. The de-facto standard algorithm used for this task is \gls{nestedsampling} within a range of different implementations \citep{Skilling:2006gxv,Ashton:2022grj,veitch:15,2019ApJS..241...27A} which, for standard \cbc signals, take of order days to weeks. The bottleneck in nested sampling is two-fold: evaluating the likelihood for \cbc signals is computationally expensive and drawing a new sample from the likelihood-constrained prior typically requires using random walks that may need up to thousands of steps, which also require evaluating the likelihood. This, in conjunction with the expected increase in the number of detected events that will result from improvements to existing detectors and next generation detectors, presents a significant computational challenge.

Multiple strategies have been suggested and demonstrated that harness the power of \ml within the existing nested sampling algorithm. Targeting the bottleneck associated with repeatedly drawing samples from the likelihood-constrained prior, it has been shown \citep{2021PhRvD.103j3006W,2021MNRAS.507.2037A,2023MLS&T...4c5011W} that incorporating \ml in the process of drawing new samples can provide significant speed-ups. In this case, a type of generative machine learning algorithm called normalising flows is used allowing a normalising flow to learn the distribution of samples within likelihood-constrained prior during sampling and, in-turn, sample from the learned distribution. This eliminates the need for random walks and improves the efficiency of drawing new samples. Comparisons of the Nessai algorithm \citep{2021PhRvD.103j3006W} with the standard nested sampler \citep{Speagle:2019ivv} used in current \lvk\ analyses \citep{LIGOScientific:2021usb,KAGRA:2021vkt} show that Bayesian posterior distributions are accurately recovered whilst requiring two-times fewer likelihood evaluations. The follow-up implementation (known as i-nessai \citep{2023MLS&T...4c5011W}) improves this factor to a range of between 2.68 and 13.3 times fewer likelihood evaluations.

The alternative speed enhancement that can be applied to nested-sampling is to optimise the likelihood calculation at the core of the algorithm. This can come in 2 forms: using new techniques to learn the likelihood function itself, or to optimise the generation of \cbc \glspl{waveform}. In the former case, much non-\ml work has been done on Reduced Order Models that allow for rapid evaluation of the likelihood by finding computationally efficient representations of the waveform model. In the latter case it has been shown that significant speed-ups can be obtained using \ml approaches. In \citet{Khan:2020fso} it was demonstrated that an order of magnitude speed-up could be achieved for \bbh signal generation over \roq techniques which was further enhanced if generating many thousands of waveforms in batches on a \gpu. This area of research has since been extended to include multi-modal precessing waveforms \citep{Thomas:2022rmc}.

An additional scientific benefit of incredibly rapid parameter estimation for \cbc events, specifically those containing a \ns component, is the possibility to perform parameter estimation quickly enough to be able to alert EM astronomers of the locations of the source. Non-\ml and \ml methods alike are being developed to achieve the goal of accurate posterior estimation within $\mathcal{O}(1)$~sec and potentially even prior to merger. One way to tackle this problem is to look at completely new approaches using solely \ml algorithms.

One of the first \ml approaches to address the issue of rapidly generating samples from a joint Bayesian posterior on \cbc source parameters \citep{1909.06296} utilized a type of neural network known as a \cvae. This CVAE implementation (known as Vitamin) and a form of simulation based inference only needs to be trained once and can be applied many times with a computational cost orders of magnitude faster than standard techniques at runtime. The process of training requires the choice of signal parameter prior distributions and assumptions on the detector network and noise properties, and so will require retraining if those assumptions change.

A standard approach to the comparison of Bayesian inference tools has been to compute the JS-divergence between 1-dimensional marginalised posterior distributions. The JS divergence between the distributions $p(x)$ and $q(x)$ is defined as
\begin{align}
    \mathrm{JS}(p,q) = \frac{1}{2}\left[\mathrm{KL}(p,m) + \mathrm{KL}(q,m)\right]
\end{align}
and has units of nats (the natural unit of information) where $m=(p+q)/2$ and the KL-divergence is 
\begin{align}
    \mathrm{KL}(p,m) = \int p(x)\frac{\log p(x)}{\log q(x)}dx.
\end{align}
Identical distributions have $\mathrm{JS}(p,p) = 0$ and in the opposite extreme with maximally differing distributions the JS becomes $\log(2)$.

Vitamin posterior distributions have been compared with existing techniques such as \mcmc and nested sampling \citep{1909.06296} when applied to inference of the type of \bbh systems detected by the \lvk. 
These comparisons have shown JS-divergences between 1-dimensional marginalised posteriors of $\mathcal{O}(10^{-2})$ nats and can be compared to the findings of \citet{2020MNRAS.499.3295R} where for different sets of samples drawn from the same Gaussian distribution, values of $> 0.002$ nats were considered statistically significant.

At present the state-of-the-art application of \ml to Bayesian parameter estimation is done using neural posterior inference \citep{2020arXiv200207656G,dax_2021,PhysRevLett.130.171403}. The first work on this application incorporated autoregressive flows within a variational \gls{autoencoder} framework but was subsequently developed into a dedicated spline coupling \nf incorporating an embedding network to compress the conditional input \gw timeseries data. This approach allows for the analysis of 8~sec of data from a network of \gw detectors compressed into 128 features to be input as the \nf conditional data. The \nf is expected to accurately model the \gw parameters posterior for any likely signal and noise realisation, and hence compression is crucial to helping the \nf by representing the conditional timeseries data in a compact and information-rich form. Results from the pipeline (known as DINGO) applied to real \ligo-\virgo\ detections provide posteriors that match incredibly well with benchmark analyses, returning JS-divergences between 1-dimensional marginalized posteriors of $\mathcal{O}(10^{-3})$~nats. 

Parallel applications of the DINGO framework have included the addition of an importance sampling component that essentially uses DINGO as a highly efficient tool for sampling from an approximate posterior distribution \citep{PhysRevLett.130.171403}. Then through comparison with the likelihood obtained from analytic models (the same models used within \mcmc and \ns algorithms) at those sample locations, can be used to obtain a corrected posterior. The motivation behind such an approach is to hedge against the known ``imperfect'' nature of current \ml applications where despite rigorous testing, the behaviour of generative models cannot be guaranteed in all regions of parameter space. The additional computational cost of using the analytic likelihood model in this scenario is minimal in comparison to traditional techniques (\mcmc and \ns) where $\mathcal{O}(10^6)$ likelihood evaluations may be required. Depending on the accuracy of the \nf approximate (proposal) distribution this can be reduced to $\mathcal{O}(10^3)$ evaluations (that can also be parallelised) therefore adding minutes to the inference latency. The disadvantages beyond the additional computational cost (minimal in comparison with traditional techniques but significant in terms of the \nf) are that in order to use an analytic likelihood, one must assume a specific form for the noise model. This limits the ways that real (non-Gaussian) detector noise can be used in training the model where no assumption is made on the mathematical form of the noise distribution which includes transient detector noise artifacts (glitches). In other words, this approach can no longer be classified as ``likelihood free'' inference.

Ref.~\citet{Yamamoto:2020rse} used \cvae for estimating quasi-normal-mode frequencies from the ringdown portion of \bbh signals.

One challenge of likelihood-free/simulation-based inference is that the \ml models trained for it (e.g., \nfs) can get quite large. In \citet{2023arXiv231207615C} it is suggested that ``Self-supervised Neural Symmetry Embeddings'' can mitigate this problem by exploiting intrinsic symmetries of the problems studied. In \citet{Mao:2023zdr} it was demonstrated that \glspl{autoencoder} and \anns can also help in calibrating the coverage of posterior credible intervals for \gw parameter estimation -- the example application was for LISA, but the methods should translate to ground-based detectors as well.

\subsection{Source parameter estimation (continuous signals)}
\label{sec:interpretation-CWPE}

Beyond simple maximum-likelihood parameter estimates
\citep{Jaranowski:1998qm,Prix:2015cfs},
the \pe problem for \cw signals has so far been less explored than for \cbcs.
As for Bayesian approaches, \gls{nestedsampling} is used for analyzing known pulsars in a time-domain pipeline \citep{Pitkin:2017qfy,Pitkin:2022gbe},
either at a single frequency-evolution template or in a very narrow band,
combining the traditionally separate detection and \pe steps
into a single analysis.
An alternative frequency-domain implementation based on the $\mathcal{F}$-statistic \citep{Jaranowski:1998qm} was recently presented \citep{Ashok:2024fts}, also using \gls{nestedsampling}.
In addition, parallel-tempered ensemble \mcmc sampling \citep{Foreman-Mackey:2012any,Vousden:2016ptemcee} is used for the hierarchical follow-up
of candidates from wide-parameter space searches with the PyFstat package \citep{Ashton:2018ure,Keitel:2021xeq,Tenorio:2021njf,Mirasola:2024lcq}.
Here, the chains are typically run only for a limited time, not necessarily reaching the level of convergence needed for robust parameter estimates,
with the main goal of excluding/confirming signal candidates for passing from one stage to the other.
An alternative implementation if Bayesian \pe for \cws has been presented by \citet{Covas:2024pam}, accessing a larger variety of samplers through the bilby package \citep{2019ApJS..241...27A}.

The first application of \ml-based \pe for \cw-like signals
has been implemented for the weakly modeled \gls{Viterbi}-based \soap pipeline \citep{Bayley:2019bcb} (see also Sect.~\ref{sec:searches-cw}),
proceeding in two steps:
First, a \cnn is used to eliminate spurious candidates caused by instrumental lines
\citep{Bayley:2020zfa};
then, a \cvae
delivers posterior estimates for the frequency-evolution (``Doppler'') parameters
of a \cw signal candidate \citep{Bayley:2022hkz}.
Like the \soap search itself, this approach so far works only for relatively high \snrs
(low search depths of around $\sqrt{S_\mathrm{n}}/h_0=10$\,Hz$^{-1/2}$, compare figure~\ref{fig:cw-spectrogram}).

Therefore, as of 2024, a full solution to \pe across the full parameter space of modeled \cw signals is still lacking, using either traditional or \ml approaches.

As discussed in Sect.~\ref{sec:searches-stochastic},
a first application of neural networks to intermittent, non-Gaussian stochastic backgrounds was presented by \citet{Yamamoto:2022kuh}
(along with detection methods).

\subsection{Applications for multi-messenger physics}
\label{sec:multimessenger}
\bnss represent a unique laboratory for probing the fundamental physics governing the dynamics of compact objects, nuclear astrophysics and the synthesis of heavy elements in the universe. The detection of \gw signals coupled with the observation of their accompanying electromagnetic counterparts, and in particular gamma-ray bursts, has opened a new era of multi-messenger astronomy \citep{LIGOScientific:2017ync,LIGOScientific:2017zic}.

For multi-messenger astronomy, \gw analyses must be able to provide information on detection candidates as quickly as possible to enable follow-up searches for electromagnetic and neutrino counterparts, and to enable multi-messenger studies of a compact binary merger. Astronomers must decide quickly whether to follow the low-latency Open Public Alerts for significant \gw candidate events. 
in \citet{Abbott_2022} GWSkyNet-Multi, an advanced \ml modelwas presented as an extension of the earlier GWSkyNet \citep{Cabero:2020eik}. It classifies potential \gw events detected by the \ligo\ and \virgo\ observatories, using the limited data from low-latency Open Public Alerts to quickly determine whether an event represents a black hole merger, a merger involving neutron stars, or simply a non-physical incident.
Specifically, GWSkyNet-Multi uses the following information from the alerts:
image representations of the skymap and three volume-projected versions of it, along with four corresponding normalization factors;
the available \gw detectors;
distance information;
and two Bayes factors for the signal-vs-noise and coherent-vs-incoherent hypothesis tests.
In \citet{Raza:2023gyv}, a study was conducted on how the complex GWSkyNet-Multi network uses input features to make a correct prediction. Factors such as the localization area of the sky maps and the computed coherence versus incoherence
Bayes factors play a key role in distinguishing between authentic events and glitches. In addition, the estimated distance to the source helps to distinguish between different types of glitches.

\begin{figure}[ht]
    \centering
    \includegraphics[width=\linewidth]{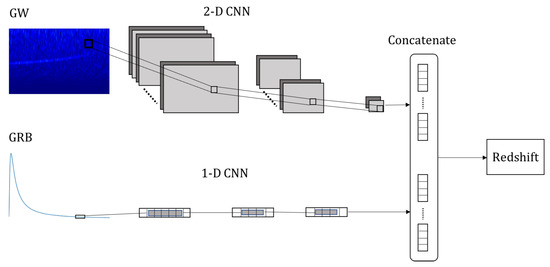}
    \caption{The analysis from \citet{natureComp} employs a multimodal \ml model to calculate the redshifts of joint \GW and gamma-ray burst (GRB) sources. This model integrates two distinct types of data -- images and time series -- to effectively address the regression problem. A schematic of the architecture of the neural network used is shown.}
    \label{fig:mmml-mma}
\end{figure}

Multimodal machine learning is a cutting-edge approach in artificial intelligence where models are designed to process and understand information from multiple modalities in input, such as text, images, audio, and sensor data.
By integrating diverse sources of information, multimodal learning enables Artificial Intelligent systems to capture a more complete understanding of the world, mimicking human-like perception \citep{10.1109/TPAMI.2018.2798607}
The multimodal approach could offer complementary information on the merger process and provide constraints on the properties of neutron stars and the nature of the resulting remnant or help in parameter inference. 
In \citet{natureComp} the authors introduced the idea of applying the multimodal machine learning (MMML) approach to multi-messenger data where there is the emission of \gw signals and electromagnetic or neutrino counterpart. in \citet{universe7110394} the authors introduced the first approach of multimodal machine learning to joint analysis of signals emitted as \gw and as gamma-ray electromagnetic signals to produce a proof of concept of this approach.

On the other hand, as already discussed in Sect.~\ref{sec:modeling-astro}, neutron star properties (especially the equation of state) can also be inferred by combining more indirectly the measurements from \gw and \elmag observations of different objects.
\ml-based examples of this rapidly growing field include, e.g., a random-forest based approach \citep{HernandezVivanco:2020cyp},
a predictive variational \glspl{autoencoder} \citep{Han:2022sxt},
or Bayesian Neural Networks \citep{Han:2021kjx,Carvalho:2023ele},
which have all been chosen with a mind to being flexible enough to combine the mass--radius constraints from the observations of \bnss and radio pulsars.

\subsection{Population inference and cosmology}
\label{sec:interpretation-pop-cosmo}

The main developments in the application of \ml to population inference and cosmology are very recent. As with a large fraction of \ml applications in \gw astrophysics, the aim in the area of cosmology is an improvement in the speed of analysis when compared to the accurate existing benchmark analyses and the correlated issue of efficiency when the number of detections and the complexity of the cosmological models is increased. Regarding population analysis, speed is of less motivation and model flexibility through the use of \nfs and \gpus appears to be the driving force. In both areas the techniques share many commonalities and all work discussed in this section falls under the category of Hierarchical Bayesian inference. This is where the input to these algorithms comes in the form of samples from Bayesian posteriors on the source parameters of individually detected events.

In \citet{Stachurski:2023ntw} the authors use a normalising flow model trained to learn the true prior distribution of \gw source parameters given knowledge of the distribution of mass in the local universe and conditioned on the Hubble constant. The results of the process then allow for the rapid calculation of the posterior distribution on the Hubble constant using posterior samples from detected \bbh events. The posterior samples are initially processed by undoing the priors applied on the \gw samples during the event \pe process. Then the Hubble constant dependent normalising flow model priors are applied allowing the extraction of the posterior probability directly from the \nf model marginalized over the event uncertainties. The paper focuses on the task of accurately comparing their results with those of the benchmark analysis \citep{O3_cosmology_2023} which only provided inference on the Hubble constant. The authors compare with the benchmark results and provide JS-divergence measurements of $\mathcal{O}(\text{few})$ nats over 42 \bbh events.

Although the \nf model in \citet{Stachurski:2023ntw} was presented as a Hubble constant inference tool, it is easily generalisable to additional cosmological and population parameters. At the heart of the analysis is the generation of training data for the model which crucially incorporates the fact that \gw sources within sky and distance regions that would likely be contained within galaxy catalogs (specifically the GLADE+ catalog; \citealt{dalya_2022}). Data generation therefore includes samples from the \gw parameter prior that follow the mass distribution described by the distribution and magnitude properties of known galaxies. Those generated samples whose location and host galaxy properties would not be within the catalog are sampled according to a uniform in comoving volume distribution. Data generation is also restricted such that samples must be detectable by the assumed network of \gw detectors. In this sense the authors require that a training data signal within an assumed detector network would achieve an \snr above a predefined threshold. An identical \snr threshold criteria must then be applied to any signals analysed (or tested) by the trained model. Ultimately these processes form a prior distribution of \gw event parameters conditional on sources being detectable but also conditional on the assumed cosmological and population parameters which are represented by random draws from their respective priors for each generated sample.

The choice to use training data conditional on being detectable naturally limits the volume of space considered. However, the evaluation of the detectability of each prior sample is costly since it requires the simulation of a gravitational waveform in order to calculate the \snr. To avoid this significant bottleneck in the speed of data generation an additional \ml tool was used. To compute the optimal \snr of each sample rapidly, a \mlp model was used to approximate the SNR function (from the {\it poplar} package developed by \citealt{Chapman-Bird:2022tvu}, and similar to \citealt{gerosa_2020}). This same tool is used to evaluate the probability of detection when processing event posterior samples through the \nf model and accounts for the selection bias imposed by the original \snr threshold\footnote{In reality an \snr threshold is not used for determining whether \gw events have been detected. This decision is made based on a false alarm rate leading to a correlated but variable threshold on the matched-filter \snr.} used for determining detections.

Work done in parallel \citep{leyde_2023} applied the same \nf technology but used a different approach to the training data. Here the authors make use of a previously discussed \ml tool used for rapid Bayesian \pe for \cbc events, DINGO \citep{dax_2021}. This allows them to sample from a prior distribution of both cosmological and population parameters from which they can then sample the prior parameters of \gw events conditional on the cosmology and population. In this first work they do not consider information from galaxy catalogs although in principle they could be incorporated in a similar way to \citet{Stachurski:2023ntw}. They can then very quickly simulate \gw signals in noise and generate $\mathcal{O}(1000s)$ of posterior samples from each event. The \nf model is designed to take as conditional input an embedded representation of event posteriors from $n$ events, each contributing a set of posterior samples. So in contrast to \citet{Stachurski:2023ntw}, where a \nf was trained to process batches of single posterior samples conditional on cosmological parameters, here the network is conditional on input batches that are composed of many events with each event represented by many posterior samples. The outputs in this case are samples drawn from the posterior on the cosmological and population parameters and results can be obtained in $\sim$minutes. 

Regarding population analyses, the first approach to apply machine learning techniques was \citet{wong_2020}, where a \nf model was used to emulate a phenomenological population model governed by 4 hyper-parameters. Each instance of the model predicts the distribution of 6 \gw observables (the primary mass, mass ratio, spin magnitudes, and spin tilts) which are modeled as being measured in the high \snr limit and therefore with no uncertainty. Their \nf method was validated against an analytic phenomenological model where the only difference is the likelihood function (learned for the \nf versus analytic). It is concluded that the \nf model can emulate the phenomenological model accurately and efficiently.

A non-parametric binned \gls{gaussianproc} approach is used in \citet{ray_2023} to model the joint mass and redshift distribution of compact binary coalescences. They account for the significant measurement uncertainties in the \gw input data (posteriors), and using a Gaussian process allows them to make very few assumptions about the functional form of the distribution model. The flexibility of their model allows them to probe the possible correlations between the mass and redshift distribution, e.g, a cosmologically evolving mass distribution. They applied their model to data from the GWTC-3 catalog \citep{KAGRA:2021vkt} but concluded that more events are required to confidently assert that correlations are present.

In \citet{ruhe_2022} \nfs were used to model the mass, redshift, and spin distributions of detected \cbc events taken from the GWTC-3 catalog \citep{KAGRA:2021vkt}. The input in this case were the posterior samples on the event parameters under the assumption of a fixed and known cosmology. Despite not including selection effects in their model, the authors claim that they have been able to recover posterior structure that agrees with existing phenomenological modeling results. 

Motivated by the growth of the observed catalog of \gw events, in \citet{cheung_2022} a Gaussian process and a \nf approach are compared in a population study mock events and a subset of events from GWTC-2 \citep{PhysRevX.11.021053}. They consider a phenomenological model of the mass and redshift parameters and use posterior samples from the simulated (and real) events as input to the analysis. They were able to conclude that the \nf model could recover the correct posterior distributions with up to 300 mock events but had a tendency to underestimate the uncertainty (or width) of the posterior for real data. The Gaussian process model struggled in all but low-dimensional cases.

We also briefly mention an application of \ml to population synthesis \citep{gerosa_98,wong_2019} where in the latter work a \gp is trained on a small but state-of-the-art, set of population-synthesis predictions of \bbh systems formed in isolation. The resultant hierarchical Bayesian analysis uses the \gp model to interpolate between simulations allowing the construction of a smooth posterior on the input hyper-parameters of the population synthesis model conditioned on \bbh data from O1 and O2 detections. In this case the inferred hyper-parameter was the strength of natal kick that black holes receive at birth. 

One aspect that is becoming increasingly clear with the development of new \ml techniques in the areas of \gw population and cosmological inference is the realistic possibility of merging the 2 areas of research. As research has developed it is clear that cosmological analyses have made fixed model assumptions about the underlying source population. Similarly, population studies have often assumed a fixed cosmology. Work specifically on the cosmology side, e.g.~\citet{leyde_2023} is leveraging the power of \ml to bridge the gap and simultaneously perform inference on a joint cosmological and population parameter-space. This will allow us to properly account for our lack of accuracy in either area and correctly account for how these uncertainties correlate between cosmological and population parameters.

\subsection{New physics}
\label{sec:new-physics}

Beyond the traditional astrophysical targets and search types,
terrestrial \gw detectors can also probe many interesting ``new physics'' scenarios
not expected under standard astrophysical scenarios,
or even beyond the standard model of particle physics.
This includes the manifold imprints of early-universe physics
on stochastic backgrounds
\citep{Renzini:2022alw,vanRemortel:2022fkb}
as well as dedicated searches for such diverse physics as
the early inspiral or full coalescence of binary \pbhs
\citep{Garcia-Bellido:2017fdg,Bird:2016dcv,LIGOScientific:2021job,Miller:2021knj,Miller:2024fpo}
(depending on their mass scale),
cosmological defects such as cosmic strings \citep{Vachaspati:1984gt,LIGOScientific:2021nrg} and domain walls \citep{Grote:2019uvn},
indirect detection of particle dark matter as emitters of \gws
from boson clouds around spinning black holes \citep{Brito:2015oca,LIGOScientific:2021rnv},
and direct detection of particle dark matter interacting with
the hardware of the detectors rather than through any propagating \gw channel
\citep{Pierce:2018xmy,Guo:2019ker,Vermeulen:2021epa,LIGOScientific:2021ffg,Miller:2023kkd,KAGRA:2024ipf}.
These references are only a limited set of examples,
as the ongoing research in these directions
would merit several dedicated review articles.
E.g. see also \citet{Maggiore:1999vm,Roshan:2024qnv} for probing the early universe with \gws
and for anything black-hole related see also \citet{Barack:2018yly}.

Most of these scenarios have only been investigated relatively recently,
and the search methods employed so far are mostly based
on direct transfer of established traditional methods
from the \cbc, \burst, \cw or stochastic domains.
This approach bears the risk of a ``searching under the streetlights'' or
``hammer in search of a nail'' effect,
where only new physics that produces signals that are qualitatively similar to known types
get searched for.
One \ml-adjacent algorithm that is already being fruitfully applied as a computationally efficient alternative to \pbh searches \citep{Alestas:2024ubs} is the \gls{Viterbi} method,
as previously discussed for \cw searches in Sect.~\ref{sec:searches-cw}.
One could in general expect that
fairly generic \ml approaches, such as those discussed in sections \ref{sec:searches-burst} and \ref{sec:searches-stochastic}
for the detection of unmodelled signals,
will generalize well to such new physics signals.
This includes e.g. anomaly detection techniques,
which by definition look for any type of unexpected signals,
and neural networks with training sets that follow simple phenomenological signal models, such as sine-Gaussians or a Taylor expansion for quasi-monochromatic \cw-like signals, but are independent of specific physical scenarios.
However, detailed studies remain to be done to see if this expectation will be borne out.
On the other hand, in the future,
more dedicated \ml methods for specific new physics scenarios
could lead to notable progress in the field,
especially in areas where fully explicit \gls{waveform} models over a broad parameter space are difficult to obtain, e.g. requiring a large new set of computationally costly numerical simulations and lots of human effort to construct an explicit model,
but \ml interpolation schemes or \sbi may be more feasible starting from a relatively sparse set of simulations and skipping the manual modelling step.
Again, it remains to be seen in the coming years if fruitful applications will be found.

\section{Citizen science \& machine learning}
\label{sec:citizenAI}

We have already mentioned the Gravity Spy and the GWitchHunters projects in Sect.~\ref{sec:noisecharact} as two examples of applications of citizen science to the characterization of detector strain data. 

A study by \citet{Soni_2021} analyses the impact of glitches on \GW searches during the O3 run of the Advanced \ligo\ detectors. Two new classes of glitches were identified, fast scattering/crowns and low-frequency blips. Gravity Spy's \ML algorithm for glitch classification was updated using training sets from detector monitoring and citizen-science volunteers. Reclassification of the data based on the updated model reveals that about 27\% of glitches at \ligo- Livingston belongs to the fast scattering class, while about 8\% belongs to the low-frequency blip class. The results underline the potential of glitch classification to improve the data quality of \gw detectors and demonstrate the value of citizen-science contributions in analyzing large datasets.

Different strategies have also been employed to involve scientists or citizens outside of the GW community, with efforts to engage the advanced data scientist community.
The approach of the G2net COST Action CA17137
\footnote{\url{https://www.g2net.eu} and \url{https://www.cost.eu/actions/CA17137}} 
was to engage external participants through data challenges for \gw data on the Kaggle platform \footnote{\url{https://www.kaggle.com/}}.

\subsection{Kaggle challenges}

Kaggle is a data science competition platform and online community of data scientists and machine learning practitioners under Google LLC. Kaggle works together with academics and business to develop self-contained data analysis challenges for this community. As of 2023 Kaggle has over 15 million registered participants who are encouraged to tackle these challenges for rewards of either cash or Kaggle credits. Example competitions currently active on the platform include the identification of text generated from large language models, predicting how small molecules change gene expression in different cell types, predicting US stocks closing movements, and helping to evaluate tackling tactics and strategy in the American National Football League. Since the founding of Kaggle in 2017 they have hosted over 600 competitions. 

In a typical challenge, participants are provided with an example (training) dataset or the means to generate artificial data, a description of the aims of the challenge, and a clearly defined metric by which to judge a submitted set of analysis results. These results are obtained through the analysis of a testing dataset, uploaded via the Kaggle competition site where the metric is evaluated. Competitions are open for $\sim3$ months (although this can vary) and participants can track their performance relative to other participants (or teams of participants) via a leaderboard that is constantly updated based on the latest submissions. 

As a challenge developer, the process of designing the challenge within the Kaggle infrastructure takes place whilst working with a small team of Kaggle employees over a number of months. During this time the feasibility of the challenge is assessed in terms of a number of factors including practical issues such as the volume of training and testing data to be made available, the difficulty of the challenge, the appropriate metric by which to judge the results, and how intrinsically interesting the challenge will be to the Kaggle community. Prior to the competition launch, documentation for participants must be provided and the datasets carefully examined by Kaggle developers to catch errors and to specifically eliminate any leakage\footnote{Leakage in this context is the use of information in the model training process which would not be expected to be available at prediction time.}. Once the challenge is launched, competition hosts are able to interact with participants through the competition online forum in order to answer any questions or address any remaining issues with the data.

In order to engage with the broader \ml community and in an effort to gain independent perspectives on how to approach \gw data analysis problems, a series of Kaggle challenges have been developed within the G2net COST action. To date two such challenges have been hosted on the Kaggle platform, each tackling a different aspect of ground based \gw data analysis. Both were classification (detection) problems with the first on the topic of compact binary coalescence of black holes (\bbhs), and the second on the detection of continuously emitted \gws (\cws) from rapidly rotating \nss.

\subsubsection{Challenge 1: Binary Black Holes}
\label{sec:kaggle-bbh}

The first G2Net Kaggle challenge \citep{kaggle1} was launched on 30th June 2021 and was the first public \gw based data analysis targeted at the machine learning community. It was designed to build upon the growing interest in applying neural network classification algorithms for the problem of transient \gw signal detection, specifically the case of stellar mass \bbhs. The task was to determine the presence or absence of a \bbh signal in simulated advanced-detector Gaussian noise and the metric used to rank submissions was the \textit{Area under the ROC curve} (AUC). This measure can be constructed from a submitted list of probability estimates for the presence of a signal in each piece of test data. At the conclusion of the challenge 1501 competitors spread between 1219 teams participated in the challenge.

In this competition, the participants were provided with a training set of time series data containing simulated \gw measurements from a network of 3 detectors ( \ligo-Hanford, \ligo-Livingston, and \virgo). Each time series contained either just detector noise or detector noise plus a simulated \gw signal. The \bbh parameters that were varied for each simulated signal were the masses, sky location, distance, black hole spins, binary orientation angle, polarisation, time of arrival, and phase at coalescence (merger). These 15 parameters were randomized according to astrophysically motivated prior distributions (not known to the participants) and used to generate the simulated signals present in the data. Each data sample contained three time series (one for each detector) and each spanned 2\,sec and was sampled at 2048\,Hz. The distribution of integrated \snr values for the data containing signals was not astrophysically motivated but tuned by varying the distances of each signal such that the challenge contained a range of easy and hard to detect cases. It was also designed so that the bulk of signals would have SNRs close to the sensitivity region of known \mf searches (SNR $\sim 8$).

At the time of writing, the results of the first challenge have yet to be published.

\subsubsection{Challenge 2: Continuous Waves}
\label{sec:kaggle-cw}

Following the success of the first challenge, which attracted strong interest from the Kaggle community, a more difficult challenge was introduced in the second G2Net competition \citep{kaggle2}. This focused on continuous gravitational wave detection, where recent \ml efforts \citep{Dreissigacker:2019edy, Dreissigacker:2020xfr} have shown promise but have yet to overcome the significant computational hurdles of traditional search methods -- see Sect.~\ref{sec:searches-cw}.

The second challenge was similar to the first in that it was a classification task to detect \cw signals in simulated advanced-detector Gaussian noise, using the Area under the ROC curve for ranking. However, the datasets were much larger, and the parameter space was more complex, making it significantly more difficult. Special efforts were made to increase the realism of the challenge by using real advanced-detector noise for the test data and simplifying the application of winning solutions to real \gw data.

In this competition, participants received time-frequency data from two gravitational wave detectors (\ligo-Hanford \& \ligo-Livingston) over a 3-month period, containing either real or simulated noise and possibly a simulated \cw signal. The data consisted of Short-time Fourier Transforms (SFTs) \citep{sfts_2022} and GPS timestamps, with realistic gaps due to detectors not always operating. Simulated signals, if present, spanned the entire dataset, and were defined by eight parameters including source location, frequency, and amplitude, drawn from astrophysical priors. The \snr distribution was tuned to include stronger signals that were easier for beginners, while more challenging signals were aimed at experienced Kaggle teams.

The results of this challenge are still pending, but like the first, it drew significant interest with 1,181 competitors across 936 teams. Surprisingly, most top-ranked teams concluded that \ml wasn't the best approach for this problem. Competitors initially tested standard \ml methods, but the weak \cw signal and large search space made them ineffective. This led many to explore more traditional data analysis techniques inspired by physical principles and g\gw literature. The winning method incorporated existing state-of-the-art \cw techniques, simplifying complex procedures and achieving faster computation using smart algorithm design and \gpu parallelism.

\section{Next-generation GW detectors}
\label{sec:nextgeneration}

Beyond the current \lvk\ network,
new third-generation ground-based detectors
like the \gls{ET} \citep{0264-9381-27-19-194002} in Europe
and \gls{CE} \citep{2017CQGra..34d4001A} in the US
are in planning.
These will be designed to bring great improvements in sensitivity,
but will also create completely new challenges in data analysis
due to the huge event rates \citep{Maggiore:2019uih}, much higher achievable \snrs which require more accurate waveform models and processing techniques \citep{Purrer:2019jcp},
and the much longer in-band duration of \cbc signals due to the lower minimum frequency.
The latter, for example, makes it necessary to include the Earth's movement in \cbc studies \citep{Zhao:2017cbb,Chen:2024kdc}, an aspect that for the \lvk\ is usually limited to \cw analyses.

\ml techniques may play crucial roles in dealing with these new challenges, and work is picking up in the community to develop new solutions. To give just a few examples,
some studies of \ml applications in this new observational regime include
\citet{10.1093/mnras/stac3797,Alhassan:2023lio} who adapted several off-the-shelf neural network for \bbh detection with \gls{ET} and found \gls{ResNet} to perform best, which was then tested on the first Einstein Telescope Mock Data Challenge \citep{Regimbau:2012ir}.
in \citet{PhysRevD.109.022006}, a deep-learning model was used to distinguish between \gw signals from cosmic string cusps and simulated blip glitches in ET data, using a realistic population of glitches for this future detector.

The sensitivity of these future ground-based \gw detectors at frequencies below approximately 10\,Hz may still be constrained by the Newtonian coupling of ground vibrations to the core optics of the detectors. This contribution of \gls{NNOISE} varies depending on the specific site and is influenced by the ambient seismic field, which, in turn, is contingent upon the geological makeup of the site and the distribution of surface and underground seismic-noise sources. in \citet{vanBeveren_2023} \ml was used for one of the candidate \gls{ET} sites to learn alongside seismic sensor networks and to predict seismic displacement noise at specific surface and underground locations. Additionally, a deep neural network has been developed to subtract Newtonian noise from the measured \gw strain data, showing its effectiveness in predicting Newtonian noise.

In \citet{9461904} the authors explored the search for a \gw background using deep neural networks, focusing on simulated astrophysical backgrounds generated by many \bbh coalescences. Specifically, the study examined the detection pipeline designed to isolate signal data from noisy detector backgrounds, utilizing three classes of deep neural network algorithms: a 1D \cnn, a 2D CNN, and a \lstm network. Results indicate that all three algorithms can effectively distinguish signals from noise with high precision for the \gls{ET} sensitivity level. 

Several space-based detectors, including \lisa~\citep{LISA:2017pwj}, Taiji~\citep{Hu:2017mde}, Tianqin~\citep{TianQin:2015yph}, and DECIGO~\citep{Kawamura:2011zz}, are being planned to access the \gw spectrum at frequencies lower than those accessible to ground-based detectors -- with \lisa\ firmly adopted by ESA for launch in the 2030s.
These next-generation detectors target a rich science case in a completely different regime than the LVK, including novel sources \citep{LISA:2022yao,LISAConsortiumWaveformWorkingGroup:2023arg,Ruan:2018tsw,TianQin:2020hid}, 
such as supermassive black hole binaries
and the very long-duration,
highly complex \emris \citep{Amaro-Seoane:2012lgq}.
\lisa\ and the analysis of its data will function very differently from the \lvk\ detectors, with each space craft receiving and sending out individual laser beams rather than them being reflected at each end point, time-delay interferometry \citep{Tinto:2020fcc} as a crucial ingredient to obtain sensitive strain measurements, and the time- and frequency-dependent detector response adding significant complications.
In addition, it is expected to operate in a signal-dominated regime, where many transients of different duration and characteristics overlap in time and a strong foreground of galactic sources dominates part of the sensitive band, leading to the ``global fit'' challenge \citep{Cornish:2005qw,Vallisneri:2008ye} of modelling and extracting all these contributions together.
And even more so than for third-generation ground-based detectors, the extremely high \snrs expected for some \lisa\ detections pose stringent requirements on the accuracy of waveform models and analysis techniques \citep{LISAConsortiumWaveformWorkingGroup:2023arg}.

So far, deep neural networks are already emerging as a popular approach
for studying the new signal types accessible with \lisa,
see for example \citet{Zhang:2022xuq,Zhao:2022qob,Yun:2023vwa,Mao:2023zdr,Sun:2023prd,Korsakova:2024sut,Xu:2024jbo,Ruan:2024qch}.
Besides evaluating individual waveforms, neural networks can also be useful to speed up population studies (like those discussed for the \lvk\ case in Sect.~\ref{sec:interpretation-pop-cosmo}), see e.g. \citet{Chapman-Bird:2022tvu} for using neural networks for \snr estimation and selection effects in studying \emri populations.
As for ground-based detectors,
glitch mitigation (see Sect.~\ref{sec:noisemitigation})
in detectors like \lisa\ 
can also be approached with \ml methods \citep{Houba:2024tyn}.
But much work remains to be done to see how \ml approaches can help with the overall challenges and fit into the overall \lisa\ data analysis pipeline.
They clearly carry great promises to deal with the complicated noise and detector properties and the more complicated signal types like \emris, but realistic end-to-end testing and integration with global fits will be crucial to realize this potential.

\ml and related methods can also be useful in the design of future detectors.
For example, \citet{Krenn:2023ljj} have used a gradient-descent optimization algorithm 
to explore the sensitivity of many possible configurations
of interferometric \gw detectors.
Such algorithms are also typically used in many training-based \ml approaches, though in this case direct optimization is performed over a large parameter space, without a training stage, so the authors categorize their work more generally as artificial intelligence.

\section{Summary and outlook}
\label{sec:summary}

The \ML revolution is rapidly changing the picture related to data analysis and, above all, the ability to provide increasingly reliable real-time answers in different fields. 
In this review, we have provided an overview of the application of \ml techniques in the field of \gw astronomy, discussing various \ml algorithms and methods used to address key challenges in \gw data analysis, including noise reduction and mitigation, signal detection, parameter estimation (\pe), classification and interpretation of astrophysical sources. We have only covered works released by the summer of 2024. More recent works will be considered in future updates of this living review. 

The main promises of \ml are twofold: greater speed, and greater flexibility and robustness where signal and noise models may be incomplete.

Many proof-of-principle studies have shown that \ml algorithms can be useful and computationally efficient instruments for identifying weak signals from noisy data and provide high-sensitivity \gw searches, supplementing classical signal processing techniques. Full end-to-end analyses and detailed comparisons against existing pipelines, to judge sensitivity and reliability in different parts of parameter space, have initially been scarce, but are now also increasingly appearing in the literature. Kaggle challenges and other mock data sets are important to this end, as are end-to-end searches on real open data.

Moreover, \ml methods can identify and suppress noise artifacts, and remove non-stationary and non-linear noise, giving thereby a better overall data quality. 
They can also facilitate the \pe process for \gw events by enabling faster determination of key parameters such as the masses and spins of \cbcs. This capability is essential for extracting detailed astrophysical information from observed signals. 
By training on labeled data sets, \ml models can distinguish between various source populations, aiding in the identification and categorization of observed events, and helping us in the classification task of different \gw signals.
\ml will also be able to help us with increasingly complex detector architectures, in identifying and reducing non-linear couplings in various noise sources.

Efficient real-time processing capabilities will be indispensable, especially for future detectors, when we will be forced to analyze many events per day and provide increasingly precise parameter estimates in a short timeframe to electromagnetic observing partners.
As the sensitivity of \gw detectors increases, particularly at low frequencies, we will also be faced with the problem of analyzing increasingly long transient events and even overlapping signals.
\ml applications in this area may be one of the solutions to untangle signals of particular interest from the astrophysical background.
At the same time, ``multimodal'' \ml methods that use multiple inputs from different observational domains to characterize the same astrophysical object, or populations of similar events or sources, may be a key advance for \gls{multimessenger} astronomy.

In addition to computational efficiency, \ml also holds the promise of making \gw data analysis more robust towards the complications of real detector noise, and to signal model limitations.
Generative noise models can help overcome the limitations of standard techniques based on Gaussian likelihoods.
Where complete and accurate signal models are difficult to obtain, \ml techniques may help with better inter- and extrapolation properties, though the details highly depend on the structure of the parameter space under study and the amount of physical intuition that can go into constructing the algorithms and, if applicable, their training data.

In general, across \gw astrophysics it is too early to say that the traditional data analysis methods have been replaced or superseded by \ml techniques. While they have demonstrated their importance and usefulness as being more efficient and faster for some applications, in others they are merely another alternative in a large toolbox with certain benefits, but not necessarily superior overall. Also, any new algorithms proposed to a large collaboration like the LVK -- not limited to \ml -- must always undergo a thorough review process before becoming an official and production-level analysis line. 
Examples where \ml is already used as a production tool by the \lvk include noise subtraction techniques and the low-latency classification of compact binary coalescence alerts.
For search pipelines and parameter estimation, various \ml methods are gearing up for production use, and this is likely to become a growing trend before the next observing run (O5).

So, we expect that in the coming years, \ml will see an increasingly deep application in production pipelines for \gw detection, classification and \pe, both inside the LVK and in the wider \gw community.
Still, such broad adaptation will also confront us with choices for the technical solutions best suited to this type of analysis.
The availability of increasingly high-performance and specific hardware for data analysis algorithms may also determine the best choices for innovative strategies, distinguishing algorithms as dedicated to real-time or offline analysis.
Many areas such as persistent \gw signals and searches with \gw detectors for new physics also are still in very early stages for exploring \ml solutions, and we expect more progress in the coming years.

On the technical side, many more recent innovations from the wider \ml community are still making their way into \gw practice. Initial studies have often focused on a few select classifier algorithms and neural network architectures, with \cnns initially particularly prominent, but a more diverse array of algorithms and network models are now being explored. Certainly, in the coming years new innovative techniques will continue to be either newly developed or adapted from other fields.

Meanwhile, it will be increasingly important to work on the the remaining conceptual and practical challenges of \ml techniques.
For example, while very computationally efficient at evaluation time, neural networks in particular can have immense computing costs for training, leading to high computing budgets overall, bottlenecks on GPUs which are still comparatively rare on many computing clusters used within the \gw community, and possibly severe climate impacts.
More fundamentally, questions have been raised about the capability of \ml-based \gw signal searches to provide reliable significance estimates on their own, or at least about how well these extrapolate from standard training/testing workflows to real applications.
Another often-raised criticism is the lacking reproducibility and explainability of many results obtained with \ML techniques, including the possibility of unforeseen failure modes. We expect in the years to come that an increasing emphasis will be placed on improving these aspects, with the adoption of best practices and novel architectures from the growing field of explainable/interpretable \ml/AI research, and also the need to make them part of the standard pipeline of \ml-based analyses. 

In summary, we look forward to an increasingly permeating role of \ml in the field of \GW research and we expect an evolution of techniques also related to the deployment of next-generation detectors. 

\bmhead{Acknowledgments}
This material is based upon work supported by NSF's LIGO Laboratory which is a major facility fully funded by the National Science Foundation.
We thank Michal Bejger, Christopher Berry, Sophie Bini, Christian Chapman-Bird, Thomas Dent, Maxime Fays, Shreejit Jadhav, Shasvath Kapadia, Jess McIver, Andrew L. Miller, Tanmaya Mishra, Carlos Palenzuela, Muhammed Saleem, Rodrigo Tenorio, and Daniel Williams for feedback on the manuscript and/or for suggesting additional references.
M.C.~was partially supported by the U.S.~National Science Foundation under awards PHY-2011334, PHY-2219212 and PHY-2308693.
D.K. was supported by
the Universitat de les Illes Balears (UIB);
the Spanish Agencia Estatal de Investigaci\'on grants
CNS2022-135440,
PID2022-138626NB-I00,
RED2022-134204-E,
RED2022-134411-T,
funded by MICIU/AEI/10.13039/501100011033,
the European Union NextGenerationEU/PRTR, and the ERDF/EU;
and the Comunitat Aut\`onoma de les Illes Balears
through the Conselleria d'Educaci\'o i Universitats
with funds from the European Union - NextGenerationEU/PRTR-C17.I1 (SINCO2022/6719)
and from the European Union - European Regional Development Fund (ERDF) (SINCO2022/18146).
This article has been assigned document numbers LIGO-P2400077 and VIR-0206A-24.

\bibliography{sn-bibliography-new}

\begin{thebibliography}{475}
\providecommand{\natexlab}[1]{#1}
\providecommand{\url}[1]{{#1}}
\providecommand{\urlprefix}{URL }
\providecommand{\doi}[1]{\url{https://doi.org/#1}}
\providecommand{\eprint}[2][]{\url{#2}}
 \bibcommenthead

\bibitem[{Aasi et~al(2015)}]{LIGOScientific:2014pky}
Aasi J, et~al (2015) {Advanced LIGO}. Class Quant Grav 32:074001. \doi{10.1088/0264-9381/32/7/074001}, {\href{https://arxiv.org/abs/1411.4547}{{arXiv:1411.4547}}} {[gr-qc]}

\bibitem[{Abac et~al(2024)}]{KAGRA:2024ipf}
Abac AG, et~al (2024) {Ultralight vector dark matter search using data from the KAGRA O3GK run}. Phys Rev D 110(4):042001. \doi{10.1103/PhysRevD.110.042001}, {\href{https://arxiv.org/abs/2403.03004}{{arXiv:2403.03004}}} {[astro-ph.CO]}

\bibitem[{{Abadie} et~al(2012)}]{2012PhRvD..85l2007A}
{Abadie} J, et~al (2012) {All-sky search for gravitational-wave bursts in the second joint LIGO-Virgo run}. \prd 85(12):122007. \doi{10.1103/PhysRevD.85.122007}, {\href{https://arxiv.org/abs/1202.2788}{{arXiv:1202.2788}}} {[gr-qc]}

\bibitem[{Abbott et~al(2016{\natexlab{a}})}]{Abbott_2016}
Abbott BP, et~al (2016{\natexlab{a}}) Characterization of transient noise in advanced ligo relevant to gravitational wave signal gw150914. Class Quantum Grav 33(13):134001. \doi{10.1088/0264-9381/33/13/134001}, {\href{https://arxiv.org/abs/1602.03844}{{arXiv:1602.03844}}} {[gr-qc]}

\bibitem[{Abbott et~al(2016{\natexlab{b}})}]{LIGOScientific:2016gtq}
Abbott BP, et~al (2016{\natexlab{b}}) {Characterization of transient noise in Advanced LIGO relevant to gravitational wave signal GW150914}. Class Quant Grav 33(13):134001. \doi{10.1088/0264-9381/33/13/134001}, {\href{https://arxiv.org/abs/1602.03844}{{arXiv:1602.03844}}} {[gr-qc]}

\bibitem[{Abbott et~al(2016{\natexlab{c}})}]{Abbott:2016blz}
Abbott BP, et~al (2016{\natexlab{c}}) {Observation of Gravitational Waves from a Binary Black Hole Merger}. Phys Rev Lett 116(6):061102. \doi{10.1103/PhysRevLett.116.061102}, {\href{https://arxiv.org/abs/1602.03837}{{arXiv:1602.03837}}} {[gr-qc]}

\bibitem[{{Abbott} et~al(2017)}]{2017CQGra..34d4001A}
{Abbott} BP, et~al (2017) {Exploring the sensitivity of next generation gravitational wave detectors}. Class Quantum Grav 34(4):044001. \doi{10.1088/1361-6382/aa51f4}, {\href{https://arxiv.org/abs/1607.08697}{{arXiv:1607.08697}}} {[astro-ph.IM]}

\bibitem[{Abbott et~al(2017{\natexlab{a}})}]{LIGOScientific:2017zic}
Abbott BP, et~al (2017{\natexlab{a}}) {Gravitational Waves and Gamma-rays from a Binary Neutron Star Merger: GW170817 and GRB 170817A}. Astrophys J Lett 848(2):L13. \doi{10.3847/2041-8213/aa920c}, {\href{https://arxiv.org/abs/1710.05834}{{arXiv:1710.05834}}} {[astro-ph.HE]}

\bibitem[{Abbott et~al(2017{\natexlab{b}})}]{LIGOScientific:2017vwq}
Abbott BP, et~al (2017{\natexlab{b}}) {GW170817: Observation of Gravitational Waves from a Binary Neutron Star Inspiral}. Phys Rev Lett 119(16):161101. \doi{10.1103/PhysRevLett.119.161101}, {\href{https://arxiv.org/abs/1710.05832}{{arXiv:1710.05832}}} {[gr-qc]}

\bibitem[{Abbott et~al(2017{\natexlab{c}})}]{LIGOScientific:2017ync}
Abbott BP, et~al (2017{\natexlab{c}}) {Multi-messenger Observations of a Binary Neutron Star Merger}. Astrophys J Lett 848(2):L12. \doi{10.3847/2041-8213/aa91c9}, {\href{https://arxiv.org/abs/1710.05833}{{arXiv:1710.05833}}} {[astro-ph.HE]}

\bibitem[{{Abbott} et~al(2018)}]{2018CQGra..35f5010A}
{Abbott} BP, et~al (2018) {Effects of data quality vetoes on a search for compact binary coalescences in Advanced LIGO's first observing run}. Class Quantum Grav 35:065010. \doi{10.1088/1361-6382/aaaafa}, {\href{https://arxiv.org/abs/1710.02185}{{arXiv:1710.02185}}} {[gr-qc]}

\bibitem[{Abbott et~al(2018)}]{KAGRA:2013rdx}
Abbott BP, et~al (2018) {Prospects for observing and localizing gravitational-wave transients with Advanced LIGO, Advanced Virgo and KAGRA}. Living Rev Rel 21(1):3. \doi{10.1007/s41114-020-00026-9}, {\href{https://arxiv.org/abs/1304.0670}{{arXiv:1304.0670}}} {[gr-qc]}

\bibitem[{Abbott et~al(2019)}]{PhysRevX.9.031040}
Abbott BP, et~al (2019) {GWTC-1: A Gravitational-Wave Transient Catalog of Compact Binary Mergers Observed by LIGO and Virgo during the First and Second Observing Runs}. Phys Rev X 9:031040. \doi{10.1103/PhysRevX.9.031040}, {\href{https://arxiv.org/abs/1811.12907}{{arXiv:1811.12907}}} {[astro-ph.HE]}

\bibitem[{{Abbott} et~al(2019)}]{2019PhRvX...9a1001A}
{Abbott} BP, et~al (2019) {Properties of the Binary Neutron Star Merger GW170817}. Physical Review X 9(1):011001. \doi{10.1103/PhysRevX.9.011001}, {\href{https://arxiv.org/abs/1805.11579}{{arXiv:1805.11579}}} {[gr-qc]}

\bibitem[{Abbott et~al(2019)}]{Abbott:2018hgk}
Abbott BP, et~al (2019) {Search for gravitational waves from a long-lived remnant of the binary neutron star merger GW170817}. ApJ 875:160. \doi{10.3847/1538-4357/ab0f3d}, {\href{https://arxiv.org/abs/1810.02581}{{arXiv:1810.02581}}} {[gr-qc]}

\bibitem[{Abbott et~al(2020{\natexlab{a}})}]{LIGOScientific:2019hgc}
Abbott BP, et~al (2020{\natexlab{a}}) {A guide to LIGO-Virgo detector noise and extraction of transient gravitational-wave signals}. Class Quant Grav 37(5):055002. \doi{10.1088/1361-6382/ab685e}, {\href{https://arxiv.org/abs/1908.11170}{{arXiv:1908.11170}}} {[gr-qc]}

\bibitem[{Abbott et~al(2020{\natexlab{b}})}]{LIGOScientific:2020aai}
Abbott BP, et~al (2020{\natexlab{b}}) {GW190425: Observation of a Compact Binary Coalescence with Total Mass $\sim 3.4 M_{\odot}$}. Astrophys J Lett 892(1):L3. \doi{10.3847/2041-8213/ab75f5}, {\href{https://arxiv.org/abs/2001.01761}{{arXiv:2001.01761}}} {[astro-ph.HE]}

\bibitem[{{Abbott} et~al(2021)}]{2021PhRvD.104l2004A}
{Abbott} R, et~al (2021) {All-sky search for short gravitational-wave bursts in the third Advanced LIGO and Advanced Virgo run}. \prd 104(12):122004. \doi{10.1103/PhysRevD.104.122004}, {\href{https://arxiv.org/abs/2107.03701}{{arXiv:2107.03701}}} {[gr-qc]}

\bibitem[{Abbott et~al(2021{\natexlab{a}})}]{LIGOScientific:2021nrg}
Abbott R, et~al (2021{\natexlab{a}}) {Constraints on Cosmic Strings Using Data from the Third Advanced LIGO-Virgo Observing Run}. Phys Rev Lett 126(24):241102. \doi{10.1103/PhysRevLett.126.241102}, {\href{https://arxiv.org/abs/2101.12248}{{arXiv:2101.12248}}} {[gr-qc]}

\bibitem[{Abbott et~al(2021{\natexlab{b}})}]{PhysRevX.11.021053}
Abbott R, et~al (2021{\natexlab{b}}) {GWTC-2: Compact Binary Coalescences Observed by LIGO and Virgo during the First Half of the Third Observing Run}. Phys Rev X 11:021053. \doi{10.1103/PhysRevX.11.021053}, {\href{https://arxiv.org/abs/2010.14527}{{arXiv:2010.14527}}} {[gr-qc]}

\bibitem[{Abbott et~al(2021{\natexlab{c}})}]{LIGOScientific:2021qlt}
Abbott R, et~al (2021{\natexlab{c}}) {Observation of Gravitational Waves from Two Neutron Star-Black Hole Coalescences}. Astrophys J Lett 915(1):L5. \doi{10.3847/2041-8213/ac082e}, {\href{https://arxiv.org/abs/2106.15163}{{arXiv:2106.15163}}} {[astro-ph.HE]}

\bibitem[{Abbott et~al(2021{\natexlab{d}})}]{LIGOScientific:2019lzm}
Abbott R, et~al (2021{\natexlab{d}}) {Open data from the first and second observing runs of Advanced LIGO and Advanced Virgo}. SoftwareX 13:100658. \doi{10.1016/j.softx.2021.100658}, {\href{https://arxiv.org/abs/1912.11716}{{arXiv:1912.11716}}} {[gr-qc]}

\bibitem[{Abbott et~al(2021{\natexlab{e}})}]{LIGOScientific:2021izm}
Abbott R, et~al (2021{\natexlab{e}}) {Search for Lensing Signatures in the Gravitational-Wave Observations from the First Half of LIGO-Virgo's Third Observing Run}. Astrophys J 923(1):14. \doi{10.3847/1538-4357/ac23db}, {\href{https://arxiv.org/abs/2105.06384}{{arXiv:2105.06384}}} {[gr-qc]}

\bibitem[{Abbott et~al(2022{\natexlab{a}})}]{LIGOScientific:2021rnv}
Abbott R, et~al (2022{\natexlab{a}}) {All-sky search for gravitational wave emission from scalar boson clouds around spinning black holes in LIGO O3 data}. Phys Rev D 105(10):102001. \doi{10.1103/PhysRevD.105.102001}, {\href{https://arxiv.org/abs/2111.15507}{{arXiv:2111.15507}}} {[astro-ph.HE]}

\bibitem[{Abbott et~al(2022{\natexlab{b}})}]{LIGOScientific:2021ffg}
Abbott R, et~al (2022{\natexlab{b}}) {Constraints on dark photon dark matter using data from LIGO's and Virgo's third observing run}. Phys Rev D 105(6):063030. \doi{10.1103/PhysRevD.105.063030}, {\href{https://arxiv.org/abs/2105.13085}{{arXiv:2105.13085}}} {[astro-ph.CO]}

\bibitem[{Abbott et~al(2022{\natexlab{c}})}]{LIGOScientific:2021job}
Abbott R, et~al (2022{\natexlab{c}}) {Search for Subsolar-Mass Binaries in the First Half of Advanced LIGO's and Advanced Virgo's Third Observing Run}. Phys Rev Lett 129(6):061104. \doi{10.1103/PhysRevLett.129.061104}, {\href{https://arxiv.org/abs/2109.12197}{{arXiv:2109.12197}}} {[astro-ph.CO]}

\bibitem[{Abbott et~al(2023{\natexlab{a}})}]{O3_cosmology_2023}
Abbott R, et~al (2023{\natexlab{a}}) {Constraints on the Cosmic Expansion History from GWTC-3}. \apj 949(2):76. \doi{10.3847/1538-4357/ac74bb}, {\href{https://arxiv.org/abs/2111.03604}{{arXiv:2111.03604}}} {[astro-ph.CO]}

\bibitem[{Abbott et~al(2023{\natexlab{b}})}]{KAGRA:2021vkt}
Abbott R, et~al (2023{\natexlab{b}}) {GWTC-3: Compact Binary Coalescences Observed by LIGO and Virgo during the Second Part of the Third Observing Run}. Phys Rev X 13(4):041039. \doi{10.1103/PhysRevX.13.041039}, {\href{https://arxiv.org/abs/2111.03606}{{arXiv:2111.03606}}} {[gr-qc]}

\bibitem[{Abbott et~al(2023{\natexlab{c}})}]{KAGRA:2023pio}
Abbott R, et~al (2023{\natexlab{c}}) {Open Data from the Third Observing Run of LIGO, Virgo, KAGRA, and GEO}. Astrophys J Suppl 267(2):29. \doi{10.3847/1538-4365/acdc9f}, {\href{https://arxiv.org/abs/2302.03676}{{arXiv:2302.03676}}} {[gr-qc]}

\bibitem[{Abbott et~al(2024{\natexlab{a}})}]{LIGOScientific:2021usb}
Abbott R, et~al (2024{\natexlab{a}}) {GWTC-2.1: Deep extended catalog of compact binary coalescences observed by LIGO and Virgo during the first half of the third observing run}. Phys Rev D 109(2):022001. \doi{10.1103/PhysRevD.109.022001}, {\href{https://arxiv.org/abs/2108.01045}{{arXiv:2108.01045}}} {[gr-qc]}

\bibitem[{Abbott et~al(2024{\natexlab{b}})}]{LIGOScientific:2023bwz}
Abbott R, et~al (2024{\natexlab{b}}) {Search for Gravitational-lensing Signatures in the Full Third Observing Run of the LIGO-Virgo Network}. Astrophys J 970(2):191. \doi{10.3847/1538-4357/ad3e83}, {\href{https://arxiv.org/abs/2304.08393}{{arXiv:2304.08393}}} {[gr-qc]}

\bibitem[{Abbott et~al(2024{\natexlab{c}})}]{LIGOScientific:2022sts}
Abbott R, et~al (2024{\natexlab{c}}) {Search for Gravitational-wave Transients Associated with Magnetar Bursts in Advanced LIGO and Advanced Virgo Data from the Third Observing Run}. Astrophys J 966(1):137. \doi{10.3847/1538-4357/ad27d3}, {\href{https://arxiv.org/abs/2210.10931}{{arXiv:2210.10931}}} {[astro-ph.HE]}

\bibitem[{Abbott et~al(2022{\natexlab{d}})Abbott, Buffaz, Vieira, Cabero, Haggard, Mahabal, and McIver}]{Abbott:2021cuf}
Abbott TC, Buffaz E, Vieira N, et~al (2022{\natexlab{d}}) {GWSkyNet-Multi: A Machine-learning Multiclass Classifier for LIGO-Virgo Public Alerts}. Astrophys J 927(2):232. \doi{10.3847/1538-4357/ac5019}, {\href{https://arxiv.org/abs/2111.04015}{{arXiv:2111.04015}}} {[astro-ph.IM]}

\bibitem[{Abbott et~al(2022{\natexlab{e}})Abbott, Buffaz, Vieira, Cabero, Haggard, Mahabal, and McIver}]{Abbott_2022}
Abbott TC, Buffaz E, Vieira N, et~al (2022{\natexlab{e}}) {GWSkyNet-Multi: A Machine-learning Multiclass Classifier for LIGO–Virgo Public Alerts}. Astrophys J 927(2):232. \doi{10.3847/1538-4357/ac5019}, {\href{https://arxiv.org/abs/2111.04015}{{arXiv:2111.04015}}} {[astro-ph]}

\bibitem[{Abe et~al(2022)}]{KAGRA:2022qtq}
Abe H, et~al (2022) {The Current Status and Future Prospects of KAGRA, the Large-Scale Cryogenic Gravitational Wave Telescope Built in the Kamioka Underground}. Galaxies 10(3):63. \doi{10.3390/galaxies10030063}

\bibitem[{Accadia et~al(2012)}]{Accadia:2012zz}
Accadia T, et~al (2012) {The NoEMi (Noise Frequency Event Miner) framework}. J Phys Conf Ser 363:012037. \doi{10.1088/1742-6596/363/1/012037}

\bibitem[{Acernese et~al(2014)}]{Acernese_2015}
Acernese F, et~al (2014) {Advanced Virgo: a second-generation interferometric gravitational wave detector}. Class Quantum Grav 32(2):024001. \doi{10.1088/0264-9381/32/2/024001}

\bibitem[{Acernese et~al(2018)}]{Virgo:2018gxa}
Acernese F, et~al (2018) {Calibration of Advanced Virgo and Reconstruction of the Gravitational Wave Signal $h(t)$ during the Observing Run O2}. Class Quant Grav 35(20):205004. \doi{10.1088/1361-6382/aadf1a}, {\href{https://arxiv.org/abs/1807.03275}{{arXiv:1807.03275}}} {[gr-qc]}

\bibitem[{Acernese et~al(2023{\natexlab{a}})}]{Acernese_2023}
Acernese F, et~al (2023{\natexlab{a}}) Virgo detector characterization and data quality: results from the o3 run. Class Quantum Grav 40(18):185006. \doi{10.1088/1361-6382/acd92d}

\bibitem[{Acernese et~al(2023{\natexlab{b}})}]{F_Acernese_2023}
Acernese F, et~al (2023{\natexlab{b}}) Virgo detector characterization and data quality: tools. Class Quantum Grav 40(18):185005. \doi{10.1088/1361-6382/acdf36}

\bibitem[{Afshordi et~al(2023)}]{LISAConsortiumWaveformWorkingGroup:2023arg}
Afshordi N, et~al (2023) {Waveform Modelling for the Laser Interferometer Space Antenna}. arXiv e-prints {\href{https://arxiv.org/abs/2311.01300}{{arXiv:2311.01300}}} {[gr-qc]}

\bibitem[{Agazie et~al(2024)}]{InternationalPulsarTimingArray:2023mzf}
Agazie G, et~al (2024) {Comparing Recent Pulsar Timing Array Results on the Nanohertz Stochastic Gravitational-wave Background}. Astrophys J 966(1):105. \doi{10.3847/1538-4357/ad36be}, {\href{https://arxiv.org/abs/2309.00693}{{arXiv:2309.00693}}} {[astro-ph.HE]}

\bibitem[{Akutsu et~al(2020)Akutsu, Ando, Arai, Arai, Araki, Araya, Aritomi, Aso, Bae, Bae et~al}]{10.1093/ptep/ptaa125}
Akutsu T, Ando M, Arai K, et~al (2020) {Overview of KAGRA: Detector design and construction history}. Progress of Theoretical and Experimental Physics 2021(5):05A101. \doi{10.1093/ptep/ptaa125}

\bibitem[{Alestas et~al(2024)Alestas, Morras, Yamamoto, Garcia-Bellido, Kuroyanagi, and Nesseris}]{Alestas:2024ubs}
Alestas G, Morras G, Yamamoto TS, et~al (2024) {Applying the Viterbi algorithm to planetary-mass black hole searches}. Phys Rev D 109(12):123516. \doi{10.1103/PhysRevD.109.123516}, {\href{https://arxiv.org/abs/2401.02314}{{arXiv:2401.02314}}} {[astro-ph.CO]}

\bibitem[{{Alfaidi} and {Messenger}(2024)}]{2024arXiv240204589A}
{Alfaidi} R, {Messenger} C (2024) {Long Short-Term Memory for Early Warning Detection of Gravitational Waves}. arXiv e-prints \doi{10.48550/arXiv.2402.04589}, {\href{https://arxiv.org/abs/2402.04589}{{arXiv:2402.04589}}} {[gr-qc]}

\bibitem[{Alhassan et~al(2022)Alhassan, Bulik, and Suchenek}]{10.1093/mnras/stac3797}
Alhassan W, Bulik T, Suchenek M (2022) {Detection of Einstein telescope gravitational wave signals from binary black holes using deep learning}. Mon Not R Astron, Soc 519(3):3843--3850. \doi{10.1093/mnras/stac3797}

\bibitem[{Alhassan et~al(2023)Alhassan, Bulik, and Suchenek}]{Alhassan:2023lio}
Alhassan W, Bulik T, Suchenek M (2023) {Einstein Telescope: binary black holes gravitational wave signals detection from three detectors combined data using deep learning}. arXiv e-prints {\href{https://arxiv.org/abs/2310.10409}{{arXiv:2310.10409}}} {[astro-ph.IM]}

\bibitem[{Allen and Romano(1999)}]{Allen:1997ad}
Allen B, Romano JD (1999) {Detecting a stochastic background of gravitational radiation: Signal processing strategies and sensitivities}. Phys Rev D 59:102001. \doi{10.1103/PhysRevD.59.102001}, {\href{https://arxiv.org/abs/gr-qc/9710117}{{arXiv:gr-qc/9710117}}}

\bibitem[{Allen et~al(2022)Allen, Goetz, Keitel, Landry, Mendell, Prix, Riles, and Wette}]{sfts_2022}
Allen B, Goetz E, Keitel D, et~al (2022) {SFT Data Format Version 2--3 Specification}. \url{https://dcc.ligo.org/T040164/public}

\bibitem[{Alvarez-Lopez(2023)}]{alvarezthesis}
Alvarez-Lopez MS (2023) {Improving our ability to distinguish gravitational-wave signals from detector transient noise for the fourth LIGO-Virgo-KAGRA observing run}. PhD thesis, {Universidad de los Andes, Bogot\'a, Colombia}, \urlprefix\url{https://dcc.ligo.org/LIGO-P2300230/public}

\bibitem[{Alvarez-Lopez et~al(2023)Alvarez-Lopez, Liyanage, Ding, Ng, and McIver}]{alvarezlopez2023gspynettree}
Alvarez-Lopez S, Liyanage A, Ding J, et~al (2023) {GSpyNetTree: A signal-vs-glitch classifier for gravitational-wave event candidates}. arXiv e-prints {\href{https://arxiv.org/abs/2304.09977}{{arXiv:2304.09977}}} {[gr-qc]}

\bibitem[{Alvey et~al(2023)Alvey, Bhardwaj, Domcke, Pieroni, and Weniger}]{Alvey:2023npw}
Alvey J, Bhardwaj U, Domcke V, et~al (2023) {Simulation-based inference for stochastic gravitational wave background data analysis}. arXiv e-prints {\href{https://arxiv.org/abs/2309.07954}{{arXiv:2309.07954}}} {[gr-qc]}

\bibitem[{Amaro-Seoane(2018)}]{Amaro-Seoane:2012lgq}
Amaro-Seoane P (2018) {Relativistic dynamics and extreme mass ratio inspirals}. Living Rev Rel 21(1):4. \doi{10.1007/s41114-018-0013-8}, {\href{https://arxiv.org/abs/1205.5240}{{arXiv:1205.5240}}} {[astro-ph.CO]}

\bibitem[{Amaro-Seoane et~al(2017)}]{LISA:2017pwj}
Amaro-Seoane P, et~al (2017) {Laser Interferometer Space Antenna}. arXiv e-prints {\href{https://arxiv.org/abs/1702.00786}{{arXiv:1702.00786}}} {[astro-ph.IM]}

\bibitem[{Andres-Carcasona et~al(2023)Andres-Carcasona, Menendez-Vazquez, Martinez, and Mir}]{Andres-Carcasona:2022prl}
Andres-Carcasona M, Menendez-Vazquez A, Martinez M, et~al (2023) {Searches for mass-asymmetric compact binary coalescence events using neural networks in the LIGO/Virgo third observation period}. Phys Rev D 107(8):082003. \doi{10.1103/PhysRevD.107.082003}, {\href{https://arxiv.org/abs/2212.02829}{{arXiv:2212.02829}}} {[gr-qc]}

\bibitem[{{Antelis} et~al(2022){Antelis}, {Cavaglia}, {Hansen}, {Morales}, {Moreno}, {Mukherjee}, {Szczepa{\'n}czyk}, and {Zanolin}}]{2022PhRvD.105h4054A}
{Antelis} JM, {Cavaglia} M, {Hansen} T, et~al (2022) {Using supervised learning algorithms as a follow-up method in the search of gravitational waves from core-collapse supernovae}. Phys Rev D 105(8):084054. \doi{10.1103/PhysRevD.105.084054}, {\href{https://arxiv.org/abs/2111.07219}{{arXiv:2111.07219}}} {[gr-qc]}

\bibitem[{Arulkumaran et~al(2017)Arulkumaran, Deisenroth, Brundage, and Bharath}]{8103164}
Arulkumaran K, Deisenroth MP, Brundage M, et~al (2017) Deep reinforcement learning: A brief survey. IEEE Signal Processing Magazine 34(6):26--38. \doi{10.1109/MSP.2017.2743240}

\bibitem[{Ashok et~al(2024)Ashok, Covas, Prix, and Papa}]{Ashok:2024fts}
Ashok A, Covas PB, Prix R, et~al (2024) {Bayesian F-statistic-based parameter estimation of continuous gravitational waves from known pulsars}. Phys Rev D 109(10):104002. \doi{10.1103/PhysRevD.109.104002}, {\href{https://arxiv.org/abs/2401.17025}{{arXiv:2401.17025}}} {[gr-qc]}

\bibitem[{Ashton and Prix(2018)}]{Ashton:2018ure}
Ashton G, Prix R (2018) {Hierarchical multistage MCMC follow-up of continuous gravitational wave candidates}. Phys Rev D 97(10):103020. \doi{10.1103/PhysRevD.97.103020}, {\href{https://arxiv.org/abs/1802.05450}{{arXiv:1802.05450}}} {[astro-ph.IM]}

\bibitem[{{Ashton} and {Talbot}(2021)}]{2021MNRAS.507.2037A}
{Ashton} G, {Talbot} C (2021) {BILBY-MCMC: an MCMC sampler for gravitational-wave inference}. \mnras 507(2):2037--2051. \doi{10.1093/mnras/stab2236}, {\href{https://arxiv.org/abs/2106.08730}{{arXiv:2106.08730}}} {[gr-qc]}

\bibitem[{{Ashton} et~al(2019){Ashton}, {H{\"u}bner}, {Lasky}, {Talbot}, {Ackley}, {Biscoveanu}, {Chu}, {Divakarla}, {Easter}, and {Goncharov}}]{2019ApJS..241...27A}
{Ashton} G, {H{\"u}bner} M, {Lasky} PD, et~al (2019) {BILBY: A user-friendly Bayesian inference library for gravitational-wave astronomy}. Astrophys J Suppl 241(2):27. \doi{10.3847/1538-4365/ab06fc}, {\href{https://arxiv.org/abs/1811.02042}{{arXiv:1811.02042}}} {[astro-ph.IM]}

\bibitem[{Ashton et~al(2022)}]{Ashton:2022grj}
Ashton G, et~al (2022) {Nested sampling for physical scientists}. Nature 2. \doi{10.1038/s43586-022-00121-x}, {\href{https://arxiv.org/abs/2205.15570}{{arXiv:2205.15570}}} {[stat.CO]}

\bibitem[{{Astone} et~al(2018){Astone}, {Cerd{\'a}-Dur{\'a}n}, {Di Palma}, {Drago}, {Muciaccia}, {Palomba}, and {Ricci}}]{2018PhRvD..98l2002A}
{Astone} P, {Cerd{\'a}-Dur{\'a}n} P, {Di Palma} I, et~al (2018) {New method to observe gravitational waves emitted by core collapse supernovae}. Phys Rev D 98:122002. \doi{10.1103/PhysRevD.98.122002}, {\href{https://arxiv.org/abs/1812.05363}{{arXiv:1812.05363}}} {[astro-ph.IM]}

\bibitem[{Attadio et~al(2024)}]{Attadio:2024otf}
Attadio F, et~al (2024) {A neural networks method to search for long transient gravitational waves}. arXiv e-prints {\href{https://arxiv.org/abs/2407.02391}{{arXiv:2407.02391}}} {[astro-ph.IM]}

\bibitem[{Aubin et~al(2021{\natexlab{a}})Aubin, Brighenti, Chierici, Estevez, Greco, Guidi, Juste, Marion, Mours, Nitoglia, Sauter, and Sordini}]{Aubin_2021}
Aubin F, Brighenti F, Chierici R, et~al (2021{\natexlab{a}}) The mbta pipeline for detecting compact binary coalescences in the third ligo–virgo observing run. Class Quantum Grav 38(9):095004. \doi{10.1088/1361-6382/abe913}

\bibitem[{Aubin et~al(2021{\natexlab{b}})}]{Aubin:2020goo}
Aubin F, et~al (2021{\natexlab{b}}) {The MBTA pipeline for detecting compact binary coalescences in the third LIGO-Virgo observing run}. Class Quant Grav 38(9):095004. \doi{10.1088/1361-6382/abe913}, {\href{https://arxiv.org/abs/2012.11512}{{arXiv:2012.11512}}} {[gr-qc]}

\bibitem[{{Auddy} et~al(2024){Auddy}, {Dey}, {Turner}, and {Basu}}]{2024MLS&T...5b5014A}
{Auddy} S, {Dey} R, {Turner} NJ, et~al (2024) {GRINN: a physics-informed neural network for solving hydrodynamic systems in the presence of self-gravity}. Machine Learning: Science and Technology 5(2):025014. \doi{10.1088/2632-2153/ad3a32}, {\href{https://arxiv.org/abs/2308.08010}{{arXiv:2308.08010}}} {[cs.LG]}

\bibitem[{Bacon et~al(2023)Bacon, Trovato, and Bejger}]{Bacon_2023}
Bacon P, Trovato A, Bejger M (2023) Denoising gravitational-wave signals from binary black holes with a dilated convolutional autoencoder. Machine Learning: Science and Technology 4(3):035024. \doi{10.1088/2632-2153/acd90f}

\bibitem[{Badaracco et~al(2020)}]{Badaracco:2020qmm}
Badaracco F, et~al (2020) {Machine learning for gravitational-wave detection: surrogate Wiener filtering for the prediction and optimized cancellation of Newtonian noise at Virgo}. Class Quant Grav 37(19):195016. \doi{10.1088/1361-6382/abab64}, {\href{https://arxiv.org/abs/2005.09289}{{arXiv:2005.09289}}} {[astro-ph.IM]}

\bibitem[{Bahaadini et~al(2017)Bahaadini, Rohani, Coughlin, Zevin, Kalogera, and Katsaggelos}]{Bahaadini:2017dqg}
Bahaadini S, Rohani N, Coughlin S, et~al (2017) {Deep Multi-view Models for Glitch Classification}. In: {42nd IEEE International Conference on Acoustics, Speech and Signal Processing (ICASSP'17)}, \eprint{1705.00034}

\bibitem[{Baltrusaitis et~al(2019)Baltrusaitis, Ahuja, and Morency}]{10.1109/TPAMI.2018.2798607}
Baltrusaitis T, Ahuja C, Morency LP (2019) Multimodal machine learning: A survey and taxonomy. IEEE Trans Pattern Anal Mach Intell 41(2):423–443. \doi{10.1109/TPAMI.2018.2798607}, \urlprefix\url{https://doi.org/10.1109/TPAMI.2018.2798607}

\bibitem[{Baltus et~al(2021)Baltus, Janquart, Lopez, Reza, Caudill, and Cudell}]{Baltus:2021nme}
Baltus G, Janquart J, Lopez M, et~al (2021) {Convolutional neural networks for the detection of the early inspiral of a gravitational-wave signal}. Phys Rev D 103:102003. \doi{10.1103/PhysRevD.103.102003}, {\href{https://arxiv.org/abs/2104.00594}{{arXiv:2104.00594}}} {[gr-qc]}

\bibitem[{{Baltus} et~al(2021){Baltus}, {Janquart}, {Lopez}, {Reza}, {Caudill}, and {Cudell}}]{2021arXiv210513664B}
{Baltus} G, {Janquart} J, {Lopez} M, et~al (2021) {Detecting the early inspiral of a gravitational-wave signal with convolutional neural networks}. arXiv e-prints arXiv:2105.13664. \doi{10.48550/arXiv.2105.13664}, {\href{https://arxiv.org/abs/2105.13664}{{arXiv:2105.13664}}} {[gr-qc]}

\bibitem[{Banagiri et~al(2019)Banagiri, Sun, Coughlin, and Melatos}]{Banagiri:2019obu}
Banagiri S, Sun L, Coughlin MW, et~al (2019) {Search strategies for long gravitational-wave transients: hidden Markov model tracking and seedless clustering}. Phys Rev D 100(2):024034. \doi{10.1103/PhysRevD.100.024034}, {\href{https://arxiv.org/abs/1903.02638}{{arXiv:1903.02638}}} {[astro-ph.IM]}

\bibitem[{Barack and Pound(2019)}]{Barack:2018yvs}
Barack L, Pound A (2019) {Self-force and radiation reaction in general relativity}. Rept Prog Phys 82(1):016904. \doi{10.1088/1361-6633/aae552}, {\href{https://arxiv.org/abs/1805.10385}{{arXiv:1805.10385}}} {[gr-qc]}

\bibitem[{Barack et~al(2019)}]{Barack:2018yly}
Barack L, et~al (2019) {Black holes, gravitational waves and fundamental physics: a roadmap}. Class Quant Grav 36(14):143001. \doi{10.1088/1361-6382/ab0587}, {\href{https://arxiv.org/abs/1806.05195}{{arXiv:1806.05195}}} {[gr-qc]}

\bibitem[{{Baron}(2019)}]{2019arXiv190407248B}
{Baron} D (2019) {Machine Learning in Astronomy: a practical overview}. arXiv e-prints arXiv:1904.07248. \doi{10.48550/arXiv.1904.07248}, {\href{https://arxiv.org/abs/1904.07248}{{arXiv:1904.07248}}} {[astro-ph.IM]}

\bibitem[{Bayley et~al(2019)Bayley, Woan, and Messenger}]{Bayley:2019bcb}
Bayley J, Woan G, Messenger C (2019) {Generalized application of the Viterbi algorithm to searches for continuous gravitational-wave signals}. Phys Rev D100(2):023006. \doi{10.1103/PhysRevD.100.023006}, {\href{https://arxiv.org/abs/1903.12614}{{arXiv:1903.12614}}} {[astro-ph.IM]}

\bibitem[{Bayley et~al(2020)Bayley, Messenger, and Woan}]{Bayley:2020zfa}
Bayley J, Messenger C, Woan G (2020) {Robust machine learning algorithm to search for continuous gravitational waves}. Phys Rev D 102(8):083024. \doi{10.1103/PhysRevD.102.083024}, {\href{https://arxiv.org/abs/2007.08207}{{arXiv:2007.08207}}} {[astro-ph.IM]}

\bibitem[{Bayley et~al(2022)Bayley, Messenger, and Woan}]{Bayley:2022hkz}
Bayley J, Messenger C, Woan G (2022) {Rapid parameter estimation for an all-sky continuous gravitational wave search using conditional varitational auto-encoders}. Phys Rev D 106(8):083022. \doi{10.1103/PhysRevD.106.083022}, {\href{https://arxiv.org/abs/2209.02031}{{arXiv:2209.02031}}} {[astro-ph.IM]}

\bibitem[{Beheshtipour and Papa(2020)}]{Beheshtipour:2020zhb}
Beheshtipour B, Papa MA (2020) {Deep learning for clustering of continuous gravitational wave candidates}. Phys Rev D 101(6):064009. \doi{10.1103/PhysRevD.101.064009}, {\href{https://arxiv.org/abs/2001.03116}{{arXiv:2001.03116}}} {[gr-qc]}

\bibitem[{Beheshtipour and Papa(2021)}]{Beheshtipour:2020nko}
Beheshtipour B, Papa MA (2021) {Deep learning for clustering of continuous gravitational wave candidates II: identification of low-SNR candidates}. Phys Rev D 103(6):064027. \doi{10.1103/PhysRevD.103.064027}, {\href{https://arxiv.org/abs/2012.04381}{{arXiv:2012.04381}}} {[gr-qc]}

\bibitem[{Benedetto et~al(2023)Benedetto, Gissi, Ciaparrone, and Troiano}]{Benedetto:2023jwn}
Benedetto V, Gissi F, Ciaparrone G, et~al (2023) {AI in Gravitational Wave Analysis, an Overview}. Appl Sciences 13(17):9886. \doi{10.3390/app13179886}

\bibitem[{Bengio et~al(2009)Bengio, Louradour, Collobert, and Weston}]{Bengio2009}
Bengio Y, Louradour J, Collobert R, et~al (2009) Curriculum learning. In: Proceedings of the 26th Annual International Conference on Machine Learning. Association for Computing Machinery, New York, NY, ICML '09, pp 41--48, \doi{10.1145/1553374.1553380}

\bibitem[{Berbel et~al(2024)Berbel, Miravet-Ten\'es, Chaudhary, Albanesi, Cavagli\`a, Zertuche, Tseneklidou, Zheng, Coughlin, and Toivonen}]{Berbel:2023vug}
Berbel M, Miravet-Ten\'es M, Chaudhary SS, et~al (2024) {Bayesian real-time classification of multi-messenger electromagnetic and gravitational-wave observations}. Class Quant Grav 41(8):085012. \doi{10.1088/1361-6382/ad3279}, {\href{https://arxiv.org/abs/2311.00045}{{arXiv:2311.00045}}} {[astro-ph.HE]}

\bibitem[{van Beveren et~al(2023)van Beveren, Bader, van~den Brand, Bulten, Campman, Koley, and Linde}]{vanBeveren_2023}
van Beveren V, Bader M, van~den Brand J, et~al (2023) {A study of deep neural networks for Newtonian noise subtraction at Terziet in Limburg—the Euregio Meuse-Rhine candidate site for Einstein Telescope}. Class Quantum Grav 40(20):205008. \doi{10.1088/1361-6382/acf3c8}

\bibitem[{{Bini} et~al(2023){Bini}, {Vedovato}, {Drago}, {Salemi}, and {Prodi}}]{2023CQGra..40m5008B}
{Bini} S, {Vedovato} G, {Drago} M, et~al (2023) {An autoencoder neural network integrated into gravitational-wave burst searches to improve the rejection of noise transients}. Class Quantum Grav 40(13):135008. \doi{10.1088/1361-6382/acd981}, {\href{https://arxiv.org/abs/2303.05986}{{arXiv:2303.05986}}} {[gr-qc]}

\bibitem[{Bini et~al(2024)Bini, Tiwari, Xu, Smith, Ebersold, Principe, Haney, Jetzer, and Prodi}]{PhysRevD.109.042009}
Bini S, Tiwari S, Xu Y, et~al (2024) Search for hyperbolic encounters of compact objects in the third ligo-virgo-kagra observing run. Phys Rev D 109:042009. \doi{10.1103/PhysRevD.109.042009}

\bibitem[{Bird et~al(2016)Bird, Cholis, Mu\~noz, Ali-Ha\"\i{}moud, Kamionkowski, Kovetz, Raccanelli, and Riess}]{Bird:2016dcv}
Bird S, Cholis I, Mu\~noz JB, et~al (2016) {Did LIGO detect dark matter?} Phys Rev Lett 116(20):201301. \doi{10.1103/PhysRevLett.116.201301}, {\href{https://arxiv.org/abs/1603.00464}{{arXiv:1603.00464}}} {[astro-ph.CO]}

\bibitem[{Biswas et~al(2020)Biswas, McIver, and Mahabal}]{Biswas_2020}
Biswas A, McIver J, Mahabal A (2020) New methods to assess and improve ligo detector duty cycle. Class Quantum Grav 37(17):175008. \doi{10.1088/1361-6382/ab8650}

\bibitem[{Biswas et~al(2024)Biswas, Smyrniotis, Liodis, and Stergioulas}]{Biswas:2023ceq}
Biswas B, Smyrniotis E, Liodis I, et~al (2024) {Bayesian investigation of the neutron star equation of state vs gravity degeneracy}. Phys Rev D 109(6):064048. \doi{10.1103/PhysRevD.109.064048}, {\href{https://arxiv.org/abs/2309.05420}{{arXiv:2309.05420}}} {[gr-qc]}

\bibitem[{Biswas et~al(2013)Biswas, Blackburn, Cao, Essick, Hodge, Katsavounidis, Kim, Kim, Le~Bigot, Lee, Oh, Oh, Son, Tao, Vaulin, and Wang}]{PhysRevD.88.062003}
Biswas R, Blackburn L, Cao J, et~al (2013) Application of machine learning algorithms to the study of noise artifacts in gravitational-wave data. Phys Rev D 88:062003. \doi{10.1103/PhysRevD.88.062003}

\bibitem[{Blackman et~al(2017{\natexlab{a}})Blackman, Field, Scheel, Galley, Hemberger, Schmidt, and Smith}]{Blackman:2017dfb}
Blackman J, Field SE, Scheel MA, et~al (2017{\natexlab{a}}) {A Surrogate Model of Gravitational Waveforms from Numerical Relativity Simulations of Precessing Binary Black Hole Mergers}. Phys Rev D95(10):104023. \doi{10.1103/PhysRevD.95.104023}, {\href{https://arxiv.org/abs/1701.00550}{{arXiv:1701.00550}}} {[gr-qc]}

\bibitem[{Blackman et~al(2017{\natexlab{b}})Blackman, Field, Scheel, Galley, Ott, Boyle, Kidder, Pfeiffer, and Szil\'agyi}]{Blackman:2017pcm}
Blackman J, Field SE, Scheel MA, et~al (2017{\natexlab{b}}) {Numerical relativity waveform surrogate model for generically precessing binary black hole mergers}. Phys Rev D96(2):024058. \doi{10.1103/PhysRevD.96.024058}, {\href{https://arxiv.org/abs/1705.07089}{{arXiv:1705.07089}}} {[gr-qc]}

\bibitem[{Blanchet(2014)}]{Blanchet:2013haa}
Blanchet L (2014) {Gravitational Radiation from Post-Newtonian Sources and Inspiralling Compact Binaries}. Living Rev Rel 17:2. \doi{10.12942/lrr-2014-2}, {\href{https://arxiv.org/abs/1310.1528}{{arXiv:1310.1528}}} {[gr-qc]}

\bibitem[{{Boccioli} and {Roberti}(2024)}]{2024Univ...10..148B}
{Boccioli} L, {Roberti} L (2024) {The Physics of Core-Collapse Supernovae: Explosion Mechanism and Explosive Nucleosynthesis}. Universe 10(3):148. \doi{10.3390/universe10030148}, {\href{https://arxiv.org/abs/2403.12942}{{arXiv:2403.12942}}} {[astro-ph.SR]}

\bibitem[{Boh\'e et~al(2017)}]{Bohe:2016gbl}
Boh\'e A, et~al (2017) {Improved effective-one-body model of spinning, nonprecessing binary black holes for the era of gravitational-wave astrophysics with advanced detectors}. Phys Rev D95(4):044028. \doi{10.1103/PhysRevD.95.044028}, {\href{https://arxiv.org/abs/1611.03703}{{arXiv:1611.03703}}} {[gr-qc]}

\bibitem[{Boudart(2023)}]{Boudart:2022apz}
Boudart V (2023) {Convolutional neural network to distinguish glitches from minute-long gravitational wave transients}. Phys Rev D 107(2):024007. \doi{10.1103/PhysRevD.107.024007}, {\href{https://arxiv.org/abs/2210.04588}{{arXiv:2210.04588}}} {[gr-qc]}

\bibitem[{Boudart and Fays(2022)}]{Boudart:2022xib}
Boudart V, Fays M (2022) {Machine learning algorithm for minute-long burst searches}. Phys Rev D 105(8):083007. \doi{10.1103/PhysRevD.105.083007}, {\href{https://arxiv.org/abs/2201.08727}{{arXiv:2201.08727}}} {[gr-qc]}

\bibitem[{Breiman(2001)}]{Breiman2001}
Breiman L (2001) Random forests. Machine Learning 45(1):5--32. \doi{10.1023/A:1010933404324}

\bibitem[{Brito et~al(2015)Brito, Cardoso, and Pani}]{Brito:2015oca}
Brito R, Cardoso V, Pani P (2015) Superradiance: Energy Extraction, Black-Hole Bombs and Implications for Astrophysics and Particle Physics, Lecture Notes in Physics, vol 906. Springer, Cham, \doi{10.1007/978-3-319-19000-6}, \eprint{1501.06570}

\bibitem[{{Burrows} and {Vartanyan}(2021)}]{2021Natur.589...29B}
{Burrows} A, {Vartanyan} D (2021) {Core-collapse supernova explosion theory}. Nature 589(7840):29--39. \doi{10.1038/s41586-020-03059-w}, {\href{https://arxiv.org/abs/2009.14157}{{arXiv:2009.14157}}} {[astro-ph.SR]}

\bibitem[{Cabero et~al(2020)Cabero, Mahabal, and McIver}]{Cabero:2020eik}
Cabero M, Mahabal A, McIver J (2020) {GWSkyNet: a real-time classifier for public gravitational-wave candidates}. Astrophys J Lett 904(1):L9. \doi{10.3847/2041-8213/abc5b5}, {\href{https://arxiv.org/abs/2010.11829}{{arXiv:2010.11829}}} {[gr-qc]}

\bibitem[{Cannon et~al(2021)Cannon, Caudill, Chan, Cousins, Creighton, Ewing, Fong, Godwin, Hanna, Hooper, Huxford, Magee, Meacher, Messick, Morisaki, Mukherjee, Ohta, Pace, Privitera, {de Ruiter}, Sachdev, Singer, Singh, Tapia, Tsukada, Tsuna, Tsutsui, Ueno, Viets, Wade, and Wade}]{CANNON2021100680}
Cannon K, Caudill S, Chan C, et~al (2021) Gstlal: A software framework for gravitational wave discovery. SoftwareX 14:100680. \doi{10.1016/j.softx.2021.100680}

\bibitem[{Carvalho et~al(2023)Carvalho, Ferreira, Malik, and Provid\^encia}]{Carvalho:2023ele}
Carvalho V, Ferreira M, Malik T, et~al (2023) {Decoding neutron star observations: Revealing composition through Bayesian neural networks}. Phys Rev D 108(4):043031. \doi{10.1103/PhysRevD.108.043031}, {\href{https://arxiv.org/abs/2306.06929}{{arXiv:2306.06929}}} {[nucl-th]}

\bibitem[{Cavaglia(2022)}]{Cavaglia:2022vlu}
Cavaglia M (2022) {Characterization of gravitational-wave detector noise with fractals}. Class Quant Grav 39(13):135012. \doi{10.1088/1361-6382/ac7325}, {\href{https://arxiv.org/abs/2201.09984}{{arXiv:2201.09984}}} {[gr-qc]}

\bibitem[{Cavagli{\`a} et~al(2018)Cavagli{\`a}, Staats, and Gill}]{CiCP-25-963}
Cavagli{\`a} M, Staats K, Gill T (2018) Finding the origin of noise transients in ligo data with machine learning. Commun Comput Phys 25(4):963--987. \doi{10.4208/cicp.OA-2018-0092}

\bibitem[{Cavaglia et~al(2020)Cavaglia, Gaudio, Hansen, Staats, Szczepa{\'n}czyk, and Zanolin}]{Cavaglia:2020qzp}
Cavaglia M, Gaudio S, Hansen T, et~al (2020) {Improving the background of gravitational-wave searches for core collapse supernovae: A machine learning approach}. Mach Learn Sci Tech 1:015005. \doi{10.1088/2632-2153/ab527d}, {\href{https://arxiv.org/abs/2002.04591}{{arXiv:2002.04591}}} {[astro-ph.IM]}

\bibitem[{{Chan} et~al(2020){Chan}, {Heng}, and {Messenger}}]{2020PhRvD.102d3022C}
{Chan} ML, {Heng} IS, {Messenger} C (2020) {Detection and classification of supernova gravitational wave signals: A deep learning approach}. Phys Rev D 102(4):043022. \doi{10.1103/PhysRevD.102.043022}, {\href{https://arxiv.org/abs/1912.13517}{{arXiv:1912.13517}}} {[astro-ph.HE]}

\bibitem[{Chapman-Bird et~al(2023)Chapman-Bird, Berry, and Woan}]{Chapman-Bird:2022tvu}
Chapman-Bird CEA, Berry CPL, Woan G (2023) {Rapid determination of LISA sensitivity to extreme mass ratio inspirals with machine learning}. Mon Not R Astron Soc 522(4):6043--6054. \doi{10.1093/mnras/stad1397}, {\href{https://arxiv.org/abs/2212.06166}{{arXiv:2212.06166}}} {[astro-ph.HE]}

\bibitem[{Chatterjee et~al(2023)Chatterjee, Kovalam, Wen, Beveridge, Diakogiannis, and Vinsen}]{Chatterjee:2022ggk}
Chatterjee C, Kovalam M, Wen L, et~al (2023) {Rapid Localization of Gravitational Wave Sources from Compact Binary Coalescences Using Deep Learning}. Astrophys J 959(1):42. \doi{10.3847/1538-4357/ad08b7}, {\href{https://arxiv.org/abs/2207.14522}{{arXiv:2207.14522}}} {[gr-qc]}

\bibitem[{Chatterjee et~al(2020{\natexlab{a}})Chatterjee, Ghosh, Brady, Kapadia, Miller, Nissanke, and Pannarale}]{Chatterjee:2019avs}
Chatterjee D, Ghosh S, Brady PR, et~al (2020{\natexlab{a}}) {A Machine Learning Based Source Property Inference for Compact Binary Mergers}. Astrophys J 896(1):54. \doi{10.3847/1538-4357/ab8dbe}, {\href{https://arxiv.org/abs/1911.00116}{{arXiv:1911.00116}}} {[astro-ph.IM]}

\bibitem[{Chatterjee et~al(2020{\natexlab{b}})Chatterjee, Ghosh, Brady, Kapadia, Miller, Nissanke, and Pannarale}]{Chatterjee_2020}
Chatterjee D, Ghosh S, Brady PR, et~al (2020{\natexlab{b}}) A machine learning-based source property inference for compact binary mergers. Astrophys J 896(1):54. \doi{10.3847/1538-4357/ab8dbe}

\bibitem[{{Chatterjee} et~al(2023){Chatterjee}, {Harris}, {Goel}, {Desai}, {Coughlin}, and {Katsavounidis}}]{2023arXiv231207615C}
{Chatterjee} D, {Harris} PC, {Goel} M, et~al (2023) {Optimizing Likelihood-free Inference using Self-supervised Neural Symmetry Embeddings}. arXiv e-prints arXiv:2312.07615. \doi{10.48550/arXiv.2312.07615}, {\href{https://arxiv.org/abs/2312.07615}{{arXiv:2312.07615}}} {[cs.LG]}

\bibitem[{{Chatterji} et~al(2004){Chatterji}, {Blackburn}, {Martin}, and {Katsavounidis}}]{2004CQGra..21S1809C}
{Chatterji} S, {Blackburn} L, {Martin} G, et~al (2004) {Multiresolution techniques for the detection of gravitational-wave bursts}. Class Quantum Grav 21(20):S1809--S1818. \doi{10.1088/0264-9381/21/20/024}, {\href{https://arxiv.org/abs/gr-qc/0412119}{{arXiv:gr-qc/0412119}}} {[gr-qc]}

\bibitem[{Chatterji et~al(2004)Chatterji, Blackburn, Martin, and Katsavounidis}]{Chatterji:2004qg}
Chatterji S, Blackburn L, Martin G, et~al (2004) {Multiresolution techniques for the detection of gravitational-wave bursts}. Class Quant Grav 21:S1809--S1818. \doi{10.1088/0264-9381/21/20/024}, {\href{https://arxiv.org/abs/gr-qc/0412119}{{arXiv:gr-qc/0412119}}}

\bibitem[{Chaudhary et~al(2024)}]{Chaudhary:2023vec}
Chaudhary SS, et~al (2024) {Low-latency gravitational wave alert products and their performance at the time of the fourth LIGO-Virgo-KAGRA observing run}. Proc Nat Acad Sci 121(18):e2316474121. \doi{10.1073/pnas.2316474121}, {\href{https://arxiv.org/abs/2308.04545}{{arXiv:2308.04545}}} {[astro-ph.HE]}

\bibitem[{Chen and Johnson-McDaniel(2024)}]{Chen:2024kdc}
Chen A, Johnson-McDaniel NK (2024) {A fast frequency-domain expression for the time-dependent detector response of ground-based gravitational-wave detectors to compact binary signals}. arXiv e-prints {\href{https://arxiv.org/abs/2407.15732}{{arXiv:2407.15732}}} {[astro-ph.IM]}

\bibitem[{Chen et~al(2020)Chen, Zhong, Feng, Li, and Li}]{Chen:2020ehw}
Chen M, Zhong Y, Feng Y, et~al (2020) {Machine Learning for Nanohertz Gravitational Wave Detection and Parameter Estimation with Pulsar Timing Array}. Sci China Phys Mech Astron 63(12):129511. \doi{10.1007/s11433-020-1609-y}, {\href{https://arxiv.org/abs/2003.13928}{{arXiv:2003.13928}}} {[astro-ph.IM]}

\bibitem[{Chen and Guestrin(2016)}]{Chen:2016btl}
Chen T, Guestrin C (2016) {XGBoost: A Scalable Tree Boosting System}. arXiv e-prints \doi{10.1145/2939672.2939785}, {\href{https://arxiv.org/abs/1603.02754}{{arXiv:1603.02754}}} {[cs.LG]}

\bibitem[{{Cheung} et~al(2022){Cheung}, {Wong}, {Hannuksela}, {Li}, and {Ho}}]{cheung_2022}
{Cheung} DHT, {Wong} KWK, {Hannuksela} OA, et~al (2022) {Testing the robustness of simulation-based gravitational-wave population inference}. \prd 106(8):083014. \doi{10.1103/PhysRevD.106.083014}, {\href{https://arxiv.org/abs/2112.06707}{{arXiv:2112.06707}}} {[astro-ph.IM]}

\bibitem[{Choudhary et~al(2023)Choudhary, More, Suyamprakasam, and Bose}]{PhysRevD.107.024030}
Choudhary S, More A, Suyamprakasam S, et~al (2023) Deep learning network to distinguish binary black hole signals from short-duration noise transients. Phys Rev D 107:024030. \doi{10.1103/PhysRevD.107.024030}

\bibitem[{Chu et~al(2022)}]{Chu:2020pjv}
Chu Q, et~al (2022) {SPIIR online coherent pipeline to search for gravitational waves from compact binary coalescences}. Phys Rev D 105(2):024023. \doi{10.1103/PhysRevD.105.024023}, {\href{https://arxiv.org/abs/2011.06787}{{arXiv:2011.06787}}} {[gr-qc]}

\bibitem[{Colgan et~al(2020)Colgan, Corley, Lau, Bartos, Wright, M\'arka, and M\'arka}]{PhysRevD.101.102003}
Colgan RE, Corley KR, Lau Y, et~al (2020) Efficient gravitational-wave glitch identification from environmental data through machine learning. Phys Rev D 101:102003. \doi{10.1103/PhysRevD.101.102003}

\bibitem[{Cornell et~al(2022)Cornell, Ncube, and Harmsen}]{Cornell:2022enn}
Cornell AS, Ncube A, Harmsen G (2022) {Using physics-informed neural networks to compute quasinormal modes}. Phys Rev D 106(12):124047. \doi{10.1103/PhysRevD.106.124047}, {\href{https://arxiv.org/abs/2205.08284}{{arXiv:2205.08284}}} {[physics.comp-ph]}

\bibitem[{Cornish(2020)}]{PhysRevD.102.124038}
Cornish NJ (2020) Time-frequency analysis of gravitational wave data. Phys Rev D 102:124038. \doi{10.1103/PhysRevD.102.124038}

\bibitem[{Cornish and Crowder(2005)}]{Cornish:2005qw}
Cornish NJ, Crowder J (2005) {LISA data analysis using MCMC methods}. Phys Rev D 72:043005. \doi{10.1103/PhysRevD.72.043005}, {\href{https://arxiv.org/abs/gr-qc/0506059}{{arXiv:gr-qc/0506059}}}

\bibitem[{Coughlin et~al(2017)Coughlin, Earle, Harms, Biscans, Buchanan, Coughlin, Donovan, Fee, Gabbard, Guy, Mukund, and Perry}]{Coughlin_2017}
Coughlin M, Earle P, Harms J, et~al (2017) Limiting the effects of earthquakes on gravitational-wave interferometers. Class Quantum Grav 34(4):044004. \doi{10.1088/1361-6382/aa5a60}

\bibitem[{Coughlin et~al(2019)}]{Coughlin:2019ref}
Coughlin SB, et~al (2019) {Classifying the unknown: discovering novel gravitational-wave detector glitches using similarity learning}. Phys Rev D 99(8):082002. \doi{10.1103/PhysRevD.99.082002}, {\href{https://arxiv.org/abs/1903.04058}{{arXiv:1903.04058}}} {[astro-ph.IM]}

\bibitem[{Covas et~al(2024)Covas, Prix, and Martins}]{Covas:2024pam}
Covas PB, Prix R, Martins J (2024) {New framework to follow up candidates from continuous gravitational-wave searches}. Phys Rev D 110(2):024053. \doi{10.1103/PhysRevD.110.024053}, {\href{https://arxiv.org/abs/2404.18608}{{arXiv:2404.18608}}} {[gr-qc]}

\bibitem[{Covas et~al(2018)}]{Covas:2018oik}
Covas PB, et~al (2018) {Identification and mitigation of narrow spectral artifacts that degrade searches for persistent gravitational waves in the first two observing runs of Advanced LIGO}. Phys Rev D97(8):082002. \doi{10.1103/PhysRevD.97.082002}, {\href{https://arxiv.org/abs/1801.07204}{{arXiv:1801.07204}}} {[astro-ph.IM]}

\bibitem[{Cranmer et~al(2020)Cranmer, Brehmer, and Louppe}]{Cranmer:2019eaq}
Cranmer K, Brehmer J, Louppe G (2020) {The frontier of simulation-based inference}. Proc Nat Acad Sci 117(48):30055--30062. \doi{10.1073/pnas.1912789117}, {\href{https://arxiv.org/abs/1911.01429}{{arXiv:1911.01429}}} {[stat.ML]}

\bibitem[{Creighton and Anderson(2011{\natexlab{a}})}]{doi:https://doi.org/10.1002/9783527636037.ch7}
Creighton J, Anderson WG (2011{\natexlab{a}}) Gravitational-Wave Data Analysis, John Wiley \& Sons, USA, chap~7, pp 269--347. \doi{10.1002/9783527636037.ch7}

\bibitem[{Creighton and Anderson(2011{\natexlab{b}})}]{doi:https://doi.org/10.1002/9783527636037.ch5}
Creighton J, Anderson WG (2011{\natexlab{b}}) Sources of Gravitational Radiation, John Wiley \& Sons, USA, chap~5, pp 149--196. \doi{10.1002/9783527636037.ch5}

\bibitem[{Cristianini and Ricci(2008)}]{Cristianini2008}
Cristianini N, Ricci E (2008) Support Vector Machines, Springer US, Boston, MA, pp 928--932. \doi{10.1007/978-0-387-30162-4_415}

\bibitem[{Cuoco et~al(2001)Cuoco, Calamai, Fabbroni, Losurdo, Mazzoni, Stanga, and Vetrano}]{Cuoco_2001}
Cuoco E, Calamai G, Fabbroni L, et~al (2001) On-line power spectra identification and whitening for the noise in interferometric gravitational wave detectors. Class Quantum Grav 18(9):1727. \doi{10.1088/0264-9381/18/9/309}

\bibitem[{Cuoco et~al(2017)Cuoco, Powell, Torres-Forn\'e, Lynch, Trifir\`o, Cavagli\`a, Siong~Heng, and A.~Font}]{Cuoco:2017evl}
Cuoco E, Powell J, Torres-Forn\'e A, et~al (2017) {Strategy for signal classification to improve data quality for Advanced Detectors gravitational-wave searches}. Nuovo Cim C 40(3):124. \doi{10.1393/ncc/i2017-17124-4}

\bibitem[{Cuoco et~al(2018)Cuoco, Razzano, and Utina}]{Cuoco2018}
Cuoco E, Razzano M, Utina A (2018) Wavelet-based classification of transient signals for gravitational wave detectors. European Signal Processing Conference 2018-September:2648--2652

\bibitem[{Cuoco et~al(2021{\natexlab{a}})Cuoco, Patricelli, Iess, and Morawski}]{universe7110394}
Cuoco E, Patricelli B, Iess A, et~al (2021{\natexlab{a}}) Multimodal analysis of gravitational wave signals and gamma-ray bursts from binary neutron star mergers. Universe 7(11):394. \doi{10.3390/universe7110394}

\bibitem[{Cuoco et~al(2022)Cuoco, Patricelli, Iess, and Morawski}]{natureComp}
Cuoco E, Patricelli B, Iess A, et~al (2022) Computational challenges for multimodal astrophysics. Nat Comput Sci 2:479--485. \doi{10.1038/s43588-022-00288-z}

\bibitem[{Cuoco et~al(2021{\natexlab{b}})}]{Cuoco:2020ogp}
Cuoco E, et~al (2021{\natexlab{b}}) {Enhancing Gravitational-Wave Science with Machine Learning}. Mach Learn Sci Tech 2(1):011002. \doi{10.1088/2632-2153/abb93a}, {\href{https://arxiv.org/abs/2005.03745}{{arXiv:2005.03745}}} {[astro-ph.HE]}

\bibitem[{{D{\'a}lya} et~al(2022){D{\'a}lya}, {D{\'\i}az}, {Bouchet}, {Frei}, {Jasche}, {Lavaux}, {Macas}, {Mukherjee}, {P{\'a}lfi}, {de Souza}, {Wandelt}, {Bilicki}, and {Raffai}}]{dalya_2022}
{D{\'a}lya} G, {D{\'\i}az} R, {Bouchet} FR, et~al (2022) {GLADE+ : an extended galaxy catalogue for multimessenger searches with advanced gravitational-wave detectors}. \mnras 514(1):1403--1411. \doi{10.1093/mnras/stac1443}, {\href{https://arxiv.org/abs/2110.06184}{{arXiv:2110.06184}}} {[astro-ph.CO]}

\bibitem[{Damour and Nagar(2016)}]{Damour:2016bks}
Damour T, Nagar A (2016) {The Effective-One-Body Approach to the General Relativistic Two Body Problem}. Lect Notes Phys 905:273--312. \doi{10.1007/978-3-319-19416-5_7}

\bibitem[{Davis et~al(2020)Davis, White, and Saulson}]{Davis:2020nyf}
Davis D, White LV, Saulson PR (2020) {Utilizing aLIGO Glitch Classifications to Validate Gravitational-Wave Candidates}. Class Quant Grav 37(14):145001. \doi{10.1088/1361-6382/ab91e6}, {\href{https://arxiv.org/abs/2002.09429}{{arXiv:2002.09429}}} {[gr-qc]}

\bibitem[{Davis et~al(2021)}]{LIGO:2021ppb}
Davis D, et~al (2021) {LIGO detector characterization in the second and third observing runs}. Class Quant Grav 38(13):135014. \doi{10.1088/1361-6382/abfd85}, {\href{https://arxiv.org/abs/2101.11673}{{arXiv:2101.11673}}} {[astro-ph.IM]}

\bibitem[{{Dax} et~al(2021){Dax}, {Green}, {Gair}, {Macke}, {Buonanno}, and {Sch{\"o}lkopf}}]{dax_2021}
{Dax} M, {Green} SR, {Gair} J, et~al (2021) {Real-Time Gravitational Wave Science with Neural Posterior Estimation}. \prl 127(24):241103. \doi{10.1103/PhysRevLett.127.241103}, {\href{https://arxiv.org/abs/2106.12594}{{arXiv:2106.12594}}} {[gr-qc]}

\bibitem[{Dax et~al(2023)Dax, Green, Gair, P\"urrer, Wildberger, Macke, Buonanno, and Sch\"olkopf}]{PhysRevLett.130.171403}
Dax M, Green SR, Gair J, et~al (2023) Neural importance sampling for rapid and reliable gravitational-wave inference. Phys Rev Lett 130:171403. \doi{10.1103/PhysRevLett.130.171403}

\bibitem[{{Demianenko} et~al(2023){Demianenko}, {Malanchev}, {Samorodova}, {Sysak}, {Shiriaev}, {Derkach}, and {Hushchyn}}]{2023A&A...677A..16D}
{Demianenko} M, {Malanchev} K, {Samorodova} E, et~al (2023) {Understanding of the properties of neural network approaches for transient light curve approximations}. Astronomy \& Astrophysics 677:A16. \doi{10.1051/0004-6361/202245189}, {\href{https://arxiv.org/abs/2209.07542}{{arXiv:2209.07542}}} {[astro-ph.IM]}

\bibitem[{Dhani et~al(2024)Dhani, V\"olkel, Buonanno, Estelles, Gair, Pfeiffer, Pompili, and Toubiana}]{Dhani:2024jja}
Dhani A, V\"olkel S, Buonanno A, et~al (2024) {Systematic Biases in Estimating the Properties of Black Holes Due to Inaccurate Gravitational-Wave Models}. arXiv e-prints {\href{https://arxiv.org/abs/2404.05811}{{arXiv:2404.05811}}} {[gr-qc]}

\bibitem[{Dhurandhar et~al(2008)Dhurandhar, Krishnan, Mukhopadhyay, and Whelan}]{Dhurandhar:2007vb}
Dhurandhar S, Krishnan B, Mukhopadhyay H, et~al (2008) {Cross-correlation search for periodic gravitational waves}. Phys Rev D 77:082001. \doi{10.1103/PhysRevD.77.082001}, {\href{https://arxiv.org/abs/0712.1578}{{arXiv:0712.1578}}} {[gr-qc]}

\bibitem[{Dieselhorst et~al(2021)Dieselhorst, Cook, Bernuzzi, and Radice}]{Dieselhorst:2021zet}
Dieselhorst T, Cook W, Bernuzzi S, et~al (2021) {Machine Learning for Conservative-to-Primitive in Relativistic Hydrodynamics}. Symmetry 13(11):2157. \doi{10.3390/sym13112157}, {\href{https://arxiv.org/abs/2109.02679}{{arXiv:2109.02679}}} {[astro-ph.IM]}

\bibitem[{Dietrich et~al(2021)Dietrich, Hinderer, and Samajdar}]{Dietrich:2020eud}
Dietrich T, Hinderer T, Samajdar A (2021) {Interpreting Binary Neutron Star Mergers: Describing the Binary Neutron Star Dynamics, Modelling Gravitational Waveforms, and Analyzing Detections}. Gen Rel Grav 53(3):27. \doi{10.1007/s10714-020-02751-6}, {\href{https://arxiv.org/abs/2004.02527}{{arXiv:2004.02527}}} {[gr-qc]}

\bibitem[{Doctor et~al(2017)Doctor, Farr, Holz, and P{\"u}rrer}]{Doctor:2017csx}
Doctor Z, Farr B, Holz DE, et~al (2017) {Statistical Gravitational Waveform Models: What to Simulate Next?} Phys Rev D96(12):123011. \doi{10.1103/PhysRevD.96.123011}, {\href{https://arxiv.org/abs/1706.05408}{{arXiv:1706.05408}}} {[astro-ph.HE]}

\bibitem[{{Drago} et~al(2023){Drago}, {Andresen}, {Di Palma}, {Tamborra}, and {Torres-Forn{\'e}}}]{2023PhRvD.108j3036D}
{Drago} M, {Andresen} H, {Di Palma} I, et~al (2023) {Multimessenger observations of core-collapse supernovae: Exploiting the standing accretion shock instability}. Phys Rev D 108(10):103036. \doi{10.1103/PhysRevD.108.103036}, {\href{https://arxiv.org/abs/2305.07688}{{arXiv:2305.07688}}} {[astro-ph.HE]}

\bibitem[{Drago et~al(2021)}]{Drago:2020kic}
Drago M, et~al (2021) {Coherent WaveBurst, a pipeline for unmodeled gravitational-wave data analysis}. SoftwareX 14:100678. \doi{10.1016/j.softx.2021.100678}, {\href{https://arxiv.org/abs/2006.12604}{{arXiv:2006.12604}}} {[gr-qc]}

\bibitem[{Dreissigacker and Prix(2020)}]{Dreissigacker:2020xfr}
Dreissigacker C, Prix R (2020) {Deep-Learning Continuous Gravitational Waves: Multiple detectors and realistic noise}. Phys Rev D 102(2):022005. \doi{10.1103/PhysRevD.102.022005}, {\href{https://arxiv.org/abs/2005.04140}{{arXiv:2005.04140}}} {[gr-qc]}

\bibitem[{Dreissigacker et~al(2018)Dreissigacker, Prix, and Wette}]{Dreissigacker:2018afk}
Dreissigacker C, Prix R, Wette K (2018) {Fast and Accurate Sensitivity Estimation for Continuous-Gravitational-Wave Searches}. Phys Rev D 98(8):084058. \doi{10.1103/PhysRevD.98.084058}, {\href{https://arxiv.org/abs/1808.02459}{{arXiv:1808.02459}}} {[gr-qc]}

\bibitem[{Dreissigacker et~al(2019)Dreissigacker, Sharma, Messenger, Zhao, and Prix}]{Dreissigacker:2019edy}
Dreissigacker C, Sharma R, Messenger C, et~al (2019) {Deep-Learning Continuous Gravitational Waves}. Phys Rev D100(4):044009. \doi{10.1103/PhysRevD.100.044009}, {\href{https://arxiv.org/abs/1904.13291}{{arXiv:1904.13291}}} {[gr-qc]}

\bibitem[{Duez and Zlochower(2019)}]{Duez:2018jaf}
Duez MD, Zlochower Y (2019) {Numerical Relativity of Compact Binaries in the 21st Century}. Rept Prog Phys 82(1):016902. \doi{10.1088/1361-6633/aadb16}, {\href{https://arxiv.org/abs/1808.06011}{{arXiv:1808.06011}}} {[gr-qc]}

\bibitem[{Dumitrescu and Irofti(2018)}]{DL_book}
Dumitrescu B, Irofti P (2018) Dictionary Learning Algorithms and Applications. Springer, Cham, \doi{10.1007/978-3-319-78674-2}

\bibitem[{Easter et~al(2019)Easter, Lasky, Casey, Rezzolla, and Takami}]{Easter2019}
Easter PJ, Lasky PD, Casey AR, et~al (2019) Computing fast and reliable gravitational waveforms of binary neutron star merger remnants. Phys Rev D 100:043005. \doi{10.1103/PhysRevD.100.043005}

\bibitem[{Edwards et~al(2024)Edwards, Wong, Lam, Coogan, Foreman-Mackey, Isi, and Zimmerman}]{Edwards:2023sak}
Edwards TDP, Wong KWK, Lam KKH, et~al (2024) {Differentiable and hardware-accelerated waveforms for gravitational wave data analysis}. Phys Rev D 110(6):064028. \doi{10.1103/PhysRevD.110.064028}, {\href{https://arxiv.org/abs/2302.05329}{{arXiv:2302.05329}}} {[astro-ph.IM]}

\bibitem[{van Engelen and Hoos(2020)}]{vanEngelen2020}
van Engelen JE, Hoos HH (2020) A survey on semi-supervised learning. Machine Learning 109(2):373--440. \doi{10.1007/s10994-019-05855-6}

\bibitem[{Essick et~al(2020{\natexlab{a}})Essick, Godwin, Hanna, Blackburn, and Katsavounidis}]{Essick_2021}
Essick R, Godwin P, Hanna C, et~al (2020{\natexlab{a}}) idq: Statistical inference of non-gaussian noise with auxiliary degrees of freedom in gravitational-wave detectors. Machine Learning: Science and Technology 2(1):015004. \doi{10.1088/2632-2153/abab5f}

\bibitem[{Essick et~al(2020{\natexlab{b}})Essick, Landry, and Holz}]{Essick:2019ldf}
Essick R, Landry P, Holz DE (2020{\natexlab{b}}) {Nonparametric Inference of Neutron Star Composition, Equation of State, and Maximum Mass with GW170817}. Phys Rev D 101(6):063007. \doi{10.1103/PhysRevD.101.063007}, {\href{https://arxiv.org/abs/1910.09740}{{arXiv:1910.09740}}} {[astro-ph.HE]}

\bibitem[{{ET steering committee, et~al.}(2020)}]{ET_design_study_2019}
{ET steering committee, et~al.} (2020) {Einstein Telescope}: Science case, design study and feasibility report. Tech. rep., {Einstein Telescope project}, \urlprefix\url{https://apps.et-gw.eu/tds/ql/?c=15662}

\bibitem[{Fan et~al(2019)Fan, Li, Li, Zhong, and Cao}]{Fan:2018vgw}
Fan X, Li J, Li X, et~al (2019) {Applying deep neural networks to the detection and space parameter estimation of compact binary coalescence with a network of gravitational wave detectors}. Sci China Phys Mech Astron 62(6):969512. \doi{10.1007/s11433-018-9321-7}, {\href{https://arxiv.org/abs/1811.01380}{{arXiv:1811.01380}}} {[astro-ph.IM]}

\bibitem[{Ferguson(2023)}]{Ferguson:2022qkz}
Ferguson D (2023) {Optimizing the placement of numerical relativity simulations using a mismatch predicting neural network}. Phys Rev D 107(2):024034. \doi{10.1103/PhysRevD.107.024034}, {\href{https://arxiv.org/abs/2209.15144}{{arXiv:2209.15144}}} {[gr-qc]}

\bibitem[{Fernandes et~al(2023)Fernandes, Vieira, Onofre, Bustillo, Torres-Forné, and Font}]{Fernandes_2023}
Fernandes T, Vieira S, Onofre A, et~al (2023) Convolutional neural networks for the classification of glitches in gravitational-wave data streams. Class Quantum Grav 40(19):195018. \doi{10.1088/1361-6382/acf26c}

\bibitem[{Field et~al(2014)Field, Galley, Hesthaven, Kaye, and Tiglio}]{Field:2013cfa}
Field SE, Galley CR, Hesthaven JS, et~al (2014) {Fast prediction and evaluation of gravitational waveforms using surrogate models}. Phys Rev X4(3):031006. \doi{10.1103/PhysRevX.4.031006}, {\href{https://arxiv.org/abs/1308.3565}{{arXiv:1308.3565}}} {[gr-qc]}

\bibitem[{Foreman-Mackey et~al(2013)Foreman-Mackey, Hogg, Lang, and Goodman}]{Foreman-Mackey:2012any}
Foreman-Mackey D, Hogg DW, Lang D, et~al (2013) {emcee: The MCMC Hammer}. Publ Astron Soc Pac 125:306--312. \doi{10.1086/670067}, {\href{https://arxiv.org/abs/1202.3665}{{arXiv:1202.3665}}} {[astro-ph.IM]}

\bibitem[{Fragkouli et~al(2023)Fragkouli, Nousi, Passalis, Iosif, Stergioulas, and Tefas}]{Fragkouli:2022lpt}
Fragkouli SC, Nousi P, Passalis N, et~al (2023) {Deep Residual Error and Bag-of-Tricks Learning for Gravitational Wave Surrogate Modeling}. {Applied Soft Computing} 147:110746. \doi{10.1016/j.asoc.2023.110746}, {\href{https://arxiv.org/abs/2203.08434}{{arXiv:2203.08434}}} {[astro-ph.IM]}

\bibitem[{Fujimoto et~al(2018)Fujimoto, Fukushima, and Murase}]{Fujimoto:2017cdo}
Fujimoto Y, Fukushima K, Murase K (2018) {Methodology study of machine learning for the neutron star equation of state}. Phys Rev D 98(2):023019. \doi{10.1103/PhysRevD.98.023019}, {\href{https://arxiv.org/abs/1711.06748}{{arXiv:1711.06748}}} {[nucl-th]}

\bibitem[{F{\"u}rnkranz(2010)}]{Fürnkranz2010}
F{\"u}rnkranz J (2010) Decision Tree, Springer US, Boston, MA, pp 263--267. \doi{10.1007/978-0-387-30164-8_204}

\bibitem[{Gabbard et~al(2018)Gabbard, Williams, Hayes, and Messenger}]{PhysRevLett.120.141103}
Gabbard H, Williams M, Hayes F, et~al (2018) Matching matched filtering with deep networks for gravitational-wave astronomy. Phys Rev Lett 120:141103. \doi{10.1103/PhysRevLett.120.141103}, \urlprefix\url{https://link.aps.org/doi/10.1103/PhysRevLett.120.141103}

\bibitem[{Gabbard et~al(2022)Gabbard, Messenger, Heng, Tonolini, and Murray-Smith}]{1909.06296}
Gabbard H, Messenger C, Heng IS, et~al (2022) {Bayesian parameter estimation using conditional variational autoencoders for gravitational-wave astronomy}. Nature Phys 18(1):112--117. \doi{10.1038/s41567-021-01425-7}, {\href{https://arxiv.org/abs/1909.06296}{{arXiv:1909.06296}}} {[astro-ph.IM]}

\bibitem[{Garc\'\i{}a-Bellido(2017)}]{Garcia-Bellido:2017fdg}
Garc\'\i{}a-Bellido J (2017) {Massive Primordial Black Holes as Dark Matter and their detection with Gravitational Waves}. J Phys Conf Ser 840(1):012032. \doi{10.1088/1742-6596/840/1/012032}, {\href{https://arxiv.org/abs/1702.08275}{{arXiv:1702.08275}}} {[astro-ph.CO]}

\bibitem[{{Gayathri} et~al(2020){Gayathri}, {Lopez}, {R.~S.}, {Heng}, {Pai}, and {Messenger}}]{2020PhRvD.102j4023G}
{Gayathri} V, {Lopez} D, {R.~S.} P, et~al (2020) {Enhancing the sensitivity of transient gravitational wave searches with Gaussian mixture models}. \prd 102(10):104023. \doi{10.1103/PhysRevD.102.104023}, {\href{https://arxiv.org/abs/2008.01262}{{arXiv:2008.01262}}} {[gr-qc]}

\bibitem[{Gebhard et~al(2019)Gebhard, Kilbertus, Harry, and Sch{\"o}lkopf}]{Gebhard:2019ldz}
Gebhard TD, Kilbertus N, Harry I, et~al (2019) {Convolutional neural networks: a magic bullet for gravitational-wave detection?} Phys Rev D 100(6):063015. \doi{10.1103/PhysRevD.100.063015}, {\href{https://arxiv.org/abs/1904.08693}{{arXiv:1904.08693}}} {[astro-ph.IM]}

\bibitem[{{George} and {Huerta}(2018)}]{2018PhLB..778...64G}
{George} D, {Huerta} EA (2018) {Deep Learning for real-time gravitational wave detection and parameter estimation: Results with Advanced LIGO data}. Physics Letters B 778:64--70. \doi{10.1016/j.physletb.2017.12.053}, {\href{https://arxiv.org/abs/1711.03121}{{arXiv:1711.03121}}} {[gr-qc]}

\bibitem[{Gerosa et~al(2018)Gerosa, Berti, O'Shaughnessy, Belczynski, Kesden, Wysocki, and Gladysz}]{gerosa_98}
Gerosa D, Berti E, O'Shaughnessy R, et~al (2018) Spin orientations of merging black holes formed from the evolution of stellar binaries. Phys Rev D 98:084036. \doi{10.1103/PhysRevD.98.084036}

\bibitem[{{Gerosa} et~al(2020){Gerosa}, {Pratten}, and {Vecchio}}]{gerosa_2020}
{Gerosa} D, {Pratten} G, {Vecchio} A (2020) {Gravitational-wave selection effects using neural-network classifiers}. \prd 102(10):103020. \doi{10.1103/PhysRevD.102.103020}, {\href{https://arxiv.org/abs/2007.06585}{{arXiv:2007.06585}}} {[astro-ph.HE]}

\bibitem[{Glanzer et~al(2023)}]{Glanzer:2022avx}
Glanzer J, et~al (2023) {Data quality up to the third observing run of advanced LIGO: Gravity Spy glitch classifications}. Class Quant Grav 40(6):065004. \doi{10.1088/1361-6382/acb633}, {\href{https://arxiv.org/abs/2208.12849}{{arXiv:2208.12849}}} {[gr-qc]}

\bibitem[{Goodfellow et~al(2016)Goodfellow, Bengio, and Courville}]{Goodfellow-et-al-2016}
Goodfellow I, Bengio Y, Courville A (2016) Deep Learning. MIT Press, Cambridge, MA, \url{http://www.deeplearningbook.org}

\bibitem[{Goyal et~al(2021)Goyal, D., Kapadia, and Ajith}]{Goyal:2021hxv}
Goyal S, D. H, Kapadia SJ, et~al (2021) {Rapid identification of strongly lensed gravitational-wave events with machine learning}. Phys Rev D 104(12):124057. \doi{10.1103/PhysRevD.104.124057}, {\href{https://arxiv.org/abs/2106.12466}{{arXiv:2106.12466}}} {[gr-qc]}

\bibitem[{Grace et~al(2023)Grace, Wette, Scott, and Sun}]{Grace:2023kqq}
Grace B, Wette K, Scott SM, et~al (2023) {Piecewise frequency model for searches for long-transient gravitational waves from young neutron stars}. Phys Rev D 108(12):123045. \doi{10.1103/PhysRevD.108.123045}, {\href{https://arxiv.org/abs/2310.12463}{{arXiv:2310.12463}}} {[gr-qc]}

\bibitem[{{Green} et~al(2020){Green}, {Simpson}, and {Gair}}]{2020arXiv200207656G}
{Green} SR, {Simpson} C, {Gair} J (2020) {Gravitational-wave parameter estimation with autoregressive neural network flows}. \prd 102(10):104057. \doi{10.1103/PhysRevD.102.104057}, {\href{https://arxiv.org/abs/2002.07656}{{arXiv:2002.07656}}} {[astro-ph.IM]}

\bibitem[{Gregory(2005)}]{gregory2005bayesian}
Gregory P (2005) {Bayesian logical data analysis for the physical sciences: a comparative approach with Mathematica support}. Cambridge University Press, Cambridge

\bibitem[{{Grespan} and {Biesiada}(2023)}]{2023Univ....9..200G}
{Grespan} M, {Biesiada} M (2023) {Strong Gravitational Lensing of Gravitational Waves: A Review}. Universe 9(5):200. \doi{10.3390/universe9050200}

\bibitem[{Grote and Stadnik(2019)}]{Grote:2019uvn}
Grote H, Stadnik YV (2019) {Novel signatures of dark matter in laser-interferometric gravitational-wave detectors}. Phys Rev Res 1(3):033187. \doi{10.1103/PhysRevResearch.1.033187}, {\href{https://arxiv.org/abs/1906.06193}{{arXiv:1906.06193}}} {[astro-ph.IM]}

\bibitem[{Gunny et~al(2022)Gunny, Rankin, Krupa, Saleem, Nguyen, Coughlin, Harris, Katsavounidis, Timm, and Holzman}]{Gunny:2021gne}
Gunny A, Rankin D, Krupa J, et~al (2022) {Hardware-accelerated Inference for Real-Time Gravitational-Wave Astronomy}. Nature Astron 6(5):529--536. \doi{10.1038/s41550-022-01651-w}, {\href{https://arxiv.org/abs/2108.12430}{{arXiv:2108.12430}}} {[gr-qc]}

\bibitem[{Guo et~al(2019)Guo, Riles, Yang, and Zhao}]{Guo:2019ker}
Guo HK, Riles K, Yang FW, et~al (2019) {Searching for Dark Photon Dark Matter in LIGO O1 Data}. Commun Phys 2:155. \doi{10.1038/s42005-019-0255-0}, {\href{https://arxiv.org/abs/1905.04316}{{arXiv:1905.04316}}} {[hep-ph]}

\bibitem[{Gurav et~al(2022)Gurav, Papalexakis, Barish, Richardson, and Vajente}]{Gurav2022IdentifyingWT}
Gurav R, Papalexakis EE, Barish BC, et~al (2022) Identifying witnesses to noise transients in ground-based gravitational-wave observations using auxiliary channels with matrix and tensor factorization techniques. In: {NeurIPS 2022 AI for Science: Progress and Promises}

\bibitem[{Haegel and Husa(2020)}]{Haegel:2019uop}
Haegel L, Husa S (2020) {Predicting the properties of black-hole merger remnants with deep neural networks}. Class Quant Grav 37(13):135005. \doi{10.1088/1361-6382/ab905c}, {\href{https://arxiv.org/abs/1911.01496}{{arXiv:1911.01496}}} {[gr-qc]}

\bibitem[{Han et~al(2021)Han, Jiang, Tang, and Fan}]{Han:2021kjx}
Han MZ, Jiang JL, Tang SP, et~al (2021) {Bayesian Nonparametric Inference of the Neutron Star Equation of State via a Neural Network}. Astrophys J 919(1):11. \doi{10.3847/1538-4357/ac11f8}, {\href{https://arxiv.org/abs/2103.05408}{{arXiv:2103.05408}}} {[hep-ph]}

\bibitem[{Han et~al(2023)Han, Tang, and Fan}]{Han:2022sxt}
Han MZ, Tang SP, Fan YZ (2023) {Nonparametric Representation of Neutron Star Equation of State Using Variational Autoencoder}. Astrophys J 950(2):77. \doi{10.3847/1538-4357/acd050}, {\href{https://arxiv.org/abs/2205.03855}{{arXiv:2205.03855}}} {[astro-ph.HE]}

\bibitem[{Haris et~al(2018)Haris, Mehta, Kumar, Venumadhav, and Ajith}]{Haris:2018vmn}
Haris K, Mehta AK, Kumar S, et~al (2018) {Identifying strongly lensed gravitational wave signals from binary black hole mergers}. arXiv e-prints {\href{https://arxiv.org/abs/1807.07062}{{arXiv:1807.07062}}} {[gr-qc]}

\bibitem[{Haskell and Bejger(2023)}]{Haskell:2023yrv}
Haskell B, Bejger M (2023) {Astrophysics with continuous gravitational waves}. Nature Astron 7(10):1160--1170. \doi{10.1038/s41550-023-02059-w}

\bibitem[{{He} et~al(2016){He}, {Zhang}, {Ren}, and {Sun}}]{2016cvpr.confE...1H}
{He} K, {Zhang} X, {Ren} S, et~al (2016) {Deep Residual Learning for Image Recognition}. In: 2016 IEEE Conference on Computer Vision and Pattern Recognition (CVPR, p~1, \doi{10.1109/CVPR.2016.90}, \eprint{1512.03385}

\bibitem[{{He} et~al(2017){He}, {Gkioxari}, {Doll{\'a}r}, and {Girshick}}]{He2017:rcnn}
{He} K, {Gkioxari} G, {Doll{\'a}r} P, et~al (2017) {Mask R-CNN}. arXiv e-prints arXiv:1703.06870. \doi{10.48550/arXiv.1703.06870}, {\href{https://arxiv.org/abs/1703.06870}{{arXiv:1703.06870}}} {[cs.CV]}

\bibitem[{Hernandez~Vivanco et~al(2020)Hernandez~Vivanco, Smith, Thrane, and Lasky}]{HernandezVivanco:2020cyp}
Hernandez~Vivanco F, Smith R, Thrane E, et~al (2020) {A scalable random forest regressor for combining neutron-star equation of state measurements: A case study with GW170817 and GW190425}. Mon Not R Astron Soc 499(4):5972--5977. \doi{10.1093/mnras/staa3243}, {\href{https://arxiv.org/abs/2008.05627}{{arXiv:2008.05627}}} {[astro-ph.HE]}

\bibitem[{{Hogg} and {Villar}(2024)}]{2024arXiv240518095H}
{Hogg} DW, {Villar} S (2024) {Is machine learning good or bad for the natural sciences?} arXiv e-prints \doi{10.48550/arXiv.2405.18095}, {\href{https://arxiv.org/abs/2405.18095}{{arXiv:2405.18095}}} {[stat.ML]}

\bibitem[{Houba et~al(2024)Houba, Ferraioli, and Giardini}]{Houba:2024tyn}
Houba N, Ferraioli L, Giardini D (2024) {Detection and mitigation of glitches in LISA data: A machine learning approach}. Phys Rev D 109(8):083027. \doi{10.1103/PhysRevD.109.083027}, {\href{https://arxiv.org/abs/2401.00846}{{arXiv:2401.00846}}} {[astro-ph.IM]}

\bibitem[{Howard et~al(2024)Howard, Thomas et~al}]{fastai}
Howard J, Thomas R, et~al (2024) fastai v 2.7.14. \urlprefix\url{https://docs.fast.ai/}

\bibitem[{Hu and Wu(2017)}]{Hu:2017mde}
Hu WR, Wu YL (2017) {The Taiji Program in Space for gravitational wave physics and the nature of gravity}. Natl Sci Rev 4(5):685--686. \doi{10.1093/nsr/nwx116}

\bibitem[{{Huang} et~al(2016){Huang}, {Liu}, {van der Maaten}, and {Weinberger}}]{Huang:2016den}
{Huang} G, {Liu} Z, {van der Maaten} L, et~al (2016) {Densely Connected Convolutional Networks}. arXiv e-prints arXiv:1608.06993. \doi{10.48550/arXiv.1608.06993}, {\href{https://arxiv.org/abs/1608.06993}{{arXiv:1608.06993}}} {[cs.CV]}

\bibitem[{Huerta et~al(2021)}]{Huerta:2020xyq}
Huerta EA, et~al (2021) {Accelerated, scalable and reproducible AI-driven gravitational wave detection}. Nature Astron 5(10):1062--1068. \doi{10.1038/s41550-021-01405-0}, {\href{https://arxiv.org/abs/2012.08545}{{arXiv:2012.08545}}} {[gr-qc]}

\bibitem[{{Huppenkothen} et~al(2023){Huppenkothen}, {Ntampaka}, {Ho}, {Fouesneau}, {Nord}, {Peek}, {Walmsley}, {Wu}, {Avestruz}, {Buck}, {Brescia}, {Finkbeiner}, {Goulding}, {Kacprzak}, {Melchior}, {Pasquato}, {Ramachandra}, {Ting}, {van de Ven}, {Villar}, {Villar}, and {Zinger}}]{2023arXiv231012528H}
{Huppenkothen} D, {Ntampaka} M, {Ho} M, et~al (2023) {Constructing Impactful Machine Learning Research for Astronomy: Best Practices for Researchers and Reviewers}. arXiv e-prints arXiv:2310.12528. \doi{10.48550/arXiv.2310.12528}, {\href{https://arxiv.org/abs/2310.12528}{{arXiv:2310.12528}}} {[astro-ph.IM]}

\bibitem[{Husa et~al(2016)Husa, Khan, Hannam, P{\"u}rrer, Ohme, Jim\'enez~Forteza, and Boh\'e}]{Husa:2015iqa}
Husa S, Khan S, Hannam M, et~al (2016) {Frequency-domain gravitational waves from nonprecessing black-hole binaries. I. New numerical waveforms and anatomy of the signal}. Phys Rev D93(4):044006. \doi{10.1103/PhysRevD.93.044006}, {\href{https://arxiv.org/abs/1508.07250}{{arXiv:1508.07250}}} {[gr-qc]}

\bibitem[{Hwang et~al(2003)Hwang, Tan, and Tsai}]{1200173}
Hwang KS, Tan SW, Tsai MC (2003) Reinforcement learning to adaptive control of nonlinear systems. IEEE Transactions on Systems, Man, and Cybernetics, Part B (Cybernetics) 33(3):514--521. \doi{10.1109/TSMCB.2003.811112}

\bibitem[{Iess et~al(2020)Iess, Cuoco, Morawski, and Powell}]{10.1088/2632-2153/ab7d31}
Iess A, Cuoco E, Morawski F, et~al (2020) Core-collapse supernova gravitational-wave search and deep learning classification. Mach Learn: Sci Technol

\bibitem[{Intini et~al(2020)Intini, Leaci, Astone, D'Antonio, Frasca, La~Rosa, Miller, Palomba, and Piccinni}]{Intini:2020blc}
Intini G, Leaci P, Astone P, et~al (2020) {A Doppler-modulation based veto to discard false continuous gravitational-wave candidates}. Class Quant Grav 37(22):225007. \doi{10.1088/1361-6382/abac43}

\bibitem[{Islam et~al(2023)Islam, Field, and Khanna}]{Islam:2023mob}
Islam T, Field SE, Khanna G (2023) {Remnant black hole properties from numerical-relativity-informed perturbation theory and implications for waveform modeling}. Phys Rev D 108(6):064048. \doi{10.1103/PhysRevD.108.064048}, {\href{https://arxiv.org/abs/2301.07215}{{arXiv:2301.07215}}} {[gr-qc]}

\bibitem[{Jackson et~al(2020)Jackson, Østerlund, Crowston, Harandi, Allen, Bahaadini, Coughlin, Kalogera, Katsaggelos, Larson, Rohani, Smith, Trouille, and Zevin}]{JACKSON2020106198}
Jackson C, Østerlund C, Crowston K, et~al (2020) Teaching citizen scientists to categorize glitches using machine learning guided training. Computers in Human Behavior 105:106198. \doi{10.1016/j.chb.2019.106198}

\bibitem[{Jacobs et~al(1991)Jacobs, Jordan, Nowlan, and Hinton}]{Jacobs:1991moe}
Jacobs RA, Jordan MI, Nowlan SJ, et~al (1991) Adaptive mixtures of local experts. Neural Computation 3(1):79--87. \doi{10.1162/neco.1991.3.1.79}

\bibitem[{Jadhav et~al(2021)Jadhav, Mukund, Gadre, Mitra, and Abraham}]{PhysRevD.104.064051}
Jadhav S, Mukund N, Gadre B, et~al (2021) Improving significance of binary black hole mergers in advanced ligo data using deep learning: Confirmation of gw151216. Phys Rev D 104:064051. \doi{10.1103/PhysRevD.104.064051}

\bibitem[{Jadhav et~al(2023)Jadhav, Shrivastava, and Mitra}]{Jadhav:2023mqx}
Jadhav S, Shrivastava M, Mitra S (2023) {Towards a robust and reliable deep learning approach for detection of compact binary mergers in gravitational wave data}. Mach Learn Sci Tech 4:045028. \doi{10.1088/2632-2153/ad0938}, {\href{https://arxiv.org/abs/2306.11797}{{arXiv:2306.11797}}} {[gr-qc]}

\bibitem[{Jaranowski et~al(1998)Jaranowski, Krolak, and Schutz}]{Jaranowski:1998qm}
Jaranowski P, Krolak A, Schutz BF (1998) {Data analysis of gravitational - wave signals from spinning neutron stars. 1. The Signal and its detection}. Phys Rev D 58:063001. \doi{10.1103/PhysRevD.58.063001}, {\href{https://arxiv.org/abs/gr-qc/9804014}{{arXiv:gr-qc/9804014}}}

\bibitem[{Jarov et~al(2023)Jarov, Thiele, Soni, Ding, McIver, Ng, Hatoya, and Davis}]{Jarov:2023qpt}
Jarov S, Thiele S, Soni S, et~al (2023) {A new method to distinguish gravitational-wave signals from detector noise transients with Gravity Spy}. arXiv e-prints {\href{https://arxiv.org/abs/2307.15867}{{arXiv:2307.15867}}} {[gr-qc]}

\bibitem[{Jaynes(2003)}]{Jaynes_2003}
Jaynes ET (2003) Probability Theory: The Logic of Science. Cambridge University Press, Cambridge

\bibitem[{Jim{\'e}nez-Forteza et~al(2017)Jim{\'e}nez-Forteza, Keitel, Husa, Hannam, Khan, and P{\"u}rrer}]{Jimenez-Forteza:2016oae}
Jim{\'e}nez-Forteza X, Keitel D, Husa S, et~al (2017) {Hierarchical data-driven approach to fitting numerical relativity data for nonprecessing binary black holes with an application to final spin and radiated energy}. Phys Rev D 95(6):064024. \doi{10.1103/PhysRevD.95.064024}, {\href{https://arxiv.org/abs/1611.00332}{{arXiv:1611.00332}}} {[gr-qc]}

\bibitem[{Jin and Han(2010)}]{Jin2010}
Jin X, Han J (2010) K-Means Clustering, Springer US, Boston, MA, pp 563--564. \doi{10.1007/978-0-387-30164-8_425}

\bibitem[{Jones et~al(2022)}]{Jones:2022fgs}
Jones D, et~al (2022) {Validating continuous gravitational-wave candidates from a semicoherent search using Doppler modulation and an effective point spread function}. Phys Rev D 106(12):123011. \doi{10.1103/PhysRevD.106.123011}, {\href{https://arxiv.org/abs/2203.14468}{{arXiv:2203.14468}}} {[gr-qc]}

\bibitem[{Joshi and Prix(2023)}]{Joshi:2023hpx}
Joshi PM, Prix R (2023) {Novel neural-network architecture for continuous gravitational waves}. Phys Rev D 108(6):063021. \doi{10.1103/PhysRevD.108.063021}, {\href{https://arxiv.org/abs/2305.01057}{{arXiv:2305.01057}}} {[gr-qc]}

\bibitem[{Joshi and Prix(2024)}]{Joshi:2024bkj}
Joshi PM, Prix R (2024) {Large-kernel Convolutional Neural Networks for Wide Parameter-Space Searches of Continuous Gravitational Waves}. arXiv e-prints {\href{https://arxiv.org/abs/2408.07070}{{arXiv:2408.07070}}} {[gr-qc]}

\bibitem[{{Kapadia} et~al(2017){Kapadia}, {Dent}, and {Dal Canton}}]{2017PhRvD..96j4015K}
{Kapadia} SJ, {Dent} T, {Dal Canton} T (2017) {Classifier for gravitational-wave inspiral signals in nonideal single-detector data}. Phys Rev D 96(10):104015. \doi{10.1103/PhysRevD.96.104015}, {\href{https://arxiv.org/abs/1709.02421}{{arXiv:1709.02421}}} {[astro-ph.IM]}

\bibitem[{{Karpov} et~al(2021){Karpov}, {Sitdikov}, {Huang}, and {Fryer}}]{2021JOSS....6.3199K}
{Karpov} P, {Sitdikov} I, {Huang} C, et~al (2021) {Sapsan: Framework for Supernovae Turbulence Modeling with Machine Learning}. The Journal of Open Source Software 6(67):3199. \doi{10.21105/joss.03199}

\bibitem[{Karpov et~al(2022)Karpov, Huang, Sitdikov, Fryer, Woosley, and Pilania}]{Karpov:2022tro}
Karpov PI, Huang C, Sitdikov I, et~al (2022) {Physics-informed Machine Learning for Modeling Turbulence in Supernovae}. Astrophys J 940(1):26. \doi{10.3847/1538-4357/ac88cc}, {\href{https://arxiv.org/abs/2205.08663}{{arXiv:2205.08663}}} {[physics.comp-ph]}

\bibitem[{Kawamura et~al(2011)}]{Kawamura:2011zz}
Kawamura S, et~al (2011) {The Japanese space gravitational wave antenna: DECIGO}. Class Quant Grav 28:094011. \doi{10.1088/0264-9381/28/9/094011}

\bibitem[{Keitel and Ashton(2018)}]{Keitel:2018pxz}
Keitel D, Ashton G (2018) {Faster search for long gravitational-wave transients: GPU implementation of the transient $\mathcal F$-statistic}. Class Quant Grav 35(20):205003. \doi{10.1088/1361-6382/aade34}, {\href{https://arxiv.org/abs/1805.05652}{{arXiv:1805.05652}}} {[astro-ph.IM]}

\bibitem[{Keitel et~al(2014)Keitel, Prix, Papa, Leaci, and Siddiqi}]{Keitel:2013wga}
Keitel D, Prix R, Papa MA, et~al (2014) {Search for continuous gravitational waves: Improving robustness versus instrumental artifacts}. Phys Rev D89(6):064023. \doi{10.1103/PhysRevD.89.064023}, {\href{https://arxiv.org/abs/1311.5738}{{arXiv:1311.5738}}} {[gr-qc]}

\bibitem[{Keitel et~al(2021)Keitel, Tenorio, Ashton, and Prix}]{Keitel:2021xeq}
Keitel D, Tenorio R, Ashton G, et~al (2021) {PyFstat: a Python package for continuous gravitational-wave data analysis}. J Open Source Softw 6(60):3000. \doi{10.21105/joss.03000}, {\href{https://arxiv.org/abs/2101.10915}{{arXiv:2101.10915}}} {[gr-qc]}

\bibitem[{Keitel et~al(2017)}]{Keitel:2016krm}
Keitel D, et~al (2017) {The most powerful astrophysical events: Gravitational-wave peak luminosity of binary black holes as predicted by numerical relativity}. Phys Rev D 96(2):024006. \doi{10.1103/PhysRevD.96.024006}, {\href{https://arxiv.org/abs/1612.09566}{{arXiv:1612.09566}}} {[gr-qc]}

\bibitem[{Keith et~al(2021)Keith, Khadse, and Field}]{Keith:2021arn}
Keith B, Khadse A, Field SE (2021) {Learning orbital dynamics of binary black hole systems from gravitational wave measurements}. Phys Rev Res 3(4):043101. \doi{10.1103/PhysRevResearch.3.043101}, {\href{https://arxiv.org/abs/2102.12695}{{arXiv:2102.12695}}} {[gr-qc]}

\bibitem[{Khan et~al(2022)Khan, Huerta, and Zheng}]{Khan:2021czv}
Khan A, Huerta EA, Zheng H (2022) {Interpretable AI forecasting for numerical relativity waveforms of quasicircular, spinning, nonprecessing binary black hole mergers}. Phys Rev D 105(2):024024. \doi{10.1103/PhysRevD.105.024024}, {\href{https://arxiv.org/abs/2110.06968}{{arXiv:2110.06968}}} {[gr-qc]}

\bibitem[{Khan and Chatterji(2009)}]{Khan:2009tx}
Khan R, Chatterji S (2009) {Enhancing the capabilities of LIGO time-frequency plane searches through clustering}. Class Quant Grav 26:155009. \doi{10.1088/0264-9381/26/15/155009}, {\href{https://arxiv.org/abs/0901.3762}{{arXiv:0901.3762}}} {[gr-qc]}

\bibitem[{Khan(2024)}]{Khan:2024whs}
Khan S (2024) {Probabilistic model for the gravitational wave signal from merging black holes}. Phys Rev D 109(10):104045. \doi{10.1103/PhysRevD.109.104045}, {\href{https://arxiv.org/abs/2403.11534}{{arXiv:2403.11534}}} {[gr-qc]}

\bibitem[{Khan and Green(2021)}]{Khan:2020fso}
Khan S, Green R (2021) {Gravitational-wave surrogate models powered by artificial neural networks}. Phys Rev D 103(6):064015. \doi{10.1103/PhysRevD.103.064015}, {\href{https://arxiv.org/abs/2008.12932}{{arXiv:2008.12932}}} {[gr-qc]}

\bibitem[{Khan et~al(2016)Khan, Husa, Hannam, Ohme, P{\"u}rrer, Jim\'enez~Forteza, and Boh\'e}]{Khan:2015jqa}
Khan S, Husa S, Hannam M, et~al (2016) {Frequency-domain gravitational waves from nonprecessing black-hole binaries. II. A phenomenological model for the advanced detector era}. Phys Rev D93(4):044007. \doi{10.1103/PhysRevD.93.044007}, {\href{https://arxiv.org/abs/1508.07253}{{arXiv:1508.07253}}} {[gr-qc]}

\bibitem[{Kim et~al(2021)Kim, Lee, Yuen, Hannuksela, and Li}]{Kim:2020xkm}
Kim K, Lee J, Yuen RSH, et~al (2021) {Identification of Lensed Gravitational Waves with Deep Learning}. Astrophys J 915(2):119. \doi{10.3847/1538-4357/ac0143}, {\href{https://arxiv.org/abs/2010.12093}{{arXiv:2010.12093}}} {[gr-qc]}

\bibitem[{Klimenko et~al(2008)Klimenko, Yakushin, Mercer, and Mitselmakher}]{Klimenko:2008fu}
Klimenko S, Yakushin I, Mercer A, et~al (2008) {Coherent method for detection of gravitational wave bursts}. Class Quant Grav 25:114029. \doi{10.1088/0264-9381/25/11/114029}, {\href{https://arxiv.org/abs/0802.3232}{{arXiv:0802.3232}}} {[gr-qc]}

\bibitem[{{Koloniari} et~al(2024){Koloniari}, {Koursoumpa}, {Nousi}, {Lampropoulos}, {Passalis}, {Tefas}, and {Stergioulas}}]{2024arXiv240707820K}
{Koloniari} AE, {Koursoumpa} EC, {Nousi} P, et~al (2024) {New Gravitational Wave Discoveries Enabled by Machine Learning}. arXiv e-prints arXiv:2407.07820. \doi{10.48550/arXiv.2407.07820}, {\href{https://arxiv.org/abs/2407.07820}{{arXiv:2407.07820}}} {[gr-qc]}

\bibitem[{Korsakova et~al(2024)Korsakova, Babak, Katz, Karnesis, Khukhlaev, and Gair}]{Korsakova:2024sut}
Korsakova N, Babak S, Katz ML, et~al (2024) {Neural density estimation for Galactic binaries in the LISA data analysis}. Phys Rev D 110(10):104069. \doi{10.1103/PhysRevD.110.104069}, {\href{https://arxiv.org/abs/2402.13701}{{arXiv:2402.13701}}} {[gr-qc]}

\bibitem[{Kovalam et~al(2022)Kovalam, Patwary, Sreekumar, Wen, Panther, and Chu}]{Kovalam_2022}
Kovalam M, Patwary MAK, Sreekumar AK, et~al (2022) Early warnings of binary neutron star coalescence using the spiir search. Astrophys J Lett 927(1):L9. \doi{10.3847/2041-8213/ac5687}

\bibitem[{Krastev(2020)}]{Krastev:2019koe}
Krastev PG (2020) {Real-Time Detection of Gravitational Waves from Binary Neutron Stars using Artificial Neural Networks}. Phys Lett B 803:135330. \doi{10.1016/j.physletb.2020.135330}, {\href{https://arxiv.org/abs/1908.03151}{{arXiv:1908.03151}}} {[astro-ph.IM]}

\bibitem[{Krastev et~al(2021)Krastev, Gill, Villar, and Berger}]{Krastev:2020skk}
Krastev PG, Gill K, Villar VA, et~al (2021) {Detection and Parameter Estimation of Gravitational Waves from Binary Neutron-Star Mergers in Real LIGO Data using Deep Learning}. Phys Lett B 815:136161. \doi{10.1016/j.physletb.2021.136161}, {\href{https://arxiv.org/abs/2012.13101}{{arXiv:2012.13101}}} {[astro-ph.IM]}

\bibitem[{Krenn et~al(2023)Krenn, Drori, and Adhikari}]{Krenn:2023ljj}
Krenn M, Drori Y, Adhikari RX (2023) {Digital Discovery of interferometric Gravitational Wave Detectors}. arXiv e-prints {\href{https://arxiv.org/abs/2312.04258}{{arXiv:2312.04258}}} {[astro-ph.IM]}

\bibitem[{Kumar et~al(2022)Kumar, Nitz, and Forteza}]{Kumar:2022tto}
Kumar S, Nitz AH, Forteza XJ (2022) {Parameter estimation with non stationary noise in gravitational waves data}. arXiv e-prints {\href{https://arxiv.org/abs/2202.12762}{{arXiv:2202.12762}}} {[astro-ph.IM]}

\bibitem[{Kyutoku et~al(2021)Kyutoku, Shibata, and Taniguchi}]{Kyutoku:2021icp}
Kyutoku K, Shibata M, Taniguchi K (2021) {Coalescence of black hole-neutron star binaries}. Living Rev Rel 24(1):5. \doi{10.1007/s41114-021-00033-4}, {\href{https://arxiv.org/abs/2110.06218}{{arXiv:2110.06218}}} {[astro-ph.HE]}

\bibitem[{Laguarta et~al(2024)}]{Laguarta:2023evo}
Laguarta P, et~al (2024) {Detection of anomalies amongst LIGO's glitch populations with autoencoders}. Class Quant Grav 41(5):055004. \doi{10.1088/1361-6382/ad1f26}, {\href{https://arxiv.org/abs/2310.03453}{{arXiv:2310.03453}}} {[astro-ph.IM]}

\bibitem[{{Lasky} et~al(2017){Lasky}, {Sarin}, and {Sammut}}]{T1700408}
{Lasky} PD, {Sarin} N, {Sammut} L (2017) Long-duration waveform models for millisecond magnetars born in binary neutron star mergers. Tech. Rep. LIGO-T1700408, {LIGO Laboratory}, \urlprefix\url{https://dcc.ligo.org/LIGO-T1700408/public}

\bibitem[{Lattimer(2021)}]{Lattimer:2021emm}
Lattimer JM (2021) {Neutron Stars and the Nuclear Matter Equation of State}. Ann Rev Nucl Part Sci 71:433--464. \doi{10.1146/annurev-nucl-102419-124827}

\bibitem[{Leaci(2015)}]{Leaci:2015iuc}
Leaci P (2015) {Methods to filter out spurious disturbances in continuous-wave searches from gravitational-wave detectors}. Phys Scripta 90(12):125001. \doi{10.1088/0031-8949/90/12/125001}

\bibitem[{Lee et~al(2021)Lee, Oh, Kim, Cho, Oh, Son, and Lee}]{Lee:2021isa}
Lee J, Oh SH, Kim K, et~al (2021) {Deep learning model on gravitational waveforms in merging and ringdown phases of binary black hole coalescences}. Phys Rev D 103(12):123023. \doi{10.1103/PhysRevD.103.123023}, {\href{https://arxiv.org/abs/2101.05685}{{arXiv:2101.05685}}} {[astro-ph.IM]}

\bibitem[{{Leyde} et~al(2023){Leyde}, {Green}, {Toubiana}, and {Gair}}]{leyde_2023}
{Leyde} K, {Green} SR, {Toubiana} A, et~al (2023) {Gravitational wave populations and cosmology with neural posterior estimation}. arXiv e-prints arXiv:2311.12093. \doi{10.48550/arXiv.2311.12093}, {\href{https://arxiv.org/abs/2311.12093}{{arXiv:2311.12093}}} {[gr-qc]}

\bibitem[{Li et~al(2023{\natexlab{a}})Li, Pei, and Li}]{LI2023110176}
Li P, Pei Y, Li J (2023{\natexlab{a}}) A comprehensive survey on design and application of autoencoder in deep learning. Applied Soft Computing 138:110176. \doi{10.1016/j.asoc.2023.110176}

\bibitem[{Li et~al(2020)Li, Babu, Yu, and Fan}]{Li:2017chi}
Li XR, Babu GJ, Yu WL, et~al (2020) {Some optimizations on detecting gravitational wave using convolutional neural network}. Front Phys (Beijing) 15(5):54501. \doi{10.1007/s11467-020-0966-4}, {\href{https://arxiv.org/abs/1712.00356}{{arXiv:1712.00356}}} {[astro-ph.IM]}

\bibitem[{Li et~al(2023{\natexlab{b}})Li, Li, and Pang}]{Li:2023eys}
Li ZH, Li CQ, Pang LG (2023{\natexlab{b}}) {Solving Einstein equations using deep learning}. arXiv e-prints {\href{https://arxiv.org/abs/2309.07397}{{arXiv:2309.07397}}} {[gr-qc]}

\bibitem[{{LIGO Scientific Collaboration, Virgo Collaboration, KAGRA Collaboration}(2023)}]{LVK:OBSWP}
{LIGO Scientific Collaboration, Virgo Collaboration, KAGRA Collaboration} (2023) {The LSC-Virgo-KAGRA Observational Science White Paper}. Tech. Rep. LIGO-T2300406, LIGO Laboratory, \urlprefix\url{https://dcc.ligo.org/LIGO-T2300406/public}

\bibitem[{Lin et~al(2020)Lin, Li, and Yu}]{Lin:2019egc}
Lin BJ, Li XR, Yu WL (2020) {Binary neutron stars gravitational wave detection based on wavelet packet analysis and convolutional neural networks}. Front Phys (Beijing) 15(2):24602. \doi{10.1007/s11467-019-0935-y}, {\href{https://arxiv.org/abs/1910.10525}{{arXiv:1910.10525}}} {[astro-ph.IM]}

\bibitem[{{Linardatos} et~al(2020){Linardatos}, {Papastefanopoulos}, and {Kotsiantis}}]{2020Entrp..23...18L}
{Linardatos} P, {Papastefanopoulos} V, {Kotsiantis} S (2020) {Explainable AI: A Review of Machine Learning Interpretability Methods}. Entropy 23(1):18. \doi{10.3390/e23010018}

\bibitem[{Liodis et~al(2024)Liodis, Smirniotis, and Stergioulas}]{Liodis:2023adg}
Liodis I, Smirniotis E, Stergioulas N (2024) {Neural-network-based surrogate model for the properties of neutron stars in 4D Einstein-Gauss-Bonnet gravity}. Phys Rev D 109(10):104008. \doi{10.1103/PhysRevD.109.104008}, {\href{https://arxiv.org/abs/2309.03991}{{arXiv:2309.03991}}} {[gr-qc]}

\bibitem[{Liu et~al(2023)Liu, Li, and Chua}]{Liu:2023oxw}
Liu M, Li XD, Chua AJK (2023) {Improving the scalability of the Gaussian-process error marginalization in gravitational-wave inference}. Phys Rev D 108(10):103027. \doi{10.1103/PhysRevD.108.103027}, {\href{https://arxiv.org/abs/2307.07233}{{arXiv:2307.07233}}} {[astro-ph.IM]}

\bibitem[{Lopez et~al(2022)Lopez, Gayathri, Pai, Heng, Messenger, and Gupta}]{Lopez:2021ikt}
Lopez D, Gayathri V, Pai A, et~al (2022) {Utilizing Gaussian mixture models in all-sky searches for short-duration gravitational wave bursts}. Phys Rev D 105(6):063024. \doi{10.1103/PhysRevD.105.063024}, {\href{https://arxiv.org/abs/2112.06608}{{arXiv:2112.06608}}} {[gr-qc]}

\bibitem[{{Lopez} et~al(2022){Lopez}, {Gayathri}, {Pai}, {Heng}, {Messenger}, and {Gupta}}]{2022PhRvD.105f3024L}
{Lopez} D, {Gayathri} V, {Pai} A, et~al (2022) {Utilizing Gaussian mixture models in all-sky searches for short-duration gravitational wave bursts}. \prd 105(6):063024. \doi{10.1103/PhysRevD.105.063024}, {\href{https://arxiv.org/abs/2112.06608}{{arXiv:2112.06608}}} {[gr-qc]}

\bibitem[{{L{\'o}pez} et~al(2021){L{\'o}pez}, {Di Palma}, {Drago}, {Cerd{\'a}-Dur{\'a}n}, and {Ricci}}]{2021PhRvD.103f3011L}
{L{\'o}pez} M, {Di Palma} I, {Drago} M, et~al (2021) {Deep learning for core-collapse supernova detection}. Phys Rev D 103(6):063011. \doi{10.1103/PhysRevD.103.063011}

\bibitem[{Lopez et~al(2022)Lopez, Boudart, Buijsman, Reza, and Caudill}]{PhysRevD.106.023027}
Lopez M, Boudart V, Buijsman K, et~al (2022) Simulating transient noise bursts in ligo with generative adversarial networks. Phys Rev D 106:023027. \doi{10.1103/PhysRevD.106.023027}

\bibitem[{Luna et~al(2023)Luna, Calder\'on~Bustillo, Mart\'\i{}nez, Torres-Forn\'e, and Font}]{Luna:2022rql}
Luna R, Calder\'on~Bustillo J, Mart\'\i{}nez JJS, et~al (2023) {Solving the Teukolsky equation with physics-informed neural networks}. Phys Rev D 107(6):064025. \doi{10.1103/PhysRevD.107.064025}, {\href{https://arxiv.org/abs/2212.06103}{{arXiv:2212.06103}}} {[gr-qc]}

\bibitem[{Luo et~al(2016)}]{TianQin:2015yph}
Luo J, et~al (2016) {TianQin: a space-borne gravitational wave detector}. Class Quant Grav 33(3):035010. \doi{10.1088/0264-9381/33/3/035010}, {\href{https://arxiv.org/abs/1512.02076}{{arXiv:1512.02076}}} {[astro-ph.IM]}

\bibitem[{Ma and Vajente(2024)}]{ma2023deep}
Ma PX, Vajente G (2024) {A deep learning technique to control the non-linear dynamics of a gravitational-wave interferometer}. Class Quant Grav 41(4):045003. \doi{10.1088/1361-6382/ad1daa}, {\href{https://arxiv.org/abs/2302.07921}{{arXiv:2302.07921}}} {[cs.LG]}

\bibitem[{Macas et~al(2022)Macas, Pooley, Nuttall, Davis, Dyer, Lecoeuche, Lyman, McIver, and Rink}]{Macas:2022afm}
Macas R, Pooley J, Nuttall LK, et~al (2022) {Impact of noise transients on low latency gravitational-wave event localization}. Phys Rev D 105(10):103021. \doi{10.1103/PhysRevD.105.103021}, {\href{https://arxiv.org/abs/2202.00344}{{arXiv:2202.00344}}} {[astro-ph.HE]}

\bibitem[{Macquet et~al(2021)Macquet, Bizouard, Christensen, and Coughlin}]{Macquet:2021ttq}
Macquet A, Bizouard MA, Christensen N, et~al (2021) {Long-duration transient gravitational-wave search pipeline}. Phys Rev D 104(10):102005. \doi{10.1103/PhysRevD.104.102005}, {\href{https://arxiv.org/abs/2108.10588}{{arXiv:2108.10588}}} {[astro-ph.IM]}

\bibitem[{Magare et~al(2024)Magare, More, and Choudary}]{Magare:2024wje}
Magare S, More A, Choudary S (2024) {SLICK: Strong Lensing Identification of Candidates Kindred in gravitational wave data}. arXiv e-prints {\href{https://arxiv.org/abs/2403.02994}{{arXiv:2403.02994}}} {[astro-ph.HE]}

\bibitem[{Magee et~al(2021)Magee, Chatterjee, Singer, Sachdev, Kovalam, Mo, Anderson, Brady, Brockill, Cannon, Canton, Chu, Clearwater, Codoreanu, Drago, Godwin, Ghosh, Greco, Hanna, Kapadia, Katsavounidis, Oloworaran, Pace, Panther, Patwary, Pietri, Piotrzkowski, Prestegard, Rei, Sreekumar, Szczepańczyk, Valsan, Viets, Wade, Wen, and Zweizig}]{Magee_2021}
Magee R, Chatterjee D, Singer LP, et~al (2021) First demonstration of early warning gravitational-wave alerts. Astrophys J Lett 910(2):L21. \doi{10.3847/2041-8213/abed54}

\bibitem[{Maggiore(2000)}]{Maggiore:1999vm}
Maggiore M (2000) {Gravitational wave experiments and early universe cosmology}. Phys Rept 331:283--367. \doi{10.1016/S0370-1573(99)00102-7}, {\href{https://arxiv.org/abs/gr-qc/9909001}{{arXiv:gr-qc/9909001}}}

\bibitem[{Maggiore et~al(2020)}]{Maggiore:2019uih}
Maggiore M, et~al (2020) {Science Case for the Einstein Telescope}. JCAP 03:050. \doi{10.1088/1475-7516/2020/03/050}, {\href{https://arxiv.org/abs/1912.02622}{{arXiv:1912.02622}}} {[astro-ph.CO]}

\bibitem[{Mao et~al(2024)Mao, Lee, Burke, Chua, Edwards, and Meyer}]{Mao:2023zdr}
Mao R, Lee JE, Burke O, et~al (2024) {Calibrating approximate Bayesian credible intervals of gravitational-wave parameters}. Phys Rev D 109(8):083002. \doi{10.1103/PhysRevD.109.083002}, {\href{https://arxiv.org/abs/2310.06321}{{arXiv:2310.06321}}} {[gr-qc]}

\bibitem[{Marianer et~al(2020)Marianer, Poznanski, and Prochaska}]{10.1093/mnras/staa3550}
Marianer T, Poznanski D, Prochaska JX (2020) {A semisupervised machine learning search for never-seen gravitational-wave sources}. Mon Not R Astron, Soc 500(4):5408--5419. \doi{10.1093/mnras/staa3550}

\bibitem[{{Markidis}(2021)}]{2021arXiv210309655M}
{Markidis} S (2021) {The Old and the New: Can Physics-Informed Deep-Learning Replace Traditional Linear Solvers?} Frontiers Big Data 4. \doi{10.3389/fdata.2021.669097}, {\href{https://arxiv.org/abs/2103.09655}{{arXiv:2103.09655}}} {[math.NA]}

\bibitem[{{Marx} et~al(2024){Marx}, {Benoit}, {Gunny}, {Omer}, {Chatterjee}, {Venterea}, {Wills}, {Saleem}, {Moreno}, {Raikman}, {Govorkova}, {Rankin}, {Coughlin}, {Harris}, and {Katsavounidis}}]{2024arXiv240318661M}
{Marx} E, {Benoit} W, {Gunny} A, et~al (2024) {A machine-learning pipeline for real-time detection of gravitational waves from compact binary coalescences}. arXiv e-prints arXiv:2403.18661. \doi{10.48550/arXiv.2403.18661}, {\href{https://arxiv.org/abs/2403.18661}{{arXiv:2403.18661}}} {[gr-qc]}

\bibitem[{Maćkiewicz and Ratajczak(1993)}]{MACKIEWICZ1993303}
Maćkiewicz A, Ratajczak W (1993) Principal components analysis (pca). Computers \& Geosciences 19(3):303--342. \doi{10.1016/0098-3004(93)90090-R}

\bibitem[{McGinn et~al(2021)McGinn, Messenger, Williams, and Heng}]{McGinn_2021}
McGinn J, Messenger C, Williams MJ, et~al (2021) Generalised gravitational wave burst generation with generative adversarial networks. Class Quantum Grav 38(15):155005. \doi{10.1088/1361-6382/ac09cc}

\bibitem[{Mei et~al(2021)}]{TianQin:2020hid}
Mei J, et~al (2021) {The TianQin project: current progress on science and technology}. PTEP 2021(5):05A107. \doi{10.1093/ptep/ptaa114}, {\href{https://arxiv.org/abs/2008.10332}{{arXiv:2008.10332}}} {[gr-qc]}

\bibitem[{Meijer et~al(2024)Meijer, Lopez, Tsuna, and Caudill}]{PhysRevD.109.022006}
Meijer Q, Lopez M, Tsuna D, et~al (2024) Gravitational-wave searches for cosmic string cusps in einstein telescope data using deep learning. Phys Rev D 109:022006. \doi{10.1103/PhysRevD.109.022006}

\bibitem[{Melatos et~al(2021)Melatos, Clearwater, Suvorova, Sun, Moran, and Evans}]{Melatos:2021mmz}
Melatos A, Clearwater P, Suvorova S, et~al (2021) {Hidden Markov model tracking of continuous gravitational waves from a binary neutron star with wandering spin. III. Rotational phase tracking}. Phys Rev D 104(4):042003. \doi{10.1103/PhysRevD.104.042003}, {\href{https://arxiv.org/abs/2107.12822}{{arXiv:2107.12822}}} {[gr-qc]}

\bibitem[{Messenger et~al(2021)Messenger, Zerafa, Cuoco, Williams, and Reade}]{kaggle1}
Messenger C, Zerafa C, Cuoco E, et~al (2021) G2net gravitational wave detection. \urlprefix\url{https://kaggle.com/competitions/g2net-gravitational-wave-detection}

\bibitem[{{M{\'e}sz{\'a}ros} et~al(2019){M{\'e}sz{\'a}ros}, {Fox}, {Hanna}, and {Murase}}]{2019NatRP...1..585M}
{M{\'e}sz{\'a}ros} P, {Fox} DB, {Hanna} C, et~al (2019) {Multi-messenger astrophysics}. Nature Reviews Physics 1(10):585--599. \doi{10.1038/s42254-019-0101-z}, {\href{https://arxiv.org/abs/1906.10212}{{arXiv:1906.10212}}} {[astro-ph.HE]}

\bibitem[{{Mezzacappa} and {Zanolin}(2024)}]{2024arXiv240111635M}
{Mezzacappa} A, {Zanolin} M (2024) {Gravitational Waves from Neutrino-Driven Core Collapse Supernovae: Predictions, Detection, and Parameter Estimation}. arXiv e-prints arXiv:2401.11635. \doi{10.48550/arXiv.2401.11635}, {\href{https://arxiv.org/abs/2401.11635}{{arXiv:2401.11635}}} {[astro-ph.HE]}

\bibitem[{Miller and Mendes(2023)}]{Miller:2023kkd}
Miller AL, Mendes L (2023) {First search for ultralight dark matter with a space-based gravitational-wave antenna: LISA Pathfinder}. Phys Rev D 107(6):063015. \doi{10.1103/PhysRevD.107.063015}, {\href{https://arxiv.org/abs/2301.08736}{{arXiv:2301.08736}}} {[gr-qc]}

\bibitem[{Miller et~al(2022)Miller, Aggarwal, Clesse, and De~Lillo}]{Miller:2021knj}
Miller AL, Aggarwal N, Clesse S, et~al (2022) {Constraints on planetary and asteroid-mass primordial black holes from continuous gravitational-wave searches}. Phys Rev D 105(6):062008. \doi{10.1103/PhysRevD.105.062008}, {\href{https://arxiv.org/abs/2110.06188}{{arXiv:2110.06188}}} {[gr-qc]}

\bibitem[{Miller et~al(2024)Miller, Aggarwal, Clesse, De~Lillo, Sachdev, Astone, Palomba, Piccinni, and Pierini}]{Miller:2024fpo}
Miller AL, Aggarwal N, Clesse S, et~al (2024) {Gravitational Wave Constraints on Planetary-Mass Primordial Black Holes Using LIGO O3a Data}. Phys Rev Lett 133(11):111401. \doi{10.1103/PhysRevLett.133.111401}, {\href{https://arxiv.org/abs/2402.19468}{{arXiv:2402.19468}}} {[gr-qc]}

\bibitem[{Miller et~al(2019)}]{Miller:2019jtp}
Miller AL, et~al (2019) {How effective is machine learning to detect long transient gravitational waves from neutron stars in a real search?} Phys Rev D100(6):062005. \doi{10.1103/PhysRevD.100.062005}, {\href{https://arxiv.org/abs/1909.02262}{{arXiv:1909.02262}}} {[astro-ph.IM]}

\bibitem[{Ming et~al(2022)Ming, Papa, Eggenstein, Machenschalk, Steltner, Prix, Allen, and Behnke}]{Ming:2021xtz}
Ming J, Papa MA, Eggenstein HB, et~al (2022) {Results From an Einstein@Home Search for Continuous Gravitational Waves From G347.3 at Low Frequencies in LIGO O2 Data}. Astrophys J 925(1):8. \doi{10.3847/1538-4357/ac35cb}, {\href{https://arxiv.org/abs/2108.02808}{{arXiv:2108.02808}}} {[gr-qc]}

\bibitem[{Mirasola and Tenorio(2024)}]{Mirasola:2024lcq}
Mirasola L, Tenorio R (2024) {Towards a computationally-efficient follow-up pipeline for blind continuous gravitational-wave searches}. arXiv e-prints {\href{https://arxiv.org/abs/2405.18934}{{arXiv:2405.18934}}} {[gr-qc]}

\bibitem[{Mishra et~al(2021)Mishra, O'Brien, Gayathri, Szczepa{\'n}czyk, Bhaumik, Bartos, and Klimenko}]{PhysRevD.104.023014}
Mishra T, O'Brien B, Gayathri V, et~al (2021) Optimization of model independent gravitational wave search for binary black hole mergers using machine learning. Phys Rev D 104:023014. \doi{10.1103/PhysRevD.104.023014}

\bibitem[{Mishra et~al(2022)Mishra, O'Brien, Szczepa{\'n}czyk, Vedovato, Bhaumik, Gayathri, Prodi, Salemi, Milotti, Bartos, and Klimenko}]{PhysRevD.105.083018}
Mishra T, O'Brien B, Szczepa{\'n}czyk M, et~al (2022) Search for binary black hole mergers in the third observing run of advanced ligo-virgo using coherent waveburst enhanced with machine learning. Phys Rev D 105:083018. \doi{10.1103/PhysRevD.105.083018}

\bibitem[{Modafferi et~al(2023)Modafferi, Tenorio, and Keitel}]{Modafferi:2023nzt}
Modafferi LM, Tenorio R, Keitel D (2023) {Convolutional neural network search for long-duration transient gravitational waves from glitching pulsars}. Phys Rev D 108(2):023005. \doi{10.1103/PhysRevD.108.023005}, {\href{https://arxiv.org/abs/2303.16720}{{arXiv:2303.16720}}} {[astro-ph.HE]}

\bibitem[{Mogushi et~al(2021)Mogushi, Quitzow-James, Cavaglià, Kulkarni, and Hayes}]{Mogushi_2021}
Mogushi K, Quitzow-James R, Cavaglià M, et~al (2021) Nnetfix: an artificial neural network-based denoising engine for gravitational-wave signals. Machine Learning: Science and Technology 2(3):035018. \doi{10.1088/2632-2153/abea69}

\bibitem[{Moore and Gair(2014)}]{Moore:2014pda}
Moore CJ, Gair JR (2014) {Novel Method for Incorporating Model Uncertainties into Gravitational Wave Parameter Estimates}. Phys Rev Lett 113:251101. \doi{10.1103/PhysRevLett.113.251101}, {\href{https://arxiv.org/abs/1412.3657}{{arXiv:1412.3657}}} {[gr-qc]}

\bibitem[{Moore et~al(2016)Moore, Berry, Chua, and Gair}]{Moore:2015sza}
Moore CJ, Berry CPL, Chua AJK, et~al (2016) {Improving gravitational-wave parameter estimation using Gaussian process regression}. Phys Rev D93(6):064001. \doi{10.1103/PhysRevD.93.064001}, {\href{https://arxiv.org/abs/1509.04066}{{arXiv:1509.04066}}} {[gr-qc]}

\bibitem[{Morawski et~al(2020)Morawski, Bejger, and Cieciel\textbackslash{}{a}g}]{Morawski:2019awi}
Morawski F, Bejger M, Cieciel\textbackslash{}{a}g P (2020) {Convolutional neural network classifier for the output of the time-domain F-statistic all-sky search for continuous gravitational waves}. Mach Learn Sci Tech 1(2):025016. \doi{10.1088/2632-2153/ab86c7}, {\href{https://arxiv.org/abs/1907.06917}{{arXiv:1907.06917}}} {[astro-ph.IM]}

\bibitem[{Morawski et~al(2021)Morawski, Bejger, Cuoco, and Petre}]{Morawski:2021kxv}
Morawski F, Bejger M, Cuoco E, et~al (2021) {Anomaly detection in gravitational waves data using convolutional autoencoders}. Mach Learn Sci Tech 2(4):045014. \doi{10.1088/2632-2153/abf3d0}, {\href{https://arxiv.org/abs/2103.07688}{{arXiv:2103.07688}}} {[astro-ph.IM]}

\bibitem[{Mozzon et~al(2020)Mozzon, Nuttall, Lundgren, Dent, Kumar, and Nitz}]{Mozzon:2020gwa}
Mozzon S, Nuttall LK, Lundgren A, et~al (2020) {Dynamic Normalization for Compact Binary Coalescence Searches in Non-Stationary Noise}. Class Quant Grav 37(21):215014. \doi{10.1088/1361-6382/abac6c}, {\href{https://arxiv.org/abs/2002.09407}{{arXiv:2002.09407}}} {[astro-ph.IM]}

\bibitem[{Mozzon et~al(2022)Mozzon, Ashton, Nuttall, and Williamson}]{Mozzon:2021wam}
Mozzon S, Ashton G, Nuttall LK, et~al (2022) {Does nonstationary noise in LIGO and Virgo affect the estimation of H0?} Phys Rev D 106(4):043504. \doi{10.1103/PhysRevD.106.043504}, {\href{https://arxiv.org/abs/2110.11731}{{arXiv:2110.11731}}} {[astro-ph.CO]}

\bibitem[{{Mukherjee} et~al(2017){Mukherjee}, {Salazar}, {Mittelstaedt}, and {Valdez}}]{2017PhRvD..96j4033M}
{Mukherjee} S, {Salazar} L, {Mittelstaedt} J, et~al (2017) {New method for enhanced efficiency in detection of gravitational waves from supernovae using coherent network of detectors}. Phys Rev D 96(10):104033. \doi{10.1103/PhysRevD.96.104033}

\bibitem[{{Mukherjee} et~al(2021){Mukherjee}, {Nurbek}, and {Valdez}}]{2021PhRvD.103j3008M}
{Mukherjee} S, {Nurbek} G, {Valdez} O (2021) {Study of efficient methods of detection and reconstruction of gravitational waves from nonrotating 3D general relativistic core collapse supernovae explosion using multilayer signal estimation method}. Phys Rev D 103(10):103008. \doi{10.1103/PhysRevD.103.103008}

\bibitem[{Mumuni and Mumuni(2022)}]{MUMUNI2022100258}
Mumuni A, Mumuni F (2022) Data augmentation: A comprehensive survey of modern approaches. Array 16:100258. \doi{10.1016/j.array.2022.100258}

\bibitem[{{Murdoch} et~al(2019){Murdoch}, {Singh}, {Kumbier}, {Abbasi-Asl}, and {Yu}}]{2019PNAS..11622071M}
{Murdoch} WJ, {Singh} C, {Kumbier} K, et~al (2019) {Definitions, methods, and applications in interpretable machine learning}. Proceedings of the National Academy of Science 116(44):22071--22080. \doi{10.1073/pnas.1900654116}

\bibitem[{Mytidis et~al(2019)Mytidis, Panagopoulos, Panagopoulos, Miller, and Whiting}]{Mytidis:2015kea}
Mytidis A, Panagopoulos AA, Panagopoulos OP, et~al (2019) {Sensitivity study using machine learning algorithms on simulated r-mode gravitational wave signals from newborn neutron stars}. Phys Rev D 99(2):024024. \doi{10.1103/PhysRevD.99.024024}, {\href{https://arxiv.org/abs/1508.02064}{{arXiv:1508.02064}}} {[astro-ph.IM]}

\bibitem[{Nagar et~al(2018)}]{Nagar:2018zoe}
Nagar A, et~al (2018) {Time-domain effective-one-body gravitational waveforms for coalescing compact binaries with nonprecessing spins, tides and self-spin effects}. Phys Rev D 98(10):104052. \doi{10.1103/PhysRevD.98.104052}, {\href{https://arxiv.org/abs/1806.01772}{{arXiv:1806.01772}}} {[gr-qc]}

\bibitem[{Neyman and Pearson(1933)}]{neyman1933}
Neyman J, Pearson E (1933) {On the Problem of the Most Efficient Tests of Statistical Hypotheses}. Phil\ Trans\ R\ Soc\ London 231:289--337. \doi{10.1098/rsta.1933.00093}

\bibitem[{Nguyen et~al(2021)}]{AdvLIGO:2021oxw}
Nguyen P, et~al (2021) {Environmental noise in advanced LIGO detectors}. Class Quant Grav 38(14):145001. \doi{10.1088/1361-6382/ac011a}, {\href{https://arxiv.org/abs/2101.09935}{{arXiv:2101.09935}}} {[astro-ph.IM]}

\bibitem[{Nielsen(2015)}]{nielsen2015neural}
Nielsen MA (2015) Neural networks and deep learning, vol~25. Determination press, San Francisco

\bibitem[{Nitz(2018)}]{Nitz:2017lco}
Nitz AH (2018) {Distinguishing short duration noise transients in LIGO data to improve the PyCBC search for gravitational waves from high mass binary black hole mergers}. Class Quant Grav 35(3):035016. \doi{10.1088/1361-6382/aaa13d}, {\href{https://arxiv.org/abs/1709.08974}{{arXiv:1709.08974}}} {[gr-qc]}

\bibitem[{Nitz et~al(2017)Nitz, Harry, Willis, Biwer, Brown, Pekowsky, Dal~Canton, Williamson, Dent, Capano, Massinger, Lenon, Nielsen, and Cabero}]{pycbc-github}
Nitz AH, Harry IW, Willis JL, et~al (2017) {PyCBC Software}. \href{https://github.com/ligo-cbc/pycbc}{github.com/ligo-cbc/pycbc}, \doi{10.5281/zenodo.344823}

\bibitem[{Nitz et~al(2018)Nitz, Dal~Canton, Davis, and Reyes}]{Nitz:2018rgo}
Nitz AH, Dal~Canton T, Davis D, et~al (2018) {Rapid detection of gravitational waves from compact binary mergers with PyCBC Live}. Phys Rev D 98(2):024050. \doi{10.1103/PhysRevD.98.024050}, {\href{https://arxiv.org/abs/1805.11174}{{arXiv:1805.11174}}} {[gr-qc]}

\bibitem[{Nitz et~al(2019)Nitz, Capano, Nielsen, Reyes, White, Brown, and Krishnan}]{Nitz_2019}
Nitz AH, Capano C, Nielsen AB, et~al (2019) 1-ogc: The first open gravitational-wave catalog of binary mergers from analysis of public advanced ligo data. Astrophys J 872(2):195. \doi{10.3847/1538-4357/ab0108}

\bibitem[{Nitz et~al(2020{\natexlab{a}})Nitz, Dent, Davies, Kumar, Capano, Harry, Mozzon, Nuttall, Lundgren, and Tápai}]{Nitz_2020}
Nitz AH, Dent T, Davies GS, et~al (2020{\natexlab{a}}) 2-ogc: Open gravitational-wave catalog of binary mergers from analysis of public advanced ligo and virgo data. Astrophys J 891(2):123. \doi{10.3847/1538-4357/ab733f}

\bibitem[{Nitz et~al(2020{\natexlab{b}})Nitz, Schäfer, and Canton}]{Nitz_2020EW}
Nitz AH, Schäfer M, Canton TD (2020{\natexlab{b}}) Gravitational-wave merger forecasting: Scenarios for the early detection and localization of compact-binary mergers with ground-based observatories. Astrophys J Lett 902(2):L29. \doi{10.3847/2041-8213/abbc10}

\bibitem[{Nousi et~al(2022)Nousi, Fragkouli, Passalis, Iosif, Apostolatos, Pappas, Stergioulas, and Tefas}]{Nousi:2021arn}
Nousi P, Fragkouli SC, Passalis N, et~al (2022) {Autoencoder-driven Spiral Representation Learning for Gravitational Wave Surrogate Modelling}. Neurocomput 491:67--77. \doi{10.1016/j.neucom.2022.03.052}, {\href{https://arxiv.org/abs/2107.04312}{{arXiv:2107.04312}}} {[cs.LG]}

\bibitem[{{Nousi} et~al(2023){Nousi}, {Koloniari}, {Passalis}, {Iosif}, {Stergioulas}, and {Tefas}}]{2023PhRvD.108b4022N}
{Nousi} P, {Koloniari} AE, {Passalis} N, et~al (2023) {Deep residual networks for gravitational wave detection}. \prd 108(2):024022. \doi{10.1103/PhysRevD.108.024022}, {\href{https://arxiv.org/abs/2211.01520}{{arXiv:2211.01520}}} {[gr-qc]}

\bibitem[{Ogata(2001)}]{10.5555/516039}
Ogata K (2001) Modern Control Engineering, 4th edn. Prentice Hall PTR, USA

\bibitem[{Olah et~al(2017)Olah, Mordvintsev, and Schubert}]{olah2017feature}
Olah C, Mordvintsev A, Schubert L (2017) Feature visualization. Distill 2(11):e7

\bibitem[{Ormiston et~al(2020)Ormiston, Nguyen, Coughlin, Adhikari, and Katsavounidis}]{PhysRevResearch.2.033066}
Ormiston R, Nguyen T, Coughlin M, et~al (2020) Noise reduction in gravitational-wave data via deep learning. Phys Rev Res 2:033066. \doi{10.1103/PhysRevResearch.2.033066}, {\href{https://arxiv.org/abs/2005.06534}{{arXiv:2005.06534}}} {[astro-ph.IM]}

\bibitem[{Osisanwo et~al(2017)Osisanwo, Akinsola, Awodele, Hinmikaiye, Olakanmi, and Akinjobi}]{osisanwo2017supervised}
Osisanwo F, Akinsola J, Awodele O, et~al (2017) Supervised machine learning algorithms: Classification and comparison. International Journal of Computer Trends and Technology (IJCTT) 48(3):128--138. \urlprefix\url{http://www.ijcttjournal.org}

\bibitem[{Ossokine et~al(2020)}]{Ossokine:2020kjp}
Ossokine S, et~al (2020) {Multipolar Effective-One-Body Waveforms for Precessing Binary Black Holes: Construction and Validation}. Phys Rev D 102(4):044055. \doi{10.1103/PhysRevD.102.044055}, {\href{https://arxiv.org/abs/2004.09442}{{arXiv:2004.09442}}} {[gr-qc]}

\bibitem[{Owen et~al(2022)Owen, Lindblom, and Pinheiro}]{Owen:2022mvu}
Owen BJ, Lindblom L, Pinheiro LS (2022) {First Constraining Upper Limits on Gravitational-wave Emission from NS 1987A in SNR 1987A}. Astrophys J Lett 935(1):L7. \doi{10.3847/2041-8213/ac84dc}, {\href{https://arxiv.org/abs/2206.01168}{{arXiv:2206.01168}}} {[gr-qc]}

\bibitem[{Papa et~al(2016)}]{Papa:2016cwb}
Papa MA, et~al (2016) {Hierarchical follow-up of subthreshold candidates of an all-sky Einstein@Home search for continuous gravitational waves on LIGO sixth science run data}. Phys Rev D 94(12):122006. \doi{10.1103/PhysRevD.94.122006}, {\href{https://arxiv.org/abs/1608.08928}{{arXiv:1608.08928}}} {[astro-ph.IM]}

\bibitem[{{Papamakarios} and {Murray}(2016)}]{2016arXiv160506376P}
{Papamakarios} G, {Murray} I (2016) {Fast $\epsilon$-free Inference of Simulation Models with Bayesian Conditional Density Estimation}. arXiv e-prints arXiv:1605.06376. \doi{10.48550/arXiv.1605.06376}, {\href{https://arxiv.org/abs/1605.06376}{{arXiv:1605.06376}}} {[stat.ML]}

\bibitem[{Patel et~al(2024)Patel, Aykutalp, and Laguna}]{Patel:2024wzo}
Patel N, Aykutalp A, Laguna P (2024) {Calculating Quasi-Normal Modes of Schwarzschild Black Holes with Physics Informed Neural Networks}. arXiv e-prints {\href{https://arxiv.org/abs/2401.01440}{{arXiv:2401.01440}}} {[gr-qc]}

\bibitem[{Pereira and Sturani(2024)}]{Pereira:2022kqn}
Pereira T, Sturani R (2024) {Deep learning waveform anomaly detector for numerical relativity catalogs}. Gen Rel Grav 56(2):24. \doi{10.1007/s10714-024-03216-w}, {\href{https://arxiv.org/abs/2210.07299}{{arXiv:2210.07299}}} {[gr-qc]}

\bibitem[{Piccinni(2022)}]{Piccinni:2022vsd}
Piccinni OJ (2022) {Status and Perspectives of Continuous Gravitational Wave Searches}. Galaxies 10(3):72. \doi{10.3390/galaxies10030072}, {\href{https://arxiv.org/abs/2202.01088}{{arXiv:2202.01088}}} {[gr-qc]}

\bibitem[{Pierce et~al(2018)Pierce, Riles, and Zhao}]{Pierce:2018xmy}
Pierce A, Riles K, Zhao Y (2018) {Searching for Dark Photon Dark Matter with Gravitational Wave Detectors}. Phys Rev Lett 121(6):061102. \doi{10.1103/PhysRevLett.121.061102}, {\href{https://arxiv.org/abs/1801.10161}{{arXiv:1801.10161}}} {[hep-ph]}

\bibitem[{Pitkin(2022)}]{Pitkin:2022gbe}
Pitkin M (2022) {CWInPy: A Python package for inference with continuous gravitational-wave signals from pulsars}. J Open Source Softw 7(77):4568. \doi{10.21105/joss.04568}

\bibitem[{Pitkin et~al(2017)Pitkin, Isi, Veitch, and Woan}]{Pitkin:2017qfy}
Pitkin M, Isi M, Veitch J, et~al (2017) {A nested sampling code for targeted searches for continuous gravitational waves from pulsars}. arXiv e-prints {\href{https://arxiv.org/abs/1705.08978}{{arXiv:1705.08978}}} {[gr-qc]}

\bibitem[{Pompili et~al(2023)}]{Pompili:2023tna}
Pompili L, et~al (2023) {Laying the foundation of the effective-one-body waveform models SEOBNRv5: Improved accuracy and efficiency for spinning nonprecessing binary black holes}. Phys Rev D 108(12):124035. \doi{10.1103/PhysRevD.108.124035}, {\href{https://arxiv.org/abs/2303.18039}{{arXiv:2303.18039}}} {[gr-qc]}

\bibitem[{Powell et~al(2015)Powell, Trifir{\`o}, Cuoco, Heng, and Cavagli{\`a}}]{Powell:2015ona}
Powell J, Trifir{\`o} D, Cuoco E, et~al (2015) {Classification methods for noise transients in advanced gravitational-wave detectors}. Class Quant Grav 32(21):215012. \doi{10.1088/0264-9381/32/21/215012}, {\href{https://arxiv.org/abs/1505.01299}{{arXiv:1505.01299}}} {[astro-ph.IM]}

\bibitem[{Powell et~al(2017)Powell, Torres-Forn\'e, Lynch, Trifir\`o, Cuoco, Cavagli\`a, Heng, and Font}]{Powell:2016rkl}
Powell J, Torres-Forn\'e A, Lynch R, et~al (2017) {Classification methods for noise transients in advanced gravitational-wave detectors II: performance tests on Advanced LIGO data}. Class Quant Grav 34(3):034002. \doi{10.1088/1361-6382/34/3/034002}, {\href{https://arxiv.org/abs/1609.06262}{{arXiv:1609.06262}}} {[astro-ph.IM]}

\bibitem[{{Powell} et~al(2023){Powell}, {Iess}, {Llorens-Monteagudo}, {Obergaulinger}, {M{\"u}ller}, {Torres-Forn{\'e}}, {Cuoco}, and {Font}}]{2023arXiv231118221P}
{Powell} J, {Iess} A, {Llorens-Monteagudo} M, et~al (2023) {Determining the core-collapse supernova explosion mechanism with current and future gravitational-wave observatories}. arXiv e-prints arXiv:2311.18221. \doi{10.48550/arXiv.2311.18221}, {\href{https://arxiv.org/abs/2311.18221}{{arXiv:2311.18221}}} {[astro-ph.HE]}

\bibitem[{Powell et~al(2023)Powell, Sun, Gereb, Lasky, and Dollmann}]{Powell_2023}
Powell J, Sun L, Gereb K, et~al (2023) Generating transient noise artefacts in gravitational-wave detector data with generative adversarial networks. Class Quantum Grav 40(3):035006. \doi{10.1088/1361-6382/acb038}

\bibitem[{Pratten et~al(2020)Pratten, Husa, Garcia-Quiros, Colleoni, Ramos-Buades, Estelles, and Jaume}]{Pratten:2020fqn}
Pratten G, Husa S, Garcia-Quiros C, et~al (2020) {Setting the cornerstone for a family of models for gravitational waves from compact binaries: The dominant harmonic for nonprecessing quasicircular black holes}. Phys Rev D 102(6):064001. \doi{10.1103/PhysRevD.102.064001}, {\href{https://arxiv.org/abs/2001.11412}{{arXiv:2001.11412}}} {[gr-qc]}

\bibitem[{Prix et~al(2011)Prix, Giampanis, and Messenger}]{Prix:2011qv}
Prix R, Giampanis S, Messenger C (2011) {Search method for long-duration gravitational-wave transients from neutron stars}. Phys Rev D 84:023007. \doi{10.1103/PhysRevD.84.023007}, {\href{https://arxiv.org/abs/1104.1704}{{arXiv:1104.1704}}} {[gr-qc]}

\bibitem[{{Prix, Reinhard}(2018)}]{Prix:2015cfs}
{Prix, Reinhard} (2018) {The F-statistic and its implementation in ComputeFstatistic\_v2}. \url{https://dcc.ligo.org/LIGO-T0900149/public}

\bibitem[{Punturo et~al(2010)}]{0264-9381-27-19-194002}
Punturo M, et~al (2010) {The Einstein Telescope: A third-generation gravitational wave observatory}. Class Quant Grav 27:194002. \doi{10.1088/0264-9381/27/19/194002}

\bibitem[{P{\"u}rrer(2014)}]{Purrer:2014fza}
P{\"u}rrer M (2014) {Frequency domain reduced order models for gravitational waves from aligned-spin compact binaries}. Class Quant Grav 31(19):195010. \doi{10.1088/0264-9381/31/19/195010}, {\href{https://arxiv.org/abs/1402.4146}{{arXiv:1402.4146}}} {[gr-qc]}

\bibitem[{P{\"u}rrer(2016)}]{Purrer:2015tud}
P{\"u}rrer M (2016) {Frequency domain reduced order model of aligned-spin effective-one-body waveforms with generic mass-ratios and spins}. Phys Rev D93(6):064041. \doi{10.1103/PhysRevD.93.064041}, {\href{https://arxiv.org/abs/1512.02248}{{arXiv:1512.02248}}} {[gr-qc]}

\bibitem[{P\"urrer and Haster(2020)}]{Purrer:2019jcp}
P\"urrer M, Haster CJ (2020) {Gravitational waveform accuracy requirements for future ground-based detectors}. Phys Rev Res 2(2):023151. \doi{10.1103/PhysRevResearch.2.023151}, {\href{https://arxiv.org/abs/1912.10055}{{arXiv:1912.10055}}} {[gr-qc]}

\bibitem[{Quitzow-James et~al(2017)Quitzow-James, Brau, Clark, Coughlin, Coughlin, Frey, Schale, Talukder, and Thrane}]{Quitzow-James:2017zeq}
Quitzow-James R, Brau J, Clark J, et~al (2017) {Exploring a search for long-duration transient gravitational waves associated with magnetar bursts}. Class Quant Grav 34(16):164002. \doi{10.1088/1361-6382/aa7d5b}, {\href{https://arxiv.org/abs/1704.03979}{{arXiv:1704.03979}}} {[astro-ph.IM]}

\bibitem[{{Raikman} et~al(2023){Raikman}, {Moreno}, {Govorkova}, {Marx}, {Gunny}, {Benoit}, {Chatterjee}, {Omer}, {Saleem}, {Rankin}, {Coughlin}, {Harris}, and {Katsavounidis}}]{2023arXiv230911537R}
{Raikman} R, {Moreno} EA, {Govorkova} E, et~al (2023) {GWAK: Gravitational-Wave Anomalous Knowledge with Recurrent Autoencoders}. arXiv e-prints arXiv:2309.11537. \doi{10.48550/arXiv.2309.11537}, {\href{https://arxiv.org/abs/2309.11537}{{arXiv:2309.11537}}} {[astro-ph.IM]}

\bibitem[{{Raissi} et~al(2017){Raissi}, {Perdikaris}, and {Karniadakis}}]{2017arXiv171110561R}
{Raissi} M, {Perdikaris} P, {Karniadakis} GE (2017) {Physics Informed Deep Learning (Part I): Data-driven Solutions of Nonlinear Partial Differential Equations}. arXiv e-prints arXiv:1711.10561. \doi{10.48550/arXiv.1711.10561}, {\href{https://arxiv.org/abs/1711.10561}{{arXiv:1711.10561}}} {[cs.AI]}

\bibitem[{Ramos-Buades et~al(2023)Ramos-Buades, Buonanno, Estell\'es, Khalil, Mihaylov, Ossokine, Pompili, and Shiferaw}]{Ramos-Buades:2023ehm}
Ramos-Buades A, Buonanno A, Estell\'es H, et~al (2023) {Next generation of accurate and efficient multipolar precessing-spin effective-one-body waveforms for binary black holes}. Phys Rev D 108(12):124037. \doi{10.1103/PhysRevD.108.124037}, {\href{https://arxiv.org/abs/2303.18046}{{arXiv:2303.18046}}} {[gr-qc]}

\bibitem[{{Rasmussen} and {Williams}(2006)}]{Rasmussen:2006gpr}
{Rasmussen} CE, {Williams} CKI (2006) Gaussian Processes for Machine Learning. MIT Press, Cambridge, MA, \doi{10.7551/mitpress/3206.001.0001}

\bibitem[{{Ray} et~al(2023){Ray}, {Hernandez}, {Mohite}, {Creighton}, and {Kapadia}}]{ray_2023}
{Ray} A, {Hernandez} IM, {Mohite} S, et~al (2023) {Nonparametric Inference of the Population of Compact Binaries from Gravitational-wave Observations Using Binned Gaussian Processes}. \apj 957(1):37. \doi{10.3847/1538-4357/acf452}, {\href{https://arxiv.org/abs/2304.08046}{{arXiv:2304.08046}}} {[gr-qc]}

\bibitem[{Raza et~al(2024)Raza, Chan, Haggard, Mahabal, McIver, Abbott, Buffaz, and Vieira}]{Raza:2023gyv}
Raza N, Chan ML, Haggard D, et~al (2024) {Explaining the GWSkyNet-Multi Machine Learning Classifier Predictions for Gravitational-wave Events}. Astrophys J 963(2):98. \doi{10.3847/1538-4357/ad13ea}, {\href{https://arxiv.org/abs/2308.12357}{{arXiv:2308.12357}}} {[astro-ph.IM]}

\bibitem[{Razzano and Cuoco(2018)}]{Razzano_2018}
Razzano M, Cuoco E (2018) Image-based deep learning for classification of noise transients in gravitational wave detectors. Class Quantum Grav 35(9):095016. \doi{10.1088/1361-6382/aab793}

\bibitem[{Razzano et~al(2023)Razzano, {Di Renzo}, Fidecaro, Hemming, and Katsanevas}]{RAZZANO2023167959}
Razzano M, {Di Renzo} F, Fidecaro F, et~al (2023) Gwitchhunters: Machine learning and citizen science to improve the performance of gravitational wave detector. Nuclear Instruments and Methods in Physics Research Section A: Accelerators, Spectrometers, Detectors and Associated Equipment 1048:167959. \doi{10.1016/j.nima.2022.167959}

\bibitem[{Reed et~al(2024)Reed, Somasundaram, De, Armstrong, Giuliani, Capano, Brown, and Tews}]{Reed:2024urq}
Reed BT, Somasundaram R, De S, et~al (2024) {Towards accelerated nuclear-physics parameter estimation from binary neutron star mergers: Emulators for the Tolman-Oppenheimer-Volkoff equations}. arXiv e-prints {\href{https://arxiv.org/abs/2405.20558}{{arXiv:2405.20558}}} {[astro-ph.HE]}

\bibitem[{Regimbau et~al(2012)}]{Regimbau:2012ir}
Regimbau T, et~al (2012) {A Mock Data Challenge for the Einstein Gravitational-Wave Telescope}. Phys Rev D 86:122001. \doi{10.1103/PhysRevD.86.122001}, {\href{https://arxiv.org/abs/1201.3563}{{arXiv:1201.3563}}} {[gr-qc]}

\bibitem[{van Remortel et~al(2023)van Remortel, Janssens, and Turbang}]{vanRemortel:2022fkb}
van Remortel N, Janssens K, Turbang K (2023) {Stochastic gravitational wave background: Methods and implications}. Prog Part Nucl Phys 128:104003. \doi{10.1016/j.ppnp.2022.104003}, {\href{https://arxiv.org/abs/2210.00761}{{arXiv:2210.00761}}} {[gr-qc]}

\bibitem[{Renzini et~al(2022)Renzini, Goncharov, Jenkins, and Meyers}]{Renzini:2022alw}
Renzini AI, Goncharov B, Jenkins AC, et~al (2022) {Stochastic Gravitational-Wave Backgrounds: Current Detection Efforts and Future Prospects}. Galaxies 10(1):34. \doi{10.3390/galaxies10010034}, {\href{https://arxiv.org/abs/2202.00178}{{arXiv:2202.00178}}} {[gr-qc]}

\bibitem[{Riles(2023)}]{Riles:2022wwz}
Riles K (2023) {Searches for continuous-wave gravitational radiation}. Living Rev Rel 26(1):3. \doi{10.1007/s41114-023-00044-3}, {\href{https://arxiv.org/abs/2206.06447}{{arXiv:2206.06447}}} {[astro-ph.HE]}

\bibitem[{{Romero-Shaw} et~al(2020){Romero-Shaw}, {Talbot}, {Biscoveanu}, {D'Emilio}, {Ashton}, {Berry}, {Coughlin}, {Galaudage}, {Hoy}, {H{\"u}bner}, {Phukon}, {Pitkin}, {Rizzo}, {Sarin}, {Smith}, {Stevenson}, {Vajpeyi}, {Ar{\`e}ne}, {Athar}, {Banagiri}, {Bose}, {Carney}, {Chatziioannou}, {Clark}, {Colleoni}, {Cotesta}, {Edelman}, {Estell{\'e}s}, {Garc{\'\i}a-Quir{\'o}s}, {Ghosh}, {Green}, {Haster}, {Husa}, {Keitel}, {Kim}, {Hernandez-Vivanco}, {Maga{\~n}a Hernandez}, {Karathanasis}, {Lasky}, {De Lillo}, {Lower}, {Macleod}, {Mateu-Lucena}, {Miller}, {Millhouse}, {Morisaki}, {Oh}, {Ossokine}, {Payne}, {Powell}, {Pratten}, {P{\"u}rrer}, {Ramos-Buades}, {Raymond}, {Thrane}, {Veitch}, {Williams}, {Williams}, and {Xiao}}]{2020MNRAS.499.3295R}
{Romero-Shaw} IM, {Talbot} C, {Biscoveanu} S, et~al (2020) {Bayesian inference for compact binary coalescences with BILBY: validation and application to the first LIGO-Virgo gravitational-wave transient catalogue}. \mnras 499(3):3295--3319. \doi{10.1093/mnras/staa2850}, {\href{https://arxiv.org/abs/2006.00714}{{arXiv:2006.00714}}} {[astro-ph.IM]}

\bibitem[{{Ronneberger} et~al(2015){Ronneberger}, {Fischer}, and {Brox}}]{2015arXiv150504597R}
{Ronneberger} O, {Fischer} P, {Brox} T (2015) {U-Net: Convolutional Networks for Biomedical Image Segmentation}. arXiv e-prints arXiv:1505.04597. \doi{10.48550/arXiv.1505.04597}, {\href{https://arxiv.org/abs/1505.04597}{{arXiv:1505.04597}}} {[cs.CV]}

\bibitem[{Roshan and White(2024)}]{Roshan:2024qnv}
Roshan R, White G (2024) {Using gravitational waves to see the first second of the Universe}. arXiv e-prints {\href{https://arxiv.org/abs/2401.04388}{{arXiv:2401.04388}}} {[hep-ph]}

\bibitem[{Rosofsky and Huerta(2023)}]{Rosofsky:2023dtc}
Rosofsky SG, Huerta EA (2023) {Magnetohydrodynamics with physics informed neural operators}. Mach Learn Sci Tech 4(3):035002. \doi{10.1088/2632-2153/ace30a}, {\href{https://arxiv.org/abs/2302.08332}{{arXiv:2302.08332}}} {[physics.comp-ph]}

\bibitem[{Rosofsky et~al(2023)Rosofsky, Majed, and Huerta}]{Rosofsky:2022lgb}
Rosofsky SG, Majed HA, Huerta EA (2023) {Applications of physics informed neural operators}. Mach Learn Sci Tech 4(2):025022. \doi{10.1088/2632-2153/acd168}, {\href{https://arxiv.org/abs/2203.12634}{{arXiv:2203.12634}}} {[physics.comp-ph]}

\bibitem[{Ruan and Guo(2024)}]{Ruan:2024qch}
Ruan WH, Guo ZK (2024) {Premerger detection of massive black hole binaries using deep learning}. Phys Rev D 109(12):123031. \doi{10.1103/PhysRevD.109.123031}, {\href{https://arxiv.org/abs/2402.16282}{{arXiv:2402.16282}}} {[astro-ph.IM]}

\bibitem[{Ruan et~al(2020)Ruan, Guo, Cai, and Zhang}]{Ruan:2018tsw}
Ruan WH, Guo ZK, Cai RG, et~al (2020) {Taiji program: Gravitational-wave sources}. Int J Mod Phys A 35(17):2050075. \doi{10.1142/S0217751X2050075X}, {\href{https://arxiv.org/abs/1807.09495}{{arXiv:1807.09495}}} {[gr-qc]}

\bibitem[{{Ruhe} et~al(2022){Ruhe}, {Wong}, {Cranmer}, and {Forr{\'e}}}]{ruhe_2022}
{Ruhe} D, {Wong} K, {Cranmer} M, et~al (2022) {Normalizing Flows for Hierarchical Bayesian Analysis: A Gravitational Wave Population Study}. arXiv e-prints arXiv:2211.09008. \doi{10.48550/arXiv.2211.09008}, {\href{https://arxiv.org/abs/2211.09008}{{arXiv:2211.09008}}} {[astro-ph.IM]}

\bibitem[{Sachdev et~al(2020)Sachdev, Magee, Hanna, Cannon, Singer, SK, Mukherjee, Caudill, Chan, Creighton, Ewing, Fong, Godwin, Huxford, Kapadia, Li, Lo, Meacher, Messick, Mohite, Nishizawa, Ohta, Pace, Reza, Sathyaprakash, Shikauchi, Singh, Tsukada, Tsuna, Tsutsui, and Ueno}]{Sachdev_2020}
Sachdev S, Magee R, Hanna C, et~al (2020) An early-warning system for electromagnetic follow-up of gravitational-wave events. Astrophys J Lett 905(2):L25. \doi{10.3847/2041-8213/abc753}

\bibitem[{Safarzadeh et~al(2022)Safarzadeh, Khan, Huerta, and Wattenberg}]{Safarzadeh:2022pij}
Safarzadeh M, Khan A, Huerta EA, et~al (2022) {Interpreting a Machine Learning Model for Detecting Gravitational Waves}. arXiv e-prints {\href{https://arxiv.org/abs/2202.07399}{{arXiv:2202.07399}}} {[gr-qc]}

\bibitem[{Saiz-Pérez et~al(2022)Saiz-Pérez, Torres-Forné, and Font}]{10.1093/mnras/stac698}
Saiz-Pérez A, Torres-Forné A, Font JA (2022) {Classification of core-collapse supernova explosions with learned dictionaries}. Mon Not R Astron, Soc 512(3):3815--3827. \doi{10.1093/mnras/stac698}

\bibitem[{Saleem et~al(2024)}]{Saleem:2023hcm}
Saleem M, et~al (2024) {Demonstration of machine learning-assisted low-latency noise regression in gravitational wave detectors}. Class Quant Grav 41(19):195024. \doi{10.1088/1361-6382/ad708a}, {\href{https://arxiv.org/abs/2306.11366}{{arXiv:2306.11366}}} {[gr-qc]}

\bibitem[{Sarin and Lasky(2021)}]{Sarin:2020gxb}
Sarin N, Lasky PD (2021) {The evolution of binary neutron star post-merger remnants: a review}. Gen Rel Grav 53(6):59. \doi{10.1007/s10714-021-02831-1}, {\href{https://arxiv.org/abs/2012.08172}{{arXiv:2012.08172}}} {[astro-ph.HE]}

\bibitem[{Sasaoka et~al(2023)Sasaoka, Koyama, Dominguez, Sakai, Somiya, Omae, and Takahashi}]{Sasaoka:2023kte}
Sasaoka S, Koyama N, Dominguez D, et~al (2023) {Visualizing convolutional neural network for classifying gravitational waves from core-collapse supernovae}. Phys Rev D 108(12):123033. \doi{10.1103/PhysRevD.108.123033}, {\href{https://arxiv.org/abs/2310.09551}{{arXiv:2310.09551}}} {[astro-ph.IM]}

\bibitem[{{Saulson}(2017)}]{Saulson_book_2017}
{Saulson} PR (2017) {Fundamentals of Interferometric Gravitational Wave Detectors}. World Scientific, \doi{10.1142/10116}

\bibitem[{{Sch{\"a}fer} et~al(2022){Sch{\"a}fer}, {Zelenka}, {Nitz}, {Ohme}, and {Br{\"u}gmann}}]{2022PhRvD.105d3002S}
{Sch{\"a}fer} MB, {Zelenka} O, {Nitz} AH, et~al (2022) {Training strategies for deep learning gravitational-wave searches}. \prd 105(4):043002. \doi{10.1103/PhysRevD.105.043002}, {\href{https://arxiv.org/abs/2106.03741}{{arXiv:2106.03741}}} {[astro-ph.IM]}

\bibitem[{Sch\"afer et~al(2023)}]{Schafer:2022dxv}
Sch\"afer MB, et~al (2023) {First machine learning gravitational-wave search mock data challenge}. Phys Rev D 107(2):023021. \doi{10.1103/PhysRevD.107.023021}, {\href{https://arxiv.org/abs/2209.11146}{{arXiv:2209.11146}}} {[astro-ph.IM]}

\bibitem[{Schmidt(2020)}]{Schmidt:2020ekt}
Schmidt P (2020) {Gravitational Waves From Binary Black Hole Mergers: Modeling and Observations}. Front Astron Space Sci 7:28. \doi{10.3389/fspas.2020.00028}

\bibitem[{Schmidt et~al(2021)Schmidt, Breschi, Gamba, Pagano, Rettegno, Riemenschneider, Bernuzzi, Nagar, and Del~Pozzo}]{Schmidt:2020yuu}
Schmidt S, Breschi M, Gamba R, et~al (2021) {Machine Learning Gravitational Waves from Binary Black Hole Mergers}. Phys Rev D 103(4):043020. \doi{10.1103/PhysRevD.103.043020}, {\href{https://arxiv.org/abs/2011.01958}{{arXiv:2011.01958}}} {[gr-qc]}

\bibitem[{{Sen} et~al(2022){Sen}, {Agarwal}, {Chakraborty}, and {Singh}}]{2022ExA....53....1S}
{Sen} S, {Agarwal} S, {Chakraborty} P, et~al (2022) {Astronomical big data processing using machine learning: A comprehensive review}. Experimental Astronomy 53(1):1--43. \doi{10.1007/s10686-021-09827-4}

\bibitem[{Seoane et~al(2023)}]{LISA:2022yao}
Seoane PA, et~al (2023) {Astrophysics with the Laser Interferometer Space Antenna}. Living Rev Rel 26(1):2. \doi{10.1007/s41114-022-00041-y}, {\href{https://arxiv.org/abs/2203.06016}{{arXiv:2203.06016}}} {[gr-qc]}

\bibitem[{Setyawati et~al(2020)Setyawati, P{\"u}rrer, and Ohme}]{Setyawati:2019xzw}
Setyawati Y, P{\"u}rrer M, Ohme F (2020) {Regression methods in waveform modeling: a comparative study}. Class Quant Grav 37(7):075012. \doi{10.1088/1361-6382/ab693b}, {\href{https://arxiv.org/abs/1909.10986}{{arXiv:1909.10986}}} {[astro-ph.IM]}

\bibitem[{Shah et~al(2023)Shah, Knee, McIver, and Stenning}]{Shah:2023twc}
Shah N, Knee AM, McIver J, et~al (2023) {Waves in a forest: a random forest classifier to distinguish between gravitational waves and detector glitches}. Class Quant Grav 40(23):235008. \doi{10.1088/1361-6382/ad0424}, {\href{https://arxiv.org/abs/2306.13787}{{arXiv:2306.13787}}} {[gr-qc]}

\bibitem[{Shakhnarovich et~al(2005)Shakhnarovich, Darrell, and Indyk}]{shakhnarovich2005nearest}
Shakhnarovich G, Darrell T, Indyk P (2005) Nearest-neighbor Methods in Learning and Vision: Theory and Practice. Neural information processing series, MIT Press, Cambridge, MA

\bibitem[{Shannon(1949)}]{Shannon1949}
Shannon C (1949) Communication in the presence of noise. Proceedings of the {IRE} 37(1):10--21. \doi{10.1109/jrproc.1949.232969}

\bibitem[{{Shibata}(2016)}]{Shibata:2016nr}
{Shibata} M (2016) Numerical Relativity. {World Scientific}, \doi{10.1142/9692}

\bibitem[{Shih et~al(2024)Shih, Freytsis, Taylor, Dror, and Smyth}]{Shih:2023jme}
Shih D, Freytsis M, Taylor SR, et~al (2024) {Fast Parameter Inference on Pulsar Timing Arrays with Normalizing Flows}. Phys Rev Lett 133(1):011402. \doi{10.1103/PhysRevLett.133.011402}, {\href{https://arxiv.org/abs/2310.12209}{{arXiv:2310.12209}}} {[astro-ph.IM]}

\bibitem[{Siemonsen et~al(2023)Siemonsen, May, and East}]{Siemonsen:2022yyf}
Siemonsen N, May T, East WE (2023) {Modeling the black hole superradiance gravitational waveform}. Phys Rev D 107(10):104003. \doi{10.1103/PhysRevD.107.104003}, {\href{https://arxiv.org/abs/2211.03845}{{arXiv:2211.03845}}} {[gr-qc]}

\bibitem[{Singh et~al(2017)Singh, Papa, Eggenstein, and Walsh}]{Singh:2017kss}
Singh A, Papa MA, Eggenstein HB, et~al (2017) {Adaptive clustering procedure for continuous gravitational wave searches}. Phys Rev D 96(8):082003. \doi{10.1103/PhysRevD.96.082003}, {\href{https://arxiv.org/abs/1707.02676}{{arXiv:1707.02676}}} {[gr-qc]}

\bibitem[{Singh et~al(2018)Singh, Li, Hannuksela, Li, and Kim}]{Singh:2018csp}
Singh AJ, Li ISC, Hannuksela OA, et~al (2018) {Classifying Lensed Gravitational Waves in the Geometrical Optics Limit with Machine Learning}. arXiv e-prints \doi{10.33697/ajur.2019.019}, {\href{https://arxiv.org/abs/1810.07888}{{arXiv:1810.07888}}} {[astro-ph.IM]}

\bibitem[{Skilling(2006)}]{Skilling:2006gxv}
Skilling J (2006) {Nested sampling for general Bayesian computation}. Bayesian Analysis 1(4):833--859. \doi{10.1214/06-BA127}

\bibitem[{Skliris et~al(2020)Skliris, Norman, and Sutton}]{Skliris:2020qax}
Skliris V, Norman MRK, Sutton PJ (2020) {Real-Time Detection of Unmodelled Gravitational-Wave Transients Using Convolutional Neural Networks}. arXiv e-prints {\href{https://arxiv.org/abs/2009.14611}{{arXiv:2009.14611}}} {[astro-ph.IM]}

\bibitem[{Slutsky et~al(2010)}]{Slutsky:2010ff}
Slutsky J, et~al (2010) {Methods for Reducing False Alarms in Searches for Compact Binary Coalescences in LIGO Data}. Class Quant Grav 27:165023. \doi{10.1088/0264-9381/27/16/165023}, {\href{https://arxiv.org/abs/1004.0998}{{arXiv:1004.0998}}} {[gr-qc]}

\bibitem[{{Smith} and {Geach}(2023)}]{2023RSOS...1021454S}
{Smith} MJ, {Geach} JE (2023) {Astronomia ex machina: a history, primer and outlook on neural networks in astronomy}. Royal Society Open Science 10(5):221454. \doi{10.1098/rsos.221454}, {\href{https://arxiv.org/abs/2211.03796}{{arXiv:2211.03796}}} {[astro-ph.IM]}

\bibitem[{Smith and Thrane(2018)}]{Smith:2017vfk}
Smith R, Thrane E (2018) {Optimal Search for an Astrophysical Gravitational-Wave Background}. Phys Rev X 8(2):021019. \doi{10.1103/PhysRevX.8.021019}, {\href{https://arxiv.org/abs/1712.00688}{{arXiv:1712.00688}}} {[gr-qc]}

\bibitem[{Soni et~al(2021)Soni, Berry, Coughlin, Harandi, Jackson, Crowston, Østerlund, Patane, Katsaggelos, Trouille, Baranowski, Domainko, Kaminski, Rodriguez, Marciniak, Nauta, Niklasch, Rote, Téglás, Unsworth, and Zhang}]{Soni_2021}
Soni S, Berry CPL, Coughlin SB, et~al (2021) Discovering features in gravitational-wave data through detector characterization, citizen science and machine learning. Class Quantum Grav 38(19):195016. \doi{10.1088/1361-6382/ac1ccb}

\bibitem[{Speagle(2020)}]{Speagle:2019ivv}
Speagle JS (2020) {dynesty: a dynamic nested sampling package for estimating Bayesian posteriors and evidences}. Mon Not R Astron Soc 493(3):3132--3158. \doi{10.1093/mnras/staa278}, {\href{https://arxiv.org/abs/1904.02180}{{arXiv:1904.02180}}} {[astro-ph.IM]}

\bibitem[{Stachurski et~al(2024)Stachurski, Messenger, and Hendry}]{Stachurski:2023ntw}
Stachurski F, Messenger C, Hendry M (2024) {Cosmological inference using gravitational waves and normalizing flows}. Phys Rev D 109(12):123547. \doi{10.1103/PhysRevD.109.123547}, {\href{https://arxiv.org/abs/2310.13405}{{arXiv:2310.13405}}} {[gr-qc]}

\bibitem[{Staley et~al(2014)Staley, Martynov, Abbott, Adhikari, Arai, Ballmer, Barsotti, Brooks, DeRosa, Dwyer, Effler, Evans, Fritschel, Frolov, Gray, Guido, Gustafson, Heintze, Hoak, Izumi, Kawabe, King, Kissel, Kokeyama, Landry, McClelland, Miller, Mullavey, O'Reilly, Rollins, Sanders, Schofield, Sigg, Slagmolen, Smith-Lefebvre, Vajente, Ward, and Wipf}]{Staley_2014}
Staley A, Martynov D, Abbott R, et~al (2014) Achieving resonance in the advanced ligo gravitational-wave interferometer. Class Quantum Grav 31(24):245010. \doi{10.1088/0264-9381/31/24/245010}

\bibitem[{Stefanou et~al(2023)Stefanou, Urb\'an, and Pons}]{Stefanou:2023jxk}
Stefanou P, Urb\'an JF, Pons JA (2023) {Solving the pulsar equation~using physics-informed neural networks}. Mon Not R Astron Soc 526(1):1504--1511. \doi{10.1093/mnras/stad2840}, {\href{https://arxiv.org/abs/2309.06410}{{arXiv:2309.06410}}} {[astro-ph.HE]}

\bibitem[{Steltner et~al(2022)Steltner, Menne, Papa, and Eggenstein}]{Steltner:2022aze}
Steltner B, Menne T, Papa MA, et~al (2022) {Density-clustering of continuous gravitational wave candidates from large surveys}. Phys Rev D 106(10):104063. \doi{10.1103/PhysRevD.106.104063}, {\href{https://arxiv.org/abs/2207.14286}{{arXiv:2207.14286}}} {[gr-qc]}

\bibitem[{Steltner et~al(2023)Steltner, Papa, Eggenstein, Prix, Bensch, Allen, and Machenschalk}]{Steltner:2023cfk}
Steltner B, Papa MA, Eggenstein HB, et~al (2023) {Deep Einstein@Home All-sky Search for Continuous Gravitational Waves in LIGO O3 Public Data}. Astrophys J 952(1):55. \doi{10.3847/1538-4357/acdad4}, {\href{https://arxiv.org/abs/2303.04109}{{arXiv:2303.04109}}} {[gr-qc]}

\bibitem[{Stergioulas(2024)}]{Stergioulas:2024jgk}
Stergioulas N (2024) {Machine Learning Applications in Gravitational Wave Astronomy}. In: {Proceedings of the 11th Aegean Summer School Recent Developments in Theory and Observations in Gravity and Cosmology}, \eprint{2401.07406}

\bibitem[{Sun and Melatos(2019)}]{Sun:2018owi}
Sun L, Melatos A (2019) {Application of hidden Markov model tracking to the search for long-duration transient gravitational waves from the remnant of the binary neutron star merger GW170817}. Phys Rev D99(12):123003. \doi{10.1103/PhysRevD.99.123003}, {\href{https://arxiv.org/abs/1810.03577}{{arXiv:1810.03577}}} {[astro-ph.IM]}

\bibitem[{Sun et~al(2018)Sun, Melatos, Suvorova, Moran, and Evans}]{Sun:2017zge}
Sun L, Melatos A, Suvorova S, et~al (2018) {Hidden Markov model tracking of continuous gravitational waves from young supernova remnants}. Phys Rev D D97(4):043013. \doi{10.1103/PhysRevD.97.043013}, {\href{https://arxiv.org/abs/1710.00460}{{arXiv:1710.00460}}} {[astro-ph.IM]}

\bibitem[{Sun et~al(2019)Sun, Melatos, and Lasky}]{Sun:2019bew}
Sun L, Melatos A, Lasky PD (2019) {Tracking continuous gravitational waves from a neutron star at once and twice the spin frequency with a hidden Markov model}. Phys Rev D 99(12):123010. \doi{10.1103/PhysRevD.99.123010}, {\href{https://arxiv.org/abs/1903.03866}{{arXiv:1903.03866}}} {[astro-ph.IM]}

\bibitem[{Sun and Li(2023)}]{Sun:2023prd}
Sun M, Li J (2023) {Parameter Estimation for Intermediate-Mass Binary Black Holes through Gravitational Waves Observed by DECIGO}. arXiv e-prints {\href{https://arxiv.org/abs/2312.07834}{{arXiv:2312.07834}}} {[gr-qc]}

\bibitem[{Suvorova et~al(2016)Suvorova, Sun, Melatos, Moran, and Evans}]{Suvorova:2016rdc}
Suvorova S, Sun L, Melatos A, et~al (2016) {Hidden Markov model tracking of continuous gravitational waves from a neutron star with wandering spin}. Phys Rev D D93(12):123009. \doi{10.1103/PhysRevD.93.123009}, {\href{https://arxiv.org/abs/1606.02412}{{arXiv:1606.02412}}} {[astro-ph.IM]}

\bibitem[{Suvorova et~al(2017)Suvorova, Clearwater, Melatos, Sun, Moran, and Evans}]{Suvorova:2017dpm}
Suvorova S, Clearwater P, Melatos A, et~al (2017) {Hidden Markov model tracking of continuous gravitational waves from a binary neutron star with wandering spin. II. Binary orbital phase tracking}. Phys Rev D96(10):102006. \doi{10.1103/PhysRevD.96.102006}, {\href{https://arxiv.org/abs/1710.07092}{{arXiv:1710.07092}}} {[astro-ph.IM]}

\bibitem[{{Szczepa{\'n}czyk} et~al(2023){Szczepa{\'n}czyk}, {Salemi}, {Bini}, {Mishra}, {Vedovato}, {Gayathri}, {Bartos}, {Bhaumik}, {Drago}, {Halim}, {Lazzaro}, {Miani}, {Milotti}, {Prodi}, {Tiwari}, and {Klimenko}}]{2023PhRvD.107f2002S}
{Szczepa{\'n}czyk} MJ, {Salemi} F, {Bini} S, et~al (2023) {Search for gravitational-wave bursts in the third Advanced LIGO-Virgo run with coherent WaveBurst enhanced by machine learning}. \prd 107(6):062002. \doi{10.1103/PhysRevD.107.062002}, {\href{https://arxiv.org/abs/2210.01754}{{arXiv:2210.01754}}} {[gr-qc]}

\bibitem[{{Szegedy} et~al(2013){Szegedy}, {Zaremba}, {Sutskever}, {Bruna}, {Erhan}, {Goodfellow}, and {Fergus}}]{2013arXiv1312.6199S}
{Szegedy} C, {Zaremba} W, {Sutskever} I, et~al (2013) {Intriguing properties of neural networks}. arXiv e-prints arXiv:1312.6199. \doi{10.48550/arXiv.1312.6199}, {\href{https://arxiv.org/abs/1312.6199}{{arXiv:1312.6199}}} {[cs.CV]}

\bibitem[{{Szegedy} et~al(2016){Szegedy}, {Ioffe}, {Vanhoucke}, and {Alemi}}]{2016arXiv160207261S}
{Szegedy} C, {Ioffe} S, {Vanhoucke} V, et~al (2016) {Inception-v4, Inception-ResNet and the Impact of Residual Connections on Learning}. arXiv e-prints arXiv:1602.07261. \doi{10.48550/arXiv.1602.07261}, {\href{https://arxiv.org/abs/1602.07261}{{arXiv:1602.07261}}} {[cs.CV]}

\bibitem[{{Tanoglidis} et~al(2023){Tanoglidis}, {Jain}, and {Qu}}]{2023arXiv231012069T}
{Tanoglidis} D, {Jain} B, {Qu} H (2023) {Transformers for scientific data: a pedagogical review for astronomers}. arXiv e-prints \doi{10.48550/arXiv.2310.12069}, {\href{https://arxiv.org/abs/2310.12069}{{arXiv:2310.12069}}} {[astro-ph.IM]}

\bibitem[{Taylor and Varma(2020)}]{Taylor:2020bmj}
Taylor A, Varma V (2020) {Gravitational wave peak luminosity model for precessing binary black holes}. Phys Rev D 102(10):104047. \doi{10.1103/PhysRevD.102.104047}, {\href{https://arxiv.org/abs/2010.00120}{{arXiv:2010.00120}}} {[gr-qc]}

\bibitem[{Tenorio et~al(2021{\natexlab{a}})Tenorio, Keitel, and Sintes}]{Tenorio:2021njf}
Tenorio R, Keitel D, Sintes AM (2021{\natexlab{a}}) {Application of a hierarchical MCMC follow-up to Advanced LIGO continuous gravitational-wave candidates}. Phys Rev D 104(8):084012. \doi{10.1103/PhysRevD.104.084012}, {\href{https://arxiv.org/abs/2105.13860}{{arXiv:2105.13860}}} {[gr-qc]}

\bibitem[{Tenorio et~al(2021{\natexlab{b}})Tenorio, Keitel, and Sintes}]{Tenorio:2021wmz}
Tenorio R, Keitel D, Sintes AM (2021{\natexlab{b}}) {Search Methods for Continuous Gravitational-Wave Signals from Unknown Sources in the Advanced-Detector Era}. Universe 7(12):474. \doi{10.3390/universe7120474}, {\href{https://arxiv.org/abs/2111.12575}{{arXiv:2111.12575}}} {[gr-qc]}

\bibitem[{Tenorio et~al(2021{\natexlab{c}})Tenorio, Keitel, and Sintes}]{Tenorio:2020cqm}
Tenorio R, Keitel D, Sintes AM (2021{\natexlab{c}}) {Time-frequency track distance for comparing continuous gravitational wave signals}. Phys Rev D 103(6):064053. \doi{10.1103/PhysRevD.103.064053}, {\href{https://arxiv.org/abs/2012.05752}{{arXiv:2012.05752}}} {[gr-qc]}

\bibitem[{Tenorio et~al(2022)Tenorio, Williams, Messenger, Reade, and Demkin}]{kaggle2}
Tenorio R, Williams MJ, Messenger C, et~al (2022) G2net detecting continuous gravitational waves. \urlprefix\url{https://kaggle.com/competitions/g2net-detecting-continuous-gravitational-waves}

\bibitem[{Theobald(2017)}]{theobald2017machine}
Theobald O (2017) Machine Learning for Absolute Beginners: A Plain English Introduction. Ai, Data Science, Python \& Statistics for Beginners, Scatterplot Press, London

\bibitem[{Thomas et~al(2022)Thomas, Pratten, and Schmidt}]{Thomas:2022rmc}
Thomas LM, Pratten G, Schmidt P (2022) {Accelerating multimodal gravitational waveforms from precessing compact binaries with artificial neural networks}. Phys Rev D 106(10):104029. \doi{10.1103/PhysRevD.106.104029}, {\href{https://arxiv.org/abs/2205.14066}{{arXiv:2205.14066}}} {[gr-qc]}

\bibitem[{Thrane and Coughlin(2013)}]{Thrane:2013bea}
Thrane E, Coughlin M (2013) {Searching for gravitational-wave transients with a qualitative signal model: seedless clustering strategies}. Phys Rev D 88(8):083010. \doi{10.1103/PhysRevD.88.083010}, {\href{https://arxiv.org/abs/1308.5292}{{arXiv:1308.5292}}} {[astro-ph.IM]}

\bibitem[{Thrane et~al(2011)}]{Thrane:2010ri}
Thrane E, et~al (2011) {Long gravitational-wave transients and associated detection strategies for a network of terrestrial interferometers}. Phys Rev D 83:083004. \doi{10.1103/PhysRevD.83.083004}, {\href{https://arxiv.org/abs/1012.2150}{{arXiv:1012.2150}}} {[astro-ph.IM]}

\bibitem[{Tinto and Dhurandhar(2021)}]{Tinto:2020fcc}
Tinto M, Dhurandhar SV (2021) {Time-delay interferometry}. Living Rev Rel 24(1):1. \doi{10.1007/s41114-020-00029-6}

\bibitem[{{Torres-Forn{\'e}} et~al(2016){Torres-Forn{\'e}}, {Marquina}, {Font}, and {Ib{\'a}{\~n}ez}}]{2016PhRvD..94l4040T}
{Torres-Forn{\'e}} A, {Marquina} A, {Font} JA, et~al (2016) {Denoising of gravitational wave signals via dictionary learning algorithms}. Phys Rev D 94(12):124040. \doi{10.1103/PhysRevD.94.124040}, {\href{https://arxiv.org/abs/1612.01305}{{arXiv:1612.01305}}} {[astro-ph.IM]}

\bibitem[{Torres-Forn\'e et~al(2020)Torres-Forn\'e, Cuoco, Font, and Marquina}]{PhysRevD.102.023011}
Torres-Forn\'e A, Cuoco E, Font JA, et~al (2020) Application of dictionary learning to denoise ligo's blip noise transients. Phys Rev D 102:023011. \doi{10.1103/PhysRevD.102.023011}

\bibitem[{Trovato et~al(2024)Trovato, Chassande-Mottin, Bejger, Flamary, and Courty}]{Trovato:2023bby}
Trovato A, Chassande-Mottin E, Bejger M, et~al (2024) {Neural network time-series classifiers for gravitational-wave searches in single-detector periods}. Class Quant Grav 41(12):125003. \doi{10.1088/1361-6382/ad40f0}, {\href{https://arxiv.org/abs/2307.09268}{{arXiv:2307.09268}}} {[gr-qc]}

\bibitem[{Urban and Macleod(2023)}]{GWDetChar}
Urban A, Macleod D (2023) {Gravitational-wave Detector Characterisation (GWDetChar), a Python package for gravitational-wave detector characterisation and data quality, v 2.2.2}. \urlprefix\url{https://gwdetchar.readthedocs.io/en/stable/#}

\bibitem[{Urb\'an et~al(2023)Urb\'an, Stefanou, Dehman, and Pons}]{Urban:2023cfk}
Urb\'an JF, Stefanou P, Dehman C, et~al (2023) {Modelling force-free neutron star magnetospheres using physics-informed neural networks}. Mon Not R Astron Soc 524(1):32--42. \doi{10.1093/mnras/stad1810}, {\href{https://arxiv.org/abs/2303.11968}{{arXiv:2303.11968}}} {[astro-ph.HE]}

\bibitem[{{Usman} et~al(2016){Usman}, {Nitz}, {Harry}, {Biwer}, {Brown}, {Cabero}, {Capano}, {Dal Canton}, {Dent}, {Fairhurst}, {Kehl}, {Keppel}, {Krishnan}, {Lenon}, {Lundgren}, {Nielsen}, {Pekowsky}, {Pfeiffer}, {Saulson}, {West}, and {Willis}}]{2016CQGra..33u5004U}
{Usman} SA, {Nitz} AH, {Harry} IW, et~al (2016) {The PyCBC search for gravitational waves from compact binary coalescence}. Class Quantum Grav 33(21):215004. \doi{10.1088/0264-9381/33/21/215004}, {\href{https://arxiv.org/abs/1508.02357}{{arXiv:1508.02357}}} {[gr-qc]}

\bibitem[{Utina et~al(2021{\natexlab{a}})Utina, Marangio, Morawski, Iess, Regimbau, Fiameni, and Cuoco}]{Utina:2021ipo}
Utina A, Marangio F, Morawski F, et~al (2021{\natexlab{a}}) Deep learning searches for gravitational wave stochastic backgrounds. In: International Conference on Content-Based Multimedia Indexing, \doi{10.1109/CBMI50038.2021.9461904}

\bibitem[{Utina et~al(2021{\natexlab{b}})Utina, Marangio, Morawski, Iess, Regimbau, Fiameni, and Cuoco}]{9461904}
Utina A, Marangio F, Morawski F, et~al (2021{\natexlab{b}}) Deep learning searches for gravitational wave stochastic backgrounds. In: 2021 International Conference on Content-Based Multimedia Indexing (CBMI), pp 1--6, \doi{10.1109/CBMI50038.2021.9461904}

\bibitem[{Vachaspati and Vilenkin(1985)}]{Vachaspati:1984gt}
Vachaspati T, Vilenkin A (1985) {Gravitational Radiation from Cosmic Strings}. Phys Rev D 31:3052. \doi{10.1103/PhysRevD.31.3052}

\bibitem[{Vajente et~al(2020)Vajente, Huang, Isi, Driggers, Kissel, Szczepa{\'n}czyk, and Vitale}]{PhysRevD.101.042003}
Vajente G, Huang Y, Isi M, et~al (2020) Machine-learning nonstationary noise out of gravitational-wave detectors. Phys Rev D 101:042003. \doi{10.1103/PhysRevD.101.042003}

\bibitem[{Vallisneri(2009)}]{Vallisneri:2008ye}
Vallisneri M (2009) {A LISA Data-Analysis Primer}. Class Quant Grav 26:094024. \doi{10.1088/0264-9381/26/9/094024}, {\href{https://arxiv.org/abs/0812.0751}{{arXiv:0812.0751}}} {[gr-qc]}

\bibitem[{Varma et~al(2019{\natexlab{a}})Varma, Field, Scheel, Blackman, Gerosa, Stein, Kidder, and Pfeiffer}]{Varma:2019csw}
Varma V, Field SE, Scheel MA, et~al (2019{\natexlab{a}}) {Surrogate models for precessing binary black hole simulations with unequal masses}. Phys Rev Research 1:033015. \doi{10.1103/PhysRevResearch.1.033015}, {\href{https://arxiv.org/abs/1905.09300}{{arXiv:1905.09300}}} {[gr-qc]}

\bibitem[{Varma et~al(2019{\natexlab{b}})Varma, Field, Scheel, Blackman, Kidder, and Pfeiffer}]{Varma:2018mmi}
Varma V, Field SE, Scheel MA, et~al (2019{\natexlab{b}}) {Surrogate model of hybridized numerical relativity binary black hole waveforms}. Phys Rev D99(6):064045. \doi{10.1103/PhysRevD.99.064045}, {\href{https://arxiv.org/abs/1812.07865}{{arXiv:1812.07865}}} {[gr-qc]}

\bibitem[{Varma et~al(2019{\natexlab{c}})Varma, Gerosa, Stein, H\'ebert, and Zhang}]{Varma:2018aht}
Varma V, Gerosa D, Stein LC, et~al (2019{\natexlab{c}}) {High-accuracy mass, spin, and recoil predictions of generic black-hole merger remnants}. Phys Rev Lett 122(1):011101. \doi{10.1103/PhysRevLett.122.011101}, {\href{https://arxiv.org/abs/1809.09125}{{arXiv:1809.09125}}} {[gr-qc]}

\bibitem[{Vaswani et~al(2017)Vaswani, Shazeer, Parmar, Uszkoreit, Jones, Gomez, Kaiser, and Polosukhin}]{Vaswani:2017lxt}
Vaswani A, Shazeer N, Parmar N, et~al (2017) {Attention Is All You Need}. In: 31st International Conference on Neural Information Processing Systems, USA, \eprint{1706.03762}

\bibitem[{{Veitch} et~al(2015){Veitch}, {Raymond}, {Farr}, {Farr}, {Graff}, {Vitale}, {Aylott}, {Blackburn}, {Christensen}, {Coughlin}, {Del Pozzo}, {Feroz}, {Gair}, {Haster}, {Kalogera}, {Littenberg}, {Mandel}, {O'Shaughnessy}, {Pitkin}, {Rodriguez}, {R{\"o}ver}, {Sidery}, {Smith}, {Van Der Sluys}, {Vecchio}, {Vousden}, and {Wade}}]{veitch:15}
{Veitch} J, {Raymond} V, {Farr} B, et~al (2015) {Parameter estimation for compact binaries with ground-based gravitational-wave observations using the LALInference software library}. Phys Rev D 91(4):042003. \doi{10.1103/PhysRevD.91.042003}, {\href{https://arxiv.org/abs/1409.7215}{{arXiv:1409.7215}}} {[gr-qc]}

\bibitem[{Venumadhav et~al(2020)Venumadhav, Zackay, Roulet, Dai, and Zaldarriaga}]{PhysRevD.101.083030}
Venumadhav T, Zackay B, Roulet J, et~al (2020) New binary black hole mergers in the second observing run of advanced ligo and advanced virgo. Phys Rev D 101:083030. \doi{10.1103/PhysRevD.101.083030}

\bibitem[{Vermeulen et~al(2021)}]{Vermeulen:2021epa}
Vermeulen SM, et~al (2021) {Direct limits for scalar field dark matter from a gravitational-wave detector}. Nature 600(7889):424--428. \doi{10.1038/s41586-021-04031-y}, {\href{https://arxiv.org/abs/2103.03783}{{arXiv:2103.03783}}} {[gr-qc]}

\bibitem[{Viets et~al(2018)}]{Viets:2017yvy}
Viets A, et~al (2018) {Reconstructing the calibrated strain signal in the Advanced LIGO detectors}. Class Quant Grav 35(9):095015. \doi{10.1088/1361-6382/aab658}, {\href{https://arxiv.org/abs/1710.09973}{{arXiv:1710.09973}}} {[astro-ph.IM]}

\bibitem[{Viterbi(1967)}]{Viterbi1967}
Viterbi A (1967) Error bounds for convolutional codes and an asymptotically optimum decoding algorithm. IEEE Transactions on Information Theory 13(2):260--269. \doi{10.1109/TIT.1967.1054010}

\bibitem[{{Vousden} et~al(2016){Vousden}, {Farr}, and {Mandel}}]{Vousden:2016ptemcee}
{Vousden} WD, {Farr} WM, {Mandel} I (2016) {Dynamic temperature selection for parallel tempering in Markov chain Monte Carlo simulations}. Mon Not R Astron Soc 455(2):1919--1937. \doi{10.1093/mnras/stv2422}, {\href{https://arxiv.org/abs/1501.05823}{{arXiv:1501.05823}}} {[astro-ph.IM]}

\bibitem[{Wainstein and Zubakov(1962)}]{Wainstein:1962vrq}
Wainstein LA, Zubakov VD (1962) Extraction of Signals from Noise. Dover books on physics and mathematical physics, Prentice-Hall, Englewood Cliffs, NJ

\bibitem[{Walsh et~al(2019)Walsh, Wette, Papa, and Prix}]{Walsh:2019nmr}
Walsh S, Wette K, Papa MA, et~al (2019) {Optimizing the choice of analysis method for all-sky searches for continuous gravitational waves with Einstein@Home}. Phys Rev D99(8):082004. \doi{10.1103/PhysRevD.99.082004}, {\href{https://arxiv.org/abs/1901.08998}{{arXiv:1901.08998}}} {[astro-ph.IM]}

\bibitem[{Wei and Huerta(2020)}]{WEI2020135081}
Wei W, Huerta E (2020) Gravitational wave denoising of binary black hole mergers with deep learning. Physics Letters B 800:135081. \doi{10.1016/j.physletb.2019.135081}

\bibitem[{Wette(2023)}]{Wette:2023dom}
Wette K (2023) {Searches for continuous gravitational waves from neutron stars: A twenty-year retrospective}. Astropart Phys 153:102880. \doi{10.1016/j.astropartphys.2023.102880}, {\href{https://arxiv.org/abs/2305.07106}{{arXiv:2305.07106}}} {[gr-qc]}

\bibitem[{Whittaker et~al(2022)Whittaker, East, Green, Lehner, and Yang}]{Whittaker:2022pkd}
Whittaker T, East WE, Green SR, et~al (2022) {Using machine learning to parametrize postmerger signals from binary neutron stars}. Phys Rev D 105(12):124021. \doi{10.1103/PhysRevD.105.124021}, {\href{https://arxiv.org/abs/2201.06461}{{arXiv:2201.06461}}} {[gr-qc]}

\bibitem[{Wiener(1949)}]{wienerbook}
Wiener N (1949) Extrapolation, Interpolation, and Smoothing of Stationary Time Series. Wiley, New York

\bibitem[{Williams et~al(2020)Williams, Heng, Gair, Clark, and Khamesra}]{Williams:2019vub}
Williams D, Heng IS, Gair J, et~al (2020) {Precessing numerical relativity waveform surrogate model for binary black holes: A Gaussian process regression approach}. Phys Rev D 101(6):063011. \doi{10.1103/PhysRevD.101.063011}, {\href{https://arxiv.org/abs/1903.09204}{{arXiv:1903.09204}}} {[gr-qc]}

\bibitem[{{Williams} et~al(2021){Williams}, {Veitch}, and {Messenger}}]{2021PhRvD.103j3006W}
{Williams} MJ, {Veitch} J, {Messenger} C (2021) {Nested sampling with normalizing flows for gravitational-wave inference}. \prd 103(10):103006. \doi{10.1103/PhysRevD.103.103006}, {\href{https://arxiv.org/abs/2102.11056}{{arXiv:2102.11056}}} {[gr-qc]}

\bibitem[{{Williams} et~al(2023){Williams}, {Veitch}, and {Messenger}}]{2023MLS&T...4c5011W}
{Williams} MJ, {Veitch} J, {Messenger} C (2023) {Importance nested sampling with normalising flows}. Machine Learning: Science and Technology 4(3):035011. \doi{10.1088/2632-2153/acd5aa}, {\href{https://arxiv.org/abs/2302.08526}{{arXiv:2302.08526}}} {[astro-ph.IM]}

\bibitem[{{Wong} and {Gerosa}(2019)}]{wong_2019}
{Wong} KWK, {Gerosa} D (2019) {Machine-learning interpolation of population-synthesis simulations to interpret gravitational-wave observations: A case study}. \prd 100(8):083015. \doi{10.1103/PhysRevD.100.083015}, {\href{https://arxiv.org/abs/1909.06373}{{arXiv:1909.06373}}} {[astro-ph.HE]}

\bibitem[{{Wong} et~al(2020){Wong}, {Contardo}, and {Ho}}]{wong_2020}
{Wong} KWK, {Contardo} G, {Ho} S (2020) {Gravitational-wave population inference with deep flow-based generative network}. \prd 101(12):123005. \doi{10.1103/PhysRevD.101.123005}, {\href{https://arxiv.org/abs/2002.09491}{{arXiv:2002.09491}}} {[astro-ph.IM]}

\bibitem[{Wong et~al(2020)Wong, Ng, and Berti}]{Wong:2020wvd}
Wong KWK, Ng KKY, Berti E (2020) {Gravitational-wave signal-to-noise interpolation via neural networks}. arXiv e-prints {\href{https://arxiv.org/abs/2007.10350}{{arXiv:2007.10350}}} {[astro-ph.HE]}

\bibitem[{Wu et~al(2024)Wu, Zevin, Berry, Crowston, \O{}sterlund, Doctor, Banagiri, Jackson, Kalogera, and Katsaggelos}]{Wu:2024tpr}
Wu Y, Zevin M, Berry CPL, et~al (2024) {Advancing Glitch Classification in Gravity Spy: Multi-view Fusion with Attention-based Machine Learning for Advanced LIGO's Fourth Observing Run}. arXiv e-prints {\href{https://arxiv.org/abs/2401.12913}{{arXiv:2401.12913}}} {[gr-qc]}

\bibitem[{Xu et~al(2024)Xu, Du, Xu, Liang, and Wang}]{Xu:2024jbo}
Xu Y, Du M, Xu P, et~al (2024) {Gravitational wave signal extraction against non-stationary instrumental noises with deep neural network}. Phys Lett B 858:139016. \doi{10.1016/j.physletb.2024.139016}, {\href{https://arxiv.org/abs/2402.13091}{{arXiv:2402.13091}}} {[gr-qc]}

\bibitem[{Yamamoto and Tanaka(2020)}]{Yamamoto:2020rse}
Yamamoto TS, Tanaka T (2020) {Use of conditional variational auto encoder to analyze ringdown gravitational waves}. arXiv e-prints {\href{https://arxiv.org/abs/2002.12095}{{arXiv:2002.12095}}} {[gr-qc]}

\bibitem[{Yamamoto and Tanaka(2021)}]{Yamamoto:2020pus}
Yamamoto TS, Tanaka T (2021) {Use of an excess power method and a convolutional neural network in an all-sky search for continuous gravitational waves}. Phys Rev D 103(8):084049. \doi{10.1103/PhysRevD.103.084049}, {\href{https://arxiv.org/abs/2011.12522}{{arXiv:2011.12522}}} {[gr-qc]}

\bibitem[{Yamamoto et~al(2022)Yamamoto, Miller, Sieniawska, and Tanaka}]{Yamamoto:2022adl}
Yamamoto TS, Miller AL, Sieniawska M, et~al (2022) {Assessing the impact of non-Gaussian noise on convolutional neural networks that search for continuous gravitational waves}. Phys Rev D 106(2):024025. \doi{10.1103/PhysRevD.106.024025}, {\href{https://arxiv.org/abs/2206.00882}{{arXiv:2206.00882}}} {[gr-qc]}

\bibitem[{Yamamoto et~al(2023)Yamamoto, Kuroyanagi, and Liu}]{Yamamoto:2022kuh}
Yamamoto TS, Kuroyanagi S, Liu GC (2023) {Deep learning for intermittent gravitational wave signals}. Phys Rev D 107(4):044032. \doi{10.1103/PhysRevD.107.044032}, {\href{https://arxiv.org/abs/2208.13156}{{arXiv:2208.13156}}} {[gr-qc]}

\bibitem[{Ying(2019)}]{Ying_2019}
Ying X (2019) An overview of overfitting and its solutions. Journal of Physics: Conference Series 1168(2):022022. \doi{10.1088/1742-6596/1168/2/022022}

\bibitem[{Yu and Adhikari(2022)}]{Yu:2021swq}
Yu H, Adhikari RX (2022) {Nonlinear Noise Cleaning in Gravitational-Wave Detectors With Convolutional Neural Networks}. Front Artif Intell 5:811563. \doi{10.3389/frai.2022.811563}, {\href{https://arxiv.org/abs/2111.03295}{{arXiv:2111.03295}}} {[astro-ph.IM]}

\bibitem[{Yu et~al(2021)Yu, Adhikari, Magee, Sachdev, and Chen}]{Yu:2021vvm}
Yu H, Adhikari RX, Magee R, et~al (2021) {Early warning of coalescing neutron-star and neutron-star-black-hole binaries from the nonstationary noise background using neural networks}. Phys Rev D 104(6):062004. \doi{10.1103/PhysRevD.104.062004}, {\href{https://arxiv.org/abs/2104.09438}{{arXiv:2104.09438}}} {[gr-qc]}

\bibitem[{Yun et~al(2025)Yun, Han, Guo, Wang, and Du}]{Yun:2023vwa}
Yun Q, Han WB, Guo YY, et~al (2025) {The detection, extraction and parameter estimation of extreme-mass-ratio inspirals with deep learning}. Sci China Phys Mech Astron 68(1):210413. \doi{10.1007/s11433-024-2500-x}, {\href{https://arxiv.org/abs/2311.18640}{{arXiv:2311.18640}}} {[gr-qc]}

\bibitem[{Zeiler(2014)}]{zeiler2014visualizing}
Zeiler M (2014) Visualizing and understanding convolutional networks. In: European conference on computer vision/arXiv

\bibitem[{{Zelenka} et~al(2024){Zelenka}, {Br{\"u}gmann}, and {Ohme}}]{2024PhRvD.110b4024Z}
{Zelenka} O, {Br{\"u}gmann} B, {Ohme} F (2024) {Convolutional neural networks for signal detection in real LIGO data}. Phys Rev D 110(2):024024. \doi{10.1103/PhysRevD.110.024024}, {\href{https://arxiv.org/abs/2402.07492}{{arXiv:2402.07492}}} {[astro-ph.IM]}

\bibitem[{Zevin et~al(2017)Zevin, Coughlin, Bahaadini, Besler, Rohani, Allen, Cabero, Crowston, Katsaggelos, Larson, Lee, Lintott, Littenberg, Lundgren, Østerlund, Smith, Trouille, and Kalogera}]{Zevin_2017}
Zevin M, Coughlin S, Bahaadini S, et~al (2017) Gravity spy: integrating advanced ligo detector characterization, machine learning, and citizen science. Class Quantum Grav 34(6):064003. \doi{10.1088/1361-6382/aa5cea}

\bibitem[{Zevin et~al(2024)}]{Zevin:2023rmt}
Zevin M, et~al (2024) {Gravity Spy: lessons learned and a path forward}. Eur Phys J Plus 139(1):100. \doi{10.1140/epjp/s13360-023-04795-4}, {\href{https://arxiv.org/abs/2308.15530}{{arXiv:2308.15530}}} {[gr-qc]}

\bibitem[{Zhang et~al(2022)Zhang, Messenger, Korsakova, Chan, Hu, and Zhang}]{Zhang:2022xuq}
Zhang XT, Messenger C, Korsakova N, et~al (2022) {Detecting gravitational waves from extreme mass ratio inspirals using convolutional neural networks}. Phys Rev D 105(12):123027. \doi{10.1103/PhysRevD.105.123027}, {\href{https://arxiv.org/abs/2202.07158}{{arXiv:2202.07158}}} {[astro-ph.HE]}

\bibitem[{Zhao et~al(2023)Zhao, Lyu, Wang, Cao, and Ren}]{Zhao:2022qob}
Zhao T, Lyu R, Wang H, et~al (2023) {Space-based gravitational wave signal detection and extraction with deep neural network}. Commun Phys 6(1):212. \doi{10.1038/s42005-023-01334-6}, {\href{https://arxiv.org/abs/2207.07414}{{arXiv:2207.07414}}} {[gr-qc]}

\bibitem[{Zhao and Wen(2018)}]{Zhao:2017cbb}
Zhao W, Wen L (2018) {Localization accuracy of compact binary coalescences detected by the third-generation gravitational-wave detectors and implication for cosmology}. Phys Rev D 97(6):064031. \doi{10.1103/PhysRevD.97.064031}, {\href{https://arxiv.org/abs/1710.05325}{{arXiv:1710.05325}}} {[astro-ph.CO]}

\bibitem[{Zhu et~al(2017)Zhu, Papa, and Walsh}]{Zhu:2017ujz}
Zhu SJ, Papa MA, Walsh S (2017) {New veto for continuous gravitational wave searches}. Phys Rev D 96(12):124007. \doi{10.1103/PhysRevD.96.124007}, {\href{https://arxiv.org/abs/1707.05268}{{arXiv:1707.05268}}} {[gr-qc]}

\end{thebibliography}

\end{document}